\documentclass[aps,prd,superscriptaddress]{revtex4}
\usepackage{epsfig,epsf}
\usepackage{amsmath}
\usepackage{amsthm}
\usepackage{amsfonts}
\usepackage{amssymb}
\usepackage{dsfont}
\usepackage{multirow}
\usepackage{appendix}
\usepackage{slashed}
\usepackage[active]{srcltx}
\usepackage{psfrag}

\begin{document}

\title{Four-quark exotic mesons}
\author{S.~S.~Agaev}
\affiliation{Institute for Physical Problems, Baku State University, Az--1148 Baku,
Azerbaijan}
\author{K.~Azizi}
\affiliation{Department of Physics, Do\v{g}u\c{s} University, Acibadem-Kadik\"{o}y, 34722
Istanbul, Turkey}
\affiliation{Department of Physics, University of Tehran, North Karegar Avenue, Tehran
14395-547, Iran}
\author{H.~Sundu}
\affiliation{Department of Physics, Kocaeli University, 41380 Izmit, Turkey}

\begin{abstract}
We review our investigations devoted to the analysis of the resonances $%
Z_{c}(3900)$, $Z_{c}(4430)$, $Z_{c}(4100)$, $X(4140)$, $X(4274)$, $a_1(1420)$%
, $Y(4660)$, $X(2100)$, $X(2239)$ and $Y(2175)$ discovered in various
processes by Belle, BaBar, BESIII, D0, CDF, CMS, LHCb and COMPASS
collaborations. These resonances are considered as serious candidates to
four-quark (tetraquark) exotic mesons. We treat all of them as
diquark-antidiquark states with relevant spin-parities, find their masses
and couplings, as well as explore their dominant strong decay channels.
Calculations are performed in the context of the QCD sum rule method. Thus,
the spectroscopic parameters of the tetraquarks are evaluated using the
two-point sum rules. For computations of the strong couplings $G_{TM_1M_2}$,
corresponding to the vertices $TM_1M_2$ and necessary to find the partial
widths of the strong decays $T \to M_1M_2$, we employ either the three-point
or full/approximate versions of the QCD light-cone sum rules methods.
Obtained results are compared with available experimental data, and with
predictions of other theoretical studies.
\end{abstract}

\maketitle


\section{Introduction}

During last five decades the Quantum Chromodynamics (QCD) as the theory of
strong interactions was successfully used to explore spectroscopic
parameters and decay channels of hadrons, to analyze features of \ numerous
exclusive and inclusive hadronic processes. The asymptotic freedom of QCD
allowed ones to employ at high momentum transfers the perturbative
methods of the quantum field theory. At relatively low momentum transfer, $Q^{2}\sim 1~\mathrm{%
GeV}^{2}$, when the coupling of the strong interactions, $\alpha _{s}(Q^{2})$,
is large enough and nonperturbative effects become important, physicists
invented and applied various models and approaches to investigate hadronic
processes. Now the QCD, appeared from merging of the parton model and
non-abelian quantum field theory of colored quarks and gluons, is a part of
the Standard Model (SM) of elementary particles. It is worth noting that despite
numerous attempts of various experimental collaborations to find particles
and interactions beyond the Standard Model, all observed experimental
processes and measured quantities can be explained within framework of this
theory.

In accordance with a contemporary paradigm, conventional mesons and baryons
have quark-antiquark $q\overline{q}$ and three-quark (antiquark) $qq^{\prime
}q^{\prime \prime }$ structures, respectively. The electromagnetic, weak and
strong interactions of these particles can be explored in the context of SM.
But fundamental principles of the QCD do not forbid existence of multiquark
hadrons, i.e., particles made of four, five, six, etc. quarks. Apart from
pure theoretical interest, multiquark systems attracted interests of
researches as possible cures to treat old standing problems of conventional
hadron spectroscopy. Actually, a hypothesis about multiquark nature some of
known particles was connected namely with evident problems of
quark-antiquark model of mesons. In fact, in the ordinary picture the nonet
of scalar mesons are $1{}^{3}P_{0}$ quark-antiquark states. But different
and independent calculations prove that $1{}^{3}P_{0}$ states are heavier
than $1~\mathrm{GeV}$. Therefore, only the isoscalar $f_{0}(1370)$ and $%
f_{0}(1710)$, isovector $a_{0}(1450)$ or isospinor $K_{0}^{\ast }(1430)$
mesons can be identified as members of the $1{}^{3}P_{0}$ multiplet. Because
the masses of mesons from the light scalar nonet are below $1\ \mathrm{GeV}$%
, during a long time the mesons $f_{0}(500)$, $f_{0}(980)$, $K_{0}^{\ast
}(800)$, and $a_{0}(980)$ were subject of controversial theoretical
hypothesis and suggestions. To describe unusual properties of light mesons
R. Jaffe assumed that they are composed of four valence quarks \cite%
{Jaffe:1976ig}. Within this paradigm problems with low masses, and a mass
hierarchy inside the light nonet seem found their solutions. The current
status of these theoretical studies can be found in Refs.\ \cite%
{Kim:2017yvd,Agaev:2017cfz,Agaev:2018sco,Agaev:2018fvz}.

Another interesting result about multiquark hadrons with important
consequences was obtained also by R.~Jaffe \cite{Jaffe:1976yi}. He
considered six-quark (dibaryon or hexaquark) states built of only light $u$,
$d$, and $s$ quarks that belong to flavor group $SU_{f}(3)$. Using for
analysis the MIT quark-bag model, Jaffe predicted existence of a $\mathrm{H}$%
-dibaryon, i.e., a flavor-singlet and neutral six-quark $uuddss$ bound state
with isospin-spin-parity $I(J^{\mathrm{P}})=0(0^{+})$. This double-strange
six-quark structure with mass $2150~\mathrm{MeV}$ lies $80~\mathrm{MeV}$
below the $2m_{\Lambda }=2230~\mathrm{MeV}\ $ threshold and is stable
against strong decays. It can transform through weak interactions, which
means that mean lifetime of $\mathrm{H}$-dibaryon, $\tau \approx 10^{-10}%
\mathrm{s}$, is considerably longer than that of most ordinary hadrons. It
is remarkable that the hexaquark $uuddss$ may be considered as a candidate
to dark matter provided its mass satisfies some constraints \cite%
{Farrar:2003qy,Farrar:2017ysn,Farrar:2018hac,Azizi:2019xla}.

Theoretical studies of stable four-quark configurations meanwhile were
continued using available methods of high energy physics. The four-quark
mesons or tetraquarks composed of heavy $bb$ or $cc$ diquarks and light
antidiquarks were considered as true candidates to such states. The class of
exotic mesons $QQ\overline{Q}\overline{Q}$ and $QQ\overline{q}\overline{q}%
^{\prime }$ was investigated in Refs.\ \cite%
{Ader:1981db,Lipkin:1986dw,Zouzou:1986qh}, where a potential model with
additive two-particle interactions was utilized to find stable tetraquarks.
In the framework of this method it was proved that tetraquarks $QQ\overline{q%
}\overline{q}$ may bind to stable states if the ratio $m_{Q}/m_{q}$ is
large. The similar conclusion was drawn in Ref.\ \cite{Carlson:1987hh},
where a restriction on the confining potential was its finiteness at small
two-particle distances. It was found there, that the isoscalar axial-vector
tetraquark $T_{bb;\overline{u}\overline{d}}^{-}$ resides below the threshold
necessary to create B mesons, and therefore can transform only through weak
decays. But the tetraquarks $T_{cc;\overline{q}\overline{q}^{\prime }}$ and $%
T_{bc;\overline{q}\overline{q}^{\prime }}$ may form both unstable or stable
compounds. The stability of structures $QQ\overline{q}\overline{q}$ in the
limit $m_{Q}\rightarrow \infty $ was studied in Ref.\ \cite{Manohar:1992nd},
as well.

Progress of those years was not limited by qualitative analysis of
four-quark bound states. Thus already at eighties of the last century
investigations of tetraquarks and hybrid hadrons were put on basis of
QCD-inspired nonperturbative methods, which allowed ones to perform
quantitative analyses and made first predictions for their masses and other
parameters \cite%
{Balitsky:1982ps,Govaerts:1984hc,Govaerts:1985fx,Balitsky:1986hf,Braun:1985ah,Braun:1988kv}%
. But achievements of these theoretical investigations then were not
accompanied by reliable experimental measurements, which negatively affected
development of the field.

Situation changed after observation of the charmoniumlike state $X(3872)$
reported in 2003 by the Belle collaboration \cite{Choi:2003ue}. Existence of
the narrow resonance $X(3872)$ was later verified by various collaborations
such as D0, CDF and BaBar \cite{Abazov:2004kp,Acosta:2003zx,Aubert:2004ns}.
Discovery of charged resonances $Z_{c}(4430)$ and $Z_{c}(3900)$ had also
important impact on physics of multiquark mesons, because they could not be
confused with neutral $\overline{c}c$ charmonia, and were candidates to
four-quark mesons. The $Z_{c}^{\pm }(4430)$ were observed in $B$ meson
decays $B\rightarrow K\psi ^{\prime }\pi ^{\pm }$ by Belle as resonances in
the $\psi ^{\prime }\pi ^{\pm }$ invariant mass distributions \cite%
{Choi:2007wga}. The resonances $Z_{c}^{+}(4430)$ and $Z_{c}^{-}(4430)$ were
fixed and investigated later again by Belle in the processes $B\rightarrow
K\psi ^{\prime }\pi ^{+}$ and $B^{0}\rightarrow K^{+}\psi ^{\prime }\pi ^{-}$
\cite{Mizuk:2009da,Chilikin:2013tch}, respectively. Evidence for $%
Z_{c}(4430)$ and its decay to $J/\psi \pi $ was found in reaction $\bar{B}%
^{0}\rightarrow J/\psi K^{-}\pi ^{+}$ by the same collaboration \cite%
{Chilikin:2014bkk}. Along with masses and widths of these states Belle fixed
also their quantum numbers $J^{\mathrm{P}}=1^{+}$ as a realistic assumption.
The parameters of $Z_{c}^{-}(4430)$ were measured in decay $B^{0}\rightarrow
K^{+}\psi ^{\prime }\pi ^{-}$ by the LHCb collaboration as well, where its
spin-parity was clearly determined to be $1^{+}$ \cite%
{Aaij:2014jqa,Aaij:2015zxa}.

Another charged tetraquarks $Z_{c}^{\pm }(3900)$ were found in the process $%
e^{+}e^{-}\rightarrow J/\psi \pi ^{+}\pi ^{-}$ by BESIII as resonances in
the $J/\psi \pi ^{\pm }$ invariant mass distributions \cite{Ablikim:2013mio}%
. These structures were seen by Belle and CLEO \cite{Liu:2013dau,Xiao:2013iha}%
, as well. The BESIII announced also detection of a neutral $%
Z_{c}^{0}(3900) $ state in the process $e^{+}e^{-}\rightarrow \pi
^{0}Z_{c}^{0}\rightarrow \pi ^{0}\pi ^{0}J/\psi $ \cite{Ablikim:2015tbp}.

An important observation of last few years was made by D0, which reported
about a structure $X(5568)$ in a chain of transformations $%
X(5568)\rightarrow B_{s}^{0}\pi ^{\pm }$, $B_{s}^{0}\rightarrow J/\psi \phi $%
, $J/\psi \rightarrow \mu ^{+}\mu ^{-}$, $\phi \rightarrow K^{+}K^{-}$ \cite%
{D0:2016mwd}. It was noted that $X(5568)$ is  first discovered exotic meson
which is composed of four different quarks. Indeed, from the decay channels $%
X(5568)\rightarrow B_{s}^{0}\pi ^{\pm }$ it is easy to conclude that $%
X(5568) $ contains $b,\,s,\,u,\,d$ quarks. The resonance $X(5568)$ is a
scalar particle with the positive charge conjugation parity $J^{\mathrm{PC}%
}=0^{++}$, its mass and width are equal to $m=5567.8\pm 2.9\mathrm{(stat)}%
_{-1.9}^{+0.9}\mathrm{(syst)}~\mathrm{MeV}$ and $\Gamma =21.9\pm 6.4\mathrm{%
(stat)}_{-2.5}^{+5.0}\mathrm{(syst)}~\mathrm{MeV}$, respectively. But, very
soon LHCb announced results of analyses of $pp$ collision data at energies $%
7~\mathrm{TeV}$ and $8~\mathrm{TeV}$ collected at CERN \cite{Aaij:2016iev}.
The LHCb could not find evidence for a resonant structure in the $%
B_{s}^{0}\pi ^{\pm }$ invariant mass distributions at the energies less than
$5700~\mathrm{MeV}$. Stated differently, a status of the resonance $X(5568)$%
, probably composed of four different quarks is controversial, and
necessitates further experimental studies. The exotic state named $X(5568)$
deserves to be looked for by other collaborations, and maybe, in other
hadronic processes.

There are new experimental results on different resonances which may be
considered as exotic mesons. Thus, recently LHCb rediscovered resonances $%
X(4140)$ and $X(4274)$ in the $J/\psi \phi $ invariant mass distribution by
analyzed the exclusive decay $B^{+}\rightarrow J/\psi \phi K^{+}$ \cite%
{Aaij:2016iza,Aaij:2016nsc}. It reported on detection of heavy resonances $%
X(4500)$ and $X(4700)$ in the same $J/\psi \phi $ channel as well. Besides,
LHCb fixed the spin-parities of these resonances. It turned out, that $%
X(4140)$ and $X(4274)$ are axial-vector states $J^{PC}=1^{++}$, whereas $%
X(4500)$ and $X(4700)$ are scalar particles with $J^{\mathrm{PC}}=0^{++}$.
The first two states were discovered already by CDF in the decays $B^{\pm
}\rightarrow J/\psi \phi K^{\pm }$ \cite{Aaltonen:2009tz}, and confirmed
later by CMS and D0 \cite{Chatrchyan:2013dma,Abazov:2013xda}, respectively.
Hence they are old members of tetraquarks' family, whereas last two heavy
states were seen for the first time. The resonances $X$ may belong to a
group of hidden-charm exotic mesons. From their decay modes, it is also
evident that as candidates to tetraquarks they have to contain a strange $s%
\overline{s}$ component. In other words, the quark content of the states $X$
is presumably $c\overline{c}s\overline{s}$.

The family of vector resonances $\{Y\}$, which are candidates to
tetraquarks, contains at least four hidden-charm particles with the quantum
numbers\ $J^{\mathrm{PC}}=1^{--}$. One of them, the resonance $Y(4660)$ for
the first time was detected by Belle via initial-state radiation in the
process $e^{+}e^{-}\rightarrow \gamma _{\mathrm{ISR}}\psi ^{\prime }$ $\pi
^{+}\pi ^{-}$ as one of two resonant structures in the $\psi ^{\prime }\pi
^{+}\pi ^{-}$ invariant mass distribution \cite{Wang:2007ea,Wang:2014hta}.
The second state observed in this experiment was labeled $Y(4360)$. The
analyses of Refs.\ \cite{Wang:2007ea,Wang:2014hta} proved that these
resonances cannot be identified with known charmonia. The state $Y(4630)$,
which is traditionally identified with $Y(4660)$, was seen in the process $%
e^{+}e^{-}\rightarrow \Lambda _{c}^{+}\Lambda _{c}^{-}$ as a peak in the $%
\Lambda _{c}^{+}\Lambda _{c}^{-}$ invariant mass distribution \cite%
{Pakhlova:2008vn}. The BaBar studied the same process $e^{+}e^{-}\rightarrow
\gamma _{\mathrm{ISR}}\psi ^{\prime }\pi ^{+}\pi ^{-}$ and independently
confirmed appearance of two resonant structures in the $\pi ^{+}\pi ^{-}\psi
^{\prime }$ invariant mass distribution \cite{Lees:2012pv}. Masses and
widths of these structures allowed BaBar to identify them with resonances $%
Y(4660)$ and $Y(4360)$, respectively. Apart from these two resonances, there
are states $Y(4260)$ and $Y(4390)$ which can also be considered as members
of $\{Y\}$ family.

Among new resonances it is worth noting the state $Z_{c}^{-}(4100)$
discovered also by LHCb in the decay $B^{0}\rightarrow K^{+}\eta _{c}\pi
^{-} $ \cite{Aaij:2018bla}. In this article it was noted that the
spin-parity of this structure is $J^{\mathrm{P}}=0^{+}$ or $J^{\mathrm{P}%
}=1^{-}$: both assignments are consistent with the data. From analysis of
the decay $Z_{c}^{-}(4100)\rightarrow \eta _{c}\pi ^{-}$ it is clear that $%
Z_{c}^{-}(4100)$ may be composed of quarks $cd\overline{c}\overline{u}$, and
is probably another member of the family of charged $Z$-resonances with the
same quark content: let us emphasize that the well-known resonances $%
Z_{c}^{\pm }(4430)$ and $Z_{c}^{\pm }(3900)$ have also the $cd\overline{c}%
\overline{u}$ or $cu\overline{c}\overline{d}$ contents.

In the present work, we review our theoretical works devoted to
investigations of these and other resonances as candidates to exotic
four-quark mesons. All investigations in our original articles were carried
out in framework of the QCD sum rule approach, which is an effective
nonperturbative method to study exclusive hadronic processes \cite%
{Shifman:1978bx,Shifman:1978by}. The spectroscopic parameters of tetraquarks
were calculated by means of the QCD two-point sum rule method. Their decays  can be explored using other versions of the sum rule method. \
It is known, that tetraquarks decay dominantly to two conventional mesons
via strong interactions. Widths of these processes are determined by strong
couplings describing vertices of initial and final particles. Therefore,
strong couplings are key components of relevant investigations, and they can
be extracted either from the QCD three-point sum rule approach or light-cone
sum rule (LCSR) method \cite{Balitsky:1989ry}.

Calculation of the strong couplings corresponding to tetraquark-meson-meson
vertices in the framework of the LCSR method requires additional technical
recipes. The reason is that a tetraquark contains four valence quarks, and
light-cone expansion of the relevant nonlocal correlation function leads to
expressions with local matrix elements of one of a final meson. Then the
four-momentum conservation in a such strong vertex can be satisfied by
setting the four-momentum of this meson equal to zero, i.e., by treating it
a "soft" particle. Difficulties appeared due to such approximation can be
evaded using a soft-meson technique of the LCSR method \cite%
{Belyaev:1994zk,Ioffe:1983ju}. Let us note, that in the case of three-meson
vertices a soft limit is an approximation to full LCSR correlation
functions, whereas for vertices with one tetraquark this approach is an only
way to compute them. For analyses of four-quark systems the soft-meson
approximation was adjusted in Ref.\ \cite{Agaev:2016dev}, and successfully
applied to study decays of various tetraquarks. The full version of the LCSR
method is restored when exploring strong vertices of two tetraquarks and a
meson \cite{Agaev:2016srl}. In the present review all of these methods will
be used to evaluate strong couplings of exotic and conventional mesons.

Detailed information on exotic resonances $XYZ$ including a history of the
problem, as well as experimental and theoretical achievements of last years
are collected in numerous interesting reviews \cite%
{Jaffe:2004ph,Swanson:2006st,Klempt:2007cp,Godfrey:2008nc,Esposito:2014rxa,
Chen:2016qju,Chen:2016spr,Esposito:2016noz,Ali:2017jda,Olsen:2017bmm,Albuquerque:2018jkn,Brambilla:2019esw}%
.

This review is organized in the following form: In Sec.\ \ref{sec:Z}, we
investigate charged axial-vector resonances $Z_{c}(3900)$ and $Z_{c}(4430)$
by treating them as exotic mesons $cu\overline{c}\overline{d}$. In our
approach we consider $Z_{c}(4430)$ as a radial excitation of the
ground-state particle $Z_{c}(3900)$. Apart from the spectroscopic parameters
of these resonances, we calculate their full widths by exploring strong
decays $Z_{c}(3900),\,Z_{c}(4430)\rightarrow J/\psi \pi ,\,\psi ^{\prime
}\pi ,\eta _{c}^{\prime }\rho ,$ and $\,\eta _{c}\rho $. In the next section
we model the resonance $Z_{c}^{-}(4100)$ as a scalar tetraquark $cd\overline{%
c}\overline{u}$, and find its mass and coupling. The full width of $%
Z_{c}^{-}(4100)$ is evaluated by taking into account the strong decays $%
Z_{c}^{-}(4100)\rightarrow \eta _{c}\pi ^{-}$ , $\eta _{c}^{\prime }\pi
^{-},D^{0}D^{-},$ and $J/\psi \rho ^{-}$. Section \ref{sec:X} is devoted to
analysis of the resonances $X(4140)$ and $X(4274)$ as tetraquarks $cs%
\overline{c}\overline{s}$ with $J^{\mathrm{PC}}=1^{++}$ and color-triplet
and color-sextet organization of constituent diquarks, respectively. We also
consider their \ decay modes $X(4140)\rightarrow J/\psi \phi $ and $%
X(4274)\rightarrow J/\psi \phi $. Our analysis demonstrate that parameters
of $X(4140)$ are compatible with LHCb data, while prediction for the full
width of $X(4274)$ exceeds experiment data. In Sec.\ \ref{sec:A1} we
investigate the tetraquark $([us][\overline{u}\overline{s}]-[ds][\overline{d}%
\overline{s}])/\sqrt{2}$ with spin-parities $J^{\mathrm{PC}}=1^{++}$ and
find its parameters. The decays of this state to final mesons $f_{0}(980)\pi
^{0}$,$\ \ K^{\ast \pm }K^{\mp }$, $K^{\ast 0}\overline{K}^{0}$ and $%
\overline{K}^{\ast 0}K^{0}$ are also investigated. Obtained results allow us
to interpret this tetraquark as the axial-vector resonance $a_{1}(1420)$.
The resonance $Y(4660)$ is explored in Sec.\ \ref{sec:Y} as a vector
tetraquark $[cs][\overline{c}\overline{s}]$ with internal structure $C\gamma
_{5}\otimes \gamma _{5}\gamma _{\mu }C$. We calculate the mass and coupling
of this state, and investigate decay channels $Y\rightarrow J/\psi
f_{0}(500) $, $\psi ^{\prime }f_{0}(500)$, $J/\psi f_{0}(980)$ and $\psi
^{\prime }f_{0}(980)$. The resonances $X(2100)$ and $X(2239)$ and their
structures, spectroscopic parameters, and decay modes are considered in
Sec.\ \ref{sec:4s}. Section \ref{sec:Y2175} is reserved for analysis of the
resonance $Y(2175)$, which is interpreted as the vector tetraquark $%
\widetilde{Y}=[su][\overline{s}\overline{u}]$. We evaluate its spectroscopic
parameters, and explore strong decays $\widetilde{Y}\rightarrow $ $\phi
f_{0}(980)$,$\ \widetilde{Y}\rightarrow $ $\phi \eta $, and $\widetilde{Y}%
\rightarrow $ $\phi \eta ^{\prime }$. The section \ref{sec:Conc} contains
our brief concluding notes. In Appendix \ref{sec:App} we provide expressions
of quark propagators which have been used in calculations.


\section{ The resonances $Z_{c}(3900)$ and $Z_{c}(4430)$}

\label{sec:Z}
The parameters of $Z^{-}(4430)$ were measured by the LHCb collaboration in
the $B^{0}\rightarrow K^{+}\psi ^{\prime }\pi ^{-}$ decay
\begin{equation}
M=(4475\pm 7_{-25}^{+15})~\mathrm{MeV},\,\Gamma =(172\pm 13_{-34}^{+37})~%
\mathrm{MeV},  \label{eq:LHCdata}
\end{equation}%
where its spin-parity was definitely fixed to be $J^{\mathrm{P}}=1^{+}$ \cite%
{Aaij:2014jqa,Aaij:2015zxa}. Another charged tetraquarks $Z_{c}^{\pm }(3900)$
were discovered by BESIII
\begin{equation}
M=(3899.0\pm 3.6\pm 4.9)~\mathrm{MeV},\,\Gamma =(46\pm 10\pm 20)~\mathrm{MeV}%
,  \label{eq:BESIIIdata}
\end{equation}%
and have the spin-parity $J^{\mathrm{P}}=1^{+}$ \cite{Ablikim:2013mio}.

Theoretical investigations of the resonances $Z_{c}(3900)$ and $Z_{c}(4430)$
(in this section $Z_{c}$ and $Z$, respectively) embrace plethora of models
and computational methods \cite{Chen:2016qju,Esposito:2016noz}. The goal of
these studies is to understand internal quark-gluon structures of the states
$Z_{c}$ and $Z$, to find their spectroscopic parameters, and partial widths
of relevant decay channels. Thus, $Z$ was examined as a diquark-antidiquark
\cite%
{Liu:2008qx,Ebert:2008kb,Bracco:2008jj,Maiani:2008zz,Wang:2010rt,Maiani:2014,Wang:2014vha,Agaev:2017tzv}
or a meson molecule state \cite%
{Lee:2007gs,Liu:2008xz,Braaten:2007xw,Branz:2010sh,Goerke:2016hxf}, a
threshold effect \cite{Rosner:2007mu}, and a hadrocharmonium composite \cite%
{Dubynskiy:2008mq}. A situation around of the resonance $Z_{c}$ does not
differ significantly from studies which try to describe properties of $Z$.
In fact, there are publications, in which $Z_{c}$ is treated as the tightly
bound diquark-antidiquark \cite%
{Dias:2013xfa,Wang:2013vex,Deng:2014gqa,Agaev:2016dev}, as a molecule built
of conventional mesons \cite%
{Wang:2013daa,Wilbring:2013cha,Dong:2013iqa,Ke:2013gia,Gutsche:2014zda,Esposito:2014hsa,Chen:2015igx,Gong:2016hlt,Ke:2016owt}%
, or as a threshold cusp \cite{Swanson:2014tra,Ikeda:2016zwx}.

The intriguing assumption was made in Ref.\ \cite{Maiani:2014}, in which the
authors interpreted $Z_{c}$ and $Z$ as the ground state and first radial
excitation of the same tetraquark. This suggestion was justified by
observation that dominant decay modes of these resonances are
\begin{equation}
Z_{c}^{\pm }\rightarrow J/\psi \pi ^{\pm },\,\,Z^{\pm }\rightarrow \psi
^{\prime }\pi ^{\pm },
\end{equation}%
and that the mass splitting $m_{\psi
^{\prime }}-m_{J/\psi }$ between $1S$ and $2S$ vector charmonia  is approximately equal to the mass gap $%
m_{Z}-m_{Z_{c}}$. This idea was realized in the diquark-antidiquark model in
Refs.\ \cite{Wang:2014vha,Agaev:2017tzv}, where the authors evaluated masses
and current couplings (pole residues) of $Z_{c}$ and $Z$. Within this
scheme decay modes of the resonances $Z_{c}$ and $Z$ were considered in
Ref.\ \cite{Agaev:2017tzv}: these processes contain  important dynamical
information on structures of particles under discussion. The analysis
performed in these works seems confirm a suggestion about their ground state
and excited natures.

The mass and decay constant (or current coupling) are parameters of ordinary
and exotic mesons, which have to be measured and evaluated primarily. As
usual, all theoretical models suggested to describe the internal
organization of tetraquarks and explain their features begin from evaluation
of these parameters. Only after successful comparison of a theoretical
result for the mass with existing experimental information a model may be
accepted and used for further analysis of a tetraquark candidate. But for
reliable conclusions on the structure of discovered resonances, one needs 
additional information. Experimental collaborations measure not only masses
of resonances, but their full widths as well. They also determine spins and
parities of these structures.

Because an overwhelming number of models predict correctly the masses of the
resonances $Z_{c}$ and $Z$, there is a necessity to compute full widths of
these structures. In all fairness, there are publications in which decays of
$Z^{\pm }$ were analyzed as well. Indeed, within a phenomenological
Lagrangian approach and a molecule picture decays $Z^{\pm }\rightarrow
J/\psi \pi ^{\pm };\,\psi ^{\prime }\pi ^{\pm }$ were studied in Ref. \cite%
{Branz:2010sh}. Unfortunately, in this article $Z^{\pm }$ were treated as
pseudoscalar or vector particles ruled out by new measurements. The decay
modes $Z^{+}\rightarrow J/\psi \pi ^{+};\,\psi ^{\prime }\pi ^{+}$ were
reanalyzed in context of the covariant quark model in Ref.\ \cite%
{Goerke:2016hxf}.

In Refs.\ \cite{Dias:2013xfa} and \cite{Agaev:2016dev} the authors studied
decays of the resonance $Z_{c}$ by modeling it as a diquark-antidiquark
state with the quantum numbers $J^{\mathrm{PC}}=1^{+-}$. In Ref. \cite%
{Dias:2013xfa} partial widths of the decays $Z_{c}^{+}\rightarrow J/\psi \pi
^{+},\,\eta _{c}\rho ,$ and $\,D^{+}\overline{D}^{\star 0}$ were computed by
employing the three-point sum rule approach. The light cone sum rule method
and a technique of soft-meson approximation were used to evaluate widths of
processes $Z_{c}^{+}\rightarrow J/\psi \pi ^{+},\,\eta _{c}\rho $ in Ref.\
\cite{Agaev:2016dev}.

Decays of the resonances $Z_{c}^{\pm }$ were also investigated in the
context of alternative approaches \cite%
{Dong:2013iqa,Gutsche:2014zda,Goerke:2016hxf}. In fact, processes $%
Z_{c}\rightarrow J/\psi \pi ,\,\psi ^{\prime }\pi ,\ h_{c}(1P)\pi $ were
considered in Ref. \cite{Dong:2013iqa} using the phenomenological Lagrangian
approach and modeling $Z_{c}$ as an axial-vector meson molecule $\overline{D}%
D$. In the context of the same model radiative and leptonic decays $%
Z_{c}^{+}\rightarrow J/\psi \pi ^{+}\gamma $ and $J/\psi \pi
^{+}l^{+}l^{-},\,l=(e,\,\mu )$ were analyzed in Ref.\ \cite{Gutsche:2014zda}%
. The covariant quark model was employed to calculate partial widths of the
channels $Z_{c}^{+}\rightarrow J/\psi \pi ^{+},\ \,\eta _{c}\rho ^{+},\
\overline{D}^{0}D^{\star +},$ and $\,\overline{D}^{\ast 0}D^{+}$ in Ref.\
\cite{Goerke:2016hxf}. Let us note also Ref. \cite{Ke:2016owt}, in which the
decay $Z_{c}\rightarrow h_{c}\pi $ was explored in the light front model.

In this section we evaluate spectroscopic parameters of the resonances $%
Z_{c} $ and $Z$, and investigate their decay channels by suggesting that $%
Z_{c}$ and $Z$ are a ground state and radial excitation of the tetraquark
with $J^{\mathrm{PC}}=1^{+-}$, respectively. In other words, we treat them
as $1S$ and $2S$ axial-vector members of the $[cu][\overline{c}\overline{d}]$
multiplet and present results of Ref. \cite{Agaev:2017tzv}.


\subsection{The masses and couplings of the tetraquarks $Z_{c}$ and $Z$}

\label{subsec:MC}
The QCD two-point sum rule method is one of best approaches to calculate the
spectroscopic parameters of the resonances $Z_{c}$ and $Z$. We find the
masses and couplings of positively charged tetraquarks $cu\overline{c}%
\overline{d}$, but due to the exact chiral limit accepted throughout this
review, parameters of resonances with negative charges do not differ from
them.

Starting point to extract the mass and coupling of the tetraquarks $Z_{c}$
and $Z$ is the correlation function
\begin{equation}
\Pi _{\mu \nu }(p)=i\int d^{4}xe^{ipx}\langle 0|\mathcal{T}\{J_{\mu
}^{Z}(x)J_{\nu }^{Z\dagger }(0)\}|0\rangle.  \label{eq:CorrF1}
\end{equation}%
Here, $J_{\mu }^{Z}(x)$ is the interpolating current for these tetraquarks:
it corresponds to axial-vector particle $J^{\mathrm{PC}}=1^{+-}$ and is
given by the expression
\begin{equation}
J_{\mu }^{Z}(x)=\frac{\epsilon \tilde{\epsilon}}{\sqrt{2}}\left\{ \left[
u_{a}^{T}(x)C\gamma _{5}c_{b}(x)\right] \left[ \overline{d}_{d}(x)\gamma
_{\mu }C\overline{c}_{e}^{T}(x)\right] -\left[ u_{a}^{T}(x)C\gamma _{\mu
}c_{b}(x)\right] \left[ \overline{d}_{d}(x)\gamma _{5}C\overline{c}%
_{e}^{T}(x)\right] \right\},  \label{eq:Curr}
\end{equation}%
where the notations $\epsilon =\epsilon _{abc}$ and $\tilde{\epsilon}%
=\epsilon _{dec}$ are introduced. In Eq.\ (\ref{eq:Curr}) $a,b,c,d,e$ are
color indices, whereas $C$ is the charge conjugation operator.

In these calculations we accept the "ground-state+radially excited
state+continuum" scheme, and carry out ordinary and well-known calculations:
we find the physical side of the sum rules by inserting into $\Pi _{\mu \nu
}(p)$ a full set of relevant states, separating contributions of the
resonances $Z_{c}$ and $Z$, and performing the integration over $x$. As a
result, for $\Pi _{\mu \nu }^{\mathrm{Phys}}(p)$ we obtain
\begin{equation}
\Pi _{\mu \nu }^{\mathrm{Phys}}(p)=\frac{\langle 0|J_{\mu
}^{Z}|Z_{c}(p)\rangle \langle Z_{c}(p)|J_{\nu }^{Z\dagger }|0\rangle }{%
m_{Z_{c}}^{2}-p^{2}}+\frac{\langle 0|J_{\mu }^{Z}|Z(p)\rangle \langle
Z(p)|J_{\nu }^{Z\dagger }|0\rangle }{m_{Z}^{2}-p^{2}}+\cdots ,  \label{Phys1}
\end{equation}%
where $m_{Z_{c}}$ and $m_{Z}$ are the masses of $Z_{c}$ and $Z$,
respectively. Contributions to the correlation function originating from
higher resonances and continuum states are denoted by dots.

In order to finish analysis of the phenomenological side, we introduce the
couplings $f_{Z_{c}}$ and $f_{Z}$ through matrix elements
\begin{equation}
\langle 0|J_{\mu }^{Z}|Z_{c}\rangle =f_{Z_{c}}m_{Z_{c}}\varepsilon _{\mu
},\,\,\langle 0|J_{\mu }^{Z}|Z\rangle =f_{Z}m_{Z}\widetilde{\varepsilon }%
_{\mu },  \label{eq:Res}
\end{equation}%
where $\varepsilon _{\mu }$ and $\widetilde{\varepsilon }_{\mu }$ are the
polarization vectors of $Z_{c}$ and $Z$ , respectively. Then the function $%
\Pi _{\mu \nu }^{\mathrm{Phys}}(p)$ can be written as
\begin{equation}
\Pi _{\mu \nu }^{\mathrm{Phys}}(p)=\frac{m_{Z_{c}}^{2}f_{Z_{c}}^{2}}{%
m_{Z_{c}}^{2}-p^{2}}\left( -g_{\mu \nu }+\frac{p_{\mu }p_{\nu }}{%
m_{Z_{c}}^{2}}\right) +\frac{m_{Z}^{2}f_{Z}^{2}}{m_{Z}^{2}-p^{2}}\left(
-g_{\mu \nu }+\frac{p_{\mu }p_{\nu }}{m_{Z}^{2}}\right) +\cdots .
\label{eq:CorM}
\end{equation}%
The Borel transformation applied to Eq.\ (\ref{eq:CorM}) yields%
\begin{equation}
\mathcal{B}\Pi _{\mu \nu }^{\mathrm{Phys}%
}(p)=m_{Z_{c}}^{2}f_{Z_{c}}^{2}e^{-m_{Z_{c}}^{2}/M^{2}}\left( -g_{\mu \nu }+%
\frac{p_{\mu }p_{\nu }}{m_{Z_{c}}^{2}}\right)
+m_{Z}^{2}f_{Z}^{2}e^{-m_{Z}^{2}/M^{2}}\left( -g_{\mu \nu }+\frac{p_{\mu
}p_{\nu }}{m_{Z}^{2}}\right) +\cdots ,
\end{equation}%
with $M^{2}$ being the Borel parameter.

The second component of the QCD sum rules is the correlation function $\Pi
_{\mu \nu }^{\mathrm{OPE}}(p)$ expressed in terms of quark propagators. It
can be found after inserting the explicit expression of $J_{\mu }^{Z}$ into
Eq.\ (\ref{eq:CorrF1}) and contracting heavy and light quark fields
\begin{eqnarray}
&&\Pi _{\mu \nu }^{\mathrm{OPE}}(p)=-\frac{i}{2}\int d^{4}xe^{ipx}\epsilon
\tilde{\epsilon}\epsilon ^{\prime }\tilde{\epsilon}^{\prime }\left\{ \mathrm{%
Tr}\left[ \gamma _{5}\widetilde{S}_{u}^{aa^{\prime }}(x)\gamma
_{5}S_{c}^{bb^{\prime }}(x)\right] \mathrm{Tr}\left[ \gamma _{\mu }%
\widetilde{S}_{c}^{e^{\prime }e}(-x)\gamma _{\nu }S_{d}^{d^{\prime }d}(-x)%
\right] \right.  \notag \\
&&-\mathrm{Tr}\left[ \gamma _{\mu }\widetilde{S}_{c}^{e^{\prime
}e}(-x)\gamma _{5}S_{d}^{d^{\prime }d}(-x)\right] \mathrm{Tr}\left[ \gamma
_{\nu }\widetilde{S}_{u}^{aa^{\prime }}(x)\gamma _{5}S_{c}^{bb^{\prime }}(x)%
\right] -\mathrm{Tr}\left[ \gamma _{5}\widetilde{S}_{u}^{a^{\prime
}a}(x)\gamma _{\mu }S_{c}^{b^{\prime }b}(x)\right]  \notag \\
&&\left. \times \mathrm{Tr}\left[ \gamma _{5}\widetilde{S}_{c}^{e^{\prime
}e}(-x)\gamma _{\nu }S_{d}^{d^{\prime }d}(-x)\right] +\mathrm{Tr}\left[
\gamma _{\nu }\widetilde{S}_{u}^{aa^{\prime }}(x)\gamma _{\mu
}S_{c}^{bb^{\prime }}(x)\right] \mathrm{Tr}\left[ \gamma _{5}\widetilde{S}%
_{c}^{e^{\prime }e}(-x)\gamma _{5}S_{d}^{d^{\prime }d}(-x)\right] \right\} .
\label{eq:CorrF2}
\end{eqnarray}%
Here
\begin{equation}
\widetilde{S}_{c(q)}^{ab}(x)=CS_{c(q)}^{ab\mathrm{T}}(x)C,
\end{equation}%
and $S_{c(q)}^{ab}(x)$ are quark propagators: their explicit expressions are
moved to Appendix \ref{sec:App}.

The function $\Pi _{\mu \nu }^{\mathrm{OPE}}(p)$ has the following
decomposition over the Lorentz structures%
\begin{equation}
\Pi _{\mu \nu }^{\mathrm{OPE}}(p)=\Pi ^{\mathrm{OPE}}(p^{2})g_{\mu \nu }+%
\widetilde{\Pi }^{\mathrm{OPE}}(p^{2})p_{\mu }p_{\nu },
\end{equation}%
where $\Pi ^{\mathrm{OPE}}(p^{2})$ and $\widetilde{\Pi }^{\mathrm{OPE}%
}(p^{2})$ are corresponding invariant amplitudes.

The QCD sum rules for the parameters of $Z$ can be found by equating
invariant amplitudes of the same structures in $\Pi _{\mu \nu }^{\mathrm{Phys%
}}(p)$ and $\Pi _{\mu \nu }^{\mathrm{OPE}}(p)$. For our purposes terms
proportional to $g_{\mu \nu }$ are convenient structures, and we employ them
in further calculations.

The invariant amplitude $\Pi ^{\mathrm{Phys}}(p^{2})$ corresponding to
structure $g_{\mu \nu }$ has a simple form. The similar function $\Pi ^{%
\mathrm{OPE}}(p^{2})$ can be written down as the dispersion integral
\begin{equation}
\Pi ^{\mathrm{OPE}}(p^{2})=\int_{4m_{c}^{2}}^{\infty }ds\frac{\rho ^{\mathrm{%
OPE}}(s)}{s-p^{2}},
\end{equation}%
where the two-point spectral density is denoted by $\rho ^{\mathrm{OPE}}(s)$%
. It is equal to the imaginary part of the correlation function $\sim g_{\mu
\nu }$, and can be obtained by means of well-known prescriptions. Let us
note that calculations have been performed by taking into account various
vacuum condensates up to dimension eight. We omit here details of
computations, and do not write down explicitly $\rho ^{\mathrm{OPE}}(s)$.

To suppress contributions of higher resonances and continuum states, we
apply the Borel transformation on the variable $p^{2}$ to both sides of QCD
sum rule's equality, and subtract them by using the assumption on the
quark-hadron duality. After some operations one gets the sum rules for the
parameters of the excited $Z$ state:
\begin{equation}
m_{Z}^{2}=\frac{\int_{4m_{c}^{2}}^{s_{0}^{\ast }}\rho ^{\mathrm{OPE}%
}(s)se^{-s/M^{2}}ds-f_{Z_{c}}^{2}m_{Z_{c}}^{4}e^{-m_{Z_{c}}^{2}/M^{2}}}{%
\int_{4m_{c}^{2}}^{s_{0}^{\ast }}\rho ^{\mathrm{OPE}%
}(s)e^{-s/M^{2}}ds-f_{Z_{c}}^{2}m_{Z_{c}}^{2}e^{-m_{Z_{c}}^{2}/M^{2}}},
\label{eq:MassEx}
\end{equation}%
and%
\begin{equation}
f_{Z}^{2}=\frac{1}{m_{Z}^{2}}\left[ \int_{4m_{c}^{2}}^{s_{0}^{\ast }}\rho ^{%
\mathrm{OPE}%
}(s)e^{(m_{Z}^{2}-s)/M^{2}}ds-f_{Z_{c}}^{2}m_{Z_{c}}^{2}e^{(m_{Z}^{2}-m_{Z_{c}}^{2})/M^{2}}%
\right],  \label{eq:CoupEx}
\end{equation}%
where $s_{0}^{\ast }$ is the continuum threshold parameter, which separates
contributions of the tetraquarks $Z_{c} + Z$ and higher resonances and
continuum states from each another.

We consider the mass and coupling of $Z_{c}$ as input parameters in Eqs.\ (%
\ref{eq:MassEx}) and (\ref{eq:CoupEx}). These parameters can be found from
the sum rules
\begin{equation}
m_{Z_{c}}^{2}=\frac{\int_{4m_{c}^{2}}^{s_{0}}ds\rho ^{\mathrm{OPE}%
}(s)se^{-s/M^{2}}}{\int_{4m_{c}^{2}}^{s_{0}}ds\rho ^{\mathrm{OPE}%
}(s)e^{-s/M^{2}}},  \label{eq:MassGS}
\end{equation}%
and
\begin{equation}
f_{Z_{c}}^{2}=\frac{1}{m_{Z_{c}}^{2}}\int_{4m_{c}^{2}}^{s_{0}}ds\rho ^{%
\mathrm{OPE}}(s)e^{(m_{Z_{c}}^{2}-s)/M^{2}}.  \label{eq:CoupGS}
\end{equation}%
The expressions (\ref{eq:MassGS}) and (\ref{eq:CoupGS}) correspond to the
"ground-state + continuum" scheme when one includes the tetraquark $Z$ into
a class of "higher resonances". It is clear, that $\rho ^{\mathrm{OPE}}(s)$
is the common spectral density, and the continuum threshold should obey $%
s_{0}<s_{0}^{\star }$. Once calculated the parameters $m_{Z_{c}}$ and $%
f_{Z_{c}}$ of the tetraquark $Z_{c}$ appear as  input information in the
sum rules (\ref{eq:MassEx}) and (\ref{eq:CoupEx}) for the tetraquark $Z$.

The sum rules obtained here depend on various vacuum condensates, which are
input parameters in numerical computations. These sum rules contain also the
mass of $c$ quark. The quark, gluon, and mixed vacuum condensates, as well
as masses of the quarks are well known%
\begin{eqnarray}
&&\langle \bar{q}q\rangle =-(0.24\pm 0.01)^{3}~\mathrm{GeV}^{3},\ \langle
\bar{s}s\rangle =0.8\ \langle \bar{q}q\rangle ,\ \langle \overline{q}%
g_{s}\sigma Gq\rangle =m_{0}^{2}\langle \overline{q}q\rangle ,  \notag \\
&&\langle \overline{s}g_{s}\sigma Gs\rangle =m_{0}^{2}\langle \bar{s}%
s\rangle ,m_{0}^{2}=(0.8\pm 0.1)~\mathrm{GeV}^{2},\langle \frac{\alpha
_{s}G^{2}}{\pi }\rangle =(0.012\pm 0.004)~\mathrm{GeV}^{4},  \notag \\
&&\langle g_{s}^{3}G^{3}\rangle =(0.57\pm 0.29)~\mathrm{GeV}^{6},\
m_{s}=93_{-5}^{+11}~\mathrm{MeV},~m_{c}=1.27\pm 0.2~\mathrm{GeV},  \notag \\
&&m_{b}=4.18_{-0.02}^{+0.03}~\mathrm{GeV}.  \label{eq:VCond}
\end{eqnarray}

The masses and couplings of the tetraquarks depend on auxiliary parameters $%
M^{2}$ and $s_{0}(s_{0}^{\star })$, which have to satisfy constraints of sum
rule computations. It means that, edges of the working windows for the Borel
parameter should be fixed by convergence of the operator product expansion
(OPE) and restriction imposed on the pole contribution ($\mathrm{PC}$).
Additionally, extracted quantities should be stable while the parameter $%
M^{2}$ is varied within this region. Analysis carried out by taking into
account these conditions allows one to extract regions of the parameters $%
M^{2}$ and $s_{0}$, where aforementioned constraints are fulfilled. Our
predictions are collected in Table \ref{tab:Results1}, where we present not
only parameters of the resonances $Z$ and $Z_{c}$, but write down also
windows for $M^{2}$ and $s_{0}(s_{0}^{\star })$ used to extract them. One
can see, that  agreement between $m_{Z_{c}}$ and experimental data is
excellent. It also confirms our previous prediction for $m_{Z_{c}}$ made in
Ref.\ \cite{Agaev:2016dev}. Result for $m_{Z}$ is less that the
corresponding LHCb datum, but still is compatible with measurements provided
one takes into account errors of calculations.
\begin{table}[tbp]
\begin{tabular}{|c|c|c|}
\hline\hline
Resonance & $Z_c$ & $Z$ \\ \hline\hline
$M^2 ~(\mathrm{GeV}^2$) & $3-6$ & $3-6$ \\ \hline
$s_0(s_0^{\star}) ~(\mathrm{GeV}^2$) & $4.2^2-4.4^2$ & $4.8^2-5.2^2$ \\
\hline
$m_{Z} ~(\mathrm{MeV})$ & $3901^{+125}_{-148} $ & $4452^{+182}_{-228} $ \\
\hline
$f_{Z}\times 10^{2} ~(\mathrm{GeV}^4)$ & $0.42^{+0.07}_{-0.09}$ & $%
1.48^{+0.31}_{-0.42}$ \\ \hline\hline
\end{tabular}%
\caption{The masses and current couplings of the resonances $Z_{c}$ and $Z$.}
\label{tab:Results1}
\end{table}


\subsection{Strong decays of the tetraquarks $Z_{c}$ and $Z$}

\label{subsec:StrongVer1}
The masses of $Z_{c}$ and $Z$ obtained above, should be employed to
distinguish from each another their kinematically allowed and forbidden
decay modes. Moreover, parameters of these resonances enter as input
information to sum rules for strong couplings corresponding to vertices $%
Z_{c}M_{h}M_{l}$ and $ZM_{h}M_{l}$, and are also embedded into formulas for
decay widths.

The tetraquarks $Z_{c}$ and $Z$ can dissociate to conventional mesons
through different ways. We consider only their decays to mesons $J/\psi \pi $%
, $\psi ^{\prime }\pi $, and $\eta _{c}\rho $, $\eta _{c}^{\prime }\rho $.
One can find masses and decay constants of these mesons in Table \ref%
{tab:Param}, and easily check that these processes are kinematically allowed
modes.

In our treatment the tetraquark $Z$ is the first radial excitation of $Z_{c}$%
. It is clear, that $\psi ^{\prime }$ and $\eta _{c}^{\prime }\equiv \eta
_{c}(2S)$ are first radial excitations of the mesons $J/\psi $ and $\eta
_{c} $, respectively. Therefore, in framework of the QCD sum rule method, we
have to analyze decays $Z_{c},\,Z\rightarrow J/\psi \pi ,\,\psi ^{\prime
}\pi $ and $Z_{c},\,Z\rightarrow \eta _{c}\rho ,\,\eta _{c}^{\prime }\rho $
in a correlated form. The reason is that, in the QCD sum rules particles are
modeled by interpolating currents which couple both to their ground states
and excitations.
\begin{table}[tbp]
\begin{tabular}{|c|c|}
\hline\hline
Parameters & Values ($\mathrm{MeV}$) \\ \hline\hline
$m_{J/\psi}$ & $3096.900 \pm 0.006$ \\
$f_{J/\psi}$ & $411 \pm 7 $ \\
$m_{\psi^{\prime}}$ & $3686.097 \pm 0.025 $ \\
$f_{\psi^{\prime}}$ & $279 \pm 8 $ \\
$m_{\eta_c}$ & $2983.4 \pm 0.5 $ \\
$f_{\eta_c}$ & $404 $ \\
$m_{\eta_{c}^{\prime}}$ & $3686.2 \pm 1.2$ \\
$f_{\eta_{c}^{\prime}}$ & $331$ \\
$m_{\pi}$ & $139.57018 \pm 0.00035$ \\
$f_{\pi}$ & $131.5 $ \\
$m_{\rho}$ & $775.26 \pm 0.25$ \\
$f_{\rho}$ & $216 \pm 3$ \\ \hline\hline
\end{tabular}%
\caption{Masses and decay constants of the conventional mesons.}
\label{tab:Param}
\end{table}


\subsubsection{Decays $Z_{c},\,Z\rightarrow J/\protect\psi \protect\pi ,\,%
\protect\psi ^{\prime }\protect\pi $}

\label{subsec:D1}
In order to calculate partial widths of the decays $Z_{c}\rightarrow J/\psi
\pi ,\,\psi ^{\prime }\pi $ and $Z\rightarrow J/\psi \pi ,\,\psi ^{\prime
}\pi $, we begin from analysis of the correlation function
\begin{equation}
\Pi _{\mu \nu }(p,q)=i\int d^{4}xe^{ipx}\langle \pi (q)|\mathcal{T}\{J_{\mu
}^{\psi }(x)J_{\nu }^{Z\dagger }(0)\}|0\rangle ,  \label{eq:CorrF3}
\end{equation}%
where%
\begin{equation}
J_{\mu }^{\psi }(x)=\overline{c}_{i}(x)\gamma _{\mu }c_{i}(x),
\label{eq:PsiCurr}
\end{equation}%
and $\psi $ is one of \ $J/\psi $ and $\psi ^{\prime }$ mesons. The current $%
J_{\nu }^{Z}(x)$ is defined by Eq.\ (\ref{eq:Curr}), and $p^{\prime }=p+q$
and $p$, $q$ are the momenta of initial and final particles, respectively.
As we have just emphasized above the interpolating currents $J_{\nu }^{Z}(x)$
and $J_{\mu }^{\psi }(x)$ couple to $Z_{c},\,Z$ and $J/\psi ,\,\psi ^{\prime
}$, respectively. Therefore, the correlation function $\Pi _{\mu \nu }^{%
\mathrm{Phys}}(p,q)$, necessary for our purposes, contains four terms
\begin{eqnarray}
&&\Pi _{\mu \nu }^{\mathrm{Phys}}(p,q)=\sum_{\psi =J/\psi ,\psi ^{\prime }}%
\left[ \frac{\langle 0|J_{\mu }^{\psi }|\psi \left( p\right) \rangle }{%
p^{2}-m_{\psi }^{2}}\langle \psi \left( p\right) \pi (q)|Z_{c}(p^{\prime
})\rangle \frac{\langle Z_{c}(p^{\prime })|J_{\nu }^{Z\dagger }|0\rangle }{%
p^{\prime 2}-m_{Z_{c}}^{2}}\right.  \notag \\
&&\left. +\frac{\langle 0|J_{\mu }^{\psi }|\psi \left( p\right) \rangle }{%
p^{2}-m_{\psi }^{2}}\langle \psi \left( p\right) \pi (q)|Z(p^{\prime
})\rangle \frac{\langle Z(p^{\prime })|J_{\nu }^{Z\dagger }|0\rangle }{%
p^{\prime 2}-m_{Z}^{2}}\right] +\cdots .  \label{eq:CorrF4}
\end{eqnarray}

To find the correlation function, we use the matrix elements
\begin{equation}
\langle 0|J_{\mu }^{\psi }|\psi \left( p\right) \rangle =f_{\psi }m_{\psi
}\varepsilon _{\mu },\,\,\langle Z_{c}(p^{\prime })|J_{\nu }^{Z\dagger
}|0\rangle =f_{Z_{c}}m_{Z_{c}}\varepsilon _{\nu }^{\prime \ast },\ \langle
Z(p^{\prime })|J_{\nu }^{Z\dagger }|0\rangle =f_{Z}m_{Z}\widetilde{%
\varepsilon }_{\nu }^{\prime \ast },  \label{eq:MelA}
\end{equation}%
with $m_{\psi }$, $f_{\psi }$, and $\,\varepsilon _{\mu }$ being the mass,
decay constant, and polarization vector of $J/\psi $ or $\psi ^{\prime }$
mesons. Accordingly, $\varepsilon _{\nu }^{\prime }$ and $\widetilde{%
\varepsilon }_{\nu }^{\prime }$ stand for the polarization vectors of the
states $Z_{c}$ and $Z$, respectively. We model the vertices in the forms
\begin{eqnarray}
\langle \psi \left( p\right) \pi (q)|Z_{c}(p^{\prime })\rangle
&=&g_{Z_{c}\psi \pi }\left[ (p\cdot p^{\prime })(\varepsilon ^{\ast }\cdot
\varepsilon ^{\prime })-(p\cdot \varepsilon ^{\prime })(p^{\prime }\cdot
\varepsilon ^{\ast })\right] ,  \notag \\
\langle \psi \left( p\right) \pi (q)|Z(p^{\prime })\rangle &=&g_{Z\psi \pi }
\left[ (p\cdot p^{\prime })(\varepsilon ^{\ast }\cdot \widetilde{\varepsilon
}^{\prime })-(p\cdot \widetilde{\varepsilon }^{\prime })(p^{\prime }\cdot
\varepsilon ^{\ast })\right] ,  \label{eq:MelB}
\end{eqnarray}%
where $g_{Z_{c}\psi \pi }$ and $g_{Z\psi \pi }$ are the strong couplings,
that have to be evaluated from the sum rules. After some transformations, we
get for $\Pi _{\mu \nu }^{\mathrm{Phys}}(p,q)$ the expression
\begin{eqnarray}
&&\Pi _{\mu \nu }^{\mathrm{Phys}}(p,q)=\sum_{\psi =J/\psi ,\psi ^{\prime }}%
\left[ \frac{f_{\psi }f_{Z_{c}}m_{Z_{c}}m_{\psi }g_{Z_{c}\psi \pi }}{\left(
p^{\prime 2}-m_{Z_{c}}^{2}\right) \left( p^{2}-m_{\psi }^{2}\right) }\left(
\frac{m_{Z_{c}}^{2}+m_{\psi }^{2}}{2}g_{\mu \nu }-p_{\mu }^{\prime }p_{\nu
}\right) \right.  \notag \\
&&\left. +\frac{f_{\psi }f_{Z}m_{Z}m_{\psi }g_{Z\psi \pi }}{\left( p^{\prime
2}-m_{Z}^{2}\right) \left( p^{2}-m_{\psi }^{2}\right) }\left( \frac{%
m_{Z}^{2}+m_{\psi }^{2}}{2}g_{\mu \nu }-p_{\mu }^{\prime }p_{\nu }\right) %
\right] +\cdots .  \label{eq:CorrF5}
\end{eqnarray}%
It is convenient to proceed by choosing structures $\sim g_{\mu \nu }$ and
corresponding invariant amplitudes.

To derive the second ingredient of the sum rule $\Pi _{\mu \nu }^{\mathrm{OPE%
}}(p,q)$, we express the correlation function \ (\ref{eq:CorrF3}) in terms
of the quark propagators, and find
\begin{eqnarray}
&&\Pi _{\mu \nu }^{\mathrm{OPE}}(p,q)=\int d^{4}xe^{ipx}\frac{\epsilon
\widetilde{\epsilon }}{\sqrt{2}}\left[ \gamma _{5}\widetilde{S}%
_{c}^{ib}(x){}\gamma _{\mu }\widetilde{S}_{c}^{ei}(-x){}\gamma _{\nu
}+\gamma _{\nu }\widetilde{S}_{c}^{ib}(x){}\gamma _{\mu }\widetilde{S}%
_{c}^{ei}(-x){}\gamma _{5}\right] _{\alpha \beta }  \notag \\
&&\times \langle \pi (q)|\overline{u}_{\alpha }^{a}(0)d_{\beta
}^{d}(0)|0\rangle,  \label{eq:CorrF6}
\end{eqnarray}%
where $\alpha $ and $\beta $ are the spinor indices.

In order to continue, we expand $\overline{u}_{\alpha }^{a}(0)d_{\beta
}^{d}(0)$ over the full set of Dirac matrices $\Gamma ^{j}$ and project them
onto the color-singlet states by employing the formula
\begin{equation}
\overline{u}_{\alpha }^{a}d_{\beta }^{d}\rightarrow \frac{1}{12}\Gamma
_{\beta \alpha }^{j}\delta _{ad}\left( \overline{u}\Gamma ^{j}d\right) ,
\label{eq:MatEx}
\end{equation}%
where $\Gamma ^{j}$
\begin{equation}
\Gamma ^{j}=\mathbf{1},\gamma _{5},\ \gamma _{\lambda },\ i\gamma _{5}\gamma
_{\lambda },\ \sigma _{\lambda \rho }/\sqrt{2}.
\end{equation}%
Then the matrix elements $\langle \pi (q)|\overline{u}_{\alpha
}^{a}(0)d_{\beta }^{d}(0)|0\rangle $ transform in accordance with the scheme%
\begin{equation}
\langle \pi (q)|\overline{u}_{\alpha }^{a}(0)d_{\beta }^{d}(0)|0\rangle
\rightarrow \frac{1}{12}\Gamma _{\beta \alpha }^{j}\delta _{ad}\langle \pi
(q)|\overline{u}(0)\Gamma ^{j}d(0)|0\rangle .  \label{eq:MatExA}
\end{equation}%
It is seen, that the correlation function $\Pi _{\mu \nu }^{\mathrm{OPE}%
}(p,q)$ depends on local matrix elements of the pion. This is typical
situation for the LCSR method when one of particles is a tetraquark. For
such tetraquark-meson-meson vertices the four-momentum conservation requires
equating a momentum one of final mesons, in the case under discussion of the
pion, to $q=0$ \cite{Agaev:2016dev}. This constraint has to be taken into
account also in the phenomenological side of the sum rule. At vertices of
ordinary two-quark mesons, in general $q\neq 0$, and only as some
approximation one sets $q$ equal to zero. A limit $q=0$ in the conventional
LCSR is known as the soft-meson approximation \cite{Belyaev:1994zk}.
Contrary, tetraquark-meson-meson vertices can be explored in the framework
of the LCSR method only if $q=0$. An important conclusion made in Ref.\ \cite%
{Belyaev:1994zk} states, that for the strong couplings of ordinary mesons
the full LCSR method and its soft-meson version lead to numerically close
predictions.

Having inserted Eq.\ (\ref{eq:MatExA}) into the correlation function, we
perform the summation over color and calculate traces over Lorentz indices.
Relevant prescriptions were explained in a detailed form in Ref.\ \cite%
{Agaev:2016dev}, hence we do not concentrate here on these questions. These
manipulations allow us to determine local matrix elements of the pion that
contribute to $\Pi _{\mu \nu }^{\mathrm{OPE}}(p,q)$, and find the spectral
density $\rho ^{\mathrm{OPE}}(s)$ as the imaginary part of $\Pi _{\mu \nu }^{%
\mathrm{OPE}}(p,q)$. It appears that the matrix element of the pion
\begin{equation}
\langle 0|\overline{d}(0)i\gamma _{5}u(0)|\pi (q)\rangle =f_{\pi }\mu _{\pi }
\label{eq:MatE2}
\end{equation}%
where $\mu _{\pi }=m_{\pi }^{2}/(m_{u}+m_{d})$, contributes to $\rho ^{%
\mathrm{OPE}}(s)$.

To calculate $\rho ^{\mathrm{OPE}}(s)$, we choose in $\Pi _{\mu \nu }^{%
\mathrm{OPE}}(p,q)$ the structure $\sim g_{\mu \nu }$, and get
\begin{equation}
\rho ^{\mathrm{OPE}}(s)=\frac{f_{\pi }\mu _{\pi }}{12\sqrt{2}}\left[ \rho ^{%
\mathrm{pert.}}(s)+\rho ^{\mathrm{n.-pert.}}(s)\right] .  \label{eq:SD3900}
\end{equation}%
The spectral density $\rho ^{\mathrm{OPE}}(s)$ consists of two components.
Thus, its perturbative part $\rho ^{\mathrm{pert.}}(s)$ has a simple form
and was computed in Ref.\ \cite{Agaev:2016dev}
\begin{equation}
\rho ^{\mathrm{OPE}}(s)=\frac{(s+2m_{c}^{2})\sqrt{s(s-4m_{c}^{2})}}{\pi ^{2}s%
}.
\end{equation}%
The $\rho ^{\mathrm{n.-pert.}}(s)$ is a nonperturbative component of the
spectral density, which includes terms up to eighth dimension: $\rho ^{%
\mathrm{n.-pert.}}(s)$ is given by the formula
\begin{equation}
\rho ^{\mathrm{n.-pert.}}(s)=\Big \langle\frac{\alpha _{s}G^{2}}{\pi }\Big
\rangle m_{c}^{2}\int_{0}^{1}f_{1}(z,s)dz+\Big \langle g_{s}^{3}G^{3}\Big
\rangle\int_{0}^{1}f_{2}(z,s)dz+\Big \langle\frac{\alpha _{s}G^{2}}{\pi }%
\Big \rangle^{2}m_{c}^{2}\int_{0}^{1}f_{3}(z,s)dz.  \label{eq:NPert}
\end{equation}%
Explicit expressions of functions $f_{1}(z,s)$, $f_{2}(z,s)$, and $%
f_{3}(z,s) $ were written down in Appendix of Ref. \cite{Agaev:2017tzv}.

Having found $\rho ^{\mathrm{OPE}}(s)$, we now are ready to calculate the
phenomenological side of the sum rule in the soft-meson approximation.
Because in the soft limit $p^{\prime }=p$, the invariant amplitude in Eq.\ (%
\ref{eq:CorrF4}) depends solely on variable $p^{2}$ and has the form
\begin{eqnarray}
&&\Pi ^{\mathrm{Phys}}(p^{2})=\frac{f_{J/\psi }f_{Z_{c}}m_{Z_{c}}m_{J/\psi
}m_{1}^{2}}{\left( p^{2}-m_{1}^{2}\right) ^{2}}g_{Z_{c}J/\psi \pi }+\frac{%
f_{\psi ^{\prime }}f_{Z_{c}}m_{Z_{c}}m_{\psi ^{\prime }}m_{2}^{2}}{\left(
p^{2}-m_{2}^{2}\right) ^{2}}g_{Z_{c}\psi ^{\prime }\pi }+\frac{f_{J/\psi
}f_{Z}m_{Z}m_{J/\psi }m_{3}^{2}}{\left( p^{2}-m_{3}^{2}\right) ^{2}}%
g_{ZJ/\psi \pi }  \notag \\
&&+\frac{f_{\psi ^{\prime }}f_{Z}m_{Z}m_{\psi ^{\prime }}m_{4}^{2}}{\left(
p^{2}-m_{4}^{2}\right) ^{2}}g_{Z\psi ^{\prime }\pi }+\ldots ,
\label{eq:PhysSide1}
\end{eqnarray}%
where
\begin{equation*}
m_{1}^{2}=(m_{Z_{c}}^{2}+m_{J/\psi }^{2})/2,m_{2}^{2}=(m_{Z_{c}}^{2}+m_{\psi
^{\prime }}^{2})/2,\ m_{3}^{2}=(m_{Z}^{2}+m_{J/\psi
}^{2})/2,m_{4}^{2}=(m_{Z}^{2}+m_{\psi ^{\prime }}^{2})/2.
\end{equation*}

In the soft-meson limit the physical side of the sum rules has  complicated
content. Thus, besides $g_{Z_{c}J/\psi \pi }$ it contains also other strong
couplings, i.e., terms that remain unsuppressed even after the Borel
transformation \cite{Belyaev:1994zk}. To exclude them from $\Pi ^{\mathrm{%
Phys}}(p^{2})$ one has to act by the operator
\begin{equation}
\mathcal{P}(M^{2},m^{2})=\left( 1-M^{2}\frac{d}{dM^{2}}\right)
M^{2}e^{m^{2}/M^{2}},  \label{eq:softop}
\end{equation}%
to both sides of sum rules \cite{Ioffe:1983ju}. In our studies, in order to
evaluate strong couplings and calculate decay widths of various tetraquarks,
we benefited from this technique (see, Ref.\ \cite{Agaev:2016dev}, as an
example). But unsuppressed terms come from vertices of excited states of
initial (final) particles, i.e., from vertices $ZJ/\psi \pi $, $Z\psi
^{\prime }\pi $ and $Z_{c}\psi ^{\prime }\pi $. In other words,
contributions considered as contaminations while one investigates a vertex
of ground-state particles become a subject of analysis in the present case.
Because, in general, $\Pi ^{\mathrm{Phys}}(p^{2})$ contains four terms and,
at the first stage of analyses, is the sum of two contributions, we do not
apply the operator $\mathcal{P}$ to present sum rules.

We proceed by following recipes of the previous subsection, i.e., we fix the
parameter $s_{0}$ below threshold for the decays $Z\rightarrow J/\psi \pi $
and $Z\rightarrow \psi ^{\prime }\pi $. Then in the considering range of $%
s\in \lbrack 0,s_{0}]$ only first two terms in Eq.\ (\ref{eq:PhysSide1})
should be explicitly taken into account: last two terms are automatically
included into a "higher resonances and continuum". The one-variable Borel
transformation applied to remaining two terms is the first step to derive a
sum rule equality. Afterwards, we equate the physical and QCD sides of the
sum rule, and in accordance with the hadron-quark duality hypothesis carry
out the continuum subtraction
\begin{equation}
f_{Z_{c}}m_{Z_{c}}\left[ f_{J/\psi }m_{J/\psi }m_{1}^{2}g_{Z_{c}J/\psi \pi
}e^{-m_{1}^{2}/M^{2}}+f_{\psi ^{\prime }}m_{\psi ^{\prime
}}m_{2}^{2}g_{Z_{c}\psi ^{\prime }\pi }e^{-m_{2}^{2}/M^{2}}\right]
=\int_{4m_{c}^{2}}^{s_{0}}dse^{-s/M^{2}}\rho ^{\mathrm{QCD}}(s).
\label{eq:SR1}
\end{equation}%
But this expression is not enough to determine two unknown variables $%
g_{Z_{c}\psi ^{\prime }\pi }$ and $g_{Z_{c}J/\psi \pi }$. The second
equality is obtained from Eq.\ (\ref{eq:SR1}) by applying the operator $%
d/d(-1/M^{2})$ to its both sides. The equality derived by this way, and the
master expression \ (\ref{eq:SR1}) allows us to extract sum rules for the
couplings $g_{Z_{c}\psi ^{\prime }\pi }$ and $g_{Z_{c}J/\psi \pi }$. They
are necessary to compute partial width of the decays $Z_{c}\rightarrow \psi
^{\prime }\pi $ and $Z_{c}\rightarrow J/\psi \pi $, and appear as input
parameters in the next sum rules.

The sum rules for the couplings $g_{Z\psi ^{\prime }\pi }$ and $g_{ZJ/\psi
\pi }$ are found by choosing $\sqrt{s_{0}^{\star }}=m_{Z}+(0.5-0.7)~\mathrm{%
GeV}$. Such choice for $s_{0}^{\star }$ is motivated by observation that a
mass splitting in a tetraquark multiplet is approximately $0.5-0.7~\mathrm{%
GeV}$. For $s\in \lbrack 0,s_{0}^{\star }]$ the processes $Z\rightarrow
J/\psi \pi $ and $Z\rightarrow \psi ^{\prime }\pi $ have to be taken into
account as well. In other words, in this step of studies all terms in Eq.\ (%
\ref{eq:PhysSide1}) have to be explicitly taken into account. We derive sum
rules for the couplings $g_{Z\psi ^{\prime }\pi }$ and $g_{ZJ/\psi \pi }$ by
repeating manipulations explained above and using two other couplings as
input parameters.

We evaluate the width of the decay $Z\rightarrow \psi \pi $ by utilizing of
the formula
\begin{equation}
\Gamma \left( Z\rightarrow \psi \pi \right) =\frac{g_{Z\psi \pi }^{2}m_{\psi
}^{2}}{24\pi }\lambda \left( m_{Z},\ m_{\psi },m_{\pi }\right) \left[ 3+%
\frac{2\lambda ^{2}\left( m_{Z},\ m_{\psi },m_{\pi }\right) }{m_{\psi }^{2}}%
\right] ,  \label{eq:DW}
\end{equation}%
where
\begin{equation}
\lambda (a,\ b,\ c)=\frac{\sqrt{a^{4}+b^{4}+c^{4}-2\left(
a^{2}b^{2}+a^{2}c^{2}+b^{2}c^{2}\right) }}{2a}.  \label{eq:Lambda}
\end{equation}%
The equation (\ref{eq:DW}) is valid for all four decay channels, where $%
Z=Z_{c}$ or $Z,$ and $\psi =$ $J/\psi $ or $\psi ^{\prime }$, respectively.

It is clear that, apart from couplings $g_{Z\psi \pi }$ the partial width of
the processes $Z\rightarrow \psi \pi $ contains parameters of initial and
final particles. The spectroscopic parameters of the tetraquarks $Z$ and $%
Z_{c}$ have been calculated in this section. Masses and decay constants of
mesons $J/\psi $, $\psi ^{\prime }$, $\pi $ are presented in Ref.\ \cite%
{Tanabashi:2018oca}. All these information are collected in Table\ \ref%
{tab:Param}, where we also write down spectroscopic parameters of the mesons
$\eta _{c}$, $\eta _{c}^{\prime }$ and $\rho $, which will be used below to
explore another decay channels of $Z$ and $Z_{c}$. Let us note that decay
constants $f_{\eta _{c}}$ and $f_{\eta _{c}^{\prime }}$ are borrowed from
Ref.\ \cite{Negash:2015rua}.

The working windows for the Borel and continuum threshold parameters used to
evaluate strong couplings do not differ from ones employed for analysis of
the masses and current couplings. Another problem, which should be
considered, is contributions to the sum rules arising from excited terms. It
is known, that dominant contribution to the sum rules is generated by a
ground-state term. In the case under analysis, besides the strong coupling
of the ground-state particles, we evaluate couplings of one or two radially
excited particles as well. The sum rules for these couplings may lead to
reliable predictions provided their effects and contributions are sizeable.
This question can be analyzed by exploring the pole contribution to the sum
rules
\begin{equation}
\mathrm{PC}=\frac{\int_{0}^{s_{0}}ds\rho ^{\mathrm{OPE}}(s)e^{-s/M^{2}}}{%
\int_{0}^{\infty }ds\rho ^{\mathrm{OPE}}(s)e^{-s/M^{2}}}.  \label{eq:PC}
\end{equation}%
Choosing $s_{0}=4.2^{2}\ \mathrm{GeV}^{2}$ and fixing $M^{2}=4.5\ \mathrm{GeV%
}^{2}$ we find $\mathrm{PC}=0.81$, which is generated by the terms
proportional to couplings $g_{Z_{c}J/\psi \pi }$ and $g_{Z_{c}\psi ^{\prime
}\pi }$. At the next phase of analysis, we fix $s_{0}\equiv s_{0}^{\star }$
and get $\mathrm{PC}=0.95$, which now embraces effects of all four terms. In
other words, contributions of terms $\sim g_{ZJ/\psi \pi }$ and $\sim
g_{Z\psi ^{\prime }\pi }$ amount to $14\%$ part of the sum rules. We see
that, effects of terms connected directly with decays of $Z$ are small,
nevertheless $g_{ZJ/\psi \pi }$ and $g_{Z\psi ^{\prime }\pi }$ are extracted
from full expressions, which contain contributions of four terms, and
therefore their evaluations are founded on reliable basis. It is seen also
that, an effect of the "higher excited states and continuum" does not exceed
$5\%$ of $\mathrm{PC}$, which means that  contaminations arising from
excited states higher than the resonance $Z$ are negligible.

Numerical values of couplings $g$ are sensitive to parameters $M^{2}$ and $%
s_{0}$, nevertheless theoretical uncertainties of $g$ generated by
variations of $M^{2}$ and $s_{0}$ remain within limits typical for sum rule
computations.These uncertainties and ones arising from other parameters form
the full theoretical errors of numerical analysis.

Our computations for $g_{Z\psi ^{\prime }\pi }$ and width of the
corresponding decay $Z\rightarrow \psi ^{\prime }\pi $ yield
\begin{equation}
g_{Z\psi ^{\prime }\pi }=(0.58\pm 0.16)~\mathrm{GeV}^{-1},\ \ \Gamma
(Z\rightarrow \psi ^{\prime }\pi )=(129.7\pm 37.6)~\mathrm{MeV}.
\label{eq:PiDW1}
\end{equation}%
The coupling $g_{ZJ/\psi \pi }$ and width of the process $Z\rightarrow
J/\psi \pi $ are found as
\begin{equation}
g_{ZJ/\psi \pi }=(0.24\pm 0.06)~\mathrm{GeV}^{-1},\ \ \Gamma (Z\rightarrow
J/\psi \pi )=(27.4\pm 7.1)~\mathrm{MeV}.  \label{eq:PiDW2}
\end{equation}%
Predictions obtained for all of strong couplings, and for the partial width
of corresponding decay channels are presented in Table\ \ref{tab:Results1A}.
\begin{table}[tbp]
\begin{tabular}{|c|c|c|c|c|}
\hline
Channels & $Z \to \psi^{\prime} \pi$ & $Z \to J/\psi \pi$ & $Z_c \to
\psi^{\prime} \pi$ & $Z_c \to J/\psi \pi$ \\ \hline\hline
$g ~(\mathrm{GeV}^{-1})$ & $0.58 \pm 0.16$ & $0.24 \pm 0.06 $ & $0.29 \pm
0.08$ & $0.38 \pm 0.11$ \\ \hline
$\Gamma ~(\mathrm{MeV})$ & $129.7 \pm 37.6$ & $27.4 \pm 7.1$ & $7.1 \pm 1.9$
& $39.9 \pm 9.3$ \\ \hline\hline
\end{tabular}%
\caption{The strong coupling $g$ and width of the $Z(Z_{c})\rightarrow
\protect\psi ^{\prime }(J/\protect\psi )\protect\pi $ decay channels.}
\label{tab:Results1A}
\end{table}


\subsubsection{Decays $Z_{c},\,Z\rightarrow \protect\eta _{c}^{\prime }%
\protect\rho ,\,\protect\eta _{c}\protect\rho $}

The $Z_{c}$ and $Z$ decay also to final mesons $\eta _{c}\rho $ and $\eta
_{c}^{\prime }\rho $. Because the decay $Z_{c}\rightarrow \eta _{c}^{\prime
}\rho $ is kinematically forbidden, in this subsection we have three
channels $Z\rightarrow \eta _{c}^{\prime }\rho $, $Z\rightarrow \eta
_{c}\rho $ and $Z_{c}\rightarrow \eta _{c}\rho $ to be studied. Let us note
that present analysis differs in some aspects from prescriptions explained
above.

As usual, we consider the correlation function
\begin{equation}
\Pi _{\nu }(p,q)=i\int d^{4}xe^{ipx}\langle \rho (q)|\mathcal{T}\{J^{\eta
_{c}}(x)J_{\nu }^{Z\dagger }(0)\}|0\rangle,
\end{equation}%
where $\eta _{c}\equiv \eta _{c},\ \eta _{c}^{\prime }$, and the current $%
J^{\eta _{c}}(x)$ is defined as%
\begin{equation}
J^{\eta _{c}}(x)=\overline{c}_{i}(x)i\gamma _{5}c_{i}(x).  \label{eq:EtaCurr}
\end{equation}

To express the correlation function in terms of involved particles' physical
parameters, we use the matrix elements
\begin{equation}
\langle 0|J^{\eta _{c}}|\eta _{c}(p)\rangle =\frac{f_{\eta _{c}}m_{\eta
_{c}}^{2}}{2m_{c}},  \label{eq:EtaCME}
\end{equation}%
with $m_{\eta _{c}}$ and $f_{\eta _{c}}$ being the mass and decay constant
of the meson $\eta _{c}$. The similar matrix element is also valid for the
meson $\eta _{c}^{\prime }$. The matrix elements of vertices are modeled in
the forms
\begin{equation}
\langle \eta _{c}\left( p\right) \rho (q)|Z(p^{\prime })\rangle =g_{Z\eta
_{c}\rho }\left[ (q\cdot \widetilde{\varepsilon }^{\prime })(p^{\prime
}\cdot \varepsilon ^{\ast })-(q\cdot p^{\prime })(\varepsilon ^{\ast }\cdot
\widetilde{\varepsilon }^{\prime })\right],
\end{equation}%
and
\begin{equation}
\langle \eta _{c}\left( p\right) \rho (q)|Z_{c}(p^{\prime })\rangle
=g_{Z_{c}\eta _{c}\rho }\left[ (q\cdot \varepsilon ^{\prime })(p^{\prime
}\cdot \varepsilon ^{\ast })-(q\cdot p^{\prime })(\varepsilon ^{\ast }\cdot
\varepsilon ^{\prime })\right],
\end{equation}%
where $q$ and $\varepsilon $ are the momentum and polarization vector of the
$\rho $-meson, respectively.

We write the phenomenological side of the sum rules $\Pi _{\nu }^{\mathrm{%
Phys}}(p,q)$ in the form
\begin{eqnarray}
&&\Pi _{\nu }^{\mathrm{Phys}}(p,q)=\frac{\langle 0|J^{\eta _{c}}|\eta
_{c}\left( p\right) \rangle }{p^{2}-m_{\eta _{c}}^{2}}\langle \eta
_{c}\left( p\right) \rho (q)|Z_{c}(p^{\prime })\rangle \frac{\langle
Z_{c}(p^{\prime })|J_{\nu }^{Z}|0\rangle }{p^{\prime 2}-m_{Z_{c}}^{2}}
\notag \\
&&+\sum_{\eta _{c}=\eta _{c},\eta _{c}^{\prime }}\frac{\langle 0|J^{\eta
_{c}}|\eta _{c}\left( p\right) \rangle }{p^{2}-m_{\eta _{c}}^{2}}\langle
\eta _{c}\left( p\right) \rho (q)|Z(p^{\prime })\rangle \frac{\langle
Z(p^{\prime })|J_{\nu }^{Z}|0\rangle }{p^{\prime 2}-m_{Z}^{2}}+\cdots .
\end{eqnarray}%
It contains three terms, which can be simplified using matrix elements
introduced above. The full expression of $\Pi _{\nu }^{\mathrm{Phys}}(p,q)$
is cumbersome, therefore we write down only the invariant amplitude
corresponding to the structure $\sim \epsilon _{\nu }^{\ast }$ in the limit $%
q\rightarrow 0$, which is employed in our analysis. This amplitude is given
by the formula
\begin{eqnarray}
&&\Pi ^{\mathrm{Phys}}(p^{2})=\frac{f_{\eta _{c}}f_{Z_{c}}m_{Z_{c}}m_{\eta
_{c}}^{2}g_{Z_{c}\eta _{c}\rho }}{4m_{c}\left( p^{2}-\widetilde{m}%
_{1}^{2}\right) ^{2}}(m_{Z_{c}}^{2}-m_{\eta _{c}}^{2})+\frac{f_{\eta
_{c}}f_{Z}m_{Z}m_{\eta _{c}}^{2}g_{Z\eta _{c}\rho }}{4m_{c}\left( p^{2}-%
\widetilde{m}_{2}^{2}\right) ^{2}}(m_{Z}^{2}-m_{\eta _{c}}^{2})  \notag \\
&&+\frac{f_{\eta _{c}^{\prime }}f_{Z}m_{Z}m_{\eta _{c}^{\prime
}}^{2}g_{Z\eta _{c}^{\prime }\rho }}{4m_{c}\left( p^{2}-\widetilde{m}%
_{3}^{2}\right) ^{2}}(m_{Z}^{2}-m_{\eta _{c}^{\prime }}^{2})+\cdots ,
\label{eq:PhysSide}
\end{eqnarray}%
where the notations $\widetilde{m}_{1}^{2}=(m_{Z_{c}}^{2}+m_{\eta
_{c}}^{2})/2$, $\widetilde{m}_{2}^{2}=(m_{Z}^{2}+m_{\eta _{c}}^{2})/2$ and $%
\widetilde{m}_{3}^{2}=(m_{Z}^{2}+m_{\eta _{c}^{\prime }}^{2})/2$ are
introduced.

Computation of the correlation function $\Pi _{\nu }^{\mathrm{OPE}}(p,q)$
using quark propagators leads to the expression
\begin{eqnarray}
&&\Pi _{\nu }^{\mathrm{OPE}}(p,q)=-i\int d^{4}xe^{ipx}\frac{\epsilon
\widetilde{\epsilon }}{\sqrt{2}}\left[ \gamma _{5}\widetilde{S}%
_{c}^{ib}(x){}\gamma _{5}\times \widetilde{S}_{c}^{ei}(-x){}\gamma _{\nu
}+\gamma _{\nu }\widetilde{S}_{c}^{ib}(x){}\gamma _{5}\widetilde{S}%
_{c}^{ei}(-x){}\gamma _{5}\right] _{\alpha \beta }  \notag \\
&&\times \langle \rho (q)|\overline{u}_{\alpha }^{d}(0)d_{\beta
}^{a}(0)|0\rangle.  \label{eq:CorrF7}
\end{eqnarray}%
In the $q\rightarrow 0$ limit the contributions to $\rho ^{\mathrm{OPE}}(s)$
come from the matrix elements \cite{Agaev:2016dev}
\begin{equation}
\langle 0|\overline{u}(0)\gamma _{\mu }d(0)|\rho (p,\lambda )\rangle
=\epsilon _{\mu }^{(\lambda )}f_{\rho }m_{\rho },  \label{eq:Melem1}
\end{equation}%
and
\begin{equation}
\langle 0|\overline{u}(0)g\widetilde{G}_{\mu \nu }\gamma _{\nu }\gamma
_{5}d(0)|\rho (p,\lambda )\rangle =f_{\rho }m_{\rho }^{3}\epsilon _{\mu
}^{(\lambda )}\zeta _{4\rho }.  \label{eq:Melem2}
\end{equation}%
These elements contain the $\rho $-meson's mass and decay constant $m_{\rho
} $, and $f_{\rho }$, and Eq.\ (\ref{eq:Melem2}) additionally depends on a
normalization factor $\zeta _{4\rho }$ of the $\rho $-meson's twist-4 matrix
element \cite{Ball:1998ff}. The numerical value of $\zeta _{4\rho }$ was
estimated in Ref.\ \cite{Ball:2007zt} at the scale $\mu =1~\mathrm{GeV}$,
and amounts to $\zeta _{4\rho }=0.07\pm 0.03$.

We derive the spectral density $\rho ^{\mathrm{OPE}}(s)$ in accordance with
known recipes, and find
\begin{equation}
\rho ^{\mathrm{OPE}}(s)=\frac{f_{\rho }m_{\rho }}{8\sqrt{2}}\left[ \frac{%
\sqrt{s(s-4m_{c}^{2})}}{\pi ^{2}}+\rho ^{\mathrm{n.-pert.}}(s)\right] .
\label{eq:SD2}
\end{equation}%
The nonperturbative component of $\rho ^{\mathrm{OPE}}(s)$ is calculated
with dimension-8 accuracy and has the following form
\begin{equation}
\rho ^{\mathrm{n.-pert.}}(s)=\frac{\zeta _{4\rho }m_{\rho }^{2}}{s}+\Big
\langle\frac{\alpha _{s}G^{2}}{\pi }\Big \rangle m_{c}^{2}\int_{0}^{1}%
\widetilde{f}_{1}(z,s)dz+\Big \langle g_{s}^{3}G^{3}\Big \rangle\int_{0}^{1}%
\widetilde{f}_{2}(z,s)dz+\Big \langle\frac{\alpha _{s}G^{2}}{\pi }\Big
\rangle^{2}m_{c}^{2}\int_{0}^{1}\widetilde{f}_{3}(z,s)dz.
\end{equation}%
Explicit expressions of the functions $\widetilde{f}_{1}(z,s)$, $\widetilde{f%
}_{2}(z,s)$, and $\widetilde{f}_{3}(z,s)$ can be found in Appendix of Ref.
\cite{Agaev:2017tzv}.

To obtain sum rules, we utilize again a prescription described above. At the
first step, i.e., for $s\in \lbrack 0,\ s_{0}]$ the physical side of the sum
rule consists of a ground-state term. At this stage, we calculate the
ground-state coupling $g_{Z_{c}\eta _{c}\rho }$, therefore to exclude
effects of excited states from the physical side of the sum rule apply the
operator $\mathcal{P}(M^{2},\widetilde{m}_{1}^{2})$. Then, we find%
\begin{equation}
g_{Z_{c}\eta _{c}\rho }=\frac{4m_{c}}{f_{\eta _{c}}f_{Z_{c}}m_{Z_{c}}m_{\eta
_{c}}^{2}(m_{Z_{c}}^{2}-m_{\eta _{c}}^{2})}\mathcal{P}(M^{2},\widetilde{m}%
_{1}^{2})\int_{4m_{c}^{2}}^{s_{0}}dse^{-s/M^{2}}\rho ^{\mathrm{OPE}}(s).
\end{equation}%
In the domain $s\in \lbrack 0,\ s_{0}^{\ast }]$ all terms from Eq.\ (\ref%
{eq:PhysSide}) should be included into analysis, and, as a result, we get
the expression with two additional couplings. Excited terms enter to this
expression explicitly, and because our goal is to determine relevant
couplings, in this situation we do not use the operator $\mathcal{P}$. The
second equality can be found by applying the operator $d/d(-1/M^{2})$ to
both sides of the first expression. Solutions of these equations are sum
rules for the couplings $g_{Z\eta _{c}}\rho $ and $g_{Z\eta _{c}^{\prime
}}\rho $. The width of the decays $Z\rightarrow \eta _{c}\rho $, $%
Z\rightarrow \eta _{c}^{\prime }\rho $ and $Z_{c}\rightarrow \eta _{c}\rho $
after replacements $m_{\pi }\rightarrow m_{\eta _{c}}(m_{\eta _{c}^{\prime
}})$ and $m_{\psi }\rightarrow m_{\rho }$ can be computed using Eq.\ (\ref%
{eq:DW}).

For the coupling $g_{Z_{c}\eta _{c}\rho }$ and width of the decay $%
Z_{c}\rightarrow \eta _{c}\rho $, we get
\begin{equation}
g_{Z_{c}\eta _{c}\rho }=(1.28\pm 0.32)~\mathrm{GeV}^{-1},\text{ }\Gamma
(Z_{c}\rightarrow \eta _{c}\rho )=(20.28\pm 5.17)~\mathrm{MeV}.
\label{eq:RhoDW1}
\end{equation}%
The strong couplings $g_{Z\eta _{c}^{\prime }\rho }$ and $g_{Z\eta _{c}\rho
} $, and width of the decays $Z\rightarrow \eta _{c}^{\prime }\rho $ and $%
Z\rightarrow \eta _{c}\rho $ are equal to
\begin{equation}
g_{Z\eta _{c}^{\prime }\rho }=(0.81\pm 0.20)~\mathrm{GeV}^{-1},\ \Gamma
(Z\rightarrow \eta _{c}^{\prime }\rho )=(1.01\pm 0.27)~\mathrm{MeV},
\label{eq:RhoDW2}
\end{equation}%
and
\begin{equation}
g_{Z\eta _{c}\rho }=(0.48\pm 0.11)~\mathrm{GeV}^{-1},\ \ \Gamma
(Z\rightarrow \eta _{c}\rho )=(11.57\pm 3.01)~\mathrm{MeV}.
\label{eq:RhoDW3}
\end{equation}

The processes $Z_{c}\rightarrow J/\psi \pi $ and $Z_{c}\rightarrow \eta
_{c}\rho $ were considered in Ref.\ \cite{Agaev:2016dev} using the QCD
light-cone sum rule method and diquark-antidiquark type interpolating
current. In Table\ \ref{tab:ZcDec}, we compare the partial widths of these
modes from Ref.\ \cite{Agaev:2016dev} with results obtained in Ref.\ \cite%
{Agaev:2017tzv}. It is clear that, these predictions are very close to each
other. Stated differently, an iterative scheme used in this section led to
results that are almost identical with predictions of Ref.\ \cite%
{Agaev:2016dev}. This fact can be treated as a serious argument in favor of
the used approach. The unessential discrepancies between two sets of results
may be explained by accuracy of the spectral densities, which here have been
calculated by taking into account condensates up to eight dimensions,
whereas in Ref.\ \cite{Agaev:2016dev} $\rho _{\pi }^{\mathrm{OPE}}(s)$ and $%
\rho _{\rho }^{\mathrm{OPE}}(s)$ contained only perturbative terms. Let us
emphasize that, we have computed also the partial width of the decay $%
Z_{c}\rightarrow \psi ^{\prime }\pi $ , which was omitted in Ref.\ \cite%
{Agaev:2016dev}.

It is evident that, $Z$ decays dominantly via the process $Z\rightarrow \psi
^{\prime }\pi $. The full width of $Z$ saturated by two channels $%
Z\rightarrow \psi ^{\prime }\pi $ and $Z\rightarrow J/\psi \pi $ equals to $%
(157.1\pm 38.3)\ {\mathrm{MeV}}$. This prediction is compatible with LHCb
information (see, Eq.\ (\ref{eq:LHCdata})), but is below the upper edge of
the experimental data $\approx 212~\mathrm{MeV}$. Experimental data on the
width of the decay $Z\rightarrow J/\psi \pi $ is limited by Belle report
about product of branching fractions%
\begin{equation}
\mathcal{B}(\overline{B}^{0}\rightarrow K^{-}Z^{+})\mathcal{B}%
(Z^{+}\rightarrow J/\psi \pi )=(5.4_{-1.0}^{+4.0}{}_{-0.9}^{+1.1})\times
10^{-6}.
\end{equation}%
By invoking similar experimental measurements for $\psi ^{\prime }$, it is
possible to estimate a ratio
\begin{equation}
R_{Z}=\Gamma (Z\rightarrow \psi ^{\prime }\pi )/\Gamma (Z\rightarrow J/\psi
\pi ),
\end{equation}%
which was carried out in Ref.\ \cite{Goerke:2016hxf}. But, we are not going
to draw strong conclusions from such computations. We think that, in the
absence of direct measurements of $\Gamma (Z\rightarrow J/\psi \pi )$, an
only reasonable way is to compute $R_{Z}$, which is equal to $R_{Z}=4.73\pm
1.84$.
\begin{table}[t]
\begin{tabular}{|c|c|c|c|}
\hline\hline
& $\Gamma(Z_c \to J/ \psi \pi)$ & $\Gamma(Z_c \to \psi^{\prime} \pi)$ & $%
\Gamma(Z_c \to \eta_c \rho)$ \\
& $(\mathrm{MeV})$ & $(\mathrm{MeV})$ & $(\mathrm{MeV})$ \\ \hline
\cite{Agaev:2017tzv} & $39.9 \pm 9.3$ & $7.1 \pm 1.9$ & $20.28 \pm 5.17$ \\
\hline
\cite{Agaev:2016dev} & $41.9\pm 9.4$ & $-$ & $23.8\pm 4.9$ \\ \hline
\cite{Dias:2013xfa} & $29.1 \pm 8.2$ & $-$ & $27.5\pm 8.5$ \\ \hline
\cite{Goerke:2016hxf}A & $27.9^{+6.3}_{-5.0}$ & $-$ & $35.7^{+6.3}_{-5.2}$
\\ \hline
\cite{Goerke:2016hxf}B & $1.8 \pm 0.3$ & $-$ & $3.2^{+0.5}_{-0.4}$ \\ \hline
\cite{Dong:2013iqa} & $10.43 - 23.89$ & $1.28-2.94$ & $-$ \\ \hline\hline
\end{tabular}%
\caption{Predictions for decays of the resonance $Z_{c}$.}
\label{tab:ZcDec}
\end{table}
\begin{table}[t]
\begin{tabular}{|c|c|c|c|c|}
\hline\hline
& $\Gamma(Z\to J/ \psi \pi)$ & $\Gamma(Z \to \psi^{\prime} \pi)$ & $\Gamma(Z
\to \eta_c \rho)$ & $\Gamma(Z \to \eta_c^{\prime} \rho)$ \\
& $(\mathrm{MeV})$ & $(\mathrm{MeV})$ & $(\mathrm{MeV})$ & $(\mathrm{MeV})$
\\ \hline
\cite{Agaev:2017tzv} & $27.4 \pm 7.1$ & $129.7 \pm 37.6$ & $11.57 \pm 3.01$
& $1.01 \pm 0.27$ \\ \hline
\cite{Goerke:2016hxf} & $26.9$ & $120.6$ & $-$ & $-$ \\ \hline\hline
\end{tabular}%
\caption{The same as in Table\ \protect\ref{tab:ZcDec}, but for the
resonance $Z$.}
\label{tab:ZDec}
\end{table}

The decays of the resonances $Z$ and $Z_{c}$ were studied in Refs.\ \cite%
{Goerke:2016hxf,Dias:2013xfa,Dong:2013iqa}: some of these predictions are
written down in Tables\ \ref{tab:ZcDec} and \ref{tab:ZDec}. Partial widths
of decay modes $Z_{c}\rightarrow J/\psi \pi $, $Z_{c}\rightarrow \eta
_{c}\rho $, $Z_{c}\rightarrow \overline{D}^{0}D^{\star }$ and $%
Z_{c}\rightarrow \overline{D}^{\star 0}D$ in the context of the three-point
sum rule method and diquark-antidiquark picture for $Z_{c}$ were calculated
in Ref.\ \cite{Dias:2013xfa}. Their predictions for first two channels are
shown in Table\ \ref{tab:ZcDec}.

The resonance $Z_{c}$ was also treated in Ref.\ \cite{Goerke:2016hxf} both
as  diquark-antidiquark and molecule-type tetraquarks. Decays $%
Z_{c}\rightarrow J/\psi \pi $, and $Z_{c}\rightarrow \eta _{c}\rho $ were
explored there using the covariant quark model. Partial widths of these
processes were evaluated in the diquark-antidiquark picture using a size
parameter $\Lambda _{Z_{c}}=2.25\pm 0.10~\mathrm{GeV}$ in their model (model
A), and in a molecular-type structure with $\Lambda _{Z_{c}}=3.3\pm 0.1~%
\mathrm{GeV}$ (model B). Obtained results are presented in Table\ \ref%
{tab:ZcDec}, as well.

In the context of the phenomenological Lagrangian method decays of the
tetraquark $Z_{c}$ were examined in Ref.\ \cite{Dong:2013iqa}. The $Z_{c}$
was considered there as hadronic molecules $\overline{D}D^{\star }$ and $%
\overline{D}^{\star }D$. In the case of the molecule's binding energy $%
\epsilon =20~\mathrm{MeV}$ the authors estimated widths of different decay
processes: some of obtained results are demonstrated in Table\ \ref%
{tab:ZcDec}.

Decays of the resonance $Z$ to $J/\psi \pi $ and $\psi ^{\prime }\pi $ were
also studied in Ref.\ \cite{Goerke:2016hxf}, where it was modeled as a
diquark-antidiquark system. Results for the partial widths of these decays
obtained at $\Lambda _{Z(4430)}=2.4\ \mathrm{GeV}$, and estimates for $%
\Gamma (Z\rightarrow J/\psi \pi )+\Gamma (Z\rightarrow \psi ^{\prime }\pi
)=147.5\ \mathrm{MeV}$ and $R_{Z}=4.48$ are close to our predictions.

We have examined the tetraquark $Z$ as first radial excitation of the
diquark-antidiquark state $Z_{c}$. We evaluated the masses and full widths
of the resonances $Z_{c}$ and $Z$, and have found: $%
m_{Z_{c}}=3901_{-148}^{+125}~\mathrm{MeV}$, $\Gamma _{Z_{c}}=(67.3\pm 10.8)~%
\mathrm{MeV}$, and $m_{Z}=4452_{-161}^{+132}~\mathrm{MeV}$, $\Gamma
_{Z}=(169.7\pm 38.4)~\mathrm{MeV}$, respectively. Predictions obtained here
seem support a suggestion about  excited nature of $Z$. But there are
problems to be considered before making viable conclusions. Namely, there is
necessity to improve our predictions for the full widths of tetraquarks $%
Z_{c}$ and $Z$ by studying their other decay modes. Experimental studies of
the $Z$ resonance's decay modes, including a direct measurement of $\Gamma
(Z\rightarrow J/\psi \pi )$ may be helpful to confirm its nature as a radial
excitation of the state $Z_{c}$.


\section{ The tetraquark $Z_{c}^{-}(4100)$}

\label{sec:Zc4100}

The tetraquark $Z_{c}^{-}(4100)$ was discovered by LHCb in $B^{0}\rightarrow
K^{+}\eta _{c}\pi ^{-}$ decays as a resonance in the $\eta _{c}\pi ^{-}$
mass distribution \cite{Aaij:2018bla}. The mass and width of this new $%
Z_{c}^{-}(4100)$ state (in this section will be denoted $\overline{Z}_{c}$)
were found equal to
\begin{equation}
m=4096\pm 20_{-22}^{+18}~\mathrm{MeV,\ }\Gamma =152\pm 58_{-35}^{+60}~%
\mathrm{MeV}.  \label{eq:SPZ4100}
\end{equation}

In Ref.\ \cite{Aaij:2018bla} the spin and parity of $Z_{c}^{-}(4100)$ were
determined as well, and it was shown that assignments $J^{\mathrm{P}}=0^{+}$
or $J^{\mathrm{P}}=1^{-}$ do not contradict to the experimental data.

The theoretical articles, as usual, consider problems connected with the
spin and possible decays of the resonance $\overline{Z}_{c}$ \cite%
{Wang:2018ntv,Wu:2018xdi,Voloshin:2018vym,Cao:2018vmv}. Thus, sum rule
calculations performed in Ref.\ \cite{Wang:2018ntv} showed that $\overline{Z}%
_{c}$ is probably a scalar tetraquark. The nature of $\overline{Z}_{c}$ as a
diquark-antidiquark state with $J^{\mathrm{PC}}=0^{++}$ was supported also
in Ref. \cite{Wu:2018xdi}. The resonances $\overline{Z}_{c}$ and $%
Z_{c}^{-}(4200)$ in the hadrocharmonium model were considered as the scalar $%
\eta _{c}$ and vector $J/\psi $ charmonia placed into a light-quark field
with pion's quantum numbers \cite{Voloshin:2018vym}. Then, due to spin
symmetry of charm quark, features of the particles $\overline{Z}_{c}$ and $%
Z_{c}^{-}(4200)$, as well as their decay modes are connected by some
relations.

Because the resonance $\overline{Z}_{c}$ was seen in the decay $\overline{Z}%
_{c}\rightarrow \eta _{c}\pi ^{-}$, it is natural to treat it as a scalar
particle with quark content $c\overline{c}d\overline{u}$. Really, the decay $%
\overline{Z}_{c}\rightarrow \eta _{c}\pi ^{-}$ is dominant $S$-wave mode for
a scalar particle, but it turns to $P$-wave decay channel in the case of a
vector tetraquark. The mass and coupling of the scalar tetraquark $\overline{%
Z}_{c}$ built of $[cd][\overline{c}\overline{u}]$ diquark-antidiquark were
computed in our paper \cite{Sundu:2018nxt}. There, we also explored decays
of $\overline{Z}_{c}$ and found its full width. The dominant strong decay of
the resonance $\overline{Z}_{c}$ is presumably the channel $Z_{c}\rightarrow
\eta _{c}\pi ^{-}$. But hidden-charm $\eta _{c}^{\prime }\pi ^{-}$, $J/\psi
\rho ^{-}$ and open-charm $D^{0}D^{-}$ and $D^{\ast 0}D^{\ast -}$ decays are
also kinematically allowed $S$-wave channels of the resonance $\overline{Z}%
_{c}$. Below we give detailed information about investigations of $\overline{%
Z}_{c}$ based on our work \cite{Sundu:2018nxt}.


\subsection{Mass and coupling of the scalar tetraquark $\overline{Z}_{c}$}

\label{sec:Mass4100}
The most stable and lower lying scalar tetraquark can be built of scalar
diquark $\epsilon ^{ijk}[c_{j}^{T}C\gamma _{5}d_{k}]$ and antidiquark $%
\epsilon ^{imn}[\overline{c}_{m}\gamma _{5}C\overline{u}_{n}^{T}]$ fields
\cite{Jaffe:2004ph}. These two-quark states are color-antitriplet and
-triplet configurations, respectively, and both are antisymmetric in flavor
indices.

For scalar particles the two-point correlation function $\Pi (p)$ has a
simple form and Lorentz structure: it is given by the following formula
\begin{equation}
\Pi (p)=i\int d^{4}xe^{ipx}\langle 0|\mathcal{T}\{J(x)J^{\dagger
}(0)\}|0\rangle .  \label{eq:CorrFZ}
\end{equation}%
In expression above, the interpolating current for the tetraquark $\overline{%
Z}_{c}$ is denoted by $J(x)$. In light of our suggestion about internal
organization of $\overline{Z}_{c}$, the current $J(x)$ can be written in the
form
\begin{equation}
J(x)=\epsilon \tilde{\epsilon}\left[ c_{j}^{T}(x)C\gamma _{5}d_{k}(x)\right] %
\left[ \overline{c}_{m}(x)\gamma _{5}C\overline{u}_{n}^{T}(x)\right] ,
\label{eq:CurrZ}
\end{equation}%
where $\epsilon =\epsilon ^{ijk}$ , $\widetilde{\epsilon }=\epsilon ^{imn}$.

The sum rules for parameters of the tetraquark $\overline{Z}_{c}$ can be
extracted using the "ground-state + continuum" scheme. First of all, we need
the phenomenological side of the sum rule $\Pi ^{\mathrm{Phys}}(p)$. For the
scalar particle relevant invariant amplitude $\Pi ^{\mathrm{Phys}}(p^{2})=$ $%
m^{2}f^{2}/(m^{2}-p^{2})$ is simple function of the mass $m$ and coupling $f$. At the next step, we have to determine the QCD side of the sum rules. In
our case, it is given by the formula
\begin{equation}
\Pi ^{\mathrm{OPE}}(p)=i\int d^{4}xe^{ipx}\epsilon \tilde{\epsilon}\epsilon
^{\prime }\tilde{\epsilon}^{\prime }\mathrm{Tr}\left[ \gamma _{5}\widetilde{S%
}_{c}^{jj^{\prime }}(x)\gamma _{5}S_{d}^{kk^{\prime }}(x)\right] \mathrm{Tr}%
\left[ \gamma _{5}\widetilde{S}_{u}^{n^{\prime }n}(-x)\gamma
_{5}S_{c}^{m^{\prime }m}(-x)\right] .  \label{eq:OPE4100}
\end{equation}%
For the mass $m$ and coupling $f$ of the tetraquark $\overline{Z}_{c}$ after
clear substitutions one can employ expressions (\ref{eq:MassGS}) and (\ref%
{eq:CoupGS}). The relevant computations are carried out by taking into
account nonperturbative terms up to dimension 10.

The sum rules for spectroscopic parameters of $\overline{Z}_{c}$ contain
various vacuum condensates, values of which have been presented in Eq.\ (\ref%
{eq:VCond}). The sum rules depend also on the Borel $M^{2}$ and continuum
threshold $s_{0}$ parameters: $M^{2}$ and $s_{0}$ are the auxiliary
parameters and should be fixed in accordance with standard restrictions of
the sum rule calculations. Thus, at the maximum of $M^{2}$ the pole
contribution (\ref{eq:PC}) should exceed some fixed value: as usual, for
four-quark systems minimum of $\mathrm{PC}$ is approximately $0.15-0.2$.

In the previous section, we have defined $\mathrm{PC}$ in terms of the
spectral density, but in a general form it can be introduced through the
ratio
\begin{equation}
\mathrm{PC}=\frac{\Pi (M^{2},s_{0})}{\Pi (M^{2},\infty )},  \label{eq:PCA}
\end{equation}%
where $\Pi (M^{2},s_{0})$ is the Borel transformed and subtracted invariant
amplitude $\Pi ^{\mathrm{OPE}}(p^{2})$. The minimum of the Borel parameter
is determined from convergence of the operator product expansion, and can be
extracted from analysis of the parameter
\begin{equation}
R(M^{2})=\frac{\Pi ^{\mathrm{DimN}}(M^{2},s_{0})}{\Pi (M^{2},s_{0})}.
\label{eq:Convergence}
\end{equation}%
Here, $\Pi ^{\mathrm{DimN}}(M^{2},s_{0})$ is a contribution of the last term
in expansion (or a sum of last few terms) to $\Pi (M^{2},s_{0})$. The
parameter $R(M^{2})$ should be small enough to guarantee a convergence of
sum rules.

The mass $m$ and coupling $f$ should not depend on the Borel parameter $%
M^{2} $. But analyses demonstrate that $m$ and $f$ are sensitive to the
choice of $M^{2}$. There are also dependence on the continuum threshold
parameter $s_{0} $, but $\sqrt{s_{0}}$ determines a position of the  first
excitation of $\overline{Z}_{c}$ and bears some information about a physical
system. Therefore, $M^{2}$ should be fixed in such a way that to minimize a
dependence of $m$ and $f$ on this parameter.

Computations demonstrate that regions for the parameters $M^{2}$ and $s_{0}$
\begin{equation}
M^{2}\in \lbrack 4,\ 6]~\mathrm{GeV}^{2},\ s_{0}\in \lbrack 19,\ 21]~\mathrm{%
GeV}^{2},  \label{eq:Wind4100}
\end{equation}%
satisfy all constraints of sum rule calculations. Indeed, at $M^{2}=6~%
\mathrm{GeV}^{2}$, we get $\mathrm{PC}=0.19$, and in the region $M^{2}\in
\lbrack 4,\ 6]~\mathrm{GeV}^{2}$ the pole contribution changes from $0.54$
till $0.19$. The low limit of the Borel parameter is fixed from Eq.\ (\ref%
{eq:Convergence}), in which we choose $\mathrm{DimN}=\mathrm{Dim(8+9+10)}$.
Then at $M^{2}=4~\mathrm{GeV}^{2}$ the parameter $R$ becomes equal to $R(4~%
\mathrm{GeV}^{2})=0.02$ which guarantees the convergence of the sum rules.
At $M^{2}=4~\mathrm{GeV}^{2}$ the perturbative contribution amounts to $83\%$
of the full result overshooting nonperturbative terms.

For the mass and coupling of the tetraquark $\overline{Z}_{c}$ our
calculations yield
\begin{equation}
m=(4080~\pm 150)~\mathrm{MeV},\ f=(0.58\pm 0.12)\times 10^{-2}~\mathrm{GeV}%
^{4}.  \label{eq:CMass4100}
\end{equation}%
One can see, that the mass of the scalar diquark-antidiquark state $%
\overline{Z}_{c}$ is in excellent agreement with LHCb data.

The scalar tetraquark $[cu][\overline{c}\overline{d}]$ with the internal
organization $C\gamma _{5}\otimes \gamma _{5}C$ was investigated in Ref.\
\cite{Wang:2017lbl} as well. Using the mass $m=(3860\pm 90)~\mathrm{MeV}$ of
this exotic state, the author interpreted it as a charged partner of the
resonance $X^{\ast }(3860)$. The charmoniumlike state $X^{\ast }(3860)$ was
seen by Belle \cite{Chilikin:2017evr} in the process $e^{+}e^{-}\rightarrow
J/\psi D\overline{D}$, where $D$ is one $D^{0}$ or $D^{+}$ mesons, and
identified there with $\chi _{c0}(2P)$ meson. Comparing our result and
prediction of Ref.\ \cite{Wang:2017lbl}, we find an overlapping region, but
a difference $200\ ~\mathrm{MeV}$ between the central values of the masses
is sizable. This difference probably stems from working windows for the
parameters $M^{2}$ and $s_{0}$ used in computations, and also may be
explained by fixed or evolved treatment of vacuum condensates.


\subsection{Decays $\overline{Z}_{c}\rightarrow \protect\eta _{c}\protect\pi %
^{-}$ and $\overline{Z}_{c}\rightarrow \protect\eta _{c}^{\prime }\protect%
\pi ^{-}$}

\label{sec:DAZ4100}

The strong decays of the resonance $\overline{Z}_{c}$ form two groups of
processes: the first of them contains decays with two pseudoscalar mesons in a
final state, whereas the second group embraces decays to two vector mesons.
The decays $\overline{Z}_{c}\rightarrow \eta _{c}\pi ^{-}$ and $\overline{Z}%
_{c}\rightarrow \eta _{c}^{\prime }\pi ^{-}$ are from the first group of
processes. The final phases of these processes are characterized by
appearance of mesons $\eta _{c}$ and $\eta _{c}^{\prime }$, where the latter
is a first radially excited state of the former one. In the QCD sum rule
method such decays are explored in a correlated way. A suitable approach to
analyze the decays $\overline{Z}_{c}\rightarrow \eta _{c}\pi ^{-}$ and $%
\overline{Z}_{c}\rightarrow \eta _{c}^{\prime }\pi ^{-}$ is the QCD
three-point sum rule method. The reason is that, this method allows one to
get for the physical side of sum rules relatively simple expression. In
fact, we are interested in extraction of sum rules for strong form factors $%
g_{\overline{Z}_{c}\eta _{ci}\pi }(q^{2})$, therefore in the context of
standard operations should apply double Borel transformation over the
momenta of particles $\overline{Z}_{c}$ and $\eta _{c}$. The Borel
transformation applied to physical side of the three-point sum rules
suppresses contributions of higher resonances in these two channels, and
eliminate contributions of pole-continuum transitions \cite%
{Belyaev:1994zk,Ioffe:1983ju}. The elimination of such terms is important
for joint treatment of the form factors $g_{\overline{Z}_{c}\eta _{ci}\pi
}(q^{2})$, because there is not a necessity to employ additional operators
to remove contaminations from the phenomenological side. Nevertheless, in
the pion channel still may survive contaminating terms corresponding to
excited states of the pion [for the $NN\pi $ vertex, see discussions in
Refs.\ \cite{Meissner:1995ra,Maltman:1997jb}]. To decrease ambiguities in
extracting of the strong couplings at the vertices, it is possible to choose
the pion on the mass shell, and consider one of remaining states ($\overline{%
Z}_{c}$ or $\eta _{c}$) as an off-shell particle. This method was employed
to investigate couplings of ordinary heavy-heavy-light mesons in Refs. \cite%
{Bracco:2006xf,Cerqueira:2015vva}. Form factors extracted by treating a
light or one of heavy mesons off-shell may differ from each other
considerably, but after extrapolating to the corresponding mass-shells give
the same or negligibly different strong couplings.

The process $\overline{Z}_{c}\rightarrow $ $J/\psi \rho ^{-}$ belongs to the
second group of $\overline{Z}_{c}$ decays. We explore this decay using the
LCSR method and soft-meson approximation. The LCSR method allows us to
determine the strong coupling by evading extrapolating prescriptions and
express $g_{\overline{Z}_{c}J/\psi \rho }$ in terms of the vacuum
condensates and matrix elements of the $\rho $ meson. The pole-continuum
contributions surviving after a single Borel transformation in the physical
side of sum rules, can be removed by employing well-known procedures \cite%
{Ioffe:1983ju}.

The strong couplings $g_{\overline{Z}_{c}\eta _{c1}\pi }$ and $g_{\overline{Z%
}_{c}\eta _{c2}\pi }$ can \ be found from analysis of the three-point
correlation function%
\begin{equation}
\Pi (p,p^{\prime })=i^{2}\int d^{4}xd^{4}ye^{-ipx}e^{ip^{\prime }y}\langle 0|%
\mathcal{T}\{J^{\eta _{c}}(y)J^{\pi }(0)J^{\dagger }(x)\}|0\rangle ,
\label{eq:CorrF1Z4100}
\end{equation}%
where $J^{\eta _{c}}(y)$ is the interpolating current for $\eta _{c}$ and $%
\eta _{c}^{\prime }$ mesons (\ref{eq:EtaCurr}), and $J^{\pi }(0)$ is the
interpolating current for the pion
\begin{equation}
\ J^{\pi }(x)=\overline{u}_{b}(x)i\gamma _{5}d_{b}(x)  \label{eq:Curr1Z4100}
\end{equation}%
at $x=0$, respectively.

The correlation function $\Pi (p,p^{\prime })$ in terms of the physical
parameters of involved particles has the form
\begin{equation}
\Pi ^{\mathrm{Phys}}(p,p^{\prime })=\sum_{i=1}^{2}\frac{\langle 0|J^{\eta
_{c}}|\eta _{ci}\left( p^{\prime }\right) \rangle }{p^{\prime 2}-m_{i}^{2}}%
\frac{\langle 0|J^{\pi }|\pi \left( q\right) \rangle }{q^{2}-m_{\pi }^{2}}%
\frac{\langle \eta _{ci}\left( p^{\prime }\right) \pi (q)|\overline{Z}%
_{c}(p)\rangle \langle \overline{Z}_{c}(p)|J^{\dagger }|0\rangle }{%
p^{2}-m^{2}}+\ldots ,  \label{eq:PhysD1}
\end{equation}%
where $m_{\pi }$ is the mass of the pion, and $m_{1}\equiv m_{\eta _{c}}$, $%
m_{2}=m_{\eta _{c}^{\prime }}$ are the masses of the mesons $\eta _{c}$ and $%
\eta _{c}^{\prime }$, respectively. Their decay constants are denoted by $%
f_{1}\equiv f_{\eta _{c}}$ and $f_{2}\equiv f_{\eta _{c}^{\prime }}$ and
together with $m_{1}$ and $m_{2}$ determine the matrix elements $\langle
0|J^{\eta _{c}}|\eta _{ci}\left( p^{\prime }\right) \rangle $ [see, Eq. (\ref%
{eq:EtaCME})]. The matrix element of the pion is also well known (\ref%
{eq:MatE2}). In addition to this information the matrix elements of the
vertices $\overline{Z}_{c}\eta _{c}\pi ^{-}$ and $\overline{Z}_{c}\eta
_{c}^{\prime }\pi ^{-}$ are required as well. For these purposes, we use
\begin{equation}
\langle \eta _{ci}\left( p^{\prime }\right) \pi (q)|\overline{Z}%
_{c}(p)\rangle =g_{\overline{Z}_{c}\eta _{ci}\pi }(p\cdot p^{\prime }).
\label{eq:ME3Z4100}
\end{equation}%
Here, the strong coupling $g_{\overline{Z}_{c}\eta _{c1}\pi }$ corresponds
to the vertex $\overline{Z}_{c}\eta _{c}\pi ^{-}$, whereas $g_{\overline{Z}%
_{c}\eta _{c2}\pi }$ describes $\overline{Z}_{c}\eta _{c}^{\prime }\pi ^{-}$.

After some manipulations for $\Pi ^{\mathrm{Phys}}(p,p^{\prime })$ we find
the following expression
\begin{equation}
\Pi ^{\mathrm{Phys}}(p,p^{\prime })=\sum_{i=1}^{2}\frac{g_{\overline{Z}%
_{c}\eta _{ci}\pi }m_{i}^{2}f_{i}mf}{2m_{c}(p^{\prime 2}-m_{i}^{2})\left(
p^{2}-m^{2}\right) }\frac{\mu _{\pi }f_{\pi }}{q^{2}-m_{\pi }^{2}}(p\cdot
p^{\prime })+\dots  \label{eq:Phys4}
\end{equation}%
The $\Pi ^{\mathrm{Phys}}(p,p^{\prime })$ has a simple Lorentz structure,
hence the invariant amplitude $\Pi ^{\mathrm{Phys}}(p^{2},p^{\prime 2})$ is
equal to the sum of two terms in Eq.\ (\ref{eq:Phys4}). The double Borel
transformation of $\Pi ^{\mathrm{Phys}}(p^{2},p^{\prime 2})$ over $p^{2}$
and $p^{\prime 2}$ with the parameters $M_{1}^{2}$ and $M_{2}^{2}$,
respectively, constitutes a physical side in a sum rule equality.

The correlation function calculated in terms of the quark propagators is:
\begin{equation}
\Pi ^{\mathrm{OPE}}(p,p^{\prime })=i^{2}\int
d^{4}xd^{4}ye^{-ipx}e^{ip^{\prime }y}\epsilon \widetilde{\epsilon }\mathrm{Tr%
}\left[ \gamma _{5}S_{c}^{aj}(y-x)\gamma _{5}\widetilde{S}%
_{d}^{bk}(-x)\gamma _{5}\widetilde{S}_{u}^{nb}(x)\gamma _{5}S_{c}^{ma}(x-y)%
\right].  \label{eq:OPEDec}
\end{equation}%
The Borel transformation $\mathcal{B}\Pi ^{\mathrm{OPE}}(p^{2},p^{\prime 2})$
of the amplitude $\Pi ^{\mathrm{OPE}}(p^{2},p^{\prime 2})$ forms the QCD
side of the sum rules.\ The first sum rule for $g_{\overline{Z}_{c}\eta _{c1}\pi }$ and $g_{\overline{Z}_{c}\eta
_{c2}\pi }$ is obtained by equating Borel transformations of  amplitudes  $\Pi ^{\mathrm{Phys}}(p^{2},p^{\prime
2}) $ and $\Pi ^{\mathrm{OPE}}(p^{2},p^{\prime 2})$ and performing the  continuum subtractions.

The Borel transformed and subtracted amplitude $\Pi ^{\mathrm{OPE}%
}(p^{2},p^{\prime 2})$ can be expressed using the spectral density $\rho _{%
\mathrm{D}}(s,s^{\prime },q^{2})$ which is determined as an imaginary part
of the correlation function $\Pi ^{\mathrm{OPE}}(p,p^{\prime })$
\begin{equation}
\Pi (\mathbf{M}^{2},\mathbf{s}_{0},q^{2})=\int_{4m_{c}^{2}}^{s_{0}}ds%
\int_{4m_{c}^{2}}^{s_{0}^{\prime }}ds^{\prime }\rho _{\mathrm{D}%
}(s,s^{\prime },q^{2})e^{-s/M_{1}^{2}}e^{-s^{\prime }/M_{2}^{2}},
\label{eq:SCoupl}
\end{equation}%
where $\mathbf{M}^{2}=(M_{1}^{2},M_{2}^{2})$ and $\mathbf{s}%
_{0}=(s_{0},s_{0}^{\prime })$ are the Borel and continuum threshold
parameters, respectively.

The second sum rule for the couplings $g_{\overline{Z}_{c}\eta _{c1}\pi }$
and $g_{\overline{Z}_{c}\eta _{c2}\pi }$ can be obtained by acting operators
$d/d(-1/M_{1}^{2})$ and/or $d/d(-1/M_{2}^{2})$ on the first expression.
These two expressions are enough to find $g_{\overline{Z}_{c}\eta _{c1}\pi }$
and $g_{\overline{Z}_{c}\eta _{c2}\pi }$. An alternative way is the master
sum rule used repeatedly to evaluate the couplings $g_{\overline{Z}_{c}\eta
_{c1}\pi }$ and $g_{\overline{Z}_{c}\eta _{c2}\pi }$. For these purposes, we
choose the continuum threshold parameter $\sqrt{s_{0}^{\prime }}$ that
corresponds to the $\eta _{c}$ channel below the mass of the radially
excited state $\eta _{c}^{\prime }$. In other words, we include $\eta
_{c}^{\prime }$ into high resonances and get sum rule for the coupling of
the ground-state meson $\eta _{c}$. At this phase of computations the
physical side of the sum rule \ (\ref{eq:Phys4}) depends only on the
coupling $g_{\overline{Z}_{c}\eta _{c1}\pi }$. This sum rule can be solved
to find the coupling $g_{Z_{c}\eta _{c1}\pi }$
\begin{equation}
g_{\overline{Z}_{c}\eta _{c1}\pi }(\mathbf{M}^{2},\mathbf{s}%
_{0}^{(1)},q^{2})=\frac{\Pi (\mathbf{M}^{2},\mathbf{s}%
_{0}^{(1)},q^{2})e^{m/M_{1}^{2}}e^{m_{1}^{2}/M_{2}^{2}}}{A_{1}},
\label{eq:StC1}
\end{equation}%
where
\begin{equation*}
A_{1}=\frac{mfm_{1}^{2}f_{1}\mu _{\pi }f_{\pi }}{4m_{c}(q^{2}-m_{\pi }^{2})}%
\left( m^{2}+m_{1}^{2}-q^{2}\right) ,
\end{equation*}%
and $\mathbf{\ s}_{0}^{(1)}=(s_{0},\ s_{0}^{\prime }\simeq m_{2}^{2}).$

At the next stage, we move the continuum threshold $\sqrt{s_{0}^{\prime }}$
to $m_{2}+(0.5-0.8)~\mathrm{GeV}$ and employ the sum rule which now includes
the ground-state meson $\eta _{c}$ and its first radial excitation $\eta
_{c}^{\prime }$. The QCD side of this sum rule is determined by $\Pi (%
\mathbf{M}^{2},\mathbf{s}_{0}^{(2)},~q^{2})$, where $\mathbf{s}%
_{0}^{(2)}=(s_{0},\ s_{0}^{\prime }\simeq \lbrack m_{2}+(0.5-0.8)]^{2})$. By
substituting the obtained expression for $g_{Z_{c}\eta _{c1}\pi }$ into this
sum rule, it is not difficult to evaluate the second coupling $g_{\overline{Z%
}_{c}\eta _{c2}\pi }$.

The couplings extracted by this manner, as usual, depend on the Borel and
continuum threshold parameters, but are functions of $q^{2}$ as well. For
simplicity of presentation, below we skip their dependence on the
parameters, and denote strong couplings obtained by substitution $%
q^{2}=-Q^{2}$ as $g_{\overline{Z}_{c}\eta _{c1}\pi }(Q^{2})$ and $g_{%
\overline{Z}_{c}\eta _{c2}\pi }(Q^{2})$. The widths of the decays under
analysis depend on values of the couplings at the pion's mass shall $%
q^{2}=m_{_{\pi }}^{2}$. This region is not accessible to sum rule
computations. The way out of this situation is to introduce extrapolating
functions $F_{1(2)}(Q^{2})$ which at $Q^{2}>0$ coincide with the sum rule's
predictions, but can be easily used in the region $Q^{2}<0$ as well.

The strong couplings depend on the masses and decay constants of the
final-state mesons, which are shown in Table \ref{tab:Param}. To perform
numerical computations the Borel $\mathbf{M}^{2}$ and continuum threshold $%
\mathbf{s}_{0}$ parameters have to be specified as well. The parameters $%
M_{2}^{2},\ s_{0}^{\prime }$ in Eq.\ (\ref{eq:StC1}) are chosen as
\begin{equation}
M_{2}^{2}\in \lbrack 3,\ 4]~\mathrm{GeV}^{2},\ s_{0}^{\prime }=13~\mathrm{GeV%
}^{2},  \label{eq:Wind2Z4100}
\end{equation}%
whereas in the sum rule for the second coupling $g_{\overline{Z}_{c}\eta
_{c2}\pi }(Q^{2})$,  we employ
\begin{equation}
M_{2}^{2}\in \lbrack 3,4]~\mathrm{GeV}^{2},\ s_{0}^{\prime }\in \lbrack
17,19]~\mathrm{GeV}^{2}.  \label{eq:Wind3Z4100}
\end{equation}

We have noted above, that at the pion mass-shell $Q^{2}=-m_{\pi }^{2}$ the
couplings can be evaluated using fit functions. For these purposes, we use
exponential-type functions
\begin{equation}
F_{i}(Q^{2})=F_{0}^{i}\mathrm{\exp }\left[ c_{1}^{i}\frac{Q^{2}}{m^{2}}%
+c_{2}^{i}\left( \frac{Q^{2}}{m^{2}}\right) ^{2}\right] ,  \label{eq:FitF}
\end{equation}%
where $F_{0}^{i}$, $c_{1}^{i}$ and $c_{2}^{i}$ are free parameters. Our
analysis allows us to fix these parameters: we get $F_{0}^{1}=0.49~\mathrm{%
GeV}^{-1}$, $c_{1}^{1}=27.64$ and $c_{1}^{2}=-34.66$. Another set reads $%
F_{0}^{2}=0.39~\mathrm{GeV}^{-1}$, $c_{2}^{1}=28.13$ and $c_{2}^{2}=-35.24$.

The strong couplings at the mass-shell are equal to
\begin{equation}
g_{\overline{Z}_{c}\eta _{c1}\pi }(-m_{\pi }^{2})=(0.47\pm 0.06)~\mathrm{GeV}%
^{-1},\ \ g_{\overline{Z}_{c}\eta _{c2}\pi }(-m_{\pi }^{2})=(0.38\pm 0.05)~%
\mathrm{GeV}^{-1}.  \label{eq:MassSC}
\end{equation}%
The widths of the decays $Z_{c}\rightarrow \eta _{c}\pi ^{-}$ and $%
Z_{c}\rightarrow \eta _{c}^{\prime }\pi ^{-}$ can be evaluated by employing
of the formula%
\begin{equation}
\Gamma \left[ \overline{Z}_{c}\rightarrow \eta _{c}(\mathrm{I}S)\pi ^{-}%
\right] =\frac{g_{\overline{Z}_{c}\eta _{ci}\pi }^{2}m_{i}^{2}}{8\pi }%
\lambda \left( m,m_{i},m_{\pi }\right) \left[ 1+\frac{\lambda ^{2}\left(
m,m_{i},m_{\pi }\right) }{m_{i}^{2}}\right],  \label{eq:DW1a}
\end{equation}%
where $\mathrm{I}\equiv i=1,2$. For the decay $\overline{Z}_{c}\rightarrow
\eta _{c}\pi ^{-}$ one has to set $g_{\overline{Z}_{c}\eta _{ci}\pi
}\rightarrow g_{\overline{Z}_{c}\eta _{c1}\pi }$ and $m_{i}\rightarrow m_{1}$%
, whereas in the case of $\overline{Z}_{c}\rightarrow \eta _{c}^{\prime }\pi
^{-}$ quantities with subscript $2$ have to be used.

Computations lead to the following predictions for the partial widths of the
decay channels%
\begin{equation}
\Gamma \left[ \overline{Z}_{c}\rightarrow \eta _{c}\pi ^{-}\right] =(81\pm
17)~\mathrm{MeV},\ \Gamma \left[ \overline{Z}_{c}\rightarrow \eta
_{c}^{\prime }\pi ^{-}\right] =(32\pm 7)~\mathrm{MeV}.  \label{eq:DW1Numeric}
\end{equation}


\subsection{Decay $\overline{Z}_{c}\rightarrow D^{0}D^{-}$}

\label{sec:Decay1a}
In this subsection we analyze $S$-wave decay of $\overline{Z}_{c}$ to a pair
of open-charm pseudoscalar mesons $\overline{Z}_{c}\rightarrow D^{0}D^{-}$.
The relevant three-point correlation function is given by the expression
\begin{equation}
\widetilde{\Pi }(p,p^{\prime })=i^{2}\int d^{4}xd^{4}ye^{-ipx}e^{ip^{\prime
}y}\langle 0|\mathcal{T}\{J^{D}(y)J^{D^{0}}(0)J^{\dagger }(x)\}|0\rangle,
\label{eq:DCF1}
\end{equation}%
where we introduce the interpolating currents for the pseudoscalar mesons $%
D^{-}$ and $D^{0}$
\begin{equation}
\ J^{D}(y)=\overline{c}_{r}(y)i\gamma _{5}d_{r}(y),\ J^{D^{0}}(0)=\overline{u%
}_{s}(0)i\gamma _{5}c_{s}(0).  \label{eq:Dcurrs}
\end{equation}

The correlation function $\Pi (p,p^{\prime })$ written down using physical
parameters of these mesons and tetraquark $\overline{Z}_{c}$ takes the form
\begin{equation}
\widetilde{\Pi }^{\mathrm{Phys}}(p,p^{\prime })=\frac{\langle
0|J^{D}|D^{-}\left( p^{\prime }\right) \rangle }{p^{\prime 2}-m_{D}^{2}}%
\frac{\langle 0|J^{D^{0}}|D^{0}\left( q\right) \rangle }{q^{2}-m_{D^{0}}^{2}}%
\frac{\langle D^{-}\left( p^{\prime }\right) D^{0}(q)|\overline{Z}%
_{c}(p)\rangle \langle \overline{Z}_{c}(p)|J^{\dagger }|0\rangle }{%
p^{2}-m^{2}}+\cdots ,  \label{eq:DCF2}
\end{equation}%
where $m_{D}$ and $m_{D^{0}}$ are masses of the mesons $D^{-}$ and $D^{0}$,
respectively.

We continue analysis by using the matrix elements
\begin{equation}
\langle 0|J^{D}|D^{-}\left( p^{\prime }\right) \rangle =\frac{f_{D}m_{D}^{2}%
}{m_{c}},\ \langle 0|J^{D^{0}}|D^{0}\left( q\right) \rangle =\frac{%
f_{D^{0}}m_{D^{0}}^{2}}{m_{c}},\ \langle D^{-}\left( p^{\prime }\right)
D^{0}(q)|\overline{Z}_{c}(p)\rangle =g_{\overline{Z}_{c}DD}(p\cdot p^{\prime
}).  \label{eq:DME1}
\end{equation}%
Simple manipulations lead to%
\begin{equation}
\widetilde{\Pi }^{\mathrm{Phys}}(p,p^{\prime })=\frac{%
f_{D^{0}}m_{D^{0}}^{2}f_{D}m_{D}^{2}}{m_{c}^{2}\left( p^{\prime
2}-m_{D}^{2}\right) \left( q^{2}-m_{D^{0}}^{2}\right) }\frac{mf}{p^{2}-m^{2}}%
(p\cdot p^{\prime })+\cdots .  \label{eq:DCF3}
\end{equation}%
The same correlation function written down in terms of the quark propagators
is%
\begin{equation}
\widetilde{\Pi }^{\mathrm{OPE}}(p,p^{\prime })=i^{2}\int
d^{4}xd^{4}ye^{-ipx}e^{ip^{\prime }y}\epsilon \widetilde{\epsilon }\mathrm{Tr%
}\left[ \gamma _{5}S_{d}^{rk}(y-x)\gamma _{5}\widetilde{S}%
_{c}^{sj}(-x)\gamma _{5}\widetilde{S}_{u}^{ns}(x)\gamma _{5}S_{c}^{mr}(x-y)%
\right] .  \label{eq:DCF4}
\end{equation}

The sum rule for the strong coupling $g_{\overline{Z}_{c}DD}$ can be
expressed in a traditional form
\begin{equation}
g_{\overline{Z}_{c}DD}(\mathbf{M}^{2},\mathbf{s}_{0},q^{2})=\frac{\widetilde{%
\Pi }(\mathbf{M}^{2},\mathbf{s}%
_{0},q^{2})e^{m/M_{1}^{2}}e^{m_{D}^{2}/M_{2}^{2}}}{B},  \label{eq:DSR}
\end{equation}%
where
\begin{equation*}
B=\frac{f_{D}m_{D}^{2}f_{D^{0}}m_{D^{0}}^{2}mf}{%
2m_{c}^{2}(q^{2}-m_{D^{0}}^{2})}\left( m^{2}+m_{D}^{2}-q^{2}\right) .
\end{equation*}%
Here, $\widetilde{\Pi }(\mathbf{M}^{2},\mathbf{s}_{0},q^{2})$ is the
amplitude $\widetilde{\Pi }^{\mathrm{OPE}}(p^{2},p^{\prime 2},q^{2})$ after
Borel transformation and subtraction procedures: it is expressible in term
of the spectral density $\widetilde{\rho }_{\mathrm{D}}(s,s^{\prime },q^{2})
$
\begin{equation}
\widetilde{\Pi }(\mathbf{M}^{2},\mathbf{s}_{0},q^{2})=%
\int_{4m_{c}^{2}}^{s_{0}}ds\int_{m_{c}^{2}}^{s_{0}^{\prime }}ds^{\prime }%
\widetilde{\rho }_{\mathrm{D}}(s,s^{\prime
},q^{2})e^{-s/M_{1}^{2}}e^{-s^{\prime }/M_{2}^{2}}.  \label{eq:DCF5}
\end{equation}

The sum rule for $g_{\overline{Z}_{c}DD}$ depends on masses and decay
constants of the mesons $D^{0}$ and $D^{-}$: for these parameters we utilize
$m_{D^{0}}=(1864.83\pm 0.05)~\mathrm{MeV}$,\ $m_{D}=(1869.65\pm 0.05)~%
\mathrm{MeV}$ and $f_{D}=f_{D^{0}}=(211.9\pm 1.1)~\mathrm{MeV}$,
respectively. Restrictions on parameters $\mathbf{M}^{2}$ and $\mathbf{s}%
_{0} $ do not differ from ones considered above and are universal for such
kind of computations. The $M_{1}^{2}$ and $s_{0}$ are varied within limits
determined in the mass calculations (\ref{eq:Wind4100}). The parameters $%
M_{2}^{2},\ s_{0}^{\prime }$ in Eq.\ (\ref{eq:DCF5}) are
\begin{equation}
M_{2}^{2}\in \lbrack 3,\ 6]~\mathrm{GeV}^{2},\ s_{0}^{\prime }\in \lbrack
7,\ 9]~\mathrm{GeV}^{2}.  \label{eq:Wind4}
\end{equation}%
Numerical computations of Eq.\ (\ref{eq:DCF5}) with regions (\ref%
{eq:Wind4100}) lead to stable results for the form factor $g_{\overline{Z}%
_{c}DD}(\mathbf{M}^{2},\mathbf{s}_{0},q^{2})$ at $q^{2}<0$. In what follows,
we denote it $g_{\overline{Z}_{c}DD}(Q^{2})$ by introducing $q^{2}=-Q^{2}$
and omit parameters $\mathbf{M}^{2}$ and $\mathbf{s}_{0}$.

The width of the decay $\overline{Z}_{c}\rightarrow D^{0}D^{-}$ depends on
the strong coupling $g_{\overline{Z}_{c}DD}$ at the mass shell of the meson $%
D^{0}$. Therefore,we utilize the fit function $\widetilde{F}(Q^{2})$ from
Eq.\ (\ref{eq:FitF}) with parameters $\widetilde{F}_{0}=0.44~\mathrm{GeV}%
^{-1}$, $\widetilde{c}_{1}=2.38$ and $\widetilde{c}_{2}=-1.61$. In Fig.\ \ref%
{fig:Fit} we depict $\widetilde{F}(Q^{2})$ and sum rule predictions for $g_{%
\overline{Z}_{c}DD}(Q^{2})$ demonstrating  very nice agreement between them.

The strong coupling at the mass shell $Q^{2}=-m_{D^{0}}^{2}$ is
\begin{equation}
g_{\overline{Z}_{c}DD}(-m_{D^{0}}^{2})=(0.25\pm 0.05)~\mathrm{GeV}^{-1}.
\label{eq:Coupl1}
\end{equation}%
The width of the decay $\overline{Z}_{c}\rightarrow D^{0}D^{-}$ is
calculated employing Eq.\ (\ref{eq:DW1a}) with necessary replacements, and
by taking into account that $\lambda \Rightarrow \lambda \left(
m,m_{D^{0}},m_{D}\right) $.

The partial width of this decay reads%
\begin{equation}
\Gamma \lbrack \overline{Z}_{c}\rightarrow D^{0}D^{-}]=(19\pm 5)~\mathrm{MeV}%
.  \label{eq:Width1}
\end{equation}%
This result will be employed to evaluate the full width of the tetraquark $%
\overline{Z}_{c}$.
\begin{figure}[h]
\includegraphics[width=8.5cm]{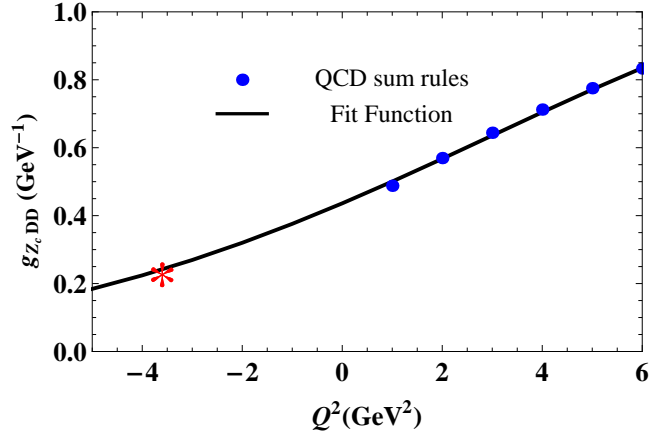}
\caption{The sum rule predictions and fit function for the strong coupling $%
g_{\overline{Z}_{c}DD}(Q^{2})$. The star marks the point $%
Q^{2}=-m_{D^{0}}^{2}$. }
\label{fig:Fit}
\end{figure}

\subsection{Decay $\overline{Z}_{c}\rightarrow J/\protect\psi \protect\rho %
^{-}$}

\label{sec:Decay2}
The scalar tetraquark $\overline{Z}_{c}$ can decay to a pair of two vector
mesons $J/\psi \rho ^{-}$. In the context of the LCSR method this decay can
be studied be means of the correlation function%
\begin{equation}
\Pi _{\mu }(p,q)=i\int d^{4}xe^{ipx}\langle \rho (q)|\mathcal{T}\{J_{\mu
}^{\psi }(x)J^{\dagger }(0)\}|0\rangle ,  \label{eq:CorrF2Z4100}
\end{equation}%
where the interpolating current for the vector meson $J/\psi $ is denoted by
$J_{\mu }^{\psi }(x)$.

The correlation function $\Pi _{\mu }^{\mathrm{Phys}}(p,q)$ in terms of the
physical parameters of the tetraquark $\overline{Z}_{c}$, and mesons $J/\psi
$ and $\rho $ is equal to
\begin{equation}
\Pi _{\mu }^{\mathrm{Phys}}(p,q)=g_{Z_{c}J/\psi \rho }\frac{m_{J/\psi
}f_{J/\psi }}{p^{2}-m_{J/\psi }^{2}}\frac{mf}{p^{\prime 2}-m^{2}}\left[
\frac{1}{2}\left( m^{2}-m_{J/\psi }^{2}-q^{2}\right) \varepsilon _{\mu
}^{\prime }-p\cdot \varepsilon ^{\prime }q_{\mu }\right] +\ldots
\label{eq:phys2A}
\end{equation}%
It contains Lorentz structures proportional to $\varepsilon _{\mu }^{\prime
} $ and $q_{\mu }$. We work with the structure $\sim \varepsilon _{\mu
}^{\prime }$ and label the corresponding invariant amplitude by $\Pi ^{%
\mathrm{Phys}}(p^{2},q^{2})$.

The second ingredient of the sum rule is the same correlation function $\Pi
_{\mu }^{\mathrm{OPE}}(p,q)$ expressed in terms of quark propagators
\begin{equation}
\Pi _{\mu }^{\mathrm{OPE}}(p,q)=i^{2}\int d^{4}xe^{ipx}\epsilon \widetilde{%
\epsilon }\left[ \gamma _{5}\widetilde{S}_{c}^{aj}(x){}\gamma _{\mu }%
\widetilde{S}_{c}^{ma}(-x){}\gamma _{5}\right] _{\alpha \beta }\langle \rho
(q)|\overline{d}_{\alpha }^{k}(0)u_{\beta }^{n}(0)|0\rangle .
\label{eq:CorrF3Z4100}
\end{equation}%
The $\Pi _{\mu }^{\mathrm{OPE}}(p,q)$ contains two- and three-particle local
matrix elements of the $\rho $-meson. Two of these elements (\ref{eq:Melem1}%
) and (\ref{eq:Melem2}) does not depend on the $\rho $ meson momentum,
whereas others are determined using momentum factors
\begin{eqnarray}
&&\langle 0|\overline{u}\sigma _{\mu \nu }d|\rho (q,\lambda )\rangle
=if_{\rho }^{T}(\epsilon _{\mu }^{(\lambda )}q_{\nu }-\epsilon _{\nu
}^{(\lambda )}q_{\mu }),\ \langle 0|\overline{u}gG_{\mu \nu }d|\rho
(q,\lambda )\rangle =if_{\rho }^{T}m_{\rho }^{3}\zeta _{4}^{T}(\epsilon
_{\mu }^{(\lambda )}q_{\nu }-\epsilon _{\nu }^{(\lambda )}q_{\mu }),  \notag
\\
&&\langle 0|\overline{u}g\widetilde{G}_{\mu \nu }i\gamma _{5}d|\rho
(q,\lambda )\rangle =if_{\rho }^{T}m_{\rho }^{3}\widetilde{\zeta }%
_{4}^{T}(\epsilon _{\mu }^{(\lambda )}q_{\nu }-\epsilon _{\nu }^{(\lambda
)}q_{\mu }).  \label{eq:HTME}
\end{eqnarray}

By substituting these matrix elements into the correlation function (\ref%
{eq:CorrF3Z4100}), carrying out the summation over color and calculating
traces over Lorentz indices, we find local matrix elements of the $\rho $
meson that contribute to $\Pi _{\mu }^{\mathrm{OPE}}(p,q)$. It appears in
the soft limit $q\rightarrow 0$ contributions to the invariant amplitude $%
\Pi ^{\mathrm{OPE}}(p^{2})$ come from the matrix elements (\ref{eq:Melem1})
and (\ref{eq:Melem2}).

The Borel transformation of the amplitude $\Pi ^{\mathrm{OPE}}(p^{2})$ is
given by the formula%
\begin{equation}
\Pi ^{\mathrm{OPE}}(M^{2})=\int_{4m_{c}^{2}}^{\infty }ds\widetilde{\rho }^{%
\mathrm{OPE}}(s)e^{-s/M^{2}}+\Pi ^{\mathrm{OPE(tw4)}}(M^{2}),
\label{eq:OPE2}
\end{equation}%
with $\widetilde{\rho }^{\mathrm{OPE}}(s)$ and $\Pi ^{\mathrm{OPE(tw4)}%
}(M^{2})$ being the spectral density and twist-4 contribution to $\Pi ^{%
\mathrm{OPE}}(M^{2})$, respectively. Computation of $\widetilde{\rho }^{%
\mathrm{OPE}}(s)$ has been performed by taking into account condensates up
to dimension six. The spectral density consists of the perturbative and
nonperturbative components
\begin{equation}
\widetilde{\rho }^{\mathrm{OPE}}(s)=\frac{f_{\rho }m_{\rho }(s+2m_{c}^{2})%
\sqrt{s(s-4m_{c}^{2})}}{24\pi ^{2}s}+\rho ^{\mathrm{n.-pert.}}(s).
\end{equation}%
The nonperturbative part of the spectral density $\rho ^{\mathrm{n.-pert.}%
}(s)$ contains terms proportional to gluon condensates $\langle \alpha
_{s}G^{2}/\pi \rangle $, $\langle \alpha _{s}G^{2}/\pi \rangle ^{2}$ and $%
\langle g_{s}^{3}G^{3}\rangle $: Here, we do not write down their
expressions explicitly. The twist-4 term in Eq.\ (\ref{eq:OPE2}) is equal to%
\begin{equation}
\Pi ^{\mathrm{OPE(tw4)}}(M^{2})=\frac{f_{\rho }m_{\rho }^{3}\zeta _{4\rho
}m_{c}^{2}}{8\pi }\int_{0}^{1}d\alpha \frac{e^{-m_{c}^{2}/M^{2}a(1-a)}}{%
a(1-a)}.  \label{eq:TW4}
\end{equation}

The sum rule for the strong coupling is given by the formula%
\begin{equation}
g_{\overline{Z}_{c}J/\psi \rho }=\frac{2}{m_{J/\psi }f_{J/\psi
}mf(m^{2}-m_{J/\psi }^{2})}\mathcal{P}(M^{2},\widetilde{m}^{2})\Pi ^{\mathrm{%
OPE}}(M^{2},s_{0}).  \label{eq:SR}
\end{equation}%
where $\mathcal{P}(M^{2},\widetilde{m}^{2})$ is the operator in Eq.\ (%
\ref{eq:softop}), and $\widetilde{m}^{2}=(m^{2}+m_{J/\psi }^{2})/2$. The
width of the decay $\overline{Z}_{c}\rightarrow J/\psi \rho ^{-}$ is
determined by the expression%
\begin{equation}
\Gamma \left( \overline{Z}_{c}\rightarrow J/\psi \rho ^{-}\right) =\frac{g_{%
\overline{Z}_{c}J/\psi \rho }^{2}m_{\rho }^{2}}{8\pi }\lambda \left( m,\
m_{J/\psi },m_{\rho }\right) \left[ 3+\frac{2\lambda ^{2}\left( m,\
m_{J/\psi },m_{\rho }\right) }{m_{\rho }^{2}}\right] .  \label{eq:DW2Z4100}
\end{equation}

Calculation of the sum rule Eq.\ (\ref{eq:SR}) is done using $M^{2}$ and $%
s_{0}$ from Eq.\ (\ref{eq:Wind4100}). For the coupling $g_{\overline{Z}%
_{c}J/\psi \rho }$, we find%
\begin{equation}
g_{\overline{Z}_{c}J/\psi \rho }=(0.56\pm 0.07)~\mathrm{GeV}^{-1}.
\label{eq:SC1}
\end{equation}%
Then the width of the decay $\overline{Z}_{c}\rightarrow J/\psi \rho ^{-}$
is
\begin{equation}
\Gamma \left[ \overline{Z}_{c}\rightarrow J/\psi \rho ^{-}\right] =(15\pm 3)~%
\mathrm{MeV.}  \label{eq:DW2num}
\end{equation}%
For the full width of the resonance $\overline{Z}_{c}$ saturated by decay
modes $\overline{Z}_{c}\rightarrow \eta _{c}\pi ^{-}$, $\eta _{c}^{\prime
}\pi ^{-}$, $D^{0}D^{-}$ and $\overline{Z}_{c}\rightarrow J/\psi \rho ^{-}$,
we get
\begin{equation}
\Gamma =(147\pm 19)~\mathrm{MeV}.  \label{eq:FW}
\end{equation}

Our predictions for the mass $m=(4080~\pm 150)~\mathrm{MeV}$ and full width $%
\Gamma =(147\pm 19)~\mathrm{MeV}$ of the resonance $\overline{Z}_{c}$ agree
with LHCb data. Therefore, it is legitimate to interpret the charged
resonance $Z_{c}^{-}(4100)$ as the scalar  diquark-antidiquark $[cd][%
\overline{c}\overline{u}]$ with $C\gamma _{5}\otimes \gamma _{5}C$
structure. It probably is a member of charged $Z$-resonance multiplets that
include also the axial-vector tetraquarks $Z_{c}^{\pm }(3900)$ and $%
Z_{c}^{\pm }(4330)$. The resonances $Z_{c}^{\pm }(4330)$ and $Z_{c}^{\pm
}(3900)$ were discovered in the $\psi ^{\prime }\pi ^{\pm }$ and $J/\psi \pi
^{\pm }$ invariant mass distributions, whereas the neutral particle $%
Z_{c}^{0}(3900)$ was seen in the process $e^{+}e^{-}\rightarrow \pi ^{0}\pi
^{0}J/\psi $. Since $J/\psi $ and $\psi ^{\prime }$ are vector mesons, and $%
\psi ^{\prime }$ is the radial excitation of $J/\psi $, it is reasonable to
treat $Z_{c}(4330)$ as first radial excitation of $Z_{c}(3900)$ (see,
section \ref{sec:Z}). Then the resonance $\overline{Z}_{c}$ fixed in the $%
\eta _{c}\pi ^{-}$ channel can be considered as a scalar partner of these
axial-vector tetraquarks. It is also meaningful to assume that a neutral
member of this family $Z_{c}^{0}(4100)$ may be seen in the process $%
e^{+}e^{-}\rightarrow \pi ^{0}\pi ^{0}\eta _{c}$ with dominantly $\pi
^{0}\pi ^{0}$ mesons at the final state rather than $D\overline{D}$ ones.


\section{The resonances $X(4140)$ and $X(4274)$}

\label{sec:X}
Recently, after analyses of exclusive decays $B^{+}\rightarrow J/\psi \phi
K^{+}$, the  LHCb confirmed existence of the resonances $X(4140)$ and $X(4274)$ in
the $J/\psi \phi $ invariant mass distribution \cite%
{Aaij:2016iza,Aaij:2016nsc}. In the same $J/\psi \phi $ channel LHCb
discovered heavy resonances $X(4500)$ and $X(4700)$ as well. The masses and
decay widths of these resonances (in this section $X(4140)\Rightarrow
X_{1},\ X(4274)\Rightarrow X_{2},\ X(4500)\Rightarrow X_{3}$ and $%
X(4700)\Rightarrow X_{4}$, respectively) in accordance with LHCb
measurements are
\begin{eqnarray}
&&X_{1}:M=4146\pm 4.5_{-2.8}^{+4.6}~\mathrm{MeV},\ \Gamma =83\pm
21_{-14}^{+21}~\mathrm{MeV},  \notag \\
&&X_{2}:M=4273\pm 8.3_{-3.6}^{+17.2}~\mathrm{MeV},\ \Gamma =56\pm
11_{-11}^{+8}~\mathrm{MeV},  \notag \\
&&X_{3}:M=4506\pm 11_{-15}^{+12}~\mathrm{MeV},\ \Gamma =92\pm 21_{-20}^{+21}~%
\mathrm{MeV},  \notag \\
&&X_{4}:M=4704\pm 10_{-24}^{+14}~\mathrm{MeV},\ \Gamma =120\pm
31_{-33}^{+42}~\mathrm{MeV}.  \label{eq:LHCb}
\end{eqnarray}%
The LHCb extracted also spins and $\mathrm{PC}$-parities of these states. It
appears, $X_{1}$ and $X_{2}$ are axial-vector resonances with $J^{\mathrm{PC}%
}=1^{++}$, whereas the $X_{3}$ and $X_{4}$ are scalar states $J^{\mathrm{PC}%
}=0^{++}$.

First experimental information on resonances $X_{1}$ and $X_{2}$ \cite%
{Aaltonen:2009tz,Chatrchyan:2013dma,Abazov:2013xda} stimulated appearance of
different models to account for their properties. Thus they were considered
as meson molecules in Refs.\ \cite%
{Liu:2008tn,Wang:2009ue,Albuquerque:2009ak,Wang:2009ry,Wang:2011uk,Liu:2010hf,He:2011ed, Finazzo:2011he,HidalgoDuque:2012pq}%
. The diquark-antidiquark picture was used in Refs.\ \cite%
{Stancu:2009ka,Patel:2014vua} to model $X_{1}$ and $X_{2}$. There are also
competing approaches which consider them as dynamically generated resonances
\cite{Molina:2009ct,Branz:2010rj} or coupled-channel effects \cite%
{Danilkin:2009hr}.

After LHCb measurements the experimental situation around the resonances $%
X_{1}$ and $X_{2}$ became more clear. The reason is that LHCb removed from
agenda an explanation of $X_{1}$ as $0^{++}$ or $2^{++}$ $D_{s}^{\ast
+}D_{s}^{\ast -}$ molecular states. The LHCb also excluded interpretation of
$X_{2}$ as a molecular bound-state and as a cusp. There were usual attempts
to interpret $X$ resonances as excitations of the ordinary charmonium or as
dynamical effects. Indeed, by studying experimental information on processes
$B\rightarrow K\chi _{c1}\pi ^{+}\pi ^{-}$ and $B\rightarrow KD\overline{D}$
by Belle and BaBar (see, Refs.\ \cite{Bhardwaj:2015rju} and \cite%
{Aubert:2008bl} ), the author of Ref.\ \cite{Chen:2016iua} identified the
resonances $X_{1}$ and $Y(4080)$ with the $P$-wave charmonia $\chi
_{c1}(3^{3}P_{1})$ and $\chi _{c0}(3^{3}P_{0})$, respectively.

Rescattering effects in the decay $B^{+}\rightarrow J/\psi \phi K^{+}$ were
investigated in Ref.\ \cite{Liu:2016onn}, where the author tried to answer
the question: can these effects simulate the discovered resonances $X_{1}$, $%
X_{2}$, $X_{3}$ and $X_{4}$ or not. In accordance with this analysis,
rescattering of $D_{s}^{\ast +}D_{s}^{-}$ and $\psi ^{\prime }\phi $ mesons
may be seen as structures $X_{1}$ and $X_{4}$, respectively. At the same
time, inclusion of $X_{2}$ and $X_{3}$ into this scheme is problematic, and
hence they maybe are genuine four-quark states. But, the author did not rule
out explanation of $X_{2}$ as the excited charmonium $\chi _{c1}(3^{3}P_{1})$%
.

The diquark-antidiquark and molecule pictures prevail over alternative
models of $X$ resonances, and constitute foundations for various studies to
explain experimental information on these states \cite%
{Chen:2010ze,Chen:2016oma,Wang:2016tzr,Wang:2016dcb,Wang:2016gxp,Agaev:2017foq}%
. Thus, the masses of the axial-vector diquark-antidiquark states $[cs][%
\overline{c}\overline{s}]$ with different spin-parities and color structures
were calculated in Ref.\ \cite{Chen:2010ze}. Results obtained there for
states $J^{\mathrm{P}}=1^{+}$ were used in Ref. \cite{Chen:2016oma} to
interpret $X_{1}$ and $X_{2}$ as tetraquarks $[cs][\overline{c}\overline{s}]$
with opposite (i.e., color-triplet or -sextet constitutent diquarks) color
organizations. Within the same approach the resonances $X_{3}$ and $X_{4}$
were interpreted as $D$-wave excited states of $X_{1}$ and $X_{2}$ \cite%
{Chen:2016oma}.

In the framework of the tetraquark model the resonances $X_{1}$ and $X_{2}$
were also explored in Refs.\ \cite{Wang:2016tzr} and \cite{Wang:2016dcb}.
Results obtained in Ref.\ \cite{Wang:2016tzr} excluded interpretation of $%
X_{1}$ as a tightly bound diquark-antidiquark state. The resonance $X_{2}$
was modeled as an octet-octet type molecule state, and it was demonstrated
that the mass of $X_{2}$ agrees with LHCb results, while its width
significantly exceeds the experimental data \cite{Wang:2016dcb}. The
resonance $X_{3}$ was examined as radial excitation of the scalar structure $%
X(3915)$, whereas $X_{4}$ was considered as the ground-state tetraquark $%
[cs][\overline{c}\overline{s}]$ composed of a vector diquark and antidiquark
\cite{Wang:2016gxp}. Let us note that the resonance $X(3915)$ was detected
by Belle in the $J/\psi \omega $ invariant mass distribution of the decay $%
B\rightarrow J/\psi \omega K$ \cite{Abe:2004zs}, and also observed in the
process $\gamma \gamma \rightarrow J/\psi \omega $ \cite{Uehara:2009tx}.
This structure was confirmed by BaBar in the same reaction $B\rightarrow
J/\psi \omega K$ \cite{Aubert:2007vj}. The $X(3915)$ was commonly considered
as the scalar charmonium $\chi _{c0}(2^{3}P_{0})$. But a lack of decay modes
$\chi _{c0}(2P)\rightarrow D\overline{D}$ stimulated other assumptions. In
fact, an alternative conjecture about the resonance $X(3915)$ was proposed
in Ref.\ \cite{Lebed:2016yvr}, where it was interpreted as the lightest
scalar diquark-antidiquark state $[cs][\bar{c}\bar{s}]$. Exactly this
structure was examined in Ref.\ \cite{Wang:2016gxp} as the ground state of $%
X_{3}$, and computations apparently support suggestions made on  nature of
the resonances $X_{3}$ and $X_{4}$.

A plethora of charmoniumlike structures seen in numerous processes
stimulated analysis of various diquark-antidiquark states, and led to
suggestions about existence of different tetraquark multiplets (see, Refs.\
\cite{Stancu:2006st,Maiani:2016wlq,Zhu:2016arf}). Thus, the resonances $X$
were included into $1S$ and $2S$ multiplets of tetraquarks $[cs]_{s=0,1}[%
\overline{c}\overline{s}]_{s=0,1}$ built of color-triplet diquarks \cite%
{Maiani:2016wlq}. The $X_{1}$ was interpreted as $J^{\mathrm{PC}}=1^{++}$
particle of the $1S$ multiplet. The $X_{2}$ resonance is probably, an
admixture of two states with the quantum numbers $J^{\mathrm{PC}}=0^{++}$
and $J^{\mathrm{PC}}=2^{++}$. The idea about mixing phenomenon is inspired
by the fact, that in the multiplet of the tetraquarks composed of
color-triplet diquarks, there is only one state with $J^{\mathrm{PC}}=1^{++}$%
. The heavy resonances $X_{3}$ and $X_{4}$ are included into the $2S$
multiplet as its $J^{\mathrm{PC}}=0^{++}$ members. But apart from the color
triplet multiplets there may exist a multiplet of tetraquarks composed of
color-sextet diquarks \cite{Stancu:2006st}, which also contains a state with
$J^{\mathrm{PC}}=1^{++}$. Stated differently, the multiplet of the
tetraquarks with color-sextet diquarks doubles a number of states \cite%
{Stancu:2006st}, and resonance $X_{2}$ may be identified with its $J^{%
\mathrm{PC}}=1^{++}$ member.

It is evident, that assumptions on internal organization of the resonances $%
X $ in the diquark-antidiquark model sometimes contradict to each other. In
most of these studies the spectroscopic parameters of these states were
calculated using the QCD two-point sum rule method. Results of these
computations obtained by employing various suggestions on interpolating
currents are in agreement with existing experimental data. In some cases
predictions of various articles coincide with each other as well. Stated
differently, the masses and current couplings of exotic states do not give
 information enough to verify supposed models by comparing them with
experimental data or/and alternative theoretical models. In such cases
additional useful information can be extracted from studies of exotic
states' decay channels. The spectroscopic parameters and strong decays of $%
X_{1}$ and $X_{2}$ were explored in Ref.\ \cite{Agaev:2017foq}, in which
they were considered as tetraquarks made of color-triplet and -sextet
diquarks, respectively. Below, we present results of this analysis.


\subsection{Parameters of the resonances $X_1$ and $X_2$}

\label{sec:Mass}
The masses and couplings of the resonances $X_{1}$ and $X_{2}$ can be
calculated by utilizing the QCD two-point sum rule method. Relevant sum
rules can be extracted from analysis of the correlation function (\ref%
{eq:CorrF1}), where $J_{\mu }(x)$ is the interpolating current of the state $%
X$ with the spin-parities $J^{\mathrm{PC}}=1^{++}$.

According to Ref.\ \cite{Chen:2016oma}, the resonances $X_{1}$ and $X_{2}$
have the same quantum numbers, but different internal color structures. This
means that colorless particles $X_{1}$ and $X_{2}$ are built of
color-triplet and color-sextet diquarks, respectively. We pursue this
suggestion and investigate $X_{1}$ and $X_{2}$ using the QCD sum rule method
and currents of different color organization. Namely, we suggest that the
current
\begin{equation}
J_{\mu }^{1}=s_{a}^{T}C\gamma _{5}c_{b}\left( \overline{s}_{a}\gamma _{\mu }C%
\overline{c}_{b}^{T}-\overline{s}_{b}\gamma _{\mu }C\overline{c}%
_{a}^{T}\right) +s_{a}^{T}C\gamma _{\mu }c_{b}\left( \overline{s}_{a}\gamma
_{5}C\overline{c}_{b}^{T}-\overline{s}_{b}\gamma _{5}C\overline{c}%
_{a}^{T}\right) ,  \label{eq:Curr1}
\end{equation}%
which has the color structure $\left[ \overline{\mathbf{3}}_{c}\right]
_{cs}\otimes \left[ \mathbf{3}_{c}\right] _{\overline{cs}}$ , presumably
describes the resonance $X_{1}$, whereas
\begin{equation}
J_{\mu }^{2}=s_{a}^{T}C\gamma _{5}c_{b}\left( \overline{s}_{a}\gamma _{\mu }C%
\overline{c}_{b}^{T}+\overline{s}_{b}\gamma _{\mu }C\overline{c}%
_{a}^{T}\right) +s_{a}^{T}C\gamma _{\mu }c_{b}\left( \overline{s}_{a}\gamma
_{5}C\overline{c}_{b}^{T}+\overline{s}_{b}\gamma _{5}C\overline{c}%
_{a}^{T}\right) ,  \label{eq:Curr2}
\end{equation}%
with the color-symmetric diquark and antidiquark $\left[ \mathbf{6}_{c}%
\right] _{cs}\otimes \left[ \overline{\mathbf{6}}_{c}\right] _{\overline{cs}%
} $ fields corresponds to the tetraquark $X_{2}$.

In order to derive required sum rules, we find an expression of the
correlator in terms of the physical parameters of the state $X$. In the case
of a single particle $X$ the Borel transformation of the phenomenological
side of the sum rules takes the simple form%
\begin{equation}
\mathcal{B}\Pi _{\mu \nu }^{\mathrm{Phys}%
}(q)=m_{X}^{2}f_{X}^{2}e^{-m_{X}^{2}/M^{2}}\left( -g_{\mu \nu }+\frac{q_{\mu
}q_{\nu }}{m_{X}^{2}}\right) +\cdots ,  \label{eq:CorBor}
\end{equation}%
with $m_{X}$ and $f_{X}$ being the mass and coupling of the state $X$.

The QCD side of the sum rule should be expressed in terms of quark
propagators. For these purposes, we contract $c$ and $s$ quark fields, and
get for the correlation function $\Pi _{\mu \nu }^{\mathrm{OPE}}(q)$ the
expression (for definiteness, we provide explicitly results for $J_{\mu
}^{1} $):
\begin{eqnarray}
&&\Pi _{\mu \nu }^{\mathrm{OPE}}(q)=-i\int d^{4}xe^{iqx}\epsilon \widetilde{%
\epsilon }\epsilon ^{\prime }\widetilde{\epsilon }^{\prime }\left\{ \mathrm{%
Tr}\left[ \gamma _{\mu }\widetilde{S}_{c}^{n^{\prime }n}(-x)\gamma _{\nu
}S_{s}^{m^{\prime }m}(-x)\right] \mathrm{Tr}\left[ \gamma _{5}\widetilde{S}%
_{s}^{aa^{\prime }}(x)\gamma _{5}S_{c}^{bb^{\prime }}(x)\right] \right.
\notag \\
&&+\mathrm{Tr}\left[ \gamma _{\mu }\widetilde{S}_{c}^{n^{\prime
}n}(-x)\gamma _{5}S_{s}^{m^{\prime }m}(-x)\right] \mathrm{Tr}\left[ \gamma
_{\nu }\widetilde{S}_{s}^{aa^{\prime }}(x)\gamma _{5}S_{c}^{bb^{\prime }}(x)%
\right] +\mathrm{Tr}\left[ \gamma _{5}\widetilde{S}_{c}^{n^{\prime
}n}(-x)\gamma _{\nu }S_{s}^{m^{\prime }m}(-x)\right]  \notag \\
&&\left. \times \mathrm{Tr}\left[ \gamma _{5}\widetilde{S}_{s}^{aa^{\prime
}}(x)\gamma _{\mu }S_{c}^{bb^{\prime }}(x)\right] +\mathrm{Tr}\left[ \gamma
_{5}\widetilde{S}_{c}^{n^{\prime }n}(-x)\gamma _{5}S_{s}^{m^{\prime }m}(-x)%
\right] \mathrm{Tr}\left[ \gamma _{\nu }\widetilde{S}_{s}^{aa^{\prime
}}(x)\gamma _{\mu }S_{c}^{bb^{\prime }}(x)\right] \right\} ,
\label{eq:CorrF8}
\end{eqnarray}%
where $\epsilon =\epsilon ^{cab},\ \widetilde{\epsilon }=\epsilon ^{cmn}$
and $\epsilon ^{\prime }=\epsilon ^{c^{\prime }a^{\prime }b^{\prime }},\
\widetilde{\epsilon }^{\prime }=\epsilon ^{c^{\prime }m^{\prime }n^{\prime
}} $.

The spectroscopic parameters of the tetraquarks $X$ can be calculated using
the sum rules (\ref{eq:MassGS}) and (\ref{eq:CoupGS}) after substituting $%
4m_{c}^{2}$,$\ m_{Z_{c}}$,$\ $and $f_{Z_{c}}$ by $4(m_{c}+m_{s})^{2}$, $%
m_{X} $ and $f_{X}$.
\begin{table}[tbp]
\begin{tabular}{|c|c|c|}
\hline\hline
$X$ & $X_1$ & $X_2$ \\ \hline\hline
$M^2 ~(\mathrm{GeV}^2$) & $4-6$ & $4-6$ \\ \hline
$s_0 ~(\mathrm{GeV}^2$) & $20-22$ & $21-23$ \\ \hline
$m_{X} ~(\mathrm{MeV})$ & $4183\pm 115$ & $4264 \pm 117$ \\ \hline
$f_{X} ~(\mathrm{GeV}^4)$ & $(0.94 \pm 0.16)\times 10^{-2}$ & $(1.51 \pm
0.21)\times 10^{-2}$ \\ \hline\hline
\end{tabular}%
\caption{The masses and couplings of the tetraquarks $X_1$ and $X_2$. }
\label{tab:Results1B}
\end{table}

\begin{widetext}

\begin{figure}[h!]
\begin{center}
\includegraphics[totalheight=6cm,width=8cm]{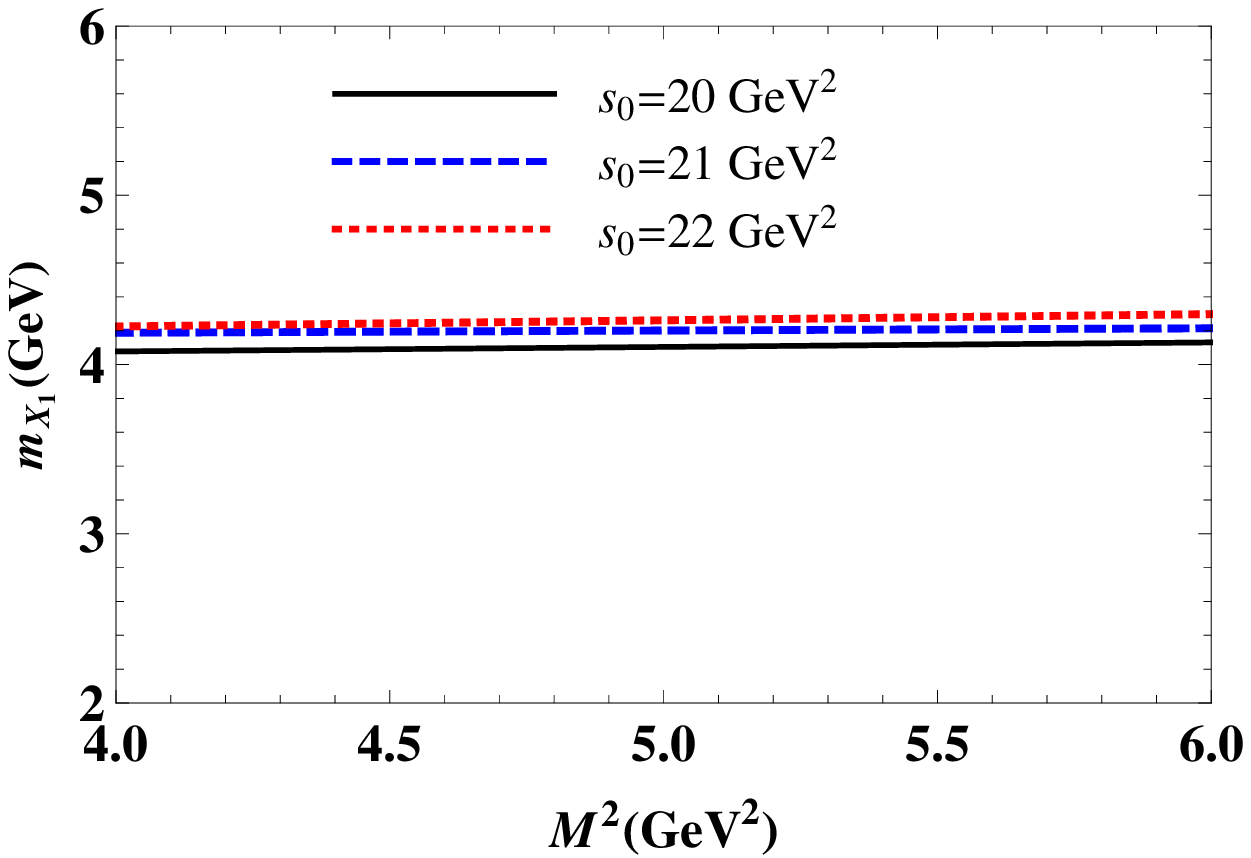}\,\, %
\includegraphics[totalheight=6cm,width=8cm]{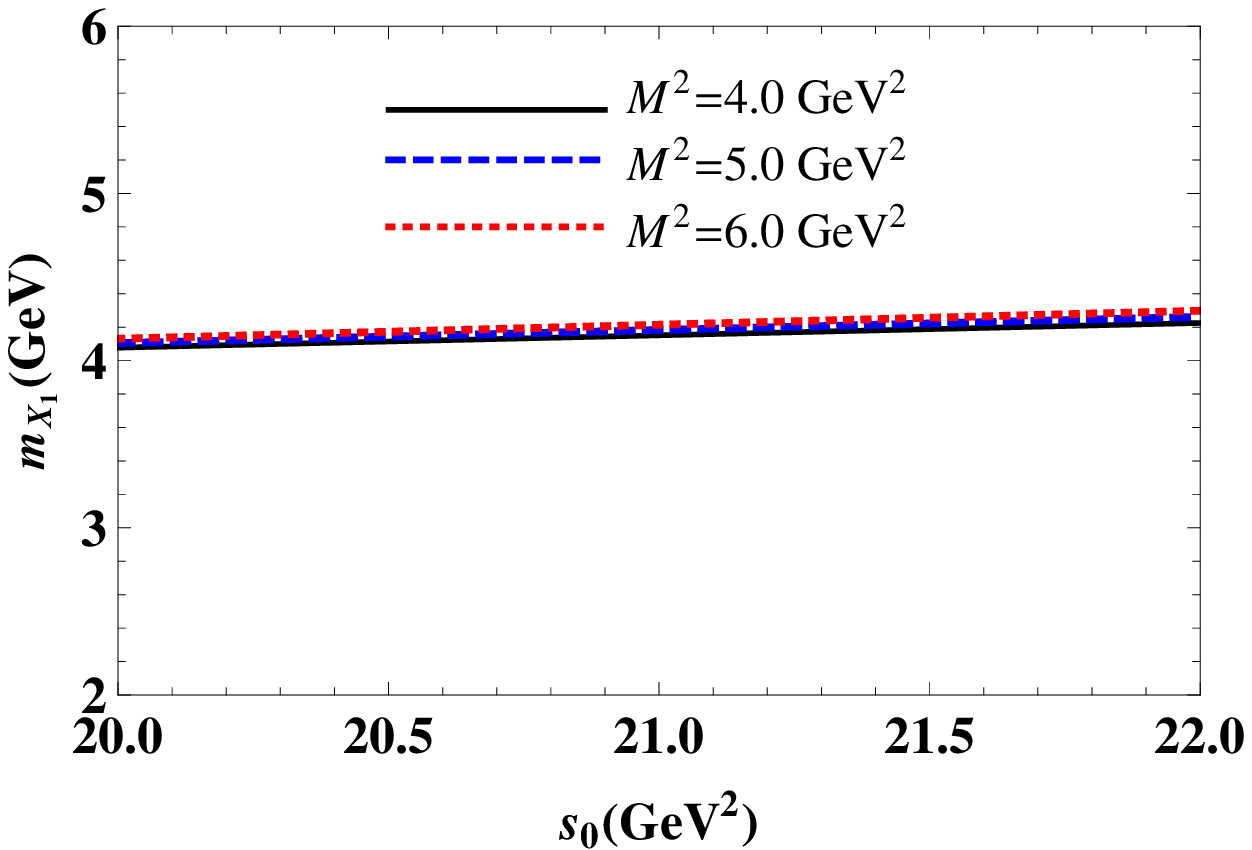}
\end{center}
\caption{ The mass of the $X_{1}$ state as a function of the Borel $M^2$ (left panel), and continuum threshold $s_0$ parameters (right panel).}
\label{fig:Mass}
\end{figure}
\begin{figure}[h!]
\begin{center}
\includegraphics[totalheight=6cm,width=8cm]{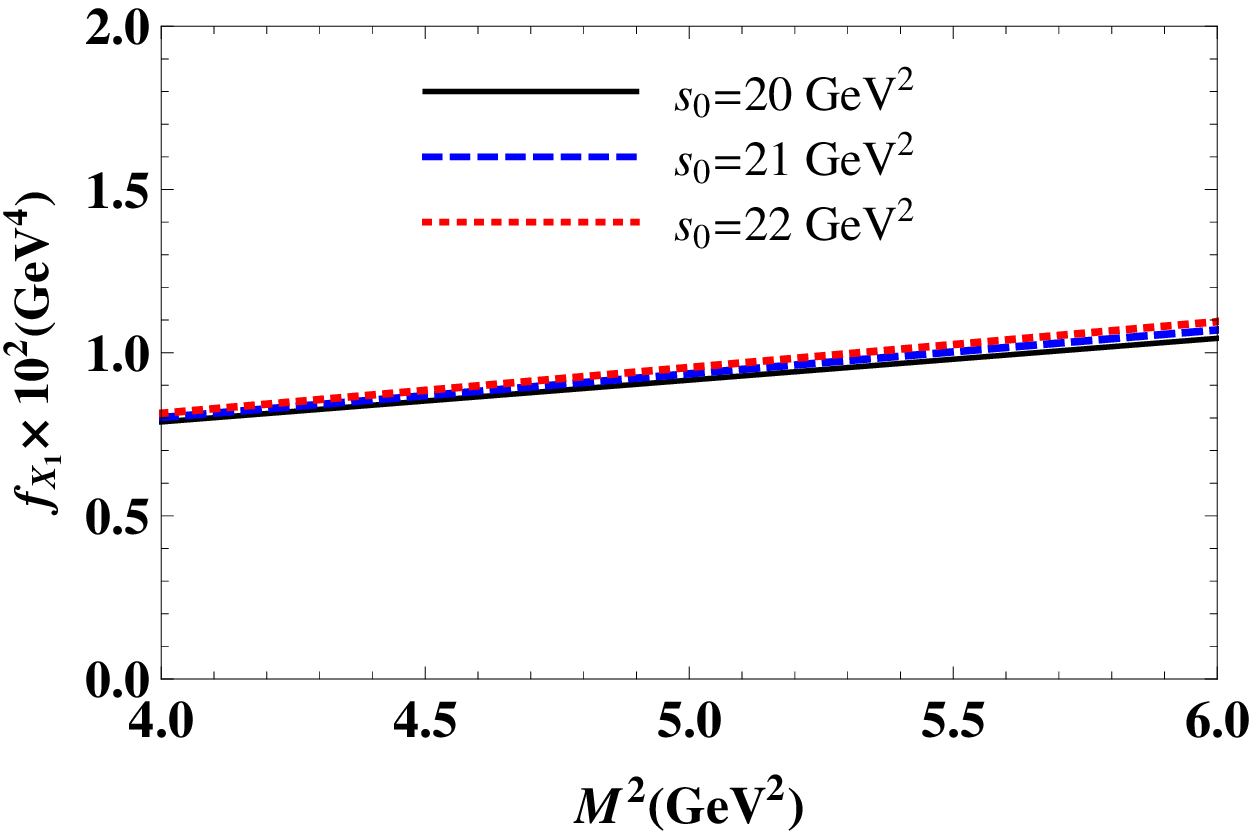}\,\, %
\includegraphics[totalheight=6cm,width=8cm]{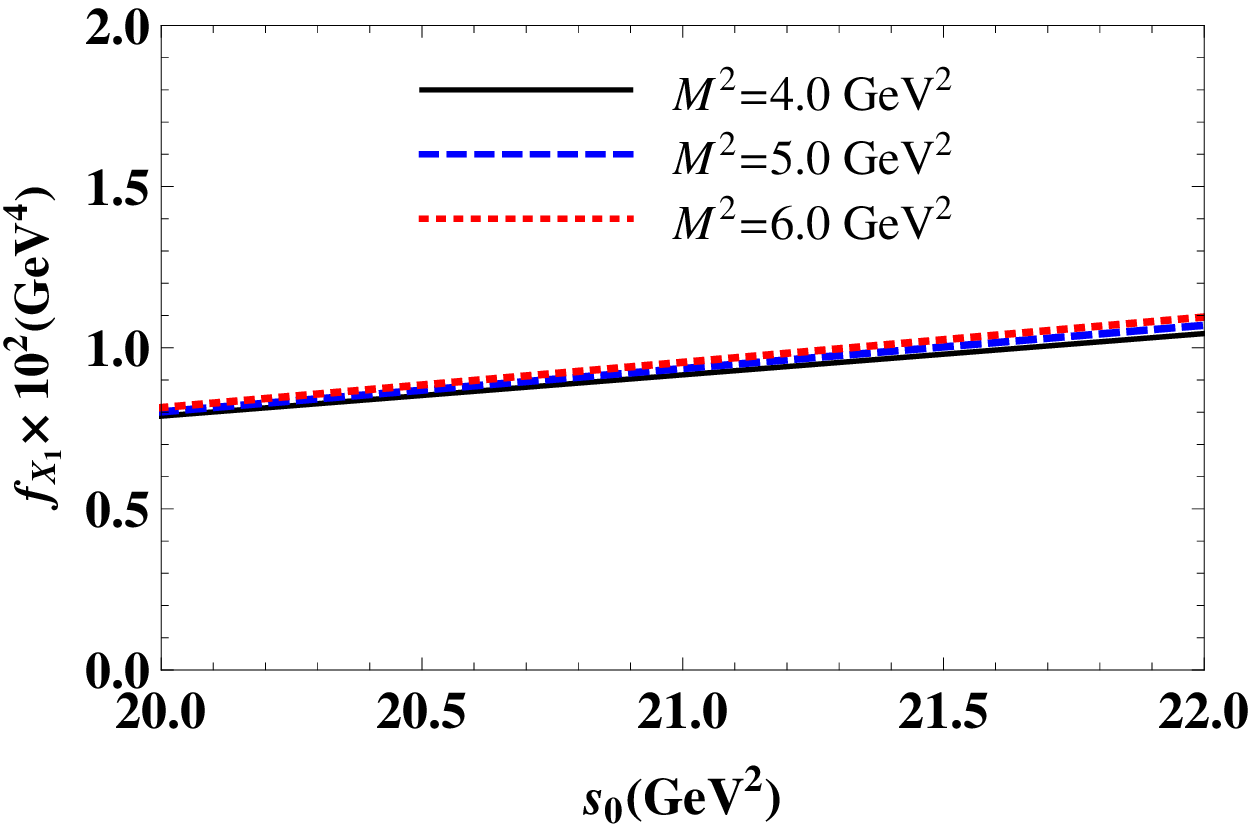}
\end{center}
\caption{ The dependence of the coupling $f_X$ of the $X_1$ resonance on the Borel parameter at chosen $s_0$ (left panel), and on $s_0$ at fixed $M^2$  (right panel).}
\label{fig:Coup}
\end{figure}
\begin{figure}[h!]
\begin{center}
\includegraphics[totalheight=6cm,width=8cm]{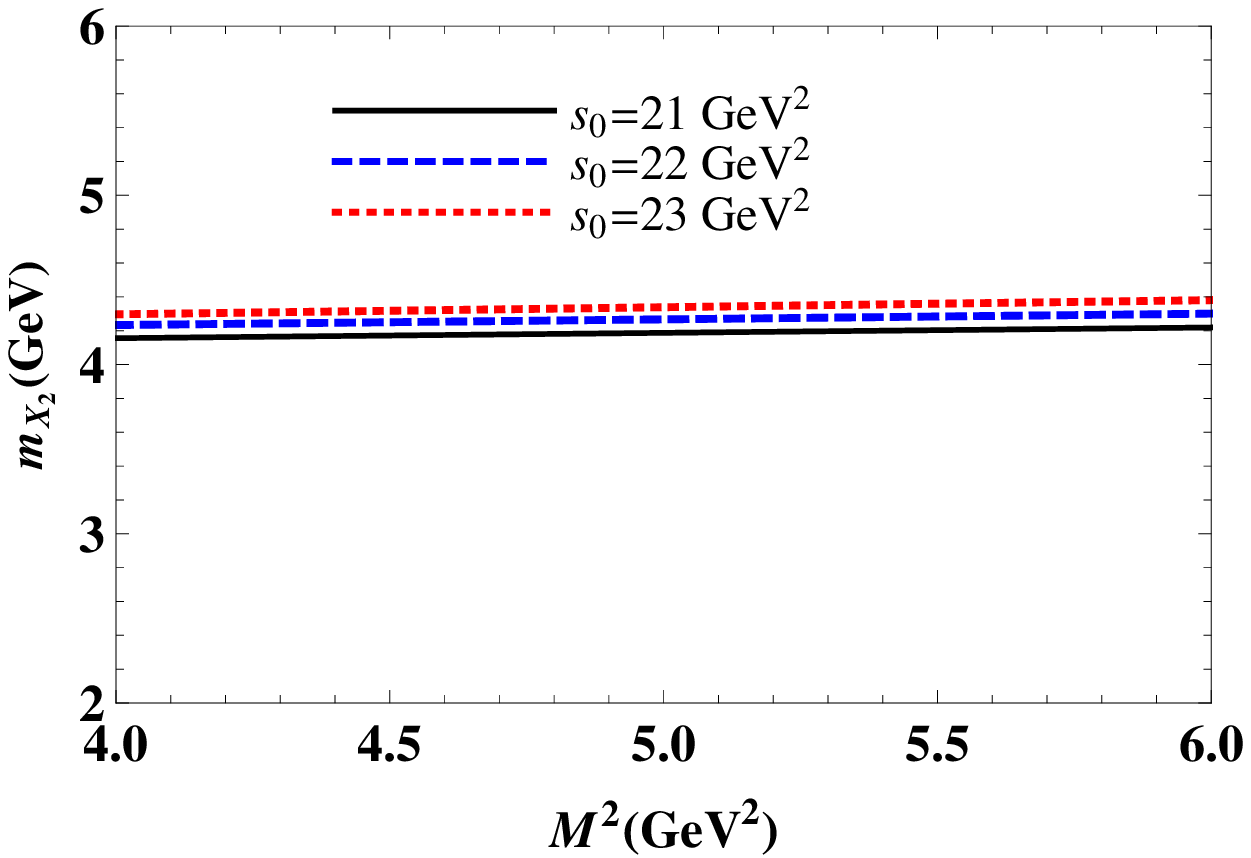}\,\, %
\includegraphics[totalheight=6cm,width=8cm]{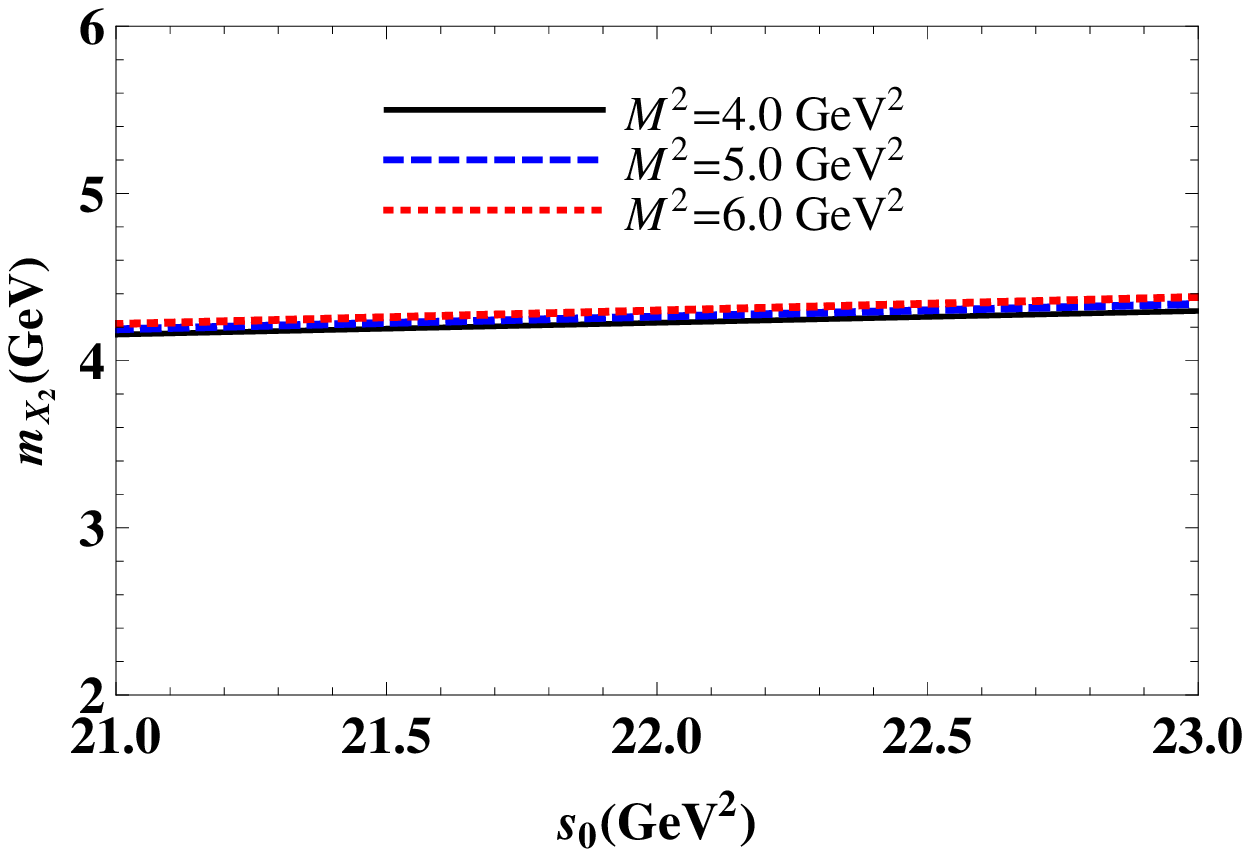}
\end{center}
\caption{ The mass of the $X_2$ resonance as a function of the Borel  $M^2$ (left panel), and continuum threshold $s_0$ parameters (right panel).}
\label{fig:Mass2}
\end{figure}
\begin{figure}[h!]
\begin{center}
\includegraphics[totalheight=6cm,width=8cm]{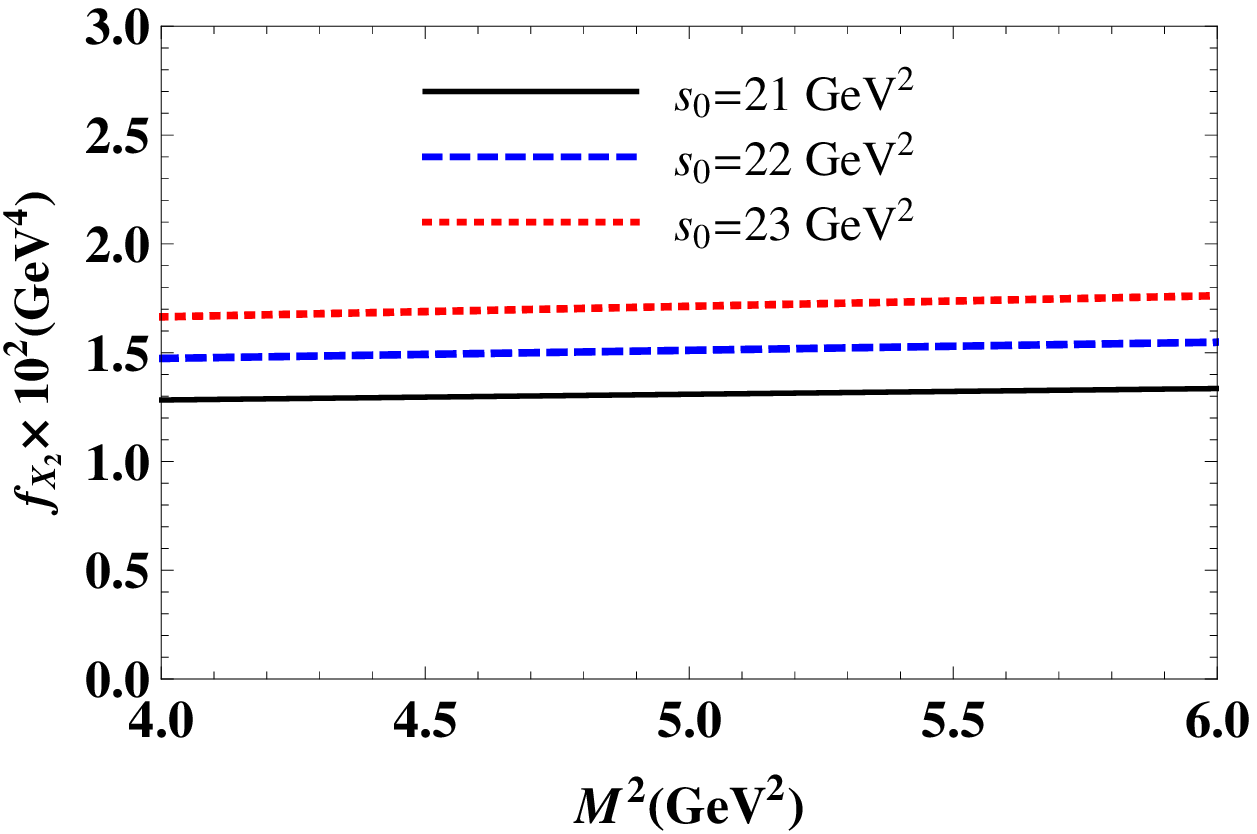}\,\, %
\includegraphics[totalheight=6cm,width=8cm]{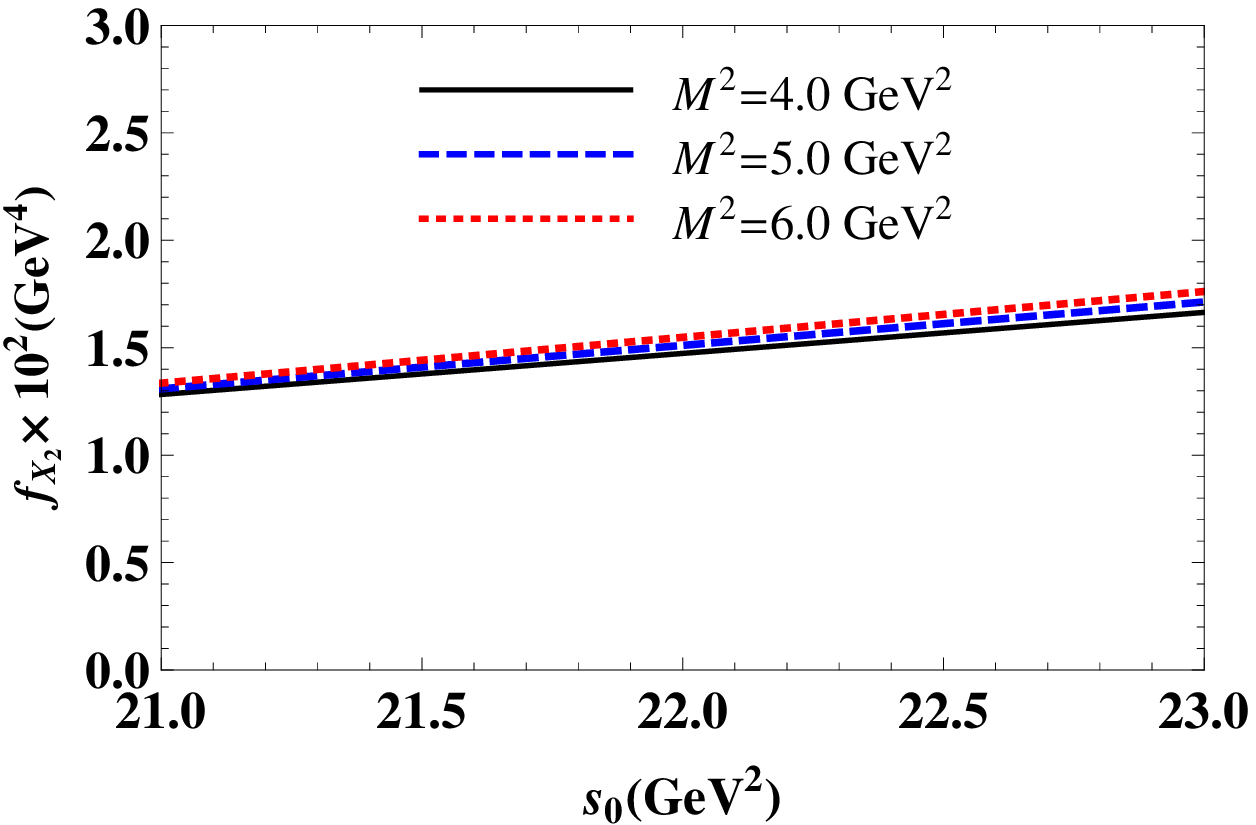}
\end{center}
\caption{ The coupling $f_X$ of the  resonance $X_2$ as a function of $M^2$ (left panel) and $s_0$ (right panel).}
\label{fig:Coup2}
\end{figure}

\end{widetext}

The two-point spectral density $\rho ^{\mathrm{OPE}}(s)$ necessary for
calculations can be derived using methods presented already in the
literature (see, for example, Ref.\ \cite{Agaev:2016dev}). Therefore, we do
not detail here these usual and routine computations. Our predictions for
parameters of the resonances $X_{1}$ and $X_{2}$ are collected in Table\ \ref%
{tab:Results1B}, where we also present working regions for $M^{2}$ and $%
s_{0} $. In the working regions of the Borel and continuum threshold
parameters the pole contribution is equal to $0.23$, which is typical for
the sum rule calculations involving tetraquarks. To keep under control
convergence of the operator product expansion, we find a contribution of
each term with fixed dimension: in the working regions convergence of OPE is
satisfied. Let us only note that a contribution of the dimension-$8$ term to
the whole result does not overshoot $1\%$.

In Figs.\ \ref{fig:Mass} and \ref{fig:Coup}, we depict the parameters of the
tetraquark $X_{1}$ as functions of $M^{2}$ and $s_{0}$. It is clear that $%
m_{X_{1}}$ and $f_{X_{1}}$ are sensitive to the choice of these parameters.
But, while effects of $M^{2}$ and $s_{0}$ on the mass $m_{X_{1}}$ are weak,
a dependence of $f_{X_{1}}$ on the Borel and continuum threshold parameters
is noticeable. These effects combined with uncertainties of other input
parameters generate errors of sum rule calculations. The theoretical errors
of calculations are presented in Table \ref{tab:Results1B} as well. The
similar analysis and conclusions are valid for the state $X_{2}$, which can
be seen in Figs.\ \ref{fig:Mass2} and \ref{fig:Coup2}.

We see, that our predictions for masses of the states $X_{1}$ and $X_{2}$
agree with the LHCb data. At this phase of studies one can conclude that the
resonances $X_{1}$ and $X_{2}$ with the spin-parities $J^{\mathrm{PC}%
}=1^{++} $ enter to multiplets of tetraquarks composed of the color-triplet
and -sextet diquarks, respectively.


\subsection{ Width of decays $X_{1}\rightarrow J/\protect\psi \protect\phi $
and $X_{2} \rightarrow J/\protect\psi \protect\phi $}

\label{subsec:Vertices}
Because $X_{1}$ and $X_{2}$ were discovered in the $J/\psi \phi $ invariant
mass distribution, processes $X_{1}\rightarrow J/\psi \phi $ and $%
X_{2}\rightarrow J/\psi \phi $ are main decay modes of these resonances. In
this subsection, we consider these two decays, and briefly explain
operations required to explore the vertex $XJ/\psi \phi $, where $X$ is one
of \ states $X_{1}$ and $X_{2}$. Below, we evaluate the strong coupling $%
g_{XJ/\psi \phi }$ and width of the corresponding process $X\rightarrow
J/\psi \phi $.

The strong coupling $g_{XJ/\psi \phi }$ can be extracted from analysis of
the correlation function
\begin{equation}
\Pi _{\mu \nu }(p,q)=i\int d^{4}xe^{ipx}\langle \phi (q)|\mathcal{T}\{J_{\mu
}^{\psi }(x)J_{\nu }^{\dag }(0)\}|0\rangle ,  \label{eq:CorrF3X}
\end{equation}%
with $J_{\nu }$ and $J_{\mu }^{\psi }$ being interpolating currents of the $%
X $ state and $J/\psi $ meson, respectively.

We calculate $\Pi _{\mu \nu }(p,q)$ using the LCSR method and the soft-meson
approximation. For these purposes, at the  first step of analysis, we express the
function $\Pi _{\mu \nu }(p,q)$ in terms of the masses, decay constants
(current couplings) of the particles $X$ and $J/\psi $, and strong coupling $%
g_{XJ/\psi \phi }$.

For $\Pi _{\mu \nu }^{\mathrm{Phys}}(p,q)$, we get
\begin{equation}
\Pi _{\mu \nu }^{\mathrm{Phys}}(p,q)=\frac{\langle 0|J_{\mu }^{\psi }|J/\psi
\left( p\right) \rangle }{p^{2}-m_{J/\psi }^{2}}\langle J/\psi \left(
p\right) \phi (q)|X(p^{\prime })\rangle \frac{\langle X(p^{\prime })|J_{\nu
}^{\dagger }|0\rangle }{p^{\prime 2}-m_{X}^{2}}+\cdots .  \label{eq:CorrF4X}
\end{equation}%
The matrix element of the $J/\psi $ meson, necessary for our calculations,
has been defined in Eq.\ (\ref{eq:MelA}), whereas for the vertex, we
introduce the matrix element
\begin{equation}
\langle J/\psi \left( p\right) \phi (q)|X(p^{\prime })\rangle =ig_{XJ/\psi
\phi }\epsilon _{\alpha \beta \gamma \delta }\varepsilon _{\alpha }^{\ast
}(p)\varepsilon _{\beta }(p^{\prime })\varepsilon _{\gamma }^{\ast
}(q)p_{\delta }.  \label{eq:Mel}
\end{equation}%
Here, $\varepsilon _{\gamma }^{\ast }(q)$ is the polarization vector of the $%
\phi $ meson. Then the contribution to $\Pi _{\mu \nu }^{\mathrm{Phys}}(p,q)$
of the ground-state particles is
\begin{equation}
\Pi _{\mu \nu }^{\mathrm{Phys}}(p,q)=i\frac{f_{J/\psi }f_{X}m_{J/\psi
}m_{X}g_{XJ/\psi \phi }}{\left( p^{\prime 2}-m_{X}^{2}\right) \left(
p^{2}-m_{J/\psi }^{2}\right) }\left( \epsilon _{\mu \nu \gamma \delta
}\varepsilon _{\gamma }^{\ast }(p)p_{\delta }-\frac{1}{m_{X}^{2}}\epsilon
_{\mu \beta \gamma \delta }\varepsilon _{\gamma }^{\ast }(p)p_{\delta
}p_{\beta }^{\prime }p_{\nu }^{\prime }\right) +\cdots .  \label{eq:CorrF5X}
\end{equation}%
In the soft limit $p=p^{\prime }$, and only the term $\sim i\epsilon _{\mu
\nu \gamma \delta }\varepsilon _{\gamma }^{\ast }(p)p_{\delta }$ survives in
Eq.\ (\ref{eq:CorrF5X}).

The correlation function $\Pi _{\mu \nu }^{\mathrm{OPE}}(p,q)$ for the
current $J_{\mu }^{1}$ is given by the expression
\begin{eqnarray}
&&\Pi _{\mu \nu }^{\mathrm{OPE}}(p,q)=i\int d^{4}xe^{ipx}\epsilon
^{ijk}\epsilon ^{imn}\left\{ \left[ \gamma _{\nu }\widetilde{S}%
_{c}^{ak}(x)\gamma _{\mu }\widetilde{S}_{c}^{na}(-x){}\gamma _{5}\right] -%
\left[ \gamma _{5}\widetilde{S}_{c}^{ak}(x){}\gamma _{\mu }\widetilde{S}%
_{c}^{na}(-x){}\gamma _{\nu }\right] \right\} _{\alpha \beta }  \notag \\
&&\times \langle \phi (q)|\overline{s}_{\alpha }^{j}s_{\beta }^{m}|0\rangle .
\label{TranCF1}
\end{eqnarray}%
In the soft-meson approximation the matrix element
\begin{equation}
\langle 0|\overline{s}(0)\gamma _{\mu }s(0)|\phi (p,\lambda )\rangle
=f_{\phi }m_{\phi }\epsilon _{\mu }^{(\lambda )},
\end{equation}%
of the $\phi $ meson contributes to the correlation function. Here, $m_{\phi
}$ and $f_{\phi }$ are the mass and decay constant of the $\phi $ meson,
respectively. The soft-meson limit reduces also possible Lorentz structures
in $\Pi _{\mu \nu }^{\mathrm{OPE}}(p,q)$ to the term $\sim i\epsilon _{\mu
\nu \gamma \delta }\varepsilon _{\gamma }^{\ast }(p)p_{\delta }$, which
should be equated to the same structure in $\Pi _{\mu \nu }^{\mathrm{Phys}%
}(p,q=0)$.
\begin{table}[tbp]
\begin{tabular}{|c|c|c|}
\hline\hline
$X$ & $X(4140)$ & $X(4274)$ \\ \hline\hline
$M^2 ~(\mathrm{GeV}^2$) & $5-7$ & $5-7$ \\ \hline
$s_0 ~(\mathrm{GeV}^2$) & $20-22$ & $21-23$ \\ \hline
$g_{XJ/\psi \phi}$ & $2.34 \pm 0.89$ & $3.41 \pm 1.21$ \\ \hline
$\Gamma (X \to J/\psi \phi) ~(\mathrm{MeV})$ & $80 \pm 29$ & $272 \pm 81$ \\
\hline\hline
\end{tabular}%
\caption{The strong coupling $g_{XJ/\protect\psi \protect\phi }$ and decay
width $\Gamma (X\rightarrow J/\protect\psi \protect\phi )$.}
\label{tab:Results2A}
\end{table}

The invariant amplitude corresponding to this Lorentz structure in $\Pi
_{\mu \nu }^{\mathrm{OPE}}(p,q=0)$ can be presented as a dispersion integral
with the spectral density $\rho _{c}^{\mathrm{OPE}}(s)$. We skip further
details of calculations, and write down the final expression for $\rho _{c}^{%
\mathrm{QCD}}(s)$, which reads
\begin{equation}
\rho _{c}^{\mathrm{OPE}}(s)=\frac{f_{\phi }m_{\phi }m_{c}}{4}\left[ \frac{%
\sqrt{s(s-4m_{c}^{2})}}{\pi ^{2}s}+\digamma ^{\mathrm{n.-pert.}}(s)\right] .
\label{eq:Sdensity}
\end{equation}%
The nonperturbative component of $\rho _{c}^{\mathrm{QCD}}(s)$, i.e., $%
\digamma ^{\mathrm{n.-pert.}}(s)$ is determined by the following formula
\begin{equation}
\digamma ^{\mathrm{n.-pert.}}(s)=\Big \langle\frac{\alpha _{s}G^{2}}{\pi }%
\Big \rangle\int_{0}^{1}f_{1}(z,s)dz+\Big \langle g_{s}^{3}G^{3}\Big \rangle%
\int_{0}^{1}f_{2}(z,s)dz+\Big \langle\frac{\alpha _{s}G^{2}}{\pi }\Big
\rangle^{2}\int_{0}^{1}f_{3}(z,s)dz.  \label{eq:NPertX}
\end{equation}%
The functions $f_{1}(z,s),\ f_{2}(z,s)$ and $f_{3}(z,s)$ are given by the
expressions
\begin{equation}
f_{1}(z,s)=\frac{1}{18r^{2}}\left\{ -\left( 2+3r(3+2r)\right) \delta
^{(1)}(s-\Phi )+(1+2r)\left[ m_{c}^{2}-sr\right] \delta ^{(2)}(s-\Phi
)\right\} ,
\end{equation}%
\begin{eqnarray}
&&f_{2}(z,s)=\frac{(1-2z)}{2^{7}\cdot 9\pi ^{2}r^{5}}\left\{ 2r\left[
3r\left( 1+rR\right) \delta ^{(2)}(s-\Phi )+\left[ 3sr^{2}(1+r)-2m_{c}^{2}%
\left( 1+rR\right) \right] \delta ^{(3)}(s-\Phi )\right] \right.  \notag \\
&&\left. +\left[ s^{2}r^{4}-2sm_{c}^{2}r^{2}(1+r)+m_{c}^{4}(1+rR)\right]
\delta ^{(4)}(s-\Phi )\right\} ,
\end{eqnarray}%
\begin{equation}
f_{3}(z,s)=\frac{m_{c}^{2}\pi ^{2}}{2^{2}\cdot 3^{4}r^{2}}\left[ \delta
^{(4)}(s-\Phi )-s\delta ^{(5)}(s-\Phi )\right] ,
\end{equation}%
where the short hand notations%
\begin{equation}
\ r=z(z-1),\ R=3+r,\ \ \Phi =\frac{m_{c}^{2}}{z(1-z)},
\end{equation}%
has been introduced. The function $\ \delta ^{(n)}(s-\Phi )$ is defined as
\begin{equation}
\delta ^{(n)}(s-\Phi )=\frac{d^{n}}{ds^{n}}\delta (s-\Phi ).
\end{equation}

For the interpolating current $J_{\mu }^{2}$ we get%
\begin{eqnarray}
&&\Pi _{\mu \nu }^{\mathrm{OPE}}(p,q)=i\int d^{4}xe^{ipx}\left\{ \left[
\gamma _{\nu }\widetilde{S}_{c}^{ib}(x)\gamma _{\mu }\widetilde{S}%
_{c}^{ai}(-x){}\gamma _{5}-\gamma _{5}\widetilde{S}_{c}^{ib}(x){}\gamma
_{\mu }\widetilde{S}_{c}^{ai}(-x){}\gamma _{\nu }\right] _{\alpha \beta
}\right.  \notag \\
&&\left. \times \langle \phi (q)|\overline{s}_{\alpha }^{a}s_{\beta
}^{b}|0\rangle +\left[ \gamma _{\nu }\widetilde{S}_{c}^{ib}(x)\gamma _{\mu }%
\widetilde{S}_{c}^{bi}(-x){}\gamma _{5}-\gamma _{5}\widetilde{S}%
_{c}^{ib}(x){}\gamma _{\mu }\widetilde{S}_{c}^{bi}(-x){}\gamma _{\nu }\right]
_{\alpha \beta }\langle \phi (q)|\overline{s}_{\alpha }^{a}s_{\beta
}^{a}|0\rangle \right\} .
\end{eqnarray}%
The corresponding spectral density is
\begin{equation}
\rho _{c}^{(2)\mathrm{OPE}}(s)=2\rho _{c}^{(1)\mathrm{OPE}}(s),
\end{equation}%
where $\rho _{c}^{(1)\mathrm{OPE}}(s)$ is given by Eq.\ (\ref{eq:Sdensity}).

The width of the decay $X\rightarrow J/\psi \phi $ can be found by means of
the formula%
\begin{eqnarray}
&&\Gamma (X\rightarrow J/\psi \phi )=\frac{\lambda (m_{X},m_{J/\psi
},m_{\phi })}{48\pi m_{X}^{4}m_{\phi }^{2}}g_{XJ/\psi \phi }^{2}\left[
\left( m_{X}^{2}+m_{\phi }^{2}\right) m_{J/\psi }^{4}+\left(
m_{X}^{2}-m_{\phi }^{2}\right) ^{2}\right.   \notag \\
&&\left. \times \left( m_{X}^{2}+m_{\phi }^{2}-2m_{J/\psi }^{2}\right)
+4m_{X}^{2}m_{J/\psi }^{2}m_{\phi }^{2}\right] ,
\end{eqnarray}%
where $\lambda (a,b,c)$ is the standard function (\ref{eq:Lambda}).

In Table\ \ref{tab:Results2A}, we have collected our results for the
couplings and decay widths. We also write down the regions for the
parameters $M^{2}$ and $s_{0}$ used in numerical calculations to evaluate
the couplings $g_{X_{1}J/\psi \phi }$ and $g_{X_{2}J/\psi \phi }$. In these
regions computations meet all standard constraints of the sum rule analysis.
\begin{table}[tbp]
\begin{tabular}{|c|c|c|c|c|}
\hline\hline
& $m_{X_1}$ & $\Gamma_{X_1}$ & $m_{X_2}$ & $\Gamma_{X_2}$ \\
& $(\mathrm{MeV})$ & $(\mathrm{MeV})$ & $(\mathrm{MeV})$ & $(\mathrm{MeV})$
\\ \hline\hline
LHCb & $4146\pm 4.5_{-2.8}^{+4.6}$ & $83\pm 21_{-14}^{+21}$ & $4273\pm
8.3_{-3.6}^{+17.2}$ & $56\pm 11_{-11}^{+8}$ \\ \hline
\cite{Agaev:2017foq} & $4183 \pm 115$ & $80 \pm 29$ & $4264 \pm 117$ & $272
\pm 81$ \\ \hline
\cite{Albuquerque:2009ak} & $4140 \pm 90$ & $-$ & $-$ & $-$ \\ \hline
\cite{Chen:2010ze} & $4070 \pm 100$ & $-$ & $4220 \pm 100$ & $-$ \\ \hline
\cite{Wang:2016tzr} & $3950 \pm 90$ & $-$ & $-$ & $-$ \\
& $5000 \pm 100$ & $-$ & $-$ & $-$ \\ \hline
\cite{Wang:2016dcb} & $-$ & $-$ & $4270 \pm 90$ & $1800$ \\ \hline\hline
\end{tabular}%
\caption{The LHCb data and theoretical predictions for the mass and width of
the resonances $X_{1}$ and $X_{2}$.}
\label{tab:Results3A}
\end{table}
In Table \ref{tab:Results3A}, we have collected the LHCb data and our
results for parameters of $X_{1}$ and $X_{2}$. \ The states $X_{1}$ and $%
X_{2}$ were explored in numerous articles \cite%
{Albuquerque:2009ak,Chen:2010ze,Chen:2016oma,Wang:2016tzr,Wang:2016dcb}:
some of their predictions are also shown. As is seen, our results for the
masses of tetraquarks $X_{1}$ and $X_{2}$, evaluated in the context of the
QCD sum rule method, are in  reasonable agreement with recent LHCb
measurements \cite{Aaij:2016iza}. We also see that width of the decay $%
X_{1}\rightarrow J/\psi \phi $ is compatible with experimental data, but $%
\Gamma (X_{2}\rightarrow J/\psi \phi )$ significantly overshoots and does
not explain them.

The resonance $X_{1}$ was considered in Ref.\ \cite{Albuquerque:2009ak} as a
molecule state $D_{s}^{\star }\overline{D}_{s}^{\star }$ with $J^{\mathrm{PC}%
}=0^{++}$. Mass of this molecule obtained by employing the QCD sum rule
method correctly describes the experimental data. But problem is that, LHCb
ruled out interpretation of the resonance $X_{1}$ as a molecule-like state.

The parameters of $X_{1}$ and $X_{2}$ in the framework of the sum rule
method were evaluated in Refs.\ \cite{Chen:2010ze,Chen:2016oma} as well.
Results obtained there, are in accord with the LHCb data. Let us emphasize
that the resonances $X_{1}$ and $X_{2}$ were considered in Refs.\ \cite%
{Chen:2010ze,Chen:2016oma} as the axial-vector states built of color-triplet
and -sextet diquarks, respectively. The studies performed in Ref.\ \cite%
{Wang:2016tzr} by means of the sum rule method and two interpolating
currents, however excluded diquark-antidiquark interpretation for $X_{1}$.
The reason is that $m_{X_{1}}$ evaluated using relevant sum rules is either
below the LHCb data or exceeds them (see, Table\ \ref{tab:Results3A}).

The $X_{2}$ was investigated as a molecule-like color-octet state \cite%
{Wang:2016dcb}, and its mass $m_{X_{2}}$ was found equal to
\begin{equation}
m_{X_{2}}=4.27\pm 0.09~\mathrm{GeV}.
\end{equation}%
But width of the decay $X_{2}\rightarrow J/\psi \phi $
\begin{equation}
\Gamma (X_{2}\rightarrow J/\psi \phi )=1.8~\mathrm{GeV}
\end{equation}%
estimated using the QCD three-point sum rule method overshoots the LHCb
value, and hence the author removed his assumption about the structure of
the state $X_{2}$ from agenda.

In this section, we explored the resonances $X_{1}$ and $X_{2}$. Our
predictions for the mass and width of the resonance $X_{1}$ permit its
interpretation as a serious candidate to a tetraquark with $J^{\mathrm{PC}%
}=1^{++}$ built of color-triplet diquark (antidiquark). But, in light of the
LHCb data, consideration of $X_{2}$ as a tetraquark with only color-sextet
diquark constituents seems is problematic. The reason is that LHCb specifies
$X_{2}$ as a relatively narrow state, while our estimate for its width
equals to a few hundred $\mathrm{MeV}$. It is quite possible that $X_{2}$ is
an admixture of a tetraquark with color-sextet ingredients and an ordinary
charmonium. But this and other assumptions on internal structure of the
resonance $X_{2}$ require additional analyses.


\section{The axial-vector resonance $a_{1}(1420)$}

\label{sec:A1}
The resonance $a_{1}(1420)$ (or $a_{1}$ throughout this section) reported by
COMPASS collaboration \cite{Adolph:2015pws} enlarged a five-member family of
axial-vector mesons with the spin-parities $J^{\mathrm{PC}}=1^{++}$. In
order to find a partner of the isosinglet $f_{1}(1420)$ meson, COMPASS
studied $J^{\mathrm{PC}}=1^{++}$ states in the diffractive process $\pi
^{-}+p\rightarrow \pi ^{-}\pi ^{-}\pi ^{+}+p_{\mathrm{recoil}}$. In the $%
f_{0}(980)\pi $ final state the collaboration discovered a resonance $1^{++}$
and identified it as $a_{1}$ meson with the mass and width
\begin{equation}
m=1414_{-13}^{+15}~\mathrm{MeV},\ \Gamma =153_{-23}^{+8}~\mathrm{MeV}.
\label{eq:COMPASS}
\end{equation}

Observation of the light axial-vector state $a_{1}$ that may be interpreted
as isovector partner of $f_{1}(1420)$ meson, stimulated theoretical studies
in the framework of numerous models and schemes. Goals of these
investigations were to reveal structure of $a_{1}$ and compute its
parameters. It is worth noting that by considering $a_{1}$ as an ordinary
axial-vector meson COMPASS, at the same time, did not rule out its possible
interpretation as an exotic state. The reason behind of this conclusion is
discovery of only $a_{1}\rightarrow f_{0}(980)\pi $ decay channel of the
meson $a_{1}$. Problems connected with identification of $a_{1}$ as a
radially excited $a_{1}(1260)$ meson also feed ideas on its exotic nature.

The meson $f_{0}(980)$ that appears in the decay $a_{1}\rightarrow
f_{0}(980)\pi $ gives  additional information on possible structure of $%
a_{1}$. It is one of the  first mesons that was considered as candidate to a
light four-quark state. The meson $f_{0}(980)$ is a member of the first
nonet of scalar particles, which already were analyzed as real candidates to
four-quark $\overline{q}^{2}q^{2}$ states \cite{Jaffe:1976ig}. Because, $%
f_{0}(980)$ has a considerable strange component it was considered also as a
$K\overline{K}$ molecule \cite{Weinstein:1990gu}. Lattice simulations and
various experiments seem confirm assumptions on four-quark structure of $%
f_{0}(980)$ and some other hadrons \cite%
{Alford:2000mm,Amsler:2004ps,Bugg:2004xu,Klempt:2007cp}. On the basis of new
theoretical analysis conclusions on a diquark-antidiquark structure of $%
f_{0}(980)$ and other light scalar mesons were also drawn in Refs.\ \cite%
{Maiani:2004uc,Hooft:2008we}.

Scalar mesons that form the first light nonet were investigated in the
framework of the QCD sum rule method as well. These studies led to
contradictory results about their internal organization \cite%
{Latorre:1985uy,Narison:1986vw,Brito:2004tv,Wang:2005cn,
Chen:2007xr,Lee:2005hs,Sugiyama:2007sg,Kojo:2008hk,Wang:2015uha}. In fact,
some computations supported the  diquark-antidiquark nature of these scalars
\cite{Brito:2004tv,Wang:2005cn,Chen:2007xr}, whereas the author of Ref.\
\cite{Lee:2005hs} could not find in the light mesons signs of diquark
components. Different models were examined to explain properties of the
light scalars and relevant experimental data. These models used various
assumptions on their structure, including mixing of diquark-antidiquarks of
different flavor structures \cite{Chen:2007xr}, and admixtures of four- and
two-quark components \cite{Sugiyama:2007sg,Kojo:2008hk,Wang:2015uha}. The
modern theoretical studies and experimental data are in favor of the
tetraquark picture for the light scalar mesons \cite%
{Kim:2017yvd,Agaev:2017cfz,Agaev:2018sco,Agaev:2018fvz,Achasov:2020fee}.

It is evident that different theoretical models consider $f_{0}(980)$ mainly
as a tetraquark state, or at least as a meson containing essential
four-quark component. These features of $f_{0}(980)$ may provide  useful
information on an internal structure of the meson $a_{1}$ itself. Indeed,
after discovery of the meson $a_{1}$, in the literature appeared various
models that considered it as an exotic state. It was modeled as an admixture
of diquark-antidiquark and two-quark components, mass of which is in accord
with the COMPASS data \cite{Wang:2014bua}. As a pure diquark-antidiquark
compound $a_{1}$ was investigated in Refs.\ \cite{Chen:2015fwa,Sundu:2017xct}%
, predictions of which also agree with the data.

Alternative confirmation of the four-quark structure of $a_{1}$ came from
investigations performed in Ref.\ \cite{Gutsche:2017oro}, where in the
soft-wall AdS/QCD approach the authors derived and solved a Schrodinger-type
equation for the tetraquark wave function. The result obtained there for the
mass of the tetraquark with $J^{\mathrm{PC}}=1^{++}$ is in agreement with
the data \cite{Adolph:2015pws}.

Explanations of $a_{1}$ as dynamical rescattering effects in $a_{1}(1260)$
meson's decays are presented in the literature by some articles \cite%
{Ketzer:2015tqa,Liu:2015taa,Aceti:2016yeb,Basdevant:2015wma,Wang:2015cis}.
Thus, a resonant structure in the $f_{0}(980)\pi $ mass distribution was
considered in Ref.\ \cite{Ketzer:2015tqa} as a triangle singularity in the
relevant decay channel of the $a_{1}(1260)$ meson. The decay of the meson $%
a_{1}(1260)$ runs in accordance with the following scheme: at the first
stage of transformations $a_{1}(1260)$ decays to $K^{\ast }\overline{K}$%
-mesons, at the second phase $K^{\ast }$ decays to $K$ and $\pi $. Finally, $%
K$ and $\overline{K}$ combine to create the $f_{0}(980)$ meson.
Investigation of these transformations and analysis of corresponding
triangle diagram shows the existence of a singularity which may be considered
as the resonance observed by COMPASS. The similar ideas were supported by
Ref.\ \cite{Liu:2015taa}, in which an anomalous triangle singularity were
considered in various processes, including $a_{1}(1260)\rightarrow
f_{0}(980)\pi $ decay.

Two-body strong decays of $a_{1}$ in the context of the covariant confined
quark model were examined in Ref.\ \cite{Gutsche:2017twh}. The meson $a_{1}$
was modeled there as a four-quark state with diquark-antidiquark and
molecule structures. In the analysis of the decay $a_{1}\rightarrow
f_{0}(980)\pi $ the meson $f_{0}(980)$ was also interpreted as a four-quark
state with molecular or diquark organizations. Partial decay widths, and
full width of the $a_{1}$ state found in this work allowed the authors to
interpret $a_{1}$ as a four-quark state with a molecule-type structure.

It is seen, that we can group theoretical studies of the axial-vector state $%
a_{1}$ into two almost equal classes: the first class contains articles, in
which it is considered as a four-quark system with different structures, the
second class encompasses works interpreting $a_{1}$ as dynamical effect
observed in the process $a_{1}(1260)\rightarrow f_{0}(980)\pi $. In this
section we present our investigation of $a_{1}$ and explain results obtained
in Ref. \cite{Sundu:2017xct}.


\subsection{Mass and current coupling of $a_{1}$}

\label{subsec:Massa1}
In the diquark picture the quark content of the neutral isovector state $I^{%
\mathrm{G}}J^{\mathrm{PC}}=1^{-}1^{++}$ has the form $([us][\overline{u}%
\overline{s}]-[ds][\overline{d}\overline{s}])/\sqrt{2}$. The isoscalar
partner of $a_{1}$, namely $f_{1}(1420)$ then should have the composition $%
([us][\overline{u}\overline{s}]+[ds][\overline{d}\overline{s}])/\sqrt{2}$.
In the chiral limit particles $a_{1}$ and $f_{1}(1420)$ have equal masses.

A next problem connected with treatment of $a_{1}$ in the framework of the
QCD sum rule method is a choice of the interpolating current. We choose the
current $J_{\mu }(x)$ in the following form \cite{Chen:2015fwa}
\begin{equation}
J_{\mu }(x)=\frac{1}{\sqrt{2}}[J_{\mu }^{u}(x)-J_{\mu }^{d}(x)].
\label{eq:Curr1a1}
\end{equation}%
Here, $J_{\mu }^{q}(x)$ is given by the expression
\begin{eqnarray}
&&J_{\mu }^{q}(x)=q_{a}^{T}(x)C\gamma _{5}s_{b}(x)\left[ \overline{q}%
_{a}(x)\gamma _{\mu }C\overline{s}_{b}^{T}(x)-\overline{q}_{b}(x)\gamma
_{\mu }C\overline{s}_{a}^{T}(x)\right]  \notag \\
&&+q_{a}^{T}(x)C\gamma _{\mu }s_{b}(x)\left[ \overline{q}_{a}(x)\gamma _{5}C%
\overline{s}_{b}^{T}(x)-\overline{q}_{b}(x)\gamma _{5}C\overline{s}%
_{a}^{T}(x)\right] ,  \label{eq:Curr2a1}
\end{eqnarray}%
with $q$ being one of the light $u$, and $d$ quarks.

After fixing $J_{\mu }(x)$, we should calculate the correlation function $%
\Pi _{\mu \nu }(p)$ given by Eq.\ (\ref{eq:CorrF1}), which allows us to
evaluate the mass $m_{a_{1}}$ and coupling $f_{a_{1}}$ of the state $a_{1}$.
The remaining manipulations are standard ones, therefore we omit further
details by emphasizing only that an invariant amplitude is calculated by
including into analysis vacuum condensates up to dimension $12$. Let us note
that contributions of terms up to dimension eight are found by using
corresponding spectral density, effects of other terms are evaluated
directly from their Borel transformations.

Sum rules depend on the auxiliary parameters $M^{2}$ and $s_{0}$, the choice
of which has to satisfy standard constraints. Our analyses allow us to find
regions, where $M^{2}$ and $s_{0}$ can be varied:
\begin{equation}
M^{2}\in \lbrack 1.4,\ 1.8]\ \mathrm{GeV}^{2},\ s_{0}\in \lbrack 2.4,\ 3.1]\
\mathrm{GeV}^{2}.  \label{eq:Wind}
\end{equation}%
Predictions for mass and coupling of the state $a_{1}$ extracted from the
sum rules are plotted in Figs.\ \ref{fig:Massa1} and \ref{fig:Coupla1}. In
these figures they are shown as functions of the Borel and continuum
threshold parameters. It is clear, that $m_{a_{1}}$ is rather stable against
varying of $M^{2}$ and $s_{0}$. The dependence of $f_{a_{1}}$ on the Borel
parameter is very weak, but its variations with $s_{0}$ are noticeable and
generate essential part of theoretical ambiguities.

For $m_{a_{1}}$ and $f_{a_{1}}$ we find:
\begin{equation}
m_{a_{1}}=1416_{-79}^{+81}~\mathrm{MeV},\
f_{a_{1}}=(1.68_{-0.26}^{+0.25})\times 10^{-3}~\mathrm{GeV}^{4}.
\label{eq:Res1}
\end{equation}%
The prediction for the mass of $a_{1}$ is in  very nice agreement with data
of the COMPASS collaboration. It is in accord also with the mass of the $%
a_{1}$ meson evaluated in Ref.\ \cite{Chen:2015fwa} in the
diquark-antidiquark model
\begin{equation}
m_{a_{1}}=(1440\pm 80)~\mathrm{MeV},\ f_{a_{1}}=(1.32\pm 0.35\ )\times
10^{-3}~\mathrm{GeV}^{4}.  \label{eq:Res2}
\end{equation}%
Our result for $f_{a_{1}}$ is compatible with prediction of Ref.\ \cite%
{Chen:2015fwa} if one takes into account errors of computations: in fact,
there is a large overlap region between (\ref{eq:Res1}) and (\ref{eq:Res2}).
A discrepancy between two sets of parameters comes presumably from
subleading terms in spectral density, which nevertheless does not change
considerably the final predictions.

\begin{widetext}

\begin{figure}[h!]
\begin{center} \includegraphics[%
totalheight=6cm,width=8cm]{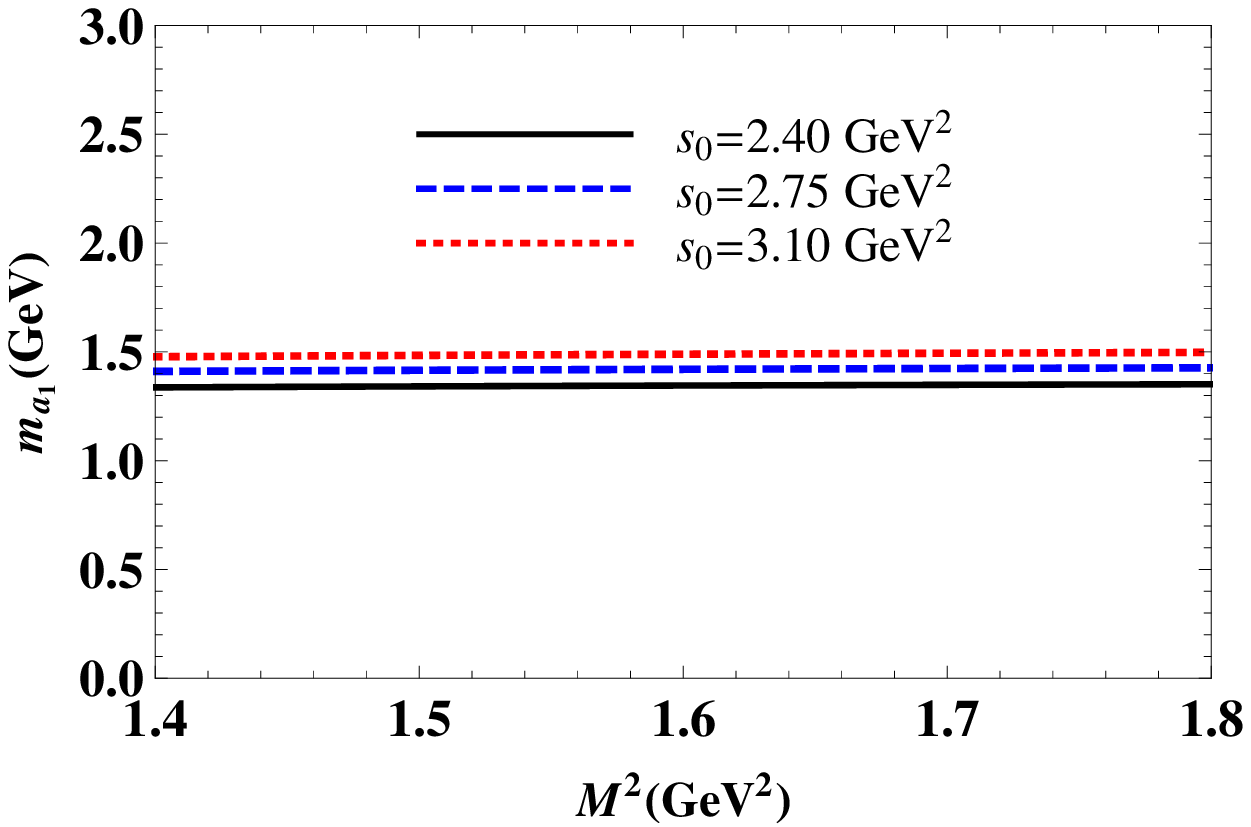}\,\,
\includegraphics[%
totalheight=6cm,width=8cm]{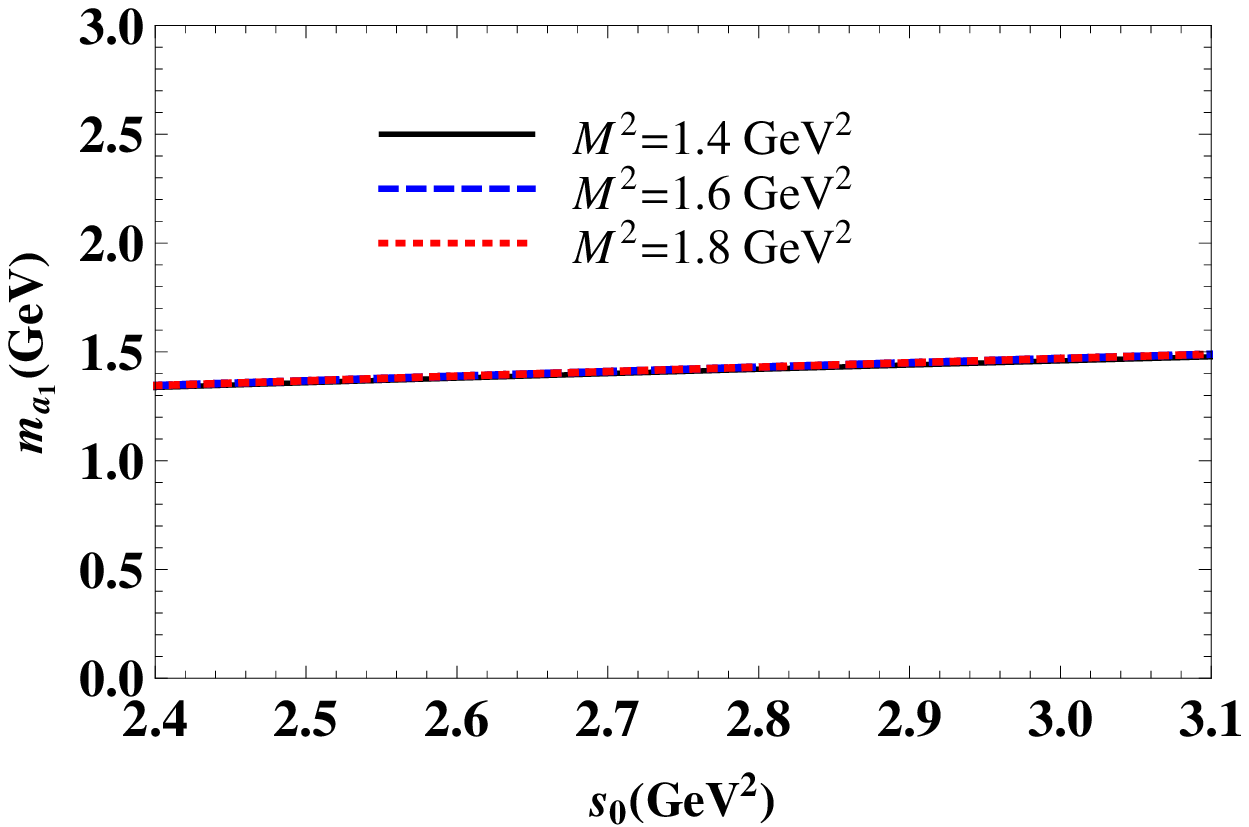}
\end{center}
\caption{ The dependence of $m_{a_{1}}$
on $M^2$
(left panel), and on $s_0$ (right panel).}
\label{fig:Massa1}
\end{figure}
\begin{figure}[h!]
\begin{%
center}
\includegraphics[totalheight=6cm,width=8cm]{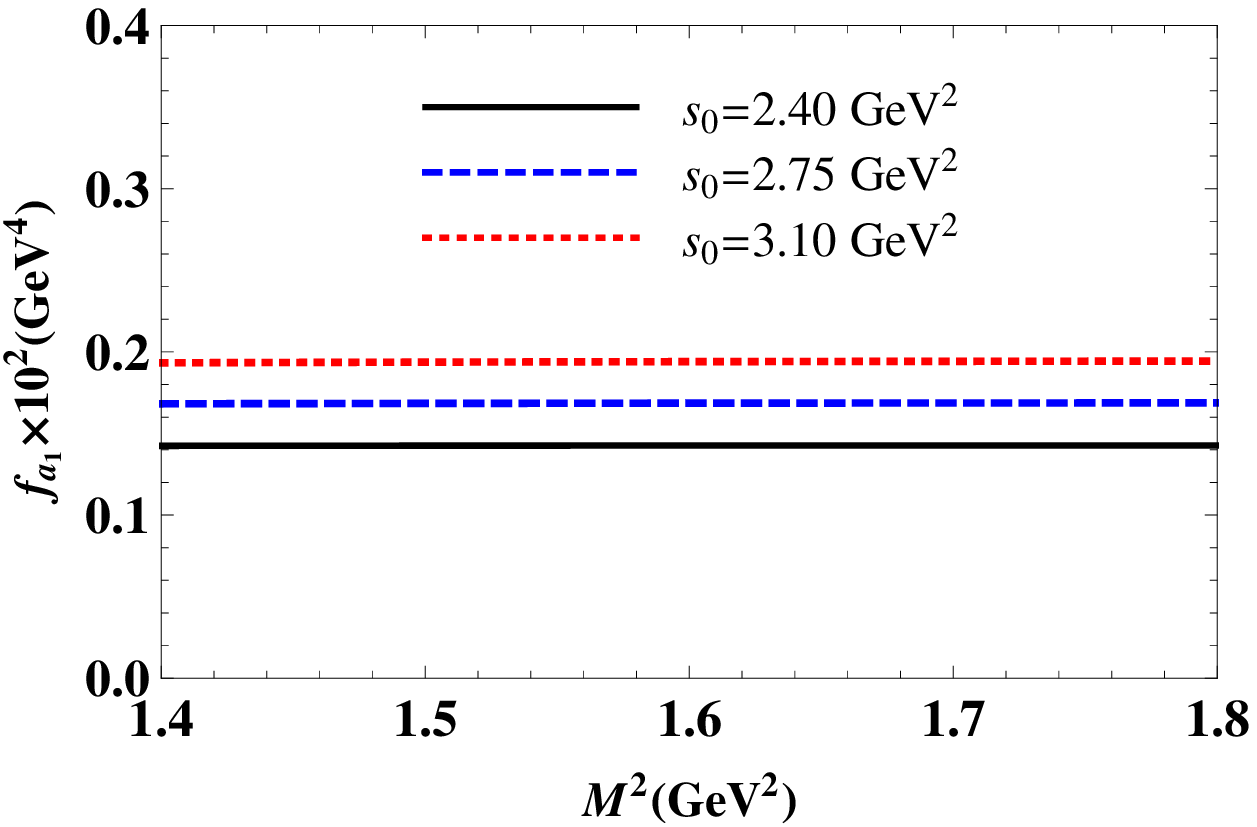}\,\,
\includegraphics[totalheight=6cm,width=8cm]{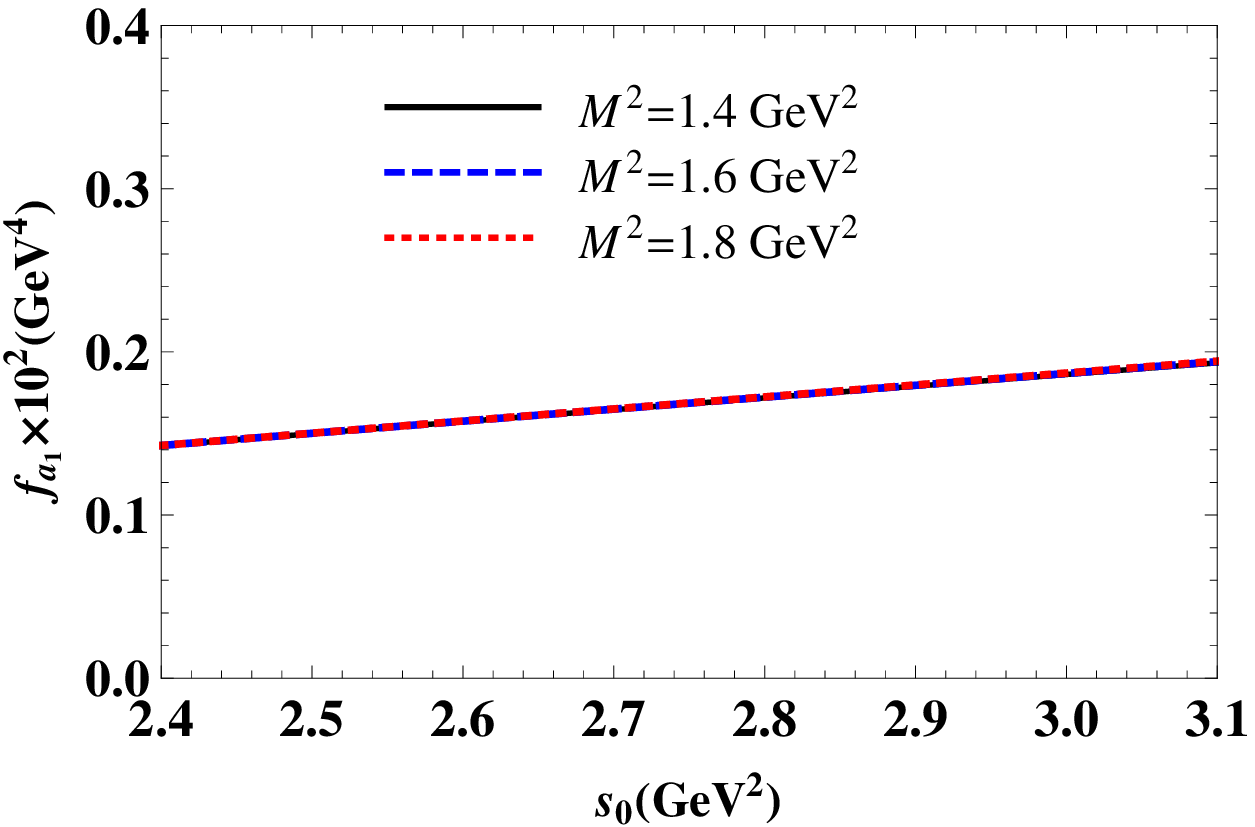} \end{center}
\caption{ The coupling $f_{a_1}$ of the $a_1$ state as a function of $%
M^2$ at fixed $s_0$ (left panel), and of $s_0$ at fixed $M^2$ (right panel).}
\label{fig:Coupla1}
\end{figure}

\end{widetext}

\subsection{The decay channel $a_{1}\rightarrow f_{0}(980)\protect\pi ^{0}$}

\label{subsec:Decf0}
The COMPASS observed the axial-vector state $a_{1}$ in the decay $%
a_{1}\rightarrow f_{0}(980)\pi ^{0}$. This process is $P$-wave decay for $%
a_{1}$, and therefore is not its dominant decay mode. Nevertheless, it has
to be analyzed in details because untill now is a solely observed decay of
the state $a_{1}$.

In the framework of the LCSR method this decay can be studied starting from
analysis of the correlation function
\begin{equation}
\Pi _{\mu }(p,q)=i\int d^{4}xe^{ipx}\langle \pi (q)|\mathcal{T}%
\{J^{f}(x)J_{\mu }^{\dagger }(0)\}|0\rangle ,  \label{eq:CorrF3a1}
\end{equation}%
where $J^{f}(x)$ is the interpolating current of $f_{0}(980)$. We treat $%
f_{0}(980)$ as the scalar diquark-antidiquark state and fix its current $%
J^{f}(x)$ in the following form
\begin{equation}
J^{f}(x)=\frac{\epsilon ^{dab}\epsilon ^{dce}}{\sqrt{2}}\left\{ \left[
u_{a}^{T}(x)C\gamma _{5}s_{b}(x)\right] \left[ \overline{u}_{c}(x)\gamma
_{5}C\overline{s}_{e}^{T}(x)\right] +\left[ d_{a}^{T}(x)C\gamma _{5}s_{b}(x)%
\right] \left[ \overline{d}_{c}(x)\gamma _{5}C\overline{s}_{e}^{T}(x)\right]
\right\} .  \label{eq:Curr3}
\end{equation}%
After adopting the currents, we should analyze the strong vertex $%
a_{1}f_{0}\pi $ that contains two tetraquarks and an ordinary meson, and
differs from vertices of a tetraquark and two conventional mesons. To find
the sum rule for the coupling $g_{a_{1}f_{0}\pi }$, we perform well-known
manipulations. Thus, at the first phase, we rewrite the correlation function
using physical parameters of involved particles and get
\begin{equation}
\Pi _{\mu }^{\mathrm{Phys}}(p,q)=\frac{\langle 0|J^{f}|f_{0}(p)\rangle }{%
p^{2}-m_{f_{0}}^{2}}\langle f_{0}\left( p\right) \pi (q)|a_{1}(p^{\prime
})\rangle \frac{\langle a_{1}(p^{\prime })|J_{\mu }^{\dagger }|0\rangle }{%
p^{\prime 2}-m_{a_{1}}^{2}}+\cdots .  \label{eq:PhysDec1}
\end{equation}%
The representation for $\Pi _{\mu }^{\mathrm{Phys}}(p,q)$ can be simplified
by means of the matrix elements of the states $a_{1}$, and $f_{0}(980)$, as
well as by introducing the strong coupling $g_{a_{1}f_{0}\pi }$ to specify
the vertex $a_{1}f_{0}\pi $
\begin{equation}
\langle f_{0}\left( p\right) \pi (q)|a_{1}(p^{\prime })\rangle
=g_{a_{1}f_{0}\pi }p\cdot \varepsilon ^{\prime \ast }.  \label{eq:Vert1}
\end{equation}%
Here $p^{\prime },\ p$ and $q$ are four-momenta of $a_{1},~f_{0}(980)$ and $%
\pi $, respectively. In Eq.\ (\ref{eq:Vert1}) $\varepsilon _{\mu }^{\prime }$
is the polarization vector of $a_{1}$. The two-variable Borel
transformations applied to $\Pi _{\mu }^{\mathrm{Phys}}(p,q)$ yield
\begin{equation}
\mathcal{B}\Pi _{\mu }^{\mathrm{Phys}}(p,q)=g_{a_{1}f_{0}\pi
}m_{f_{0}}m_{a_{1}}f_{f_{0}}f_{a_{1}}e^{-m_{f_{0}}^{2}/M_{1}^{2}-m_{a_{1}}^{2}/M_{2}^{2}}%
\left[ \frac{1}{2}\left( -1+\frac{m_{f_{0}}^{2}}{m_{a_{1}}^{2}}\right)
p_{\mu }+\frac{1}{2}\left( 1+\frac{m_{f_{0}}^{2}}{m_{a_{1}}^{2}}\right)
q_{\mu }\right] ,
\end{equation}%
where $m_{f_{0}}$ and $f_{f_{0}}$ are the mass and coupling of $f_{0}(980)$,
and $M_{1}^{2}$, $M_{2}^{2}$ are Borel parameters which correspond to
variables $p^{2}$ and $p^{\prime 2}$, respectively. The $\Pi _{\mu }^{%
\mathrm{Phys}}(p,q)$ and its Borel transformation contains structures
proportional to $p_{\mu }$ and $q_{\mu }$. In our studies, we use the
invariant amplitude that correspond to the structure $\sim p_{\mu }$
\begin{equation}
\Pi ^{\mathrm{Phys}}\left( M_{1}^{2},\ M_{2}^{2}\right) =g_{a_{1}f_{0}\pi
}m_{f_{0}}m_{a_{1}}f_{f_{0}}f_{a_{1}}\frac{1}{2}%
e^{-m_{f_{0}}^{2}/M_{1}^{2}-m_{a_{1}}^{2}/M_{2}^{2}}\left( -1+\frac{%
m_{f_{0}}^{2}}{m_{a_{1}}^{2}}\right) .
\end{equation}%
The sum rule can be derived after calculation of its second component. This
means that the correlation function $\Pi _{\mu }(p,q)$ should be expressed
in terms of quark propagators and of the pion's distribution amplitudes.
After inserting currents into Eq.\ (\ref{eq:CorrF3a1}) and contracting quark
fields, we get
\begin{eqnarray}
&&\Pi _{\mu }^{\mathrm{OPE}}(p,q)=i\int d^{4}x\epsilon \widetilde{\epsilon }%
\epsilon ^{\prime }\widetilde{\epsilon }^{\prime }e^{ipx}\left\{ \mathrm{Tr}%
\left[ \gamma _{\mu }\widetilde{S}_{u}^{a^{\prime }a}(x){}\gamma _{5}%
\widetilde{S}_{s}^{b^{\prime }b}(x){}\right] \left[ \gamma _{5}\widetilde{S}%
_{s}^{ee^{\prime }}(-x)\gamma _{5}{}\right] _{\alpha \beta }\langle \pi (q)|%
\overline{u}_{\alpha }^{c^{\prime }}(x)u_{\beta }^{c}(0)|0\rangle \right.
\notag \\
&&-\mathrm{Tr}\left[ \gamma _{5}\widetilde{S}_{s}^{ee^{\prime }}(-x){}\gamma
_{5}\widetilde{S}_{u}^{cc^{\prime }}(-x){}\right] \left[ \gamma _{\mu }%
\widetilde{S}_{s}^{b^{\prime }b}(x)\gamma _{5}{}\right] _{\alpha \beta
}\langle \pi (q)|\overline{u}_{\alpha }^{a}(x)u_{\beta }^{a^{\prime
}}(0)|0\rangle +\mathrm{Tr}\left[ \gamma _{5}\widetilde{S}_{u}^{a^{\prime
}a}(x){}\gamma _{5}\widetilde{S}_{s}^{b^{\prime }b}(x)\right]  \notag \\
&&\times \left[ \gamma _{5}\widetilde{S}_{s}^{ee^{\prime }}(-x)\gamma _{\mu }%
\right] _{\alpha \beta }\langle \pi (q)|\overline{u}_{\alpha }^{c^{\prime
}}(x)u_{\beta }^{c}(0)|0\rangle +\mathrm{Tr}\left[ \gamma _{5}\widetilde{S}%
_{s}^{ee^{\prime }}(-x)\gamma _{\mu }\widetilde{S}_{u}^{cc^{\prime }}(-x){}%
\right] \left[ \gamma _{5}\widetilde{S}_{s}^{b^{\prime }b}(x)\gamma _{5}{}%
\right] _{\alpha \beta }  \notag \\
&&\times \langle \pi (q)|\overline{u}_{\alpha }^{a}(x)u_{\beta }^{a^{\prime
}}(0)|0\rangle ,  \label{eq:CorrF4a1}
\end{eqnarray}%
where $\epsilon \widetilde{\epsilon }\epsilon ^{\prime }\widetilde{\epsilon }%
^{\prime }=\epsilon ^{dab}\epsilon ^{dce}\epsilon ^{d^{\prime }a^{\prime
}b^{\prime }}\epsilon ^{d^{\prime }c^{\prime }e^{\prime }}$.

Let us note that Eq.\ (\ref{eq:CorrF4a1}) is a full expression for $\Pi
_{\mu }^{\mathrm{OPE}}(p,q)$, that encompasses contributions due to both $u$
and $d$ components of the interpolating currents $J_{\mu }(x)$ and $J^{f}(x)$%
: this form of the correlation function is convenient for our analysis.
Apart from propagators, the function $\Pi _{\mu }^{\mathrm{OPE}}(p,q)$
contains also nonlocal quark operators sandwiched between the vacuum and
pion states, which can be transformed in accordance with the prescription \ (%
\ref{eq:MatEx}).

The matrix elements of operators $\overline{u}(x)\Gamma ^{j}u(0)$ can be
expanded over $x^{2}$ and written down using the pion's two-particle DAs of
different twist \cite{Braun:1989iv,Ball:1998je}. For example, in the case of
$\Gamma =\ i\gamma _{\mu }\gamma _{5}$ and $\gamma _{5}$ one obtains%
\begin{eqnarray}
\sqrt{2}\langle \pi ^{0}(q)|\overline{u}(x)i\gamma _{\mu }\gamma
_{5}u(0)|0\rangle  &=&f_{\pi }q_{\mu }\int_{0}^{1}due^{i\overline{u}qx}\left[
\phi _{\pi }(u)+\frac{m_{\pi }^{2}x^{2}}{16}\mathbb{A}_{4}(u)\right]   \notag
\\
&&+\frac{f_{\pi }m_{\pi }^{2}}{2}\frac{x_{\mu }}{qx}\int_{0}^{1}due^{i%
\overline{u}qx}\mathbb{B}_{4}(u),  \label{eq:LTDA}
\end{eqnarray}%
and%
\begin{equation}
\sqrt{2}\langle \pi ^{0}(q)|\overline{u}(x)i\gamma _{5}u(0)|0\rangle =\frac{%
f_{\pi }m_{\pi }^{2}}{m_{u}+m_{d}}\int_{0}^{1}due^{iuqx}\phi _{3;\pi
}^{p}(u).  \label{eq:TW3}
\end{equation}%
Above, the twist-2 (or leading twist) DA of the pion is denoted by $\phi
_{\pi }(u)$. The $\mathbb{A}_{4}(u)$ and $\mathbb{B}_{4}(u)$ are higher
twist functions which can be rewritten in terms of the pion two-particle
twist-4 distributions. The matrix element given by Eq.\ (\ref{eq:TW3}) is
determined by one of two-particle twist-3 distribution amplitudes of the
pion $\phi _{3;\pi }^{p}(u)$. Another two-particle twist-3 DA $\phi _{3;\pi
}^{\sigma }(u)$ corresponds to matrix element (\ref{eq:TW3}) with $i\gamma
_{5}\rightarrow \sigma _{\mu \nu }$ replacement. The matrix elements which
appear due to insertion into $\overline{u}(x)\Gamma ^{j}u(0)$ of the gluon
field strength tensor $G_{\mu \nu }(ux)$ can be expressed in terms of
three-particle DAs of the pion. Their definitions and further details were
presented in Refs.\ \cite{Braun:1989iv,Ball:1998je}.

Because the correlation function written down in terms pion's various DAs is
rather cumbersome, we do not provide it here. The $\Pi _{\mu }^{\mathrm{OPE}%
}(p,q)$ contains Lorentz structures proportional to $p_{\mu }$ and $q_{\mu }$%
. We use the invariant amplitude $\Pi ^{\mathrm{OPE}}\left( p^{2},\
p^{\prime 2}\right) $ that corresponds to a structure $\sim p_{\mu }$ and
equate it to similar amplitude from $\Pi _{\mu }^{\mathrm{Phys}}(p,q)$. The
Borel transform of the invariant amplitude $\Pi ^{\mathrm{OPE}}\left(
p^{2},\ p^{\prime 2}\right) $ can be computed in a manner described in Ref.\
\cite{Belyaev:1994zk}. Afterwards, we carry out the continuum subtraction,
which simplifies when two Borel parameters are equal to each other $%
M_{1}^{2}=M_{2}^{2}$. In the case under discussion we assume that a choice $%
M_{1}^{2}=M_{2}^{2}$ does not lead to essential ambiguities in sum rules,
and introduce $M^{2}$ through
\begin{equation}
\frac{1}{M^{2}}=\frac{1}{M_{1}^{2}}+\frac{1}{M_{2}^{2}}.  \label{eq:M1M2}
\end{equation}%
Continuum subtraction is carried out by means of recipes explained in Ref.\
\cite{Belyaev:1994zk}. Some of formulas used during these manipulations were
presented in Appendix B of Ref.\ \cite{Agaev:2016srl}.

Then, the strong coupling $g_{a_{1}f_{0}\pi }$ can be evaluated using the
sum rule
\begin{equation}
g_{a_{1}f_{0}\pi }=\frac{2m_{a_{1}}^{2}}{m_{f_{0}}^{2}-m_{a_{1}}^{2}}\frac{%
e^{(m_{f_{0}}^{2}+m_{a_{1}}^{2})/2M^{2}}}{%
m_{f_{0}}m_{a_{1}}f_{f_{0}}f_{a_{1}}}\Pi ^{\mathrm{OPE}}\left(
M^{2},s_{0}\right) ,  \label{eq:StrCoup1}
\end{equation}%
where $\Pi ^{\mathrm{OPE}}\left( M^{2},s_{0}\right) $ is the invariant
amplitude after Borel transformation and subtaction procedures. The partial
width of decay $a_{1}\rightarrow f_{0}(980)\pi ^{0}$ is given by the formula%
\begin{equation}
\Gamma (a_{1}\rightarrow f_{0}\pi ^{0})=g_{a_{1}f_{0}\pi }^{2}\frac{\lambda
^{3}(m_{a_{1}},m_{f_{0}},m_{\pi })}{24\pi m_{a_{1}}^{2}}.
\end{equation}

Important nonperturbative information in $\Pi ^{\mathrm{OPE}}\left( M^{2},\
s_{0}\right) $ is included into DAs of the pion. A considerable part of $\Pi
^{\mathrm{OPE}}\left( M^{2},s_{0}\right) $ is generated by two-particle DAs
of the pion at $u_{0}=1/2$. The leading twist DA $\phi _{\pi }(u)$
contributes to $\Pi ^{\mathrm{OPE}}\left( M^{2},s_{0}\right) $ not only
directly, but also via higher-twist DAs with which it is connected by
equations of motion. Therefore, $\phi _{\pi }(u)$ deserves a detailed
analysis.

The DA $\phi _{\pi }(u)$ can be expanded over the Gegenbauer polynomials $%
C_{2n}^{3/2}(\varsigma )$
\begin{equation}
\phi _{\pi }(u,\mu ^{2})=6u\overline{u}\left[ 1+\sum_{n=1,2 \ldots
}a_{2n}(\mu ^{2})C_{2n}^{3/2}(u-\overline{u})\right] ,  \label{eq:PionDALT}
\end{equation}%
where $\overline{u}=1-u$. It depends not only on a longitudinal momentum
fraction $u$ carried by a quark, but due to $a_{2n}(\mu ^{2})$ also on a
scale $\mu $. The Gegenbauer moments $a_{2n}(\mu _{0}^{2})$ at a
normalization scale $\mu =\mu _{0}$ fix an initial shape of the distribution
amplitude, and should be determined by some nonperturbative method or
extracted from experiment.

Here, we use two models for $\phi _{\pi }(u,\mu ^{2}=1\ \mathrm{GeV}^{2})$.
One of these models was obtained from LCSR analysis of the pion's
electromagnetic transition form factor \cite{Agaev:2010aq,Agaev:2012tm}. The
shape of this DA is fixed by the coefficients
\begin{equation}
a_{2}=0.1,\ a_{4}=0.1,\ a_{6}=0.1,\ a_{8}=0.034.  \label{eq:OurGM}
\end{equation}%
At the middle point it equals to $\phi _{\pi }(1/2)\simeq 1.354$, which is
not far from $\phi _{\mathrm{asy}}(1/2)=3/2$, where $\phi _{\mathrm{asy}%
}(u)=6u\overline{u}$ is the asymptotic DA. We use also the lattice model for
$\phi _{\pi }(u)$ \ \cite{Braun:2015axa}, which contains only one
nonasymptotic term
\begin{equation}
\phi _{\pi }(u,\mu ^{2})=6u\overline{u}\left[ 1+a_{2}(\mu ^{2})C_{2}^{3/2}(u-%
\overline{u})\right] .  \label{eq:PionDALT1}
\end{equation}%
The second Gegenbauer moment $a_{2}(\mu ^{2})$ of this DA at $\mu =2\
\mathrm{GeV}$ was estimated $a_{2}=0.1364\pm 0.021$, and evolved to
\begin{equation}
a_{2}(1\ \mathrm{GeV}^{2})=0.1836\pm 0.0283  \label{eq:BRLat}
\end{equation}%
at the scale $\mu =1~\mathrm{GeV}$.

The sum rule (\ref{eq:StrCoup1}) contains the spectroscopic parameters of
the particles $a_{1}$ and $f_{0}$. The mass $m_{a_{1}}$ and coupling $%
f_{a_{1}}$ have been evaluated in the previous subsection. For $m_{f_{0}}$
we use experimental data \ \cite{Tanabashi:2018oca}
\begin{equation}
m_{f_{0}}=(990\pm 20)~\mathrm{MeV},
\end{equation}%
whereas the coupling of the meson $f_{0}(980)$ is borrowed from Ref.\ \cite%
{Brito:2004tv}%
\begin{equation}
f_{f_{0}}=(1.51\pm 0.14)\times 10^{-3}~\mathrm{GeV}^{4}.  \label{eq:f0coupl}
\end{equation}%
In Ref.\ \cite{Brito:2004tv} $f_{f_{0}}$ was extracted from the QCD sum rule
analysis using the interpolating current (\ref{eq:Curr3}), and hence is
appropriate for our goals. Here, we take into account a difference between
definitions of $f_{f_{0}}$ employed in Ref.\ \cite{Brito:2004tv}, and
accepted in the present review.

Numerical computations are carried out by utilizing the following regions
for the Borel and continuum threshold parameters
\begin{equation}
M^{2}\in \lbrack 1.5,\ 2.0]\ \mathrm{GeV}^{2},\ s_{0}\in \lbrack 2.4,\ 3.1]\
\mathrm{GeV}^{2},
\end{equation}%
where all standard restrictions on $M^{2}$, and $s_{0}$ imposed by the sum
rules are satisfied.

For the pion DA (\ref{eq:OurGM}) the strong coupling $g_{a_{1}f_{0}\pi }$
and width of the decay $a_{1}\rightarrow f_{0}\pi ^{0}$ are equal to
\begin{equation}
g_{a_{1}f_{0}\pi }=3.41\pm 0.97,\ \ \Gamma (a_{1}\rightarrow f_{0}\pi
^{0})=(3.14\pm 0.96)~\mathrm{MeV},
\end{equation}%
respectively. For the DA from Eq.\ (\ref{eq:BRLat}), we find
\begin{equation}
g_{a_{1}f_{0}\pi }=3.38\pm 0.93,\ \ \Gamma (a_{1}\rightarrow f_{0}\pi
^{0})=(3.09\pm 0.91)~\mathrm{MeV}.
\end{equation}%
It is seen that an effect of different twist-2 DAs of the pion on final
results is small.


\subsection{The decay channels $a_{1}\rightarrow K^{\ast \pm }K^{\mp }$, $%
K^{\ast 0}\overline{K}^{0}$ and $\ \overline{K}^{\ast 0}K^{0}$}

\label{subsec:DecK}

Here, we consider S-wave decays of the exotic meson $a_{1}$. For these
purposes, we compute strong couplings $g_{a_{1}K^{\ast }K^{-}}$ and $%
g_{a_{1}K^{\ast }K^{+}}$ of the vertices $a_{1}K^{\ast +}K^{-}$ and $%
a_{1}K^{\ast -}K^{+}$, as well as find other two couplings corresponding to
vertices $a_{1}K^{\ast 0}\overline{K}^{0}$ and $\ a_{1}\overline{K}^{\ast
0}K^{0}$. These vertices contain a tetraquark and two ordinary mesons. For
their investigation, the LCSR method should be used in connection with the
soft-meson approximation. In other words, to satisfy the four-momentum
conservation at these vertices momentum of a final meson, for example, a
momentum of $K^{-}$ in $a_{1}K^{\ast +}K^{-}$ has to be set $q=0$.

We start from the decay channel $a_{1}\rightarrow K^{\ast +}K^{-}$ which can
be studied using of the correlation function
\begin{equation}
\Pi _{\mu \nu }(p,q)=i\int d^{4}xe^{ipx}\langle K^{-}(q)|\mathcal{T}\{J_{\mu
}^{K^{\ast +}}(x)J_{\nu }^{\dagger }(0)\}|0\rangle ,
\end{equation}%
where $J_{\mu }^{K^{\ast +}}(x)$ is the interpolating current of the $%
K^{\ast +}$ meson
\begin{equation}
J_{\mu }^{K^{\ast +}}(x)=\overline{s}(x)\gamma _{\mu }u(x).
\end{equation}%
Following standard prescriptions, we write $\Pi _{\mu \nu }(p,q)$ in terms
of physical parameters of the particles $a_{1},\ K^{\ast +}$ and $K^{-}$
\begin{equation}
\Pi _{\mu \nu }^{\mathrm{Phys}}(p,q)=\frac{\langle 0|J_{\mu }^{K^{\ast
+}}|K^{\ast +}(p)\rangle }{p^{2}-m_{K^{\ast }}^{2}}\langle K^{\ast +}\left(
p\right) K^{-}(q)|a_{1}(p^{\prime })\rangle \frac{\langle a_{1}(p^{\prime
})|J_{\nu }^{\dagger }|0\rangle }{p^{\prime 2}-m_{a_{1}}^{2}}+...,
\label{eq:PhysDec2}
\end{equation}%
where $p^{\prime }$ and $p$, $q$ are momenta of the initial and final
particles, respectively.

Further simplification of $\Pi _{\mu \nu }^{\mathrm{Phys}}(p,q)$ is achieved
by replacing the matrix elements with their explicit formulas
\begin{eqnarray}
&&\langle 0|J_{\mu }^{K^{\ast +}}|K^{\ast +}(p)\rangle =f_{K^{\ast
}}m_{K^{\ast }}\varepsilon _{\mu },  \notag \\
&&\langle K^{\ast +}\left( p\right) K^{-}(q)|a_{1}(p^{\prime })\rangle
=g_{a_{1}K^{\ast }K^{-}}\left[ (p\cdot p^{\prime })(\varepsilon ^{\ast
}\cdot \varepsilon ^{\prime })-(p\cdot \varepsilon ^{\prime })(p^{\prime
}\cdot \varepsilon ^{\ast })\right] .
\end{eqnarray}%
First of them, i.e., $\langle 0|J_{\mu }^{K^{\ast +}}|K^{\ast +}(p)\rangle $
is written in terms of the mass $m_{K^{\ast }}$ , decay constant $f_{K^{\ast
}}$ and polarization vector $\varepsilon _{\mu }$ of the $K^{\ast +}$ meson.
The second matrix element is expressed by employing the strong coupling $%
g_{a_{1}K^{\ast }K^{-}}$ that should be evaluated from a sum rule. In the
soft approximation $q\rightarrow 0$ and $p^{\prime }=p$: As a result, we
should perform one-variable Borel transformation, which leads to
\begin{equation}
\mathcal{B}\Pi _{\mu \nu }^{\mathrm{Phys}}(p)=g_{a_{1}K^{\ast
}K^{-}}m_{K^{\ast }}m_{a_{1}}f_{K^{\ast }}f_{a_{1}}\frac{e^{-m^{2}/M^{2}}}{%
M^{2}}\left( m^{2}g_{\mu \nu }-p_{\nu }p_{\mu }^{\prime }\right) +\cdots ,
\label{eq:BorelPhys}
\end{equation}%
where $m^{2}=(m_{K^{\ast }}^{2}+m_{a_{1}}^{2})/2$

We preserve in Eq.\ (\ref{eq:BorelPhys}) $p_{\nu }\neq p_{\mu }^{\prime }$
to show explicitly the Lorentz structures of $\mathcal{B}\Pi _{\mu \nu }^{%
\mathrm{Phys}}(p)$. It is known that in the soft limit there are
contributions in Eq.\ (\ref{eq:BorelPhys}) denoted by dots, which remain
unsuppressed even after Borel transformation. They correspond to strong
vertices of higher excitations of particles involved into a decay process.
These terms appear as contaminations in the physical side of the sum rules
and should be removed using well-known recipes \cite{Belyaev:1994zk}.

In the soft-meson approximation the correlation function $\Pi _{\mu \nu }^{%
\mathrm{OPE}}(p)$ is determined by the formula
\begin{eqnarray}
&&\Pi _{\mu \nu }^{\mathrm{OPE}}(p)=i\int d^{4}xe^{ipx}\frac{\epsilon
\widetilde{\epsilon }}{\sqrt{2}}\left\{ \left[ \gamma _{5}\widetilde{S}%
_{u}^{ic}(x){}\gamma _{\mu }\widetilde{S}_{s}^{bi}(-x){}\gamma _{\nu }\right]
+\left[ \gamma _{\nu }\widetilde{S}_{u}^{ic}(x)\gamma _{\mu }\widetilde{S}%
_{s}^{bi}(-x)\gamma _{5}\right] \right\} _{\alpha \beta }  \notag \\
&&\times \langle K^{-}(q)|\overline{s}_{\alpha }^{e}(0)u_{\beta
}^{a}(0)|0\rangle,  \label{eq:OPE1}
\end{eqnarray}%
where $\epsilon \widetilde{\epsilon }=\epsilon ^{dab}\epsilon ^{dec}$ .

It turns out, that the matrix element of the $K$ meson that contributes to
this correlation function is
\begin{equation}
\langle 0|\overline{u}(0)i\gamma _{5}s(0)|K\rangle =f_{K}\mu _{K},
\end{equation}%
where $\mu _{K}=m_{K}^{2}/(m_{s}+m_{u})$. The function $\Pi _{\mu \nu }^{%
\mathrm{OPE}}(p)$ contains the same Lorentz structures as its
phenomenological counterpart (\ref{eq:BorelPhys}). To derive the sum rule
for $g_{a_{1}K^{\ast }K^{-}}$, we choose the invariant amplitude
proportional to $g_{\mu \nu }$. The Borel transform of this amplitude reads%
\begin{eqnarray}
&&\Pi ^{\mathrm{OPE}}(M^{2})=\int_{4m_{s}^{2}}^{\infty }ds\rho ^{\mathrm{%
pert.}}(s)e^{-s/M^{2}}+\frac{f_{K}\mu _{K}}{\sqrt{2}}\left[ \frac{m_{s}}{6}%
\left( 2\langle \overline{u}u\rangle -\langle \overline{s}s\rangle \right) +%
\frac{1}{72}\langle \frac{\alpha _{s}G^{2}}{\pi }\rangle \right.  \notag \\
&&\left. +\frac{m_{s}}{36M^{2}}\langle \overline{s}g_{s}\sigma Gs\rangle -%
\frac{g_{s}^{2}}{243M^{2}}\left( \langle \overline{s}s\rangle ^{2}+\langle
\overline{u}u\rangle ^{2}\right) \right] ,  \label{eq:Borel2}
\end{eqnarray}%
where
\begin{equation*}
\rho ^{\mathrm{pert.}}(s)=\frac{f_{K}\mu _{K}}{12\sqrt{2}\pi ^{2}}s.
\end{equation*}%
The function $\Pi ^{\mathrm{OPE}}(M^{2})$ contains nonperturbative terms up
to dimension six, and has a simple form. It is worth emphasizing that, the
spectral density $\rho ^{\mathrm{pert.}}(s)$ in Eq.\ (\ref{eq:Borel2}) is
calculated as imaginary part of the correlation function, whereas Borel
transforms of other terms are extracted directly from $\Pi ^{\mathrm{OPE}%
}(p^{2})$.

The sum rule for the strong coupling $g_{a_{1}K^{\ast }K^{-}}$ is given by
the equality
\begin{equation}
g_{a_{1}K^{\ast }K^{-}}m_{K^{\ast }}m_{a_{1}}f_{K^{\ast }}f_{a_{1}}m^{2}%
\frac{e^{-m^{2}/M^{2}}}{M^{2}}+...=\Pi ^{\mathrm{OPE}}(M^{2}).
\label{eq:SRraw}
\end{equation}%
But before to carry out the continuum subtraction, we need to exclude
unsuppressed terms from the physical side of this expression. To this end,
we act on both sides of Eq.\ (\ref{eq:SRraw}) by the operator $\mathcal{P}%
(M^{2},m^{2})$, which singles out the ground-state term and cancel
contaminations. Remaining contributions can be subtracted in a standard way,
which requires replacing $\infty \rightarrow s_{0}$ in the first term of $%
\Pi ^{\mathrm{OPE}}(M^{2})$ while leaving components $\sim (M^{2})^{0}$ and $%
\sim 1/M^{2}$ in their original forms \cite{Belyaev:1994zk}. The width of
the decay $a_{1}\rightarrow K^{\ast +}K^{-}$ after replacements $g_{Z\psi
\pi },m_{\psi },\lambda \left( m_{Z},\ m_{\psi },m_{\pi }\right) \rightarrow
g_{a_{1}K^{\ast }K^{-}},m_{K^{\ast }},\lambda \left( m_{a_{1}},m_{K^{\ast
}},m_{K}\right) $ can be calculated using Eq.\ (\ref{eq:DW}).

The sum rule for $g_{a_{1}K^{\ast }K^{-}}$ can be easily used for numerical
calculations. The regions for parameters $M^{2}$ and $s_{0}$ employed in the
decay $a_{1}\rightarrow f_{0}(980)\pi $ are suitable for the process $%
a_{1}\rightarrow K^{\ast +}K^{-}$ as well. For the masses and decay
constants of the mesons $K^{\ast +}$ and $K^{-}$ we use
\begin{equation}
m_{K^{\pm }}=(493.677\pm 0.016)~\mathrm{MeV},\ m_{K^{\ast \pm }}=(891.76\pm
0.25)~\mathrm{MeV},
\end{equation}%
and
\begin{equation}
f_{K^{\pm }}=(155.72\pm 0.51)~\mathrm{MeV,\ }f_{K^{\ast 0(\pm )}}=225~%
\mathrm{MeV},  \label{eq:Kdecays}
\end{equation}%
respectively.

Results of calculations are presented below%
\begin{equation}
g_{a_{1}K^{\ast }K^{-}}=(2.84\pm 0.79)~\mathrm{GeV}^{-1},\ \Gamma
(a_{1}\rightarrow K^{\ast +}K^{-})=(37.84\pm 10.97)~\mathrm{MeV}.
\label{eq:DW2}
\end{equation}%
The width of the decay $\Gamma (a_{1}\rightarrow K^{\ast -}K^{+})$ are also
given by Eq.\ (\ref{eq:DW2}).

The analysis of the decays $a_{1}\rightarrow K^{\ast 0}\overline{K}%
^{0}\left( \overline{K}^{\ast 0}K^{0}\right) $ does not differ from one
presented above. Let us write down only masses of the $K^{0}(\overline{K}%
^{0})$ and $K^{\ast 0}(\overline{K}^{\ast 0})$ mesons
\begin{equation}
m_{K^{0}}=(497.611\pm 0.013)~\mathrm{MeV},\ m_{K^{\ast 0}}=(895.55\pm 0.20)~%
\mathrm{MeV},
\end{equation}%
employed in numerical calculations. The decay constants of these
pseudoscalar and vector mesons are presented in Eq.\ (\ref{eq:Kdecays}). We
skip further details and write down final sum rule predictions for one of
these channels
\begin{equation}
g_{a_{1}K^{\ast 0}\overline{K}^{0}}=(2.85\pm 0.82)~\mathrm{GeV}^{-1},\
\Gamma (a_{1}\rightarrow K^{\ast 0}\overline{K}^{0})=(33.35\pm 9.76)~\mathrm{%
MeV}.  \label{eq:DW3a1}
\end{equation}%
Parameters of the process $a_{1}\rightarrow \overline{K}^{\ast 0}K^{0}$ are
identical to ones of the decay $a_{1}\rightarrow K^{\ast 0}\overline{K}^{0}$
presented  in Eq.\ (\ref{eq:DW3a1}).

Predictions for decays of the state $a_{1}$ obtained in this section allow
us to find its full width $\Gamma $
\begin{equation}
\Gamma =(145.52\pm 20.79)~\mathrm{MeV,}
\end{equation}%
which is compatible with the COMPASS data, if we take into account
ambiguities of computations.

We have treated the meson $a_{1}$ as a diquark-antidiquark state, and
calculated its mass and widths of five decay modes \cite{Sundu:2017xct}. Our
prediction for the mass $m_{a_{1}}=1416_{-79}^{+81}~\mathrm{MeV}$ of the $%
a_{1}$ is in  nice agreement with the experimental result. Within
small computational errors, it is also in accord with the result of Ref.\
\cite{Chen:2015fwa}. The full width of the meson $a_{1}$ calculated
utilizing five decay channels led to prediction $\Gamma =(145.52\pm 20.79)~%
\mathrm{MeV}$. By taking into account errors of theoretical calculations and
experimental measurements, it is consistent with the COMPASS data $\Gamma
=153_{-23}^{+8}~\mathrm{MeV}$ as well. Analysis performed in Ref. \cite%
{Sundu:2017xct} proved that the axial-vector meson $a_{1}(1420)$ can be
considered as a viable candidate to a diquark-antidiquark state.


\section{The resonance $Y(4660)$}

\label{sec:Y}
The resonances $Y(4660)$ (in a short form $Y$) and $Y(4360)$ were seen by
the Belle collaboration through initial-state radiation (ISR) in the
electron-positron annihilation $e^{+}e^{-}\rightarrow \gamma _{\mathrm{ISR}%
}\psi ^{\prime }\pi ^{+}\pi ^{-}$ : they were fixed as resonant structures
in the $\psi ^{\prime }\pi ^{+}\pi ^{-}$ invariant mass distribution \cite%
{Wang:2007ea,Wang:2014hta}. The mass and full width of the resonance $Y$
measured by Belle are \cite{Wang:2014hta}
\begin{equation}
m_{Y}=4652\pm 10\pm 8\ \mathrm{MeV},\ \Gamma _{Y}=68\pm 11\pm 1\ \mathrm{MeV}%
.  \label{eq:SpPar}
\end{equation}

It is interesting, that there are theoretical papers in the literature
claiming to interpret $Y$ and $Y(4360)$ in the contexts of different models
and schemes of the high-energy physics. In fact, the resonance $Y$ was
considered as the excited charmonia $5{}^{3}S_{1}$ and $6{}^{3}S_{1}$ in
Refs.\ \cite{Ding:2007rg} and \cite{Li:2009zu}, respectively. To account for
collected experimental data, $Y$ was analyzed as a bound state of the scalar
$f_{0}(980)$ and vector $\psi ^{\prime }$ mesons \cite%
{Guo:2008zg,Wang:2009hi,Albuquerque:2011ix}, or as a baryonium \cite%
{Qiao:2007ce,Cotugno:2009ys}. The hadrocharmonium model for the resonances $%
Y $ and $Y(4360)$ was proposed in Ref.\ \cite{Dubynskiy:2008mq}.

The diquark-antidiquark picture is among widely used models of $Y(4360)$ and
$Y$, in which one assumes they are composition of a diquark and an
antidiquark with certain features. Thus, computations performed in the
context of the relativistic diquark model allowed the authors of Ref.\ \cite%
{Ebert:2008kb} to interpret the resonance $Y(4360)$ as an excited $1P$
tetraquark composed of an axial-vector diquark and antidiquark. In this
picture the resonance $Y$ is $2P$ excitation of a scalar diquark-antidiquark
state. As a radial excitation of the tetraquark $Y(4008)$ the resonance $%
Y(4360)$ was examined in Ref.\ \cite{Maiani:2014}. In the framework of the
QCD sum rule method $Y$ was analyzed as the $P$-wave tetraquark with $%
C\gamma _{5}\otimes D_{\mu }\gamma _{5}C$ type structure and $[cs][\overline{%
c}\overline{s}]$ content in Ref.\ \cite{Zhang:2010mw}. It was also modeled
in Ref.\ \cite{Albuquerque:2008up} as the tetraquark $[cs][\overline{c}%
\overline{s}]$ with the interpolating current $C\gamma _{5}\otimes \gamma
_{5}\gamma _{\mu }C$. The mass of such compound computed using the sum rule
method agrees with experimental data.

Strictly speaking, there are some options to construct a tetraquark with
required $P$ and $C$ parities from the five independent diquark fields
without derivatives, which bear spins $0$ or $1$ and have different $P$%
-parities \cite{Chen:2010ze}. This means that there are numerous tetraquarks
with different diquark-antidiquark structures, but the same quantum numbers $%
J^{PC}=1^{--}$. In the context of the QCD sum rule method such currents,
excluding ones with derivatives, were employed in Ref.\ \cite{Chen:2010ze}
to compute masses of \ tetraquarks with $J^{PC}=1^{-+},1^{--},1^{++},\
1^{+-} $ and quark contents $[cs][\overline{c}\overline{s}]$ and $[cq][%
\overline{c}\overline{q}]$. All examined currents for the tetraquark $[cq][%
\overline{c}\overline{q}]$ with $J^{PC}=1^{--}$ predicted $m\sim 4.6-4.7\
\mathrm{GeV}$, which implies a possible tetraquark nature of $Y$ as well.
But these results do not exclude interpretation of $Y$ as a state $[cs][%
\overline{c}\overline{s}]$ with $J^{PC}=1^{--}$ and structure $C\gamma ^{\nu
}\otimes \sigma _{\mu \nu }C-C\sigma _{\mu \nu }\otimes \gamma ^{\nu }C$,
because the mass of such state $m=4.64\pm 0.09\ \mathrm{GeV}$ is also
consistent with the mass of the $Y$ resonance. The sum rule method was
utilized in Refs.\ \cite{Wang:2013exa,Wang:2016mmg,Wang:2018rfw} as well, in
which the resonance $Y$ was modeled as a tetraquark with $[cq][\overline{c}%
\overline{q}]$ or $[cs][\overline{c}\overline{s}]$ contents, and $C\gamma
_{\mu }\otimes \gamma _{\nu }C-C\gamma _{v}\otimes \gamma _{\mu }C$ and $%
C\otimes \gamma _{\mu }C$ type structures.


\subsection{Mass and coupling of the vector tetraquark $Y$}

\label{subsec:MassY}

We consider $Y$ as the $[cs][\overline{c}\overline{s}]$ tetraquark made of a
scalar diquark and vector antidiquark with the $C\gamma _{5}\otimes \gamma
_{5}\gamma _{\mu }C$ type structure \cite{Sundu:2018toi}. In our
calculations we take into account condensates up to dimension $10$ by
including into consideration the gluon condensate $\langle
g_{s}^{3}G^{3}\rangle $ omitted in aforementioned works, and improve an
accuracy of the predictions obtained there.

We start from consideration of the correlation function (\ref{eq:CorrF1}),
where the interpolating current $J_{\mu }(x)$ is
\begin{equation}
J_{\mu }(x)=\epsilon \widetilde{\epsilon }\left[ s_{b}^{T}(x)C\gamma
_{5}c_{c}(x)\overline{s}_{d}(x)\gamma _{5}\gamma _{\mu }C\overline{c}%
_{e}^{T}(x)+s_{b}^{T}(x)C\gamma _{\mu }\gamma _{5}c_{c}(x)\overline{s}%
_{d}(x)\gamma _{5}C\overline{c}_{e}^{T}(x)\right] .  \label{eq:Curr1Y}
\end{equation}%
Remaining operations are standard, and the invariant amplitude proportional
to a structure $g_{\mu \nu }$ in the physical side of the sum rule is equal
to
\begin{equation}
\Pi ^{\mathrm{Phys}}(p^{2})=\frac{m_{Y}^{2}f_{Y}^{2}}{m_{Y}^{2}-p^{2}}
\label{eq:Phys2}
\end{equation}%
The QCD side of the sum rule $\Pi _{\mu \nu }^{\mathrm{OPE}}(p)$ should be
expressed in terms of the quark propagators, and has the form
\begin{eqnarray}
&&\Pi _{\mu \nu }^{\mathrm{OPE}}(p)=i\int d^{4}xe^{ipx}\epsilon \widetilde{%
\epsilon }\epsilon ^{\prime }\widetilde{\epsilon }^{\prime }\left\{ \mathrm{%
Tr}\left[ \gamma _{5}\widetilde{S}_{s}^{bb^{\prime }}(x)\gamma
_{5}S_{c}^{cc^{\prime }}(x)\right] \mathrm{Tr}\left[ \gamma _{5}\gamma _{\mu
}\widetilde{S}_{c}^{e^{\prime }e}(-x)\gamma _{\nu }\gamma
_{5}S_{s}^{d^{\prime }d}(-x)\right] \right.  \notag \\
&&+\mathrm{Tr}\left[ \gamma _{5}\gamma _{\mu }\widetilde{S}_{c}^{ee^{\prime
}}(-x)\gamma _{5}S_{s}^{d^{\prime }d}(-x)\right] \mathrm{Tr}\left[ \gamma
_{5}\gamma _{\nu }\widetilde{S}_{s}^{bb^{\prime }}(x)\gamma
_{5}S_{c}^{cc^{\prime }}(x)\right] +\mathrm{Tr}\left[ \gamma _{5}\widetilde{S%
}_{c}^{ee^{\prime }}(-x)\gamma _{\nu }\gamma _{5}S_{s}^{d^{\prime }d}(-x)%
\right]  \notag \\
&&\left. \times \mathrm{Tr}\left[ \gamma _{5}\widetilde{S}_{s}^{bb^{\prime
}}(x)\gamma _{\mu }\gamma _{5}S_{c}^{cc^{\prime }}(x)\right] +\mathrm{Tr}%
\left[ \gamma _{5}\gamma _{\nu }\widetilde{S}_{s}^{bb^{\prime }}(x)\gamma
_{\mu }\gamma _{5}S_{c}^{cc^{\prime }}(x)\right] \mathrm{Tr}\left[ \gamma
_{5}\widetilde{S}_{c}^{ee^{\prime }}(-x)\gamma _{5}S_{s}^{d^{\prime }d}(-x%
\right] \right\} .  \label{eq:OPE1Y}
\end{eqnarray}

The analysis performed by taking into account all usual restrictions of sum
rule computations permits us to find
\begin{equation}
M^{2}\in \lbrack 4.9,\ 6.8]\ \mathrm{GeV}^{2},\ s_{0}\in \lbrack 23.2,\
25.2]\ \mathrm{GeV}^{2},  \label{eq:WindY}
\end{equation}%
as working windows for $M^{2}$ and $s_{0}$. Really, at $M^{2}=4.9\ \mathrm{%
GeV}^{2}$ the convergence of the OPE is satisfied with high accuracy and $%
R(4.8~\mathrm{GeV}^{2})=0.017$ [$R$ is evaluated using $\mathrm{DimN\equiv
Dim(8+9+10)}$ in Eq.\ (\ref{eq:Convergence})]. At $\ M^{2}=6.8\ \mathrm{GeV}%
^{2}$ the pole contribution is $\mathrm{PC}=0.16$, and at $M^{2}=4.9\
\mathrm{GeV}^{2}$ reaches its maximum $\mathrm{PC}=0.78$. Moreover, at
minimum of $M^{2}$ the perturbative contribution constitutes more than $74\%$
of the result and significantly exceeds nonperturbative effects.

\begin{widetext}

\begin{figure}[h!]
\begin{center} \includegraphics[%
totalheight=6cm,width=8cm]{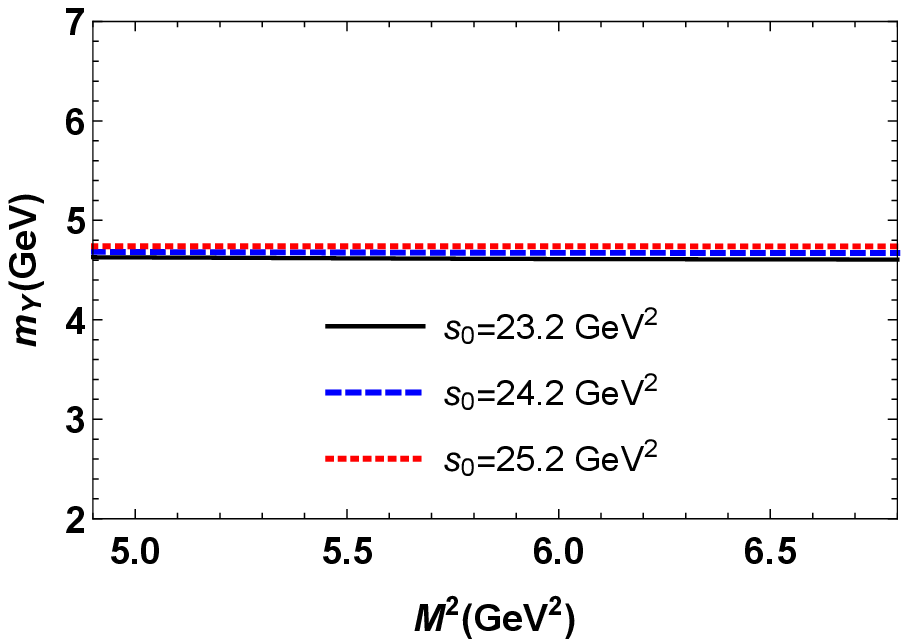}\,\,
\includegraphics[
totalheight=6cm,width=8cm]{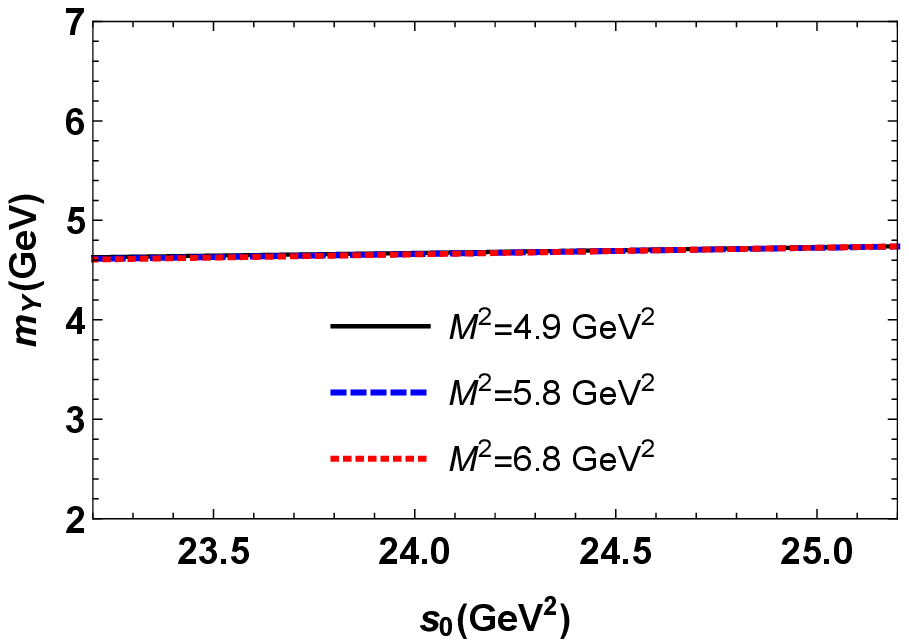}
\end{center}
\caption{ The dependence of the $Y(4660)$ resonance's mass on the Borel (left) and continuum threshold
(right) parameters.}
\label{fig:MassY}
\end{figure}
\begin{figure}[h!]
\begin{center}
\includegraphics[%
totalheight=6cm,width=8cm]{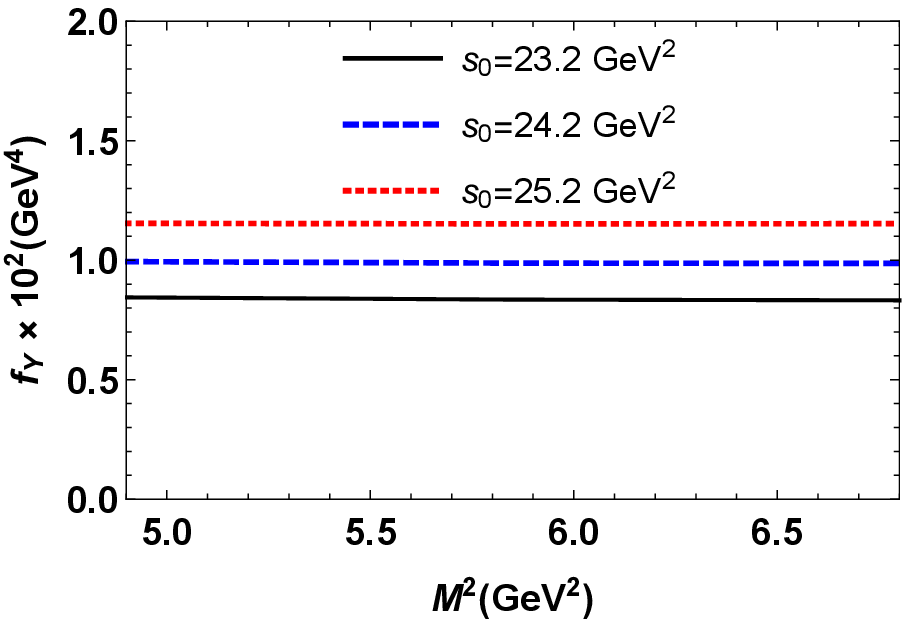}\,\,
\includegraphics[
totalheight=6cm,width=8cm]{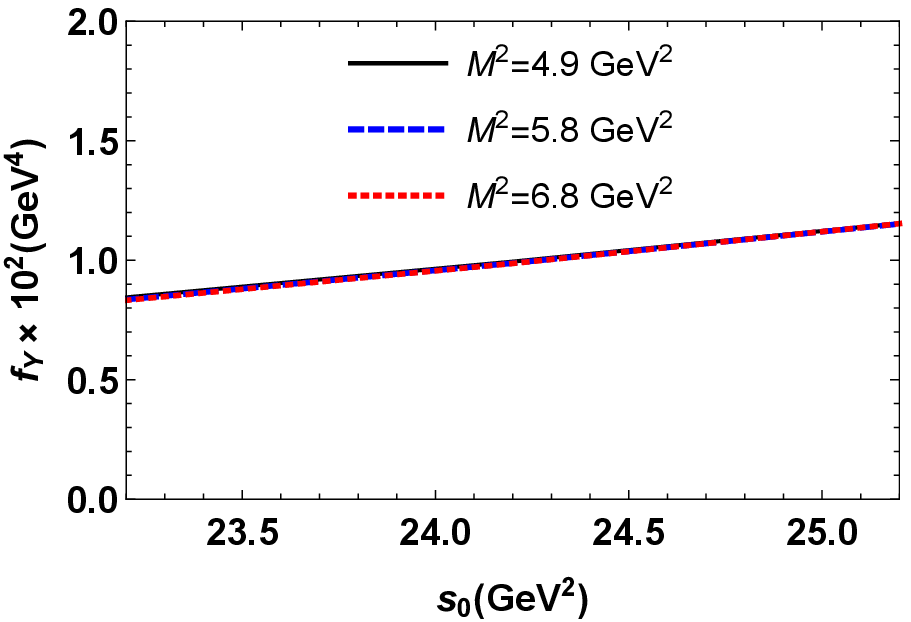}
\end{center}
\caption{ The same as in Fig.\ \ref{fig:MassY} but for the coupling $f_{Y}$.}
\label{fig:CouplingY}
\end{figure}

\end{widetext}

In Figs.\ \ref{fig:MassY} and \ref{fig:CouplingY} we depict $m_{Y}$ and $%
f_{Y}$ as functions of the parameters $M^{2}$ and $s_{0}$. It is evident
that variations of the mass and coupling on the Borel parameter are very
weak: predictions for $m_{Y}$ and $f_{Y}$ are stable against changes of $%
M^{2}$ within limits of the working region. But $m_{Y}$ and $f_{Y}$
demonstrate a sensitivity to the continuum threshold parameter $s_{0}$: this
dependence forms an essential part of ambiguities in obtained predictions,
which, however are within limits traditional for sum rule calculations. Then
for the mass and coupling of the resonance $Y(4660)$, we get
\begin{equation}
m_{Y}=4677_{-63}^{+71}~\mathrm{MeV},\ f_{Y}=(0.99\pm 0.16)\times 10^{-2}~%
\mathrm{GeV}^{4}.  \label{eq:CMass}
\end{equation}

The result for $m_{Y}$ is compatible with experimental data \cite%
{Wang:2014hta}. It is interesting to confront $m_{Y}$ with results of other
theoretical studies. We have noted above, that the mass of the resonance $Y$
was estimated using QCD sum rule method in various publications. Thus, in
Ref.\ \cite{Zhang:2010mw}, in which the authors examined $Y$ as $P$-wave
excitation of the scalar tetraquark $[cs][\overline{c}\overline{s}]$, its
mass was found equal to $m_{Y}=(4.69\pm 0.36)~\mathrm{GeV}$. As the vector
tetraquark $[cs][\overline{c}\overline{s}]$ the resonance $Y$ was considered
also in Ref.\ \cite{Albuquerque:2008up}, with the prediction
\begin{equation}
m_{Y}=(4.65\pm 0.10)~\mathrm{GeV}.
\end{equation}%
These results agree with experimental data, and, by taking into account
errors of calculations, are in accord also with our prediction.

Vector tetraquarks with $[cq][\overline{c}\overline{q}]$ or $[cs][\overline{c%
}\overline{s}]$ contents and charge conjugation parities $C=\pm $ were
studied in Ref.\ \cite{Chen:2010ze} as well. In this article sum rules for
the mass were calculated by including into analysis vacuum condensates up to
dimension $8$. For the tetraquark $[cs][\overline{c}\overline{s}]$ built of
the scalar diquark and vector antidiquark, the authors employed two currents
denoted there by $J_{1\mu }$ and $J_{3\mu }$, respectively. The prediction
obtained using the first current exceeds the mass of the resonance $Y$
\begin{equation}
m_{J_{1}}=(4.92\pm 0.10)~\mathrm{GeV},
\end{equation}%
whereas the second one underestimates it, and leads to
\begin{equation}
m_{J_{3}}=(4.52\pm 0.10)~\mathrm{GeV}.
\end{equation}%
These results do not coincide with data, and agree neither with our result
nor with prediction of Ref.\ \cite{Albuquerque:2008up} made by employing the
current Eq.\ (\ref{eq:Curr1Y}).

The $Y$ was assigned in Ref.\ \cite{Wang:2018rfw} to be the $C\otimes \gamma
_{\mu }C$-type vector tetraquark with the mass $m_{Y}=(4.66\pm 0.09)~\mathrm{%
GeV}$ and pole residue $\lambda _{Y}=(6.74\pm 0.88)\times 10^{-2}~\mathrm{GeV%
}^{5}$, which for the coupling $f_{Y}$ leads to $f_{Y}=(1.45\pm 0.19)\times
10^{-2}~\mathrm{GeV}^{4}$. The difference between this result and our
prediction (\ref{eq:CMass}) for $f_{Y}$ can be explained by assumptions on
the internal structure of the vector resonance $Y$. In fact, we treat $Y$ a
state built of a scalar diquark and vector antidiquark, whereas in Ref.\
\cite{Wang:2018rfw} it was considered as a bound state of a pseudoscalar
diquark and axial-vector antidiquark.

It is evident, that one can interpret the resonance $Y$ as vector
tetraquarks with the same content $[cs][\overline{c}\overline{s}]$, but
distinct internal organizations and interpolating currents. Therefore, there
is a necessity to analyze decay channels of the state $Y$ to make a choice
between existing models. In the next subsection we investigate strong decay
modes of $Y$, where $m_{Y}$ and $f_{Y}$ appear as input information.


\subsection{Strong decays of the tetraquark $Y$}

\label{subsec:Decays}

The strong decays of the tetraquark $Y$ can be determined by employing a
kinematical constraint which is evident from Eq.\ (\ref{eq:CMass}). We
consider $S$-wave decays of $Y$, therefore the spin and parity in these
processes are conserved. Performed analysis allows us to see that processes $%
Y\rightarrow J/\psi f_{0}(980),$ $\psi ^{\prime }f_{0}(980)$, $J/\psi
f_{0}(500),$\ and $\psi ^{\prime }f_{0}(500)$ are among important decay
modes of $Y$.

These decays in the final state contain scalar mesons $f_{0}(980)$ and $%
f_{0}(500)$, which will be considered as diquark-antidiquark states. A model
proposed in Ref.\ \cite{Kim:2017yvd} treats $f_{0}(980)$ and $f_{0}(500)$ as
superpositions of the basic states $\mathbf{L}=[ud][\overline{u}\overline{d}%
] $ and $\mathbf{H}=([su][\overline{s}\overline{u}]+[ds][\overline{d}s])/%
\sqrt{2}$. $\mathbf{\ }$Calculations carried out by employing this model
predicted the mass and full width of mesons $f_{0}(980)$ and $f_{0}(500)$
\cite{Agaev:2017cfz,Agaev:2018sco}, which are in reasonable agreement with
existing experimental data.

We consider here in a detailed form decays of the tetraquark $Y$ to mesons $%
J/\psi f_{0}(980)$ and $\psi ^{\prime }f_{0}(980)$, and compute the strong
couplings $g_{Y\psi f_{0}(980)}$ and $g_{Y\psi ^{\prime }f_{0}(980)}$
corresponding to the vertices $YJ/\psi f_{0}(980)$ and $Y\psi ^{\prime
}f_{0}(980)$, respectively. To this end, we use the LCSR method and analyze
the correlation function
\begin{equation}
\Pi _{\mu \nu }(p,q)=i\int d^{4}xe^{ipx}\langle f_{0}(q)|\mathcal{T}\{J_{\mu
}^{\psi }(x)J_{\nu }^{\dagger }(0)\}|0\rangle ,  \label{eq:CF2Y4600}
\end{equation}%
where $J_{\mu }^{\psi }(x)$ is the interpolating currents to $J/\psi $, and $%
\psi ^{\prime }$.

To extract from Eq.\ (\ref{eq:CF2Y4600}) the sum rules for $g_{Y\psi
f_{0}(980)}$ and $g_{Y\psi ^{\prime }f_{0}(980)}$, we first find $\Pi _{\mu
\nu }(p,q)$ in terms of the physical parameters of involved particles. \
After standard manipulations discussed in this review, we get
\begin{eqnarray}
&&\Pi _{\mu \nu }^{\mathrm{Phys}}(p,q)=\frac{g_{Y\psi f_{0}(980)}f_{\psi
}m_{\psi }f_{Y}m_{Y}}{\left( p^{\prime 2}-m_{Y}^{2}\right) \left(
p^{2}-m_{\psi }^{2}\right) }\left( -p_{\mu }^{\prime }p_{\nu }+\frac{%
m_{Y}^{2}+m_{\psi }^{2}}{2}g_{\mu \nu }\right)  \notag \\
&&+\frac{g_{Y\psi ^{\prime }f_{0}(980)}f_{\psi ^{\prime }}m_{\psi ^{\prime
}}f_{Y}m_{Y}}{\left( p^{\prime 2}-m_{Y}^{2}\right) \left( p^{2}-m_{\psi
^{\prime }}^{2}\right) }\left( -p_{\mu }^{\prime }p_{\nu }+\frac{%
m_{Y}^{2}+m_{\psi ^{\prime }}^{2}}{2}g_{\mu \nu }\right) +\cdots ,
\label{eq:CF3Y4600}
\end{eqnarray}%
where $m_{\psi }$, and $m_{\psi ^{\prime }}$ are the mass of the mesons $%
J/\psi $ and $\psi ^{\prime }$, respectively. The decay constants of these
mesons are denoted by $f_{\psi }$ and $f_{\psi ^{\prime }}$. In order to
derive sum rules for couplings $g_{Y\psi f_{0}(980)}$ and $g_{Y\psi ^{\prime
}f_{0}(980)}$, we use structures proportional $g_{\mu \nu }$ and
corresponding invariant amplitudes.

At the next stage of calculations, we express the correlation function using
the quark propagators, and obtain
\begin{eqnarray}
&&\Pi _{\mu \nu }^{\mathrm{OPE}}(p,q)=\int d^{4}xe^{ipx}\epsilon \widetilde{%
\epsilon }\left[ \gamma _{5}\widetilde{S}_{c}^{ic}(x){}\gamma _{\mu }%
\widetilde{S}_{c}^{ei}(-x){}\gamma _{\nu }\gamma _{5}-\gamma _{\nu }\gamma
_{5}\widetilde{S}_{c}^{ic}(x){}\gamma _{\mu }\widetilde{S}%
_{c}^{ei}(-x){}\gamma _{5}\right] _{\alpha \beta }  \notag \\
&&\times \langle f_{0}(q)|\overline{s}_{\alpha }^{b}(0)s_{\beta
}^{d}(0)|0\rangle ,  \label{eq:CF4Y4600}
\end{eqnarray}%
Our computations for the Borel transformed correlation function $\Pi ^{%
\mathrm{OPE}}(M^{2})$ give%
\begin{equation}
\Pi ^{\mathrm{OPE}}(M^{2})=\frac{\lambda _{f^{\prime }}}{24\pi ^{2}}%
\int_{4m_{c}^{2}}^{\infty }\frac{ds}{s}\sqrt{s(s-4m_{c}^{2})}%
(s+8m_{c}^{2})+\lambda _{f^{\prime }}\int_{0}^{1}d\alpha
e^{-m_{c}^{2}/M^{2}Z}F(\alpha ,M^{2}).  \label{eq:CF5Y4600}
\end{equation}%
\ Here, $\lambda _{f^{\prime }}$ is the matrix element
\begin{equation}
\langle f_{0}(980)(q)|\overline{s}(0)s(0)|0\rangle =\lambda _{f^{\prime }},
\end{equation}%
which has been calculated by employing the QCD two-point sum rule method in
Ref.\ \cite{Sundu:2018toi}. In Eq.\ (\ref{eq:CF5Y4600}) $F(\alpha ,M^{2})$
is a function that contains all nonperturbative contributions to $\Pi ^{%
\mathrm{OPE}}(M^{2})$ up to dimension $8$
\begin{eqnarray}
&&F(\alpha ,M^{2})=-\frac{\left\langle \alpha _{s}G^{2}/\pi \right\rangle
m_{c}^{2}}{72M^{4}}\frac{1}{Z}\left[ m_{c}^{2}\left( 1-2Z\right)
-M^{2}Z\left( 3-7Z\right) \right]  \notag \\
&&+\frac{\langle g_{s}^{3}G^{3}\rangle }{45\cdot 2^{9}\pi ^{2}M^{8}Z^{5}}%
\left\{ m_{c}^{6}(1-2\alpha
)^{2}(9-11Z)+2m_{c}^{2}M^{4}Z^{2}[-42+Z(122+9Z)]\right.  \notag \\
&&\left. -2M^{6}Z^{3}\left[ 6-Z(22-9Z)\right] +m_{c}^{4}M^{2}Z\left(
-11+119Z-190Z^{2}\right) \right\}  \notag \\
&&+\frac{\left\langle \alpha _{s}G^{2}/\pi \right\rangle ^{2}m_{c}^{2}\pi
^{2}}{648M^{10}Z^{3}}\left[ m_{c}^{4}-m_{c}^{2}M^{2}(1+4Z)+2M^{4}Z(2-Z)%
\right] ,  \label{eq:Func}
\end{eqnarray}%
where $Z=\alpha (1-\alpha )$.

After equating the Borel transform of the invariant amplitudes $\Pi ^{%
\mathrm{Phys}}(p^{2})$ and $\Pi ^{\mathrm{OPE}}(M^{2})$, and carrying out
continuum subtraction, we obtain an expression that contains two unknown
variables $g_{YJf_{0}(980)}$ and $g_{Y\Psi f_{0}(980)}$. It is worth noting
that continuum subtraction in the perturbative part has to done by $\infty
\rightarrow s_{0\text{ }}$ replacement. As is seen, all terms in Eq.\ (\ref%
{eq:Func}) are proportional to inverse powers of the parameter $M^{2}$,
therefore the nonperturbative contributions should be left in an
unsubtracted form \cite{Belyaev:1994zk}. The second equality can be found by
applying the operator $d/d(-1/M^{2})$ to the first expression. By solving
these equalities it is possible to extract sum rules for $g_{YJf_{0}(980)}$
and $g_{Y\Psi f_{0}(980)}$.

The similar analysis is valid also for decays of $Y$ to $J/\psi f_{0}(500)$,
and $\psi ^{\prime }f_{0}(500)$. The width of the decays under analysis can
be evaluated by means of the formula (\ref{eq:DW}), where instead of
parameters $g_{Z\psi \pi }$, $m_{\psi }$, and $\lambda \left( m_{Z},\
m_{\psi },m_{\pi }\right) $ one has to use $g_{Y\psi f_{0}}$,$\ m_{\psi }$,
and $\lambda (m_{Y},m_{\psi },m_{f_{0}})$: here, $\psi $ and $f_{0}$ are one
of the mesons $J/\psi $,$\ \psi ^{\prime }$ and $f_{0}(500)$, $\ f_{0}(980)$%
, respectively.

The numerical computations are fulfilled by employing the vacuum condensates
given in Eq.\ (\ref{eq:VCond}), and using the mass and decay constant of the
mesons $J/\psi $ and $\psi ^{\prime }$ (see, Table \ref{tab:Param}). The
parameters of the resonance $Y$ have been found in the previous subsection,
and for the mass of the $f_{0}(980)$ meson we use its experimental value $%
m_{f_{0}}=(990\pm 20)$ $\mathrm{MeV}$.

The parameters $M^{2}$ and $s_{0}$ are changed in the regions $M^{2}\in
\lbrack 4.9,\ 6.8]~\mathrm{GeV}^{2}$ and $s_{0}\in \lbrack 23.2,\ 25.2]~%
\mathrm{GeV}^{2}$. The results obtained for the strong couplings read%
\begin{equation}
|g_{YJf_{0}(980)}|=(0.22\pm 0.07)~\mathrm{GeV}^{-1},\ g_{Y\Psi
f_{0}(980)}=(1.22\pm 0.33)~\mathrm{GeV}^{-1}.  \label{eq:SCoupl1}
\end{equation}%
Then partial widths of the decay modes under analysis become equal to (in
units of $\mathrm{MeV}$):%
\begin{equation}
\Gamma (Y\rightarrow J/\psi f_{0}(980))=18.8\pm 5.4,\ \Gamma (Y\rightarrow
\psi ^{\prime }f_{0}(980))=30.2\pm 8.5.  \label{eq:DW1Y}
\end{equation}%
Exploration of the next decays does not differ from previous analysis and
gives the  following predictions
\begin{equation}
g_{YJf_{0}(500)}=(0.07\pm 0.02)~\mathrm{GeV}^{-1},\ |g_{Y\Psi
f_{0}(500)}|=(0.18\pm 0.05)~\mathrm{GeV}^{-1},  \label{eq:SCoupl2}
\end{equation}%
and (in $\mathrm{MeV}$)
\begin{equation}
\Gamma (Y\rightarrow J/\psi f_{0}(500))=2.7\pm 0.7,\ \Gamma (Y\rightarrow
\psi ^{\prime }f_{0}(500))=13.1\pm 3.7.  \label{eq:DW2Y}
\end{equation}%
The full width of the resonance $Y$ evaluated by taking into account these
four strong decay modes%
\begin{equation}
\Gamma _{Y}=(64.8\pm 10.8)~\mathrm{MeV}  \label{eq:TWD}
\end{equation}%
agrees with the experimental result $(68\pm 11\pm 1)~\mathrm{MeV}$ from Eq.\
(\ref{eq:SpPar}). The Particle Data Group for the full width of $Y$ provides
the world averaged estimate $\Gamma _{Y}=(72\pm 11)~\mathrm{MeV}$ \cite%
{Tanabashi:2018oca}, which exceeds (\ref{eq:SpPar}). But the result Eq.\ (%
\ref{eq:TWD}) within errors of computations and experimental measurements is
consistent also with the world average. We also take into account that in
the diquark-antidiquark model there are other $S$-wave decay channels $%
Y\rightarrow D_{s}^{\pm }D_{s1}^{\mp }(2460)$ and $Y\rightarrow D_{s}^{\ast
\pm }D_{s0}^{\mp }(2317)$ of the resonance $Y$ which contribute to $\Gamma
_{Y}$ and may improve this agreement.

We have calculated the full width of the resonance $Y$ by taking into
account the strong decays $Y\rightarrow J/\psi f_{0}(500)$, $\psi ^{\prime
}f_{0}(500)$, $J/\psi f_{0}(980)$ and $\psi ^{\prime }f_{0}(980)$. However,
only the process $Y\rightarrow \psi ^{\prime }\pi ^{+}\pi ^{-}$ was observed
experimentally. It is known that the dominant decays of $\ $the $f_{0}(500)$
and $f_{0}(980)$ mesons are processes $f_{0}\rightarrow \pi ^{+}\pi ^{-}$
and $f_{0}\rightarrow \pi ^{0}\pi ^{0}$. Therefore, the chains $Y\rightarrow
\psi ^{\prime }f_{0}(980)\rightarrow \psi ^{\prime }\pi ^{+}\pi ^{-}$ and $%
Y\rightarrow \psi ^{\prime }f_{0}(500)\rightarrow \psi ^{\prime }\pi ^{+}\pi
^{-}$ explain a dominance of the observed $\psi ^{\prime }\pi ^{+}\pi ^{-}$
final state in the decay of the resonance $Y$. In the diquark-antidiquark
model the width of the mode $Y\rightarrow J/\psi f_{0}(980)$ is
considerable. Besides, decays to mesons $\psi ^{\prime }\pi ^{0}\pi ^{0}$
and $J/\psi \pi ^{0}\pi ^{0}$ have to be detected as well. But neither $%
J/\psi \pi ^{+}\pi ^{-}$ nor $\pi ^{0}\pi ^{0}$ were seen in decays of $Y$.
It is interesting that aforementioned final-state particles were observed in
decays of the vector resonance $Y(4260)$: its decays to $J/\psi \pi ^{+}\pi
^{-}$ and $J/\psi \pi ^{0}\pi ^{0}$ as well as to $J/\psi K^{+}K^{-}$ were
discovered in experiments. Therefore, more accurate measurements may fix
these modes in decays of the resonance $Y$ as well.


\section{The light resonances $X(2100)$ and $X(2239)$}

\label{sec:4s}

In previous sections we have explored heavy resonances which are candidates
to exotic four-quark mesons. They are heavy particles and contain a pair of $%
c\overline{c}$ quarks. Only small number of resonances observed
experimentally may be interpreted as multiquark mesons composed exclusively
of light quarks. The famous resonance $Y(2175)$ seen by BaBar in the process
$e^{+}e^{-}\rightarrow \gamma _{\mathrm{ISR}}\phi f_{0}(980)$ \cite%
{Aubert:2006bu} is one of such states. It was fixed as a resonant structure
in the $\phi f_{0}(980)$ invariant mass spectrum. The BESII, Belle, and
BESIII collaborations observed this state as well \cite%
{Ablikim:2007ab,Shen:2009zze,Ablikim:2014pfc}. The mass and width of the
resonance $Y(2175)$ with spin-parities $J^{\mathrm{PC}}=1^{--}$ are $%
m=(2175\pm 10\pm 15)~\mathrm{MeV}$ and $\Gamma =(58\pm 16\pm 20)~\mathrm{MeV}
$, respectively.

Some other light resonances that can be considered as four-quark states were
discovered recently by BESIII. One of them, i.e., $X(2239)$ was fixed in the
cross section's lineshape of the process $e^{+}e^{-}\rightarrow K^{+}K^{-}$
\cite{Ablikim:2018iyx}. The mass and width of this resonant structure are
equal to $m=(2239.2\pm 7.1\pm 11.3)~\mathrm{MeV}$ and $\Gamma =(139.8\pm
12.3\pm 20.6)~\mathrm{MeV}$, respectively. The $X(2100)$ was discovered in
the process $J/\psi \rightarrow \phi \eta \eta ^{\prime }$ as a resonance in
the $\phi \eta ^{\prime }$ mass spectrum \cite{Ablikim:2018xuz}. The BESIII
explored angular distribution of $J/\psi \rightarrow X(2100)\eta $, but
because of limited statistics could not distinguish $1^{+}$ and $1^{-}$
options for the spin-parity $J^{\mathrm{P}}$ of the $X(2100)$. \ Therefore,
the mass and width of this state were extracted by employing both of these
options. For $J^{P}=1^{-}$ the mass and width of the $X(2100)$ were
extracted to be $m=(2002.1\pm 27.5\pm 21.4)~\mathrm{MeV}$ and $\Gamma
=(129\pm 17\pm 9)~\mathrm{MeV}$. In the case $J^{\mathrm{P}}=1^{+}$ BESIII
found $m=(2062.8\pm 13.1\pm 7.2)~\mathrm{MeV}$ and $\Gamma =(177\pm 36\pm
35)~\mathrm{MeV}$.

Almost all models and methods of the high energy physics were used to
understand structures of these light resonances. Because $Y(2175)$ was
observed for the first time more than ten years ago, there are various
articles in literature, in which it was investigated thoroughly. The $%
Y(2175) $ was interpreted as $2{}^{3}D_{1}$ excited state of the ordinary
meson $\overline{s}s$ \cite{Ding:2007pc,Wang:2012wa}. It was considered as a
dynamically generated $\phi K\overline{K}$ system \cite%
{MartinezTorres:2008gy}, or a system appeared due to self-interaction
between $\phi $ and $f_{0}(980)$ mesons \cite{AlvarezRuso:2009xn}. Other
models suggest to explain the resonance $Y(2175)$ as a hybrid meson $%
\overline{s}sg$, or a baryon-antibaryon $qqs\overline{q}\overline{q}%
\overline{s}$ state that couples strongly to the $\Lambda \overline{\Lambda }
$ channel (see Ref.\ \cite{Ablikim:2018iyx} for other models).

As a vector tetraquark with $s\overline{s}s\overline{s}$ or $ss\overline{s}%
\overline{s}$ contents $Y(2175)$ was examined in Refs.\ \cite{Wang:2006ri}
and \cite{Chen:2008ej,Chen:2018kuu}, respectively. An alternative suggestion
on nature of this state was made in Ref.\ \cite{Agaev:2019coa}, where it was
interpreted as a vector diquark-antidiquark system with the content $sq%
\overline{s}\overline{q}$. The resonances $X(2100)$ and $X(2239)$ (hereafter
$X_{1}$ and $X_{2}$, respectively) were explored as vector or axial-vector
tetraquarks as well. Indeed, in Ref.\ \cite{Lu:2019ira} the $ss\overline{s}%
\overline{s}$ four-quark compounds were studied within the relativized quark
model. The authors made a conclusion that the resonance $X_{2}$ can be
considered as a $P$-wave $ss\overline{s}\overline{s}$ tetraquark. The $X_{1}$
was investigated within framework the QCD sum rule method in Refs.\ \cite%
{Cui:2019roq,Wang:2019nln}. Results of these analyses can be explained by
interpreting $X_{1}$ as the axial-vector $ss\overline{s}\overline{s}$
tetraquark with $J^{PC}=1^{+-}$. An alternative explanation of $X_{1}$ as
the second radial excitation of the meson $h_{1}(1380)$ was suggested in
Ref.\ \cite{Wang:2019qyy}.

In our article \cite{Azizi:2019ecm}, we explored the light axial-vector $T_{%
\mathrm{AV}}$ and vector $T_{\mathrm{V}}$ tetraquarks $ss\overline{s}%
\overline{s}$ and calculated their spectroscopic parameters. It appears, the
resonance $X_{1}$ may be considered as a axial-vector tetraquark: we
identified $X_{1}$ with $T_{\mathrm{AV}}$. Among the vector particles $%
Y(2175)$ and $X_{2}$, which we treated as different resonances, parameters
of the latter are closer to our predictions. Therefore, we interpreted the
resonance $X_{2}$ as the tetraquark $T_{\mathrm{V}}$. We evaluated also
width of the decays $X_{1}\rightarrow \phi \eta ^{\prime }$ and $%
X_{1}\rightarrow \phi \eta $ which are essential for our interpretation of $%
X_{1}$. Presentation in this section is based on Ref. \cite{Azizi:2019ecm}.


\subsection{Mass and coupling of the axial-vector tetraquark $ss\overline{s}%
\overline{s}$}

\label{subsec:Mass1}
In this subsection, we calculate the spectroscopic parameters of the
axial-vector tetraquark $T_{\mathrm{AV}}=ss\overline{s}\overline{s}$ using
the QCD sum rule method. We consider the two-point correlation function $\Pi
_{\mu \nu }(p)$ given by Eq.\ (\ref{eq:CorrF1}), with $J_{\mu }(x)$ being
the interpolating current for the axial-vector tetraquark $T_{\mathrm{AV}}$.
The choice of $J_{\mu }(x)$ is a main decision in our analysis, because $T_{%
\mathrm{AV}}$ with spin-parities $J^{\mathrm{PC}}=1^{+-}$ can be modeled by
employing various currents. The current that leads to credible results for
parameters of $T_{\mathrm{AV}}$ is given by the expression \cite{Cui:2019roq}
\begin{equation}
J_{\mu }(x)=\left[ s_{a}^{T}(x)C\gamma ^{\nu }s_{b}(x)\right] \left[
\overline{s}_{a}(x)\sigma _{\mu \nu }\gamma _{5}C\overline{s}_{b}^{T}(x)%
\right] -\left[ s_{a}^{T}(x)C\sigma _{\mu \nu }\gamma _{5}s_{b}(x)\right] %
\left[ \overline{s}_{a}(x)\gamma ^{\nu }C\overline{s}_{b}^{T}(x)\right] .
\label{eq:Curr1X1}
\end{equation}

The sum rules for the mass $m$ and coupling $f$ of the tetraquark $T_{%
\mathrm{AV}}$ can be obtained in accordance with prescriptions of the
method. First, we should rewrite the correlation function $\Pi _{\mu \nu
}(p) $ by utilizing the physical parameters of $T_{\mathrm{AV}}$. After some
operations for the physical side of the sum rules, we obtain
\begin{equation}
\Pi _{\mu \nu }^{\mathrm{Phys}}(p)=\frac{m^{2}f^{2}}{m^{2}-p^{2}}\left(
-g_{\mu \nu }+\frac{p_{\mu }p_{\nu }}{m^{2}}\right) +\cdots .
\label{eq:CorF1}
\end{equation}%
The correlation function $\Pi _{\mu \nu }(p)$ calculated using the quark
propagators forms the QCD side of the sum rules. It is determined by the
formula
\begin{eqnarray}
&&\Pi _{\mu \nu }^{\mathrm{OPE}}(p)=\frac{i}{4}\int d^{4}xe^{ipx}\left\{
\mathrm{Tr}\left[ \gamma ^{\alpha }\widetilde{S}^{a^{\prime }b}(-x)\gamma
^{\beta }S^{b^{\prime }a}(-x)\right] \mathrm{Tr}\left[ S^{ab^{\prime
}}(x)\gamma _{\nu }\gamma _{\beta }\gamma _{5}\widetilde{S}^{ba^{\prime
}}(x)\gamma _{5}\gamma _{\mu }\gamma _{\alpha }\right] \right.  \notag \\
&&\left. -\mathrm{Tr}\left[ \gamma ^{\alpha }\widetilde{S}^{bb^{\prime
}}(-x)\gamma ^{\beta }S^{a^{\prime }a}(-x)\right] \mathrm{Tr}\left[
S^{ab^{\prime }}(x)\gamma _{\nu }\gamma _{\beta }\gamma _{5}\widetilde{S}%
^{ba^{\prime }}(x)\gamma _{5}\gamma _{\mu }\gamma _{\alpha }\right] +62\
\mathrm{similar\ terms}\right\} .  \label{eq:CF3}
\end{eqnarray}%
In these computations, we employ the $x$-space light-quark propagator
\begin{equation}
S^{ab}(x)\Rightarrow S^{ab}(x)+\frac{m_{s}g_{s}}{32\pi ^{2}}G_{ab}^{\mu \nu
}\sigma _{\mu \nu }\left[ \ln \left( \frac{-x^{2}\Lambda ^{2}}{4}\right)
+2\gamma _{E}\right] +\cdots ,  \label{eq:QProp}
\end{equation}%
where $\gamma _{E}\simeq 0.577$ is the Euler constant, and $\Lambda $ is the
QCD scale parameter. The scale parameter $\Lambda $ can be fixed inside of
the region $[0.5,1]~\mathrm{GeV}$; we use the central value $\Lambda =0.75\
\mathrm{GeV}$. We introduce also the notation $G_{ab}^{\mu \nu }\equiv
G_{A}^{\mu \nu }t_{ab}^{A},\ A=1,2,\ldots 8$, and $t^{A}=\lambda ^{A}/2$,
with $\lambda ^{A}$ being the Gell-Mann matrices.

At the next phase, we compute four-$x$ Fourier integrals appeared in $\Pi
_{\mu \nu }^{\mathrm{OPE}}(p)$ due to propagators. The correlation function $%
\Pi _{\mu \nu }^{\mathrm{OPE}}(p)$ found by this manner contains two Lorentz
structures. To extract the sum rules, we work with terms proportional to $%
g_{\mu \nu }$, because they do not receive contributions from scalar
particles. By equating the relevant invariant amplitudes $\Pi ^{\mathrm{Phys}%
}(p^{2})$ and $\Pi ^{\mathrm{OPE}}(p^{2})$, we get an expression in momentum
space, which after applying the Borel transformation and subtracting
continuum effects becomes the first sum rule equality. An expression
obtained after these operations depends on the Borel $M^{2}$ and continuum
subtraction $s_{0}$ parameters.

A second equality which is necessary to get sum rules for $m$ and $f$, can
be derived by applying\ the operator $d/d\left( -1/M^{2}\right) $ to the
first expression. These two equalities are enough to fix the sum rules for $%
m $ and $f$ . Obtained expressions for the mass and coupling of $T_{\mathrm{%
AV}}$ contains perturbative and nonperturbative parts. In numerical
computations, we take into account nonperturbative terms up to dimension-$20$%
, and bear in mind that higher dimensional contributions appear due to the
factorization hypothesis as product of basic condensates, and do not embrace
all dimension-$20$ terms.
\begin{figure}[h]
\includegraphics[width=8.8cm]{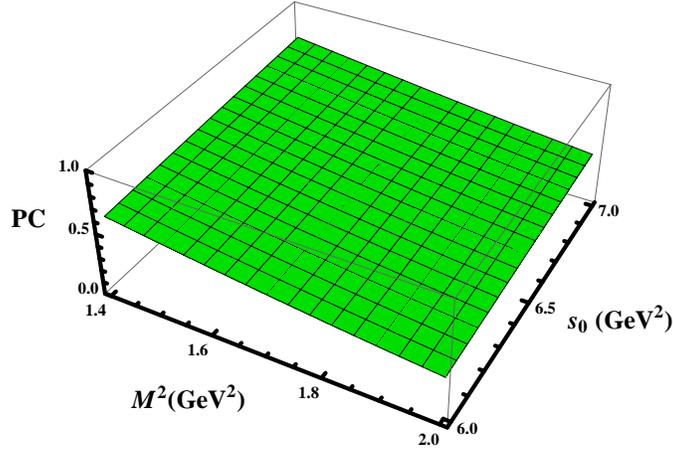}
\caption{The pole contribution vs $M^{2}$ and $s_{0}$. }
\label{fig:PC}
\end{figure}
\begin{figure}[h]
\includegraphics[width=8.8cm]{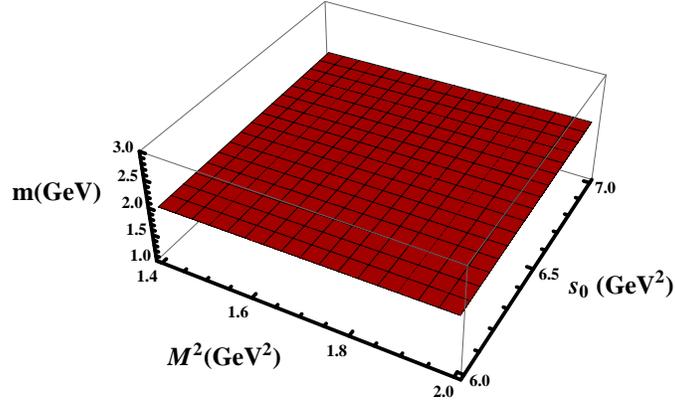}
\caption{ The dependence of the mass $m$ on $M^{2}$ and $s_{0}$.}
\label{fig:MassAV}
\end{figure}

Traditionally an important question is a choice of regions for the Borel $%
M^{2}$ and continuum threshold $s_{0}$ parameters. These parameters should
meet some known requirements. Our investigations prove that the regions
\begin{equation}
M^{2}\in \lbrack 1.4,\ 2]~\mathrm{GeV}^{2},\ s_{0}\in \lbrack 6,\ 7]~\mathrm{%
GeV}^{2},  \label{eq:Wind1}
\end{equation}%
obey required constraints. Indeed, at $M^{2}=2~\mathrm{GeV}^{2}$ the pole
contribution is equal to $0.39$, and reaches $\mathrm{PC=}$ $0.68$ at the
minimum of $M^{2}=1.4~\mathrm{GeV}^{2}$. In Fig.\ \ref{fig:PC} we plot the
pole contribution as functions of $M^{2}$ and $s_{0}$, where these effects
are seen explicitly. Convergence of the sum rules is also satisfied $R(1.4~%
\mathrm{GeV}^{2})<0.01$. The predictions for the mass and coupling of the
tetraquark $T_{\mathrm{AV}}$ are
\begin{equation}
m=\left( 2067\pm 84\right) ~\mathrm{MeV},\ \ f=\left( 0.89\pm 0.11\right)
\times 10^{-2}~\mathrm{GeV}^{4}.  \label{eq:Result1}
\end{equation}

Obtained results should not depend on the parameter $M^{2}$. But $m$ and $f$
evaluated from relevant sum rules are sensitive to a choice of $M^{2}$.
Theoretical errors in computations appear namely due to choices of $M^{2}$
and $s_{0}$. Errors generated by ambiguities of $m_{s}$ and vacuum
condensates are not considerable. Varying of $m_{s}$, for example, inside
boundaries $88~\mathrm{MeV}\leq m_{s}\leq 104~\mathrm{MeV}\ $generates
corrections $\binom{+2~}{-1}~\mathrm{MeV}$ to $m$ and $\binom{+0.0002}{%
-0.0001}$ $~\mathrm{GeV}^{4}$ to $f$. Of course, these errors and others
connected with condensates are included into Eq.\ (\ref{eq:Result1}). The
mass $m$ is depicted in Fig.\ \ref{fig:MassAV}, where one sees its weak
dependence on $M^{2}$ and $s_{0}$.

The prediction for the mass of $T_{\mathrm{AV}}$ agrees with the mass of the
resonance $X_{1}$ measured by BESIII. Therefore, it is reasonable to
identify $T_{\mathrm{AV}}$ with the resonance $X_{1}$. This is in line with
existing theoretical predictions for $m$ extracted from the QCD sum rule
computations. In fact, the mass of $X_{1}$ was found in Refs.\ \cite%
{Cui:2019roq,Wang:2019nln} equal to
\begin{equation}
m=2000_{-90}^{+100}~\mathrm{MeV},\ m=\left( 2080\pm 120\right) ~\mathrm{MeV},
\label{eq:Previous}
\end{equation}%
where computations were performed by taking into account condensates untill
dimensions $12$ and $10$, respectively. There are differences between
results (\ref{eq:Result1}) and (\ref{eq:Previous}), but all of them support
a suggestion about an diquark-antidiquark structure of the axial-vector
resonance $X_{1}$. But to unveil a whole picture, we should explore decay
channels $X_{1}\rightarrow \phi \eta ^{\prime }$ and $X_{1}\rightarrow \phi
\eta $ to find width of $X_{1}$ and compare it with experimental
information: only after successful outcome, we will be able to make firm
decision about structure of $X_{1}$. In this section, we are going to
analyze this problem as well.


\subsection{ Spectroscopic parameters of the vector tetraquark $ss\overline{s%
}\overline{s}$}

\label{subsec:Mass2}

We have investigated the axial-vector tetraquark $T_{\mathrm{AV}}$ and
classified it as a candidate to the resonance $X_{1}$. But there are light
resonances $Y(2175)$ and $X_{2}$ which may be interpreted as four-quark
states. Here, we study the vector tetraquark $T_{\mathrm{V}}=ss\overline{s}%
\overline{s}$ with spin-parities $J^{\mathrm{PC}}=1^{--}$ and compute its
mass. After confronting our result with the experimental data of the BaBar
and BESIII collaborations, we can identify $Y(2175)$ or $X_{2}$ as the state
$T_{\mathrm{V}}$.

\begin{widetext}

\begin{figure}[h!]
\begin{center}
\includegraphics[%
totalheight=6cm,width=8cm]{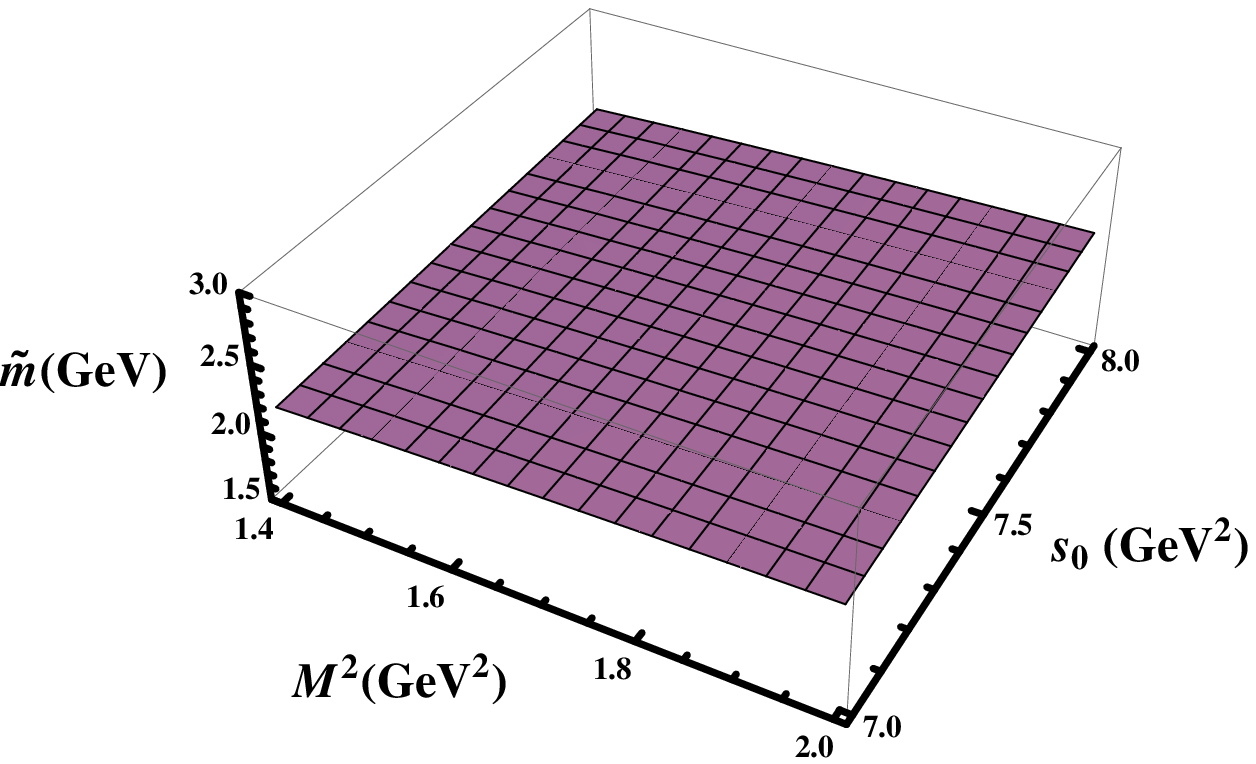}
\includegraphics[
totalheight=6cm,width=8cm]{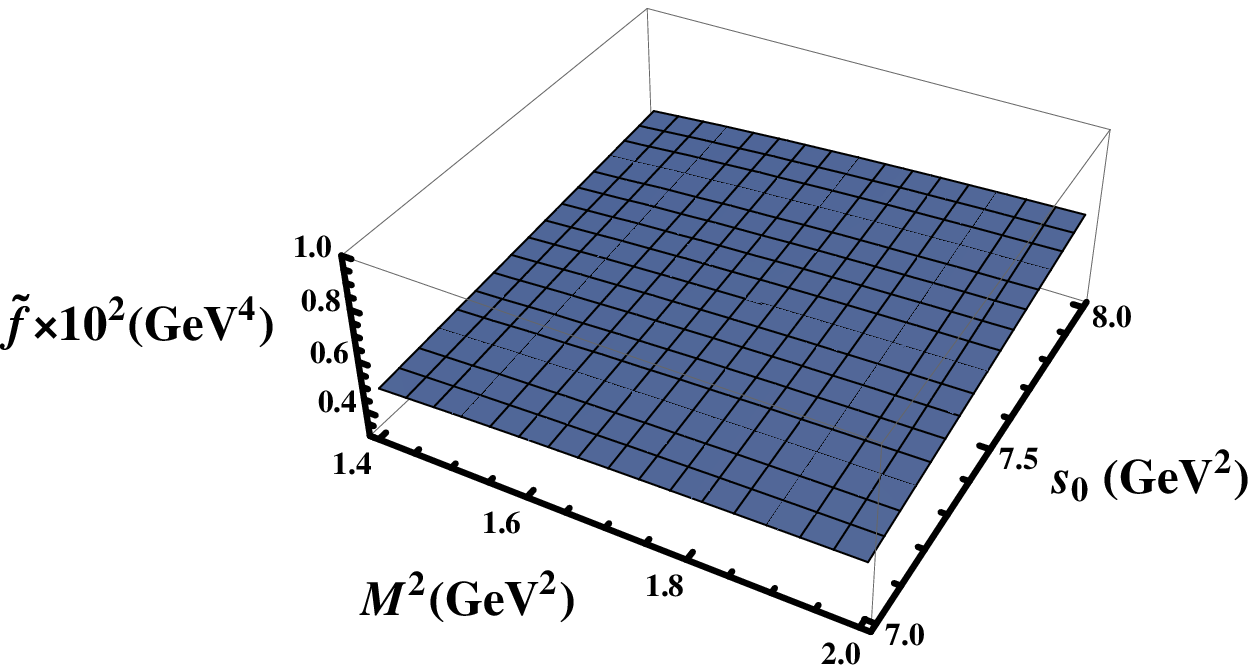}
\end{center}
\caption{ The $\widetilde{m}$ (left panel) and $\widetilde{f}$ (right panel) as functions
of the Borel and continuum threshold  parameters.}
\label{fig:MassCoupl2}
\end{figure}

\end{widetext}

The mass $\widetilde{m}$ and coupling $\widetilde{f}$ of the tetraquark $T_{%
\mathrm{V}}$ can be found using standard tools of the sum rule method. This
analysis does not differ significantly from operations performed above. A
difference in the case under consideration stems from a choice of the
interpolating current $\widetilde{J}_{\mu }(x)$, which for the vector
tetraquark is given by the formula \cite{Chen:2018kuu}
\begin{equation}
\widetilde{J}_{\mu }(x)=\left[ s_{a}^{T}(x)C\gamma _{5}s_{b}(x)\right] \left[
\overline{s}_{a}(x)\gamma _{\mu }\gamma _{5}C\overline{s}_{b}^{T}(x)\right] -%
\left[ s_{a}^{T}(x)C\gamma _{\mu }\gamma _{5}s_{b}(x)\right] \left[
\overline{s}_{a}(x)\gamma _{5}C\overline{s}_{b}^{T}(x)\right] .
\label{eq:Curr2X2}
\end{equation}%
We should determine both sides of the sum rule equalities. The physical side
of the sum rule is fixed by Eq.\ (\ref{eq:CorF1}) with evident replacements.
The QCD side of the sum rule $\widetilde{\Pi }_{\mu \nu }^{\mathrm{OPE}}(p)$
has the form
\begin{eqnarray}
&&\widetilde{\Pi }_{\mu \nu }^{\mathrm{OPE}}(p)=i\int d^{4}xe^{ipx}\left\{
\mathrm{Tr}\left[ \gamma _{5}\widetilde{S}^{b^{\prime }b}(-x)\gamma
_{5}\gamma _{\nu }S^{a^{\prime }a}(-x)\right] \mathrm{Tr}\left[
S^{aa^{\prime }}(x)\gamma _{5}\widetilde{S}^{bb^{\prime }}(x)\gamma
_{5}\gamma _{\mu }\right] \right.  \notag \\
&&\left. -\mathrm{Tr}\left[ \gamma _{5}\widetilde{S}^{a^{\prime
}b}(-x)\gamma _{\nu }\gamma _{5}S^{b^{\prime }a}(-x)\right] \mathrm{Tr}\left[
S^{aa^{\prime }}(x)\gamma _{5}\widetilde{S}^{bb^{\prime }}(x)\gamma
_{5}\gamma _{\mu }\right] +14\ \mathrm{similar\ terms}\right\} .
\label{eq:CF4}
\end{eqnarray}%
The regions for the Borel and continuum threshold parameters $M^{2}$ and $%
s_{0}$ are given by the intervals
\begin{equation}
M^{2}\in \lbrack 1.4,\ 2]~\mathrm{GeV}^{2},\ s_{0}\in \lbrack 7,\ 8]~\mathrm{%
GeV}^{2}.  \label{eq:Wind2}
\end{equation}%
These regions differ from working windows of the axial-vector state (\ref%
{eq:Wind1}) by a small shift in $s_{0}$. The regions (\ref{eq:Wind2}) meet
all restrictions on the $\mathrm{PC}$ and convergence of OPE imposed by the
sum rule method. In fact, at $M^{2}=1.4~\mathrm{GeV}^{2}$ the $\mathrm{PC}$
amounts to $0.6$, and at $M^{2}=2~\mathrm{GeV}^{2}$ is equal to $0.3$.
Convergence of OPE is also fulfilled. The spectroscopic parameters of the
vector tetraquark $T_{\mathrm{V}}$ are
\begin{equation}
\widetilde{m}=\left( 2283\pm 114\right) ~\mathrm{MeV},\ \ \widetilde{f}%
=\left( 0.57\pm 0.10\right) \times 10^{-2}~\mathrm{GeV}^{4}.
\label{eq:Result2}
\end{equation}%
In Fig.\ \ref{fig:MassCoupl2} we depict $\widetilde{m}$ and $\widetilde{f}$
as functions of $M^{2}$ and $s_{0}$.

Confronting now $\widetilde{m}$ with collected data on the resonances $%
Y(2175)$ and $X_{2}$, we see that $T_{\mathrm{V}}$ can be interpreted as the
resonance $X_{2}$. Indeed, a mass gap $T_{\mathrm{V}}-$ $X_{2}$ is
approximately $60~\mathrm{MeV}$ smaller than mass splitting of $T_{\mathrm{V}%
}$ and $Y(2175)$. The mass $m_{X_{2}}=2227~\mathrm{MeV}$ of the vector
tetraquark $ss\overline{s}\overline{s}$ found in Ref.\ \cite{Lu:2019ira}
also agrees with BESIII data for $X_{2}$. This fact forced the authors to
draw the similar conclusion about internal organization of $X_{2}$.

Parameters of the vector tetraquark $ss\overline{s}\overline{s}$ were also
calculated in the context the sum rule method in Refs.\ \cite{Wang:2019nln}
and \cite{Chen:2018kuu}. The result for the mass of this exotic meson $%
\widetilde{m}=(3080\pm 110)~\mathrm{MeV}$ obtained in Ref.\ \cite%
{Wang:2019nln} disfavors interpreting it as the resonance $Y(2175)$.
Confronting this prediction with the BESIII data, we see that it is also
difficult to classify this structure as the resonance $X_{2}$. In Ref.\ \cite%
{Chen:2018kuu} the authors employed two interpolating currents to study
the vector tetraquark $ss\overline{s}\overline{s}$. For $\widetilde{m}$ these
currents led to different values
\begin{equation}
\widetilde{m}_{1}=(2410\pm 250)~\mathrm{MeV},\ \widetilde{m}_{2}=(2340\pm
170)~\mathrm{MeV}.  \label{eq:PreviousA}
\end{equation}%
The first tetraquark was interpreted as a structure at around $2.4~\mathrm{%
GeV}$ in the invariant mass spectrum $\phi f_{0}(980)$ \cite{Chen:2018kuu}.
The second structure was identified with the resonance $Y(2175)$. We think
that it is closer to $X_{2}$, which was discovered later and the authors did
not know about existence of this resonance.


\subsection{Decays $X_{1}\rightarrow \protect\phi \protect\eta ^{\prime }$
and $X_{1}\rightarrow \protect\phi \protect\eta $}

\label{subsec:Decay}

In this subsection we investigate decays $X_{1}\rightarrow \phi \eta
^{\prime }$ and $X_{1}\rightarrow \phi \eta $ of the resonance $X_{1}$. We
concentrate on the process $X_{1}\rightarrow \phi \eta ^{\prime }$ and
present our studies in a detailed form. The second mode $X_{1}\rightarrow
\phi \eta $ can be considered in the same way.

We begin from analysis of the correlation function
\begin{equation}
\widehat{\Pi }_{\mu \nu }(p,q)=i\int d^{4}xe^{ipx}\langle \eta ^{\prime }(q)|%
\mathcal{T}\{J_{\mu }^{\phi }(x)J_{\nu }^{\dagger }(0)\}|0\rangle ,
\label{eq:CF5}
\end{equation}%
which is necessary to study the decay $X_{1}\rightarrow \phi \eta ^{\prime }$%
. In Eq.\ (\ref{eq:CF5}) $J_{\mu }^{\phi }(x)$ is the interpolating current
of the $\phi $ meson
\begin{equation}
J_{\mu }^{\phi }(x)=\overline{s}_{i}(x)\gamma _{\mu }s_{i}(x).
\label{eq:Curr3Phi}
\end{equation}%
Following the standard recipes, we write down $\widehat{\Pi }_{\mu \nu
}(p,q) $ in terms of the physical parameters of the particles $X_{1},\ \phi $
and $\eta ^{\prime }$%
\begin{equation}
\widehat{\Pi }_{\mu \nu }^{\mathrm{Phys}}(p,q)=\frac{\langle 0|J_{\mu
}^{\phi }(x)|\phi (p)\rangle }{p^{2}-m_{\phi }^{2}}\langle \phi (p)\eta
^{\prime }(q)|X_{1}(p^{\prime })\rangle \frac{\langle X_{1}(p^{\prime
})|J_{\nu }^{\dagger }|0\rangle }{p^{\prime 2}-m^{2}}+\cdots ,
\label{eq:CF5A}
\end{equation}%
where the momenta of the initial and final particles are denoted by $%
p^{\prime }$ and $p$, $q$, respectively. By utilizing the matrix elements
\begin{equation}
\langle 0|J_{\mu }^{\phi }(x)|\phi (p)\rangle =f_{\phi }m_{\phi }\varepsilon
_{\mu },\ \ \langle \phi (p)\eta ^{\prime }(q)|X_{1}(p^{\prime })\rangle
=g_{X_{1}\phi \eta ^{\prime }}\left[ (p\cdot p^{\prime })(\varepsilon ^{\ast
}\cdot \varepsilon ^{\prime })-(p\cdot \varepsilon ^{\prime })(p^{\prime
}\cdot \varepsilon ^{\ast })\right] ,  \label{eq:MelX}
\end{equation}%
one can simplify $\widehat{\Pi }_{\mu \nu }^{\mathrm{Phys}}(p,q)$. The
matrix element $\langle 0|J_{\mu }^{\phi }(x)|\phi (p)\rangle $ is
determined by the mass $m_{\phi }$, decay constant $f_{\phi }$ and
polarization vector $\varepsilon _{\mu }$ of $\phi $ meson. The vertex $%
X_{1}\phi \eta ^{\prime }$ is modeled using the strong coupling $%
g_{X_{1}\phi \eta ^{\prime }}$, that should be extracted from the sum rule.
In the soft-meson approximation $q\rightarrow 0$ and $p^{\prime }=p$, we
have to perform one-variable Borel transformation which gives
\begin{equation}
\mathcal{B}\widehat{\Pi }_{\mu \nu }^{\mathrm{Phys}}(p)=g_{X_{1}\phi \eta
^{\prime }}m_{\phi }mf_{\phi }f\frac{e^{-\overline{m}^{2}/M^{2}}}{M^{2}}%
\left( \overline{m}^{2}g_{\mu \nu }-p_{\nu }p_{\mu }^{\prime }\right)
+\cdots ,  \label{eq:CF5B}
\end{equation}%
where $\overline{m}^{2}=(m_{\phi }^{2}+m^{2})/2.$ To make explicit Lorentz
structures of $\mathcal{B}\widehat{\Pi }_{\mu \nu }^{\mathrm{Phys}}(p)$, we
keep in Eq.\ (\ref{eq:CF5B}) $p_{\nu }\neq p_{\mu }^{\prime }$. The sum rule
for $g_{X_{1}\phi \eta ^{\prime }}$ will be derived by using a structure
proportional to $g_{\mu \nu }$. In the soft limit we also act on both the
physical and QCD sides of the sum rule by the operator $\mathcal{P}(M^{2},%
\overline{m}^{2})$, that cancels unsuppressed terms in $\mathcal{B}\widehat{%
\Pi }_{\mu \nu }^{\mathrm{Phys}}(p)$.

In the soft limit $\widehat{\Pi }_{\mu \nu }^{\mathrm{OPE}}(p)$ is given by
the formula
\begin{eqnarray}
&&\widehat{\Pi }_{\mu \nu }^{\mathrm{OPE}}(p)=2i\int d^{4}xe^{ipx}\left\{ %
\left[ \sigma _{\mu \rho }\gamma _{5}\widetilde{S}^{ib}(x){}\gamma _{\nu }%
\widetilde{S}^{bi}(-x)\gamma ^{\rho }-\gamma ^{\rho }\widetilde{S}%
^{ib}(x){}\gamma _{\nu }\widetilde{S}^{bi}(-x){}\gamma _{5}\sigma _{\mu \rho
}\right] _{\alpha \beta }\langle \eta ^{\prime }(q)|\overline{s}_{\alpha
}^{a}(0)s_{\beta }^{a}(0)|0\rangle \right.  \notag \\
&&\left. +\left[ {}\gamma ^{\rho }\widetilde{S}^{ia}(x)\gamma _{\nu }%
\widetilde{S}^{bi}(-x)\gamma _{5}\sigma _{\mu \rho }-\gamma _{5}\sigma _{\mu
\rho }\widetilde{S}^{ia}(x)\gamma ^{\rho }\widetilde{S}^{bi}(-x){}\gamma
_{\nu }\right] _{\alpha \beta }\langle \eta ^{\prime }(q)|\overline{s}%
_{\alpha }^{b}(0)s_{\beta }^{a}(0)|0\rangle \right\}.  \label{eq:CF6}
\end{eqnarray}%
The local matrix element of the $\eta ^{\prime }$ meson $\langle \eta
^{\prime }(q)|\overline{s}_{\alpha }^{b}s_{\beta }^{a}|0\rangle $ can be
transformed in accordance with Eq.\ (\ref{eq:MatEx}). After this
transformation operators $\overline{s}\Gamma ^{j}s$, and ones generated due
to $G_{\mu \nu }$ insertions from quark propagators, form local matrix
elements of the $\eta ^{\prime }$ meson. Applying Eq.\ (\ref{eq:MatEx}) to
the correlation function, performing summation over color and calculating
traces of over spinor indices, we determine local matrix elements of the $%
\eta ^{\prime }$ meson that contribute to $\widehat{\Pi }_{\mu \nu }^{%
\mathrm{OPE}}(p)$.

Our studies show that in the soft limit the twist-3 matrix element $\langle
\eta ^{\prime }|\overline{s}i\gamma _{5}s|0\rangle $ contributes to the
correlation function $\widehat{\Pi }_{\mu \nu }^{\mathrm{OPE}}(p)$. The
matrix elements of the $\eta $ and $\eta ^{\prime }$ mesons have some
peculiarities connected with mixing in the $\eta -\eta ^{\prime }$ system.
In fact, \ through the mixing both the $\eta ^{\prime }$ and $\eta $ mesons
contains $\overline{s}s$ components. It is clear that in the $\eta ^{\prime
} $ meson dominant is a strange component, but it plays some role also in
the $\eta $ meson. Due to existence of strange components both the decays $%
X_{1}\rightarrow \phi \eta ^{\prime }$ and $X_{1}\rightarrow \phi \eta $ can
be realized.

The $\eta -\eta ^{\prime }$ mixing can be described using two basics. For
our analysis a suitable is the quark-flavor basis, \ which was employed to
investigate various exclusive processes with $\eta ^{\prime }$ and $\eta $
mesons \cite{Agaev:2014wna,Agaev:2015faa,Agaev:2016dsg}. In this basis the
twist-3 matrix element $\langle \eta ^{\prime }|\overline{s}i\gamma
_{5}s|0\rangle $ is given by the formula%
\begin{equation}
2m_{s}\langle \eta ^{\prime }|\overline{s}i\gamma _{5}s|0\rangle =h_{\eta
^{\prime }}^{s},  \label{eq:MEl1}
\end{equation}%
where the parameter $h_{\eta ^{\prime }}^{s}$ is defined by the expression
\begin{equation}
h_{\eta ^{\prime }}^{s}=m_{\eta ^{\prime }}^{2}f_{\eta ^{\prime
}}^{s}-A_{\eta ^{\prime }},\ A_{\eta ^{\prime }}=\langle 0|\frac{\alpha _{s}%
}{4\pi }G_{\mu \nu }^{a}\widetilde{G}^{a,\mu \nu }|\eta ^{\prime }\rangle .
\label{eq:MEleta}
\end{equation}%
In Eq.\ (\ref{eq:MEleta}) $m_{\eta ^{\prime }}$ and $f_{\eta ^{\prime }}^{s}$
are the mass and $s$-component of the $\eta ^{\prime }$ meson decay
constant. Here, $A_{\eta ^{\prime }}$ is the matrix element which appear due
to $U(1)$ axial-anomaly. The parameter $h_{\eta ^{\prime }}^{s}$ can be
calculated using Eqs.\ (\ref{eq:MEl1}) and (\ref{eq:MEleta}). It is also
possible to use the phenomenological value
\begin{equation}
h_{\eta ^{\prime }}^{s}=h_{s}\cos \varphi ,\ h_{s}=(0.087\pm 0.006)\ \mathrm{%
GeV}^{3},
\end{equation}%
where $\varphi =39^{\circ }.3\pm 1^{\circ }.0$ is the mixing angle in the
quark-flavor basis.

The Borel transform of the invariant function $\widehat{\Pi }^{\mathrm{OPE}%
}(p^{2})$ which is related to a structure $\sim g_{\mu \nu }$ reads%
\begin{eqnarray}
\widehat{\Pi }^{\mathrm{OPE}}(M^{2}) &=&\int_{16m_{s}^{2}}^{\infty }ds\rho ^{%
\mathrm{pert.}}(s)e^{-s/M^{2}}-h_{\eta ^{\prime }}^{s}\langle \overline{s}%
s\rangle -\langle \frac{\alpha _{s}G^{2}}{\pi }\rangle \frac{h_{\eta
^{\prime }}^{s}}{8m_{s}}-\frac{h_{\eta ^{\prime }}^{s}}{6M^{2}}\langle
\overline{s}g_{s}\sigma Gs\rangle +\frac{2g_{s}^{2}h_{\eta ^{\prime }}^{s}}{%
81m_{s}M^{2}}\langle \overline{s}s\rangle ^{2},  \notag \\
&&  \label{eq:Borel2X}
\end{eqnarray}%
where the perturbative contribution is given in terms of the spectral
density
\begin{equation}
\rho ^{\mathrm{pert.}}(s)=-\frac{h_{\eta ^{\prime }}^{s}}{4m_{s}\pi ^{2}}%
(s+3m_{s}^{2}).  \label{eq:Pert}
\end{equation}%
Other components of $\widehat{\Pi }^{\mathrm{OPE}}(M^{2})$ are
nonperturbative contributions calculated by including terms up to dimension $%
6$. To carry out the continuum subtraction, we need to apply the operator $%
\mathcal{P}(M^{2},\overline{m}^{2})$ to $\widehat{\Pi }^{\mathrm{OPE}}(M^{2})
$. Afterwards, we should replace in the first term $\infty $ by $s_{0}$, and
leave in original forms contributions $\sim (M^{2})^{0}$ and $\sim 1/M^{2}$.
The width of the decay $X_{1}\rightarrow \phi \eta ^{\prime }$ is determined
by the formula (\ref{eq:DW}), where substitutions $g_{Z\psi \pi }$, $m_{\psi
}$, $\lambda \left( m_{Z},\ m_{\psi },m_{\pi }\right) \rightarrow
g_{X_{1}\phi \eta ^{\prime }}$, $m_{\phi },$, $\lambda (m,m_{\phi },m_{\eta
^{\prime }})$ must be done.

The parameters $M^{2}$ and $s_{0}$ in numerical analysis are varied inside
limits
\begin{equation}
M^{2}\in \lbrack 1.4,\ 2]~\mathrm{GeV}^{2},\ s_{0}\in \lbrack 6.2,\ 7.2]~%
\mathrm{GeV}^{2}.  \label{eq:Wind3}
\end{equation}%
The masses of the mesons $\phi $ and $\eta ^{\prime }$ are borrowed from
Ref. \cite{Tanabashi:2018oca}
\begin{equation}
m_{\phi }=(1019.461\pm 0.019)~\mathrm{MeV},\ m_{\eta ^{\prime }}=(957.78\pm
0.06)~\mathrm{MeV},\ f_{\phi }=(215\pm 5)~\mathrm{MeV}.
\label{eq:Finalmesons}
\end{equation}%
We get the following results%
\begin{equation}
g_{X_{1}\phi \eta ^{\prime }}=(2.82\pm 0.54)~\mathrm{GeV}^{-1},\ \Gamma
(X_{1}\rightarrow \phi \eta ^{\prime })=(105.3\pm 28.6)~\mathrm{MeV}.
\label{eq:DWX1}
\end{equation}

The $X_{1}\rightarrow \phi \eta ^{\prime }$ is the dominant decay mode of
the tetraquark $X_{1}$. The width of the decay $X_{1}\rightarrow \phi \eta $
can be computed using formulas derived in the present subsection. The
distinctions between two decays of $X_{1}$ are connected with the twist-3
matrix element
\begin{equation}
2m_{s}\langle \eta |\overline{s}i\gamma _{5}s|0\rangle =-h_{s}\sin \varphi ,
\end{equation}%
and the mass of the $\eta $ meson $m_{\eta }=(547.862\pm 0.018)~\mathrm{MeV}$%
. Computations lead to results%
\begin{equation}
|g_{X_{1}\phi \eta }|=(0.85\pm 0.22)~\mathrm{GeV}^{-1},\ \Gamma
(X_{1}\rightarrow \phi \eta )=(24.9\pm 9.5)~\mathrm{MeV}.  \label{eq:DWX2}
\end{equation}%
It is worth emphasizing that $|g_{X_{1}\phi \eta }|$ has been evaluated
using the region $s_{0}\in \lbrack 5.8,\ 6.8]~\mathrm{GeV}^{2}$.

The full width of the resonance $X_{1}$ saturated by two decays is equal to%
\begin{equation}
\Gamma =(130.2\pm 30.1)~\mathrm{MeV}.  \label{eq:FullW}
\end{equation}%
This result for $\Gamma $ is comparable with experimentally measured width
of the resonance $X_{1}$.

In this section we have investigated the axial-vector and vector states $ss%
\overline{s}\overline{s}$. The mass $m=\left( 2067\pm 84\right) ~\mathrm{MeV}
$ of $T_{\mathrm{AV}}$ evaluated here agrees with results of BESIII. The
width of $T_{\mathrm{AV}}$ which has been found equal to $\Gamma =\left(
130.2\pm 30.1\right) ~\mathrm{MeV}$ is consistent with these data as well.
This information has allowed us to interpret the resonance $X_{1}$ as an
axial-vector exotic meson $ss\overline{s}\overline{s}$.

Another conclusion that can be made is that the vector tetraquark $ss%
\overline{s}\overline{s}$ may be considered as the structure $X_{2}$ rather
than the resonance $Y(2175)$. Let us note that we have treated the
resonances $X_{2}$ and $Y(2175)$ as different particles, though their masses
are close to each other. This picture is typical for a family of heavy
vector resonances $Y$ as well \cite{Sundu:2018toi}. Some of these states may
be treated as the same resonances, but even in this situation the mass range
$4-5$ $\mathrm{GeV}$ is overpopulated by $J^{PC}=1^{--}$ mesons. A similar
picture seems persists also in a light sector of $J^{PC}=1^{--}$ particles.
Hence, more detailed experimental analyses are necessary to differentiate
these resonances, and determine reliably their parameters.


\section{The resonance $Y(2175)$}

\label{sec:Y2175}

As we have noted above, the vector resonance $Y(2175)$ (in this section, $%
\widetilde{Y}$) is one of a few light particles which can be considered as a
serious candidate to an exotic meson. Because it was observed in $\phi
f_{0}(980)$ invariant mass distribution, usually was treated as a state
containing exclusively strange quarks and antiquarks $ss\overline{s}%
\overline{s}$. The reason for such interpretation of $\widetilde{Y}$ is
quite natural. Indeed, in the conventional model both $\phi $ and $f_{0}(980)
$ are mesons with $\overline{s}s$ structure, a difference being only in
their quantum numbers: While $\phi $ is the vector particle, $f_{0}(980)$ is
the scalar meson. But, the light meson $f_{0}(980)$, as a member of the
first scalar nonet, can also be treated as a four-quark state. In this
picture $f_{0}(980)$ is a superposition of diquark-antidiquark states $%
\mathbf{L}=[ud][\overline{u}\overline{d}]$ and $\mathbf{H}=([su][\overline{s}%
\overline{u}]+[ds][\overline{d}s])/\sqrt{2}$. Then it appears that $%
\widetilde{Y}$ can be interpreted as a tetraquark with $sq\overline{s}%
\overline{q}$ content. In this section, we provide results of our analysis,
obtained in Ref. \cite{Agaev:2019coa} by treating $\widetilde{Y}$ as a
vector tetraquark $[su][\overline{s}\overline{u}]$.


\subsection{Spectroscopic parameters of the tetraquark $\widetilde{Y}$: the
mass $m_{Y}$ and current coupling $f_{Y}$}

\label{subsec:M}

To evaluate the mass $m_{Y}$ and coupling\ $f_{Y}$ of the vector tetraquark $%
\widetilde{Y}$, we use the QCD two-point sum rule method and start our
calculations from analysis of the correlation function (\ref{eq:CorrF1}),
where use the current
\begin{equation}
J_{\mu }^{Y}(x)=[u_{a}^{T}(x)C\gamma _{5}s_{b}(x)][\overline{u}_{a}(x)\gamma
_{\mu }\gamma _{5}C\overline{s}_{b}^{T}(x)]-[u_{a}^{T}(x)C\gamma _{\mu
}\gamma _{5}s_{b}(x)][\overline{u}_{a}(x)\gamma _{5}C\overline{s}%
_{b}^{T}(x)].  \label{eq:CurrY2175}
\end{equation}%
The current $J_{\mu }^{Y}$ consists of two pieces and each of them describes
a vector $J^{\mathrm{P}}=1^{-}$ tetraquark. This is evident from quantum
numbers of the diquark-antidiquark fields: the first term is built of the
scalar diquark $u^{T}C\gamma _{5}s$ and vector antidiquark $\overline{u}%
\gamma _{\mu }\gamma _{5}C\overline{s}^{T}$, whereas in the second term the
diquark and antidiquark are vector and scalar states, respectively. The $%
J_{\mu }^{Y}$ corresponds to the vector tetraquark with a definite
charge-conjugation parity $J^{\mathrm{PC}}=1^{--}$. Indeed, because the
charge-conjugation transforms diquarks to antidiquarks (and antidiquarks to
diquarks) the minus sign between two currents in $J_{\mu }^{Y}$ implies $%
C=-1 $.

The analysis of the phenomenological side of the sum rules $\Pi _{\mu \nu }^{%
\mathrm{Phys}}(p)$ does not differ from similar expression (\ref{eq:CorF1}),
where now one should use the mass $m_{Y}$ and coupling $f_{Y}$ of the state $%
\widetilde{Y}$. Because a part of $\Pi _{\mu \nu }^{\mathrm{Phys}}(p)$
proportional to $g_{\mu \nu }$ is formed due to contributions of vector
states, we work with this term and corresponding invariant amplitude $\Pi ^{%
\mathrm{Phys}}(p^{2})$.

To get the sum rules' QCD side, we compute $\Pi _{\mu \nu }(p)$ using
quark-gluon degrees of freedom, and find
\begin{eqnarray}
&&\Pi _{\mu \nu }^{\mathrm{OPE}}(p)=i\int d^{4}xe^{ipx}\left\{ \mathrm{Tr}%
\left[ \gamma _{5}\widetilde{S}_{s}^{b^{\prime }b}(-x)\gamma _{5}\gamma
_{\nu }S_{u}^{a^{\prime }a}(-x)\right] \mathrm{Tr}\left[ S_{u}^{aa^{\prime
}}(x)\gamma _{5}\widetilde{S}_{s}^{bb^{\prime }}(x)\gamma _{5}\gamma _{\mu }%
\right] \right.   \notag \\
&&+\mathrm{Tr}\left[ \gamma _{\mu }\gamma _{5}\widetilde{S}_{s}^{b^{\prime
}b}(-x)\gamma _{5}S_{u}^{a^{\prime }a}(-x)\right] \mathrm{Tr}\left[
S_{u}^{aa^{\prime }}(x)\gamma _{\nu }\gamma _{5}\widetilde{S}%
_{s}^{bb^{\prime }}(x)\gamma _{5}\right] +\mathrm{Tr}\left[
S_{u}^{aa^{\prime }}(x)\gamma _{5}\widetilde{S}_{s}^{bb^{\prime }}(x)\gamma
_{5}\right]   \notag \\
&&\left. \times \mathrm{Tr}\left[ \gamma _{\mu }\gamma _{5}\widetilde{S}%
_{s}^{b^{\prime }b}(-x)\gamma _{5}\gamma _{\nu }S_{u}^{a^{\prime }a}(-x)%
\right] +\mathrm{Tr}\left[ \gamma _{5}\widetilde{S}_{s}^{b^{\prime
}b}(-x)\gamma _{5}S_{u}^{a^{\prime }a}(-x)\right] \mathrm{Tr}\left[
S_{u}^{aa^{\prime }}(x)\gamma _{\nu }\gamma _{5}\widetilde{S}%
_{s}^{bb^{\prime }}(x)\gamma _{5}\gamma _{\mu }\right] \right\} .
\label{eq:CF3Y}
\end{eqnarray}%
The required sum rules for the mass and coupling of the tetraquark $%
\widetilde{Y}$ can be obtained by extracting the invariant amplitude $\Pi ^{%
\mathrm{OPE}}(p^{2})$ related to a structure $g_{\mu \nu }$ in Eq.\ (\ref%
{eq:CF3Y}), and equating it to $\Pi ^{\mathrm{Phys}}(p^{2})$. Afterwards,
one should apply to this equality the Borel transformation and perform
continuum subtraction. These operations generate a dependence of sum rules
on the Borel $M^{2}$ and continuum threshold $s_{0}$ parameters. Next steps
to get sum rules for $m_{Y}$ and $f_{Y}$ were described many times in this
review, therefore we omit further details. Let us only note that calculation
of $\Pi ^{\mathrm{OPE}}(p^{2})$ is carried out by including into analysis
nonperturbative terms up to dimension $15$.

The quantities $m_{Y}$ and $f_{Y}$ should be stable against variations of
the Borel parameter $M^{2}$. But in actual computations one can minimize
these effects by fixing a plateau where dependence of physical quantities on
$M^{2}$ is minimal. The continuum threshold parameter $s_{0}$ separates
contributions of ground-state particles from ones due to higher resonances and
continuum states. In other words, $s_{0}$ should be below the first excited
state of the particle under discussion $\widetilde{Y}$. In the case of
ordinary hadrons, masses of excited states are known either from
experimental measurements or from alternative theoretical studies. For
exotic particles a situation is more complicated: there is not information
on their radial and/or orbital excitations. For tetraquarks this problem was
addressed only in a  few publications \cite%
{Maiani:2014,Wang:2014vha,Agaev:2017tzv}. Therefore, one chooses $s_{0}$ by
demanding maximum for the pole contribution, and a stability of extracting
physical quantity. In this situation a self-consistency of the prediction
for $m_{Y}$, and $s_{0}$ used for its computation is very important: $\sqrt{%
s_{0}}$ may exceed $m_{Y}$ approximately $[0.3,0.5]~\mathrm{MeV}$, then a
first excited state of $\widetilde{Y}$ \ is above $\sqrt{s_{0}}$.

Computations show that the regions
\begin{equation}
M^{2}\in \lbrack 1.2,\ 1.7]~\mathrm{GeV}^{2},\ s_{0}\in \lbrack 6,\ 6.5]~%
\mathrm{GeV}^{2}  \label{eq:Wind1Y}
\end{equation}%
satisfy all restrictions imposed on $M^{2}$ and $s_{0}$ by the sum rule
analysis. Predictions for $m_{Y}$ and $f_{Y}$ extracted from this analysis
read
\begin{equation}
m_{Y}=\left( 2173\pm 85\right) ~\mathrm{MeV},\ \ f_{Y}=\left( 2.8\pm
0.5\right) \times 10^{-3}~\mathrm{GeV}^{4}.  \label{eq:Result1Y}
\end{equation}%
Comparing $m_{Y}$ with $\sqrt{s_{0}}$,  we see that $\sqrt{s_{0}}%
-m_{Y}=[0.28,0.38]~\mathrm{MeV}$ is a reasonable mass gap to separate $%
\widetilde{Y}$ from its excitations.

Our result for $m_{Y}$ is in good  agreement with the BaBar datum $%
(2175\pm 10\pm 15)~\mathrm{MeV}$, but is below new result of BESIII $%
(2200\pm 6\pm 5)~\mathrm{MeV}$. Nevertheless, if one takes into account
theoretical errors of computations, and errors of the experiment $m_{Y}$ is
consistent with BESIII data as well. In this situation, when there are
different models for $\widetilde{Y}$, a prediction for full width of this
tetraquark and its confrontation with data can shed light on internal
structure of $\widetilde{Y}$.


\subsection{The decay $\widetilde{Y}\rightarrow \protect\phi f_{0}(980)$}

\label{subsec:Decay1}
The process $\widetilde{Y}\rightarrow \phi f_{0}(980)$ is an important decay
channel of the tetraquark $\widetilde{Y}$. The partial width of this mode
can be expressed in term of the strong coupling $G_{Y\phi f}$ describing the
vertex $\widetilde{Y}\phi f_{0}(980)$. In its turn, the coupling $G_{Y\phi
f} $ can be evaluated in the context of the LCSR method and expressed using
various vacuum condensates and distribution amplitudes of the $\phi $ meson.

We extract the  sum rule for $G_{Y\phi f}$ by computing the correlation
function
\begin{equation}
\Pi _{\mu }(p,q)=i\int d^{4}xe^{ipx}\langle \phi (q)|\mathcal{T}%
\{J^{f}(x)J_{\mu }^{Y\dagger }(0)\}|0\rangle.  \label{eq:CF4Y}
\end{equation}

We treat the scalar meson $f_{0}(980)$ [hereafter in expressions $%
f=f_{0}(980)$] as a pure $\mathbf{H}$ state, interpolating current of which
has been presented in Eq.\ (\ref{eq:Curr3}). The phenomenological side of
the sum rule is equal to the expression
\begin{eqnarray}
&&\Pi _{\mu }^{\mathrm{Phys}}(p,q)=G_{Y\phi f}\frac{m_{Y}f_{Y}m_{f}F_{f}}{%
2\left( p^{\prime 2}-m_{Y}^{2}\right) \left( p^{2}-m_{f}^{2}\right) }\left[
\left( m_{f}^{2}-m_{Y}^{2}-m_{\phi }^{2}\right) \varepsilon _{\mu }^{\ast }+%
\frac{m_{Y}^{2}+m_{f}^{2}-m_{\phi }^{2}}{m_{Y}^{2}}p\cdot \varepsilon ^{\ast
}q_{\mu }\right].  \notag \\
&&  \label{eq:Phys2Y}
\end{eqnarray}%
The function $\Pi _{\mu }^{\mathrm{Phys}}(p,q)$ is a sum of two terms with
different Lorentz structures. We choose a structure $\sim \varepsilon _{\mu
}^{\ast }$ to extract the sum rule necessary for our purposes.

The QCD side of the sum rule $\Pi _{\mu }^{\mathrm{OPE}}(p,q)$ is derived by
inserting interpolating currents into Eq.\ (\ref{eq:CF4Y}). After
contracting quark fields, and rewriting an obtained expression using the
quarks' light-cone propagators $\mathcal{S}_{q}(x)$, we see that the matrix
element in Eq.\ (\ref{eq:CF4Y}) is a sum of terms
\begin{equation}
\left[ A(x)\right] _{\alpha \beta }^{ab}\langle \phi (q)|\overline{s}%
_{\alpha }^{a}(x)s_{\beta }^{b}(0)|0\rangle ,\ \ \left[ B(x)\right] _{\alpha
\beta }^{ab}\langle \phi (q)|\overline{s}_{\alpha }^{a}(0)s_{\beta
}^{b}(x)|0\rangle.  \label{eq:AB}
\end{equation}%
Here $A(x)$ and $B(x)$ are composed of the propagators $\mathcal{S}_{q}(\pm
x)$,$\ \widetilde{\mathcal{S}}_{q}(\pm x)=C\mathcal{S}_{q}^{T}(\pm x)C$, and
$\gamma _{5(\sigma )}$ matrices. Explicit expression of $\mathcal{S}_{q}(x)$
is moved to Appendix.

Besides propagators, the function $\Pi _{\mu }^{\mathrm{OPE}}(p,q)$ depends
on nonlocal matrix elements of operator $\overline{s}s$ placed between the
vacuum and $\phi $ meson. These matrix elements, after using Eq.\ (\ref%
{eq:MatEx}), can be expressed via the $\phi $ meson's distribution
amplitudes. In fact, after performed operations $A(x)$ and $B(x)$ depend on
colorless operators $\overline{s}(x)\Gamma ^{j}s(0)$ which can be expanded
over $x^{2}$ and expressed in terms of the $\phi $ meson's DAs of different
twist. For $\ \Gamma ^{j}=\mathbf{1}$ and $\ i\gamma _{\mu }\gamma _{5}$ we
employ the following formulas%
\begin{equation}
\langle 0|\overline{s}(x)s(0)|\phi (q)\rangle =-if_{\phi }^{\perp
}\varepsilon \cdot xm_{\phi }^{2}\int_{0}^{1}due^{i\overline{u}qx}\psi
_{3}^{\parallel }(u),  \label{eq:DA1}
\end{equation}%
and
\begin{equation}
\langle 0|\overline{s}(x)\gamma _{\mu }\gamma _{5}s(0)|\phi (q)\rangle =%
\frac{1}{2}f_{\phi }^{\parallel }m_{\phi }\epsilon _{\mu \nu \alpha \beta
}\varepsilon ^{\nu }q^{\alpha }x^{\beta }\int_{0}^{1}due^{i\overline{u}%
qx}\psi _{3}^{\perp }(u).  \label{eq:DA1A}
\end{equation}%
For the structures $\Gamma ^{j}=\gamma _{\mu }\ $and $\sigma _{\mu \nu }$ we
have
\begin{eqnarray}
&&\langle 0|\overline{s}(x)\gamma _{\mu }s(0)|\phi (q)\rangle =f_{\phi
}^{\parallel }m_{\phi }\left\{ \frac{\varepsilon \cdot x}{q\cdot x}q_{\mu
}\int_{0}^{1}due^{i\overline{u}qx}\left[ \phi _{2}^{\parallel }(u)+\frac{%
m_{\phi }^{2}x^{2}}{4}\phi _{4}^{\parallel }(u)\right] \right.   \notag \\
&&\left. +\left( \varepsilon _{\mu }-q_{\mu }\frac{\varepsilon \cdot x}{%
q\cdot x}\right) \int_{0}^{1}due^{i\overline{u}qx}\phi _{3}^{\perp }(u)-%
\frac{1}{2}x_{\mu }\frac{\varepsilon \cdot x}{(q\cdot x)^{2}}m_{\phi
}^{2}\int_{0}^{1}due^{i\overline{u}qx}C(u)+\cdots \right\} ,  \label{eq:DA2}
\end{eqnarray}%
and
\begin{eqnarray}
&&\langle 0|\overline{s}(x)\sigma _{\mu \nu }s(0)|\phi (q)\rangle =if_{\phi
}^{\perp }\left\{ \left( \varepsilon _{\mu }q_{\nu }-\varepsilon _{\nu
}q_{\mu }\right) \int_{0}^{1}due^{i\overline{u}qx}\left[ \phi _{2}^{\perp
}(u)+\frac{m_{\phi }^{2}x^{2}}{4}\phi _{4}^{\perp }(u)\right] \right.
\notag \\
&&+\frac{1}{2}\left( \varepsilon _{\mu }x_{\nu }-\varepsilon _{\nu }x_{\mu
}\right) \frac{m_{\phi }^{2}}{q\cdot x}\int_{0}^{1}due^{i\overline{u}qx}%
\left[ \psi _{4}^{\perp }(u)-\phi _{2}^{\perp }(u)\right] +\left( q_{\mu
}x_{\nu }-q_{\nu }x_{\mu }\right)   \notag \\
&&\left. \times \frac{\varepsilon \cdot x}{(q\cdot x)^{2}}m_{\phi
}^{2}\int_{0}^{1}due^{i\overline{u}qx}D(u)+\cdots \right\} .  \label{eq:DA3}
\end{eqnarray}%
In equations above $u$ is a longitudinal momentum fraction carried by a
quark, and $\overline{u}=1-u$ is a momentum of an antiquark. The mass and
polarization vector of the $\phi $ meson are denoted respectively by $%
m_{\phi }$ and $\varepsilon _{\mu }$. Combinations of two-particle DAs $C(u)$
and $D(u)$ are given by the following expressions
\begin{equation}
C(u)=\psi _{4}^{\parallel }(u)+\phi _{2}^{\parallel }(u)-2\phi _{3}^{\perp
}(u),\ \ D(u)=\phi _{3}^{\parallel }(u)-\frac{1}{2}\phi _{2}^{\perp }(u)-%
\frac{1}{2}\psi _{4}^{\perp }(u),  \label{eq:DA4}
\end{equation}%
where subscripts in functions denote their twists. Expressions of matrix
elements $\langle 0|\overline{s}(x)\Gamma ^{J}G_{\mu \nu }(vx)s(0)|\phi
(q)\rangle $ in terms of the $\phi $ meson higher twist DAs, and detailed
information on features of these DAs themselves can be found in Refs.\ \cite%
{Ball:1996tb,Ball:1998sk,Ball:1998ff,Ball:2007rt,Ball:2007zt}.

\begin{figure}[h]
\includegraphics[width=7.8cm]{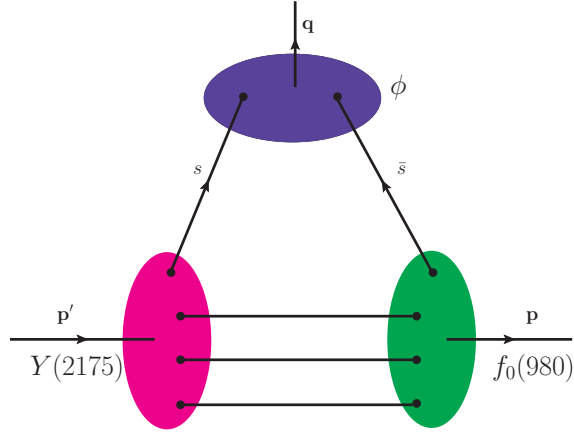}
\caption{The leading order diagram contributing to $\Pi _{\protect\mu }^{%
\mathrm{OPE}}(p,q)$.}
\label{fig:Diag1}
\end{figure}

\begin{figure}[h]
\includegraphics[width=14.8cm]{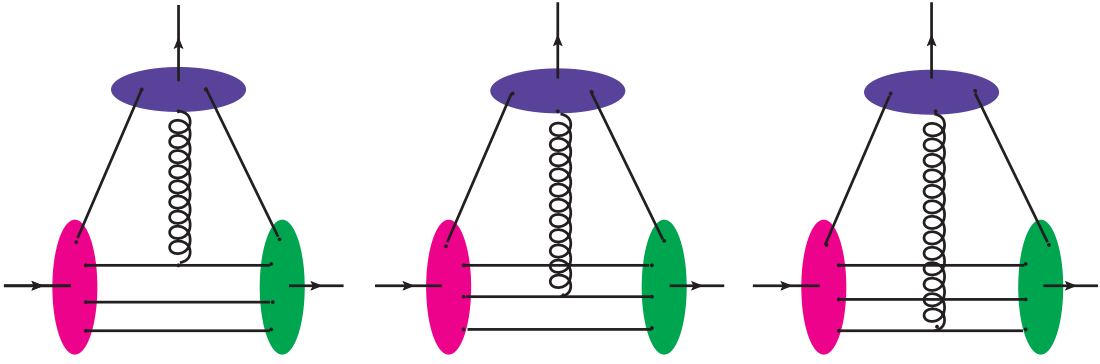}
\caption{The one-gluon exchange diagrams connected to three-particle DAs of $%
\protect\phi $ meson.}
\label{fig:Diag2}
\end{figure}

The main contribution to $\Pi _{\mu }^{\mathrm{OPE}}(p,q)$ comes from the
terms (\ref{eq:AB}), where only perturbative components of the propagators
are used (see, Fig.\ \ref{fig:Diag1}). Contribution of this diagram can be
evaluated by employing the $\phi $ meson's two-particle distribution
amplitudes. The one-gluon exchange diagrams shown in Fig.\ \ref{fig:Diag2}
are corrections, which can be expressed and calculated using relevant
three-particle DAs. An analytic expression of the $\Pi _{\mu }^{\mathrm{OPE}%
}(p,q)$ in terms of the $\phi $ meson's DAs is lengthy enough, hence we
refrain from providing it here.

In calculations, we use the amplitude $\Pi ^{\mathrm{OPE}}(p^{\prime
2},p^{2})$ extracted from a term proportional to $\varepsilon _{\mu }^{\ast
} $, and equate it to relevant amplitude from $\Pi _{\mu }^{\mathrm{Phys}%
}(p,q) $. The invariant amplitudes depend on $p^{\prime 2}$ and $p^{2}$,
therefore one has to perform double Borel transformation over $p^{\prime 2}$
and $p^{2}$
\begin{equation}
\Pi ^{\mathrm{OPE}}(M_{1}^{2},M_{2}^{2})=\mathcal{B}_{p^{\prime
2}}^{M_{1}^{2}}\mathcal{B}_{p^{2}}^{M_{2}^{2}}\Pi ^{\mathrm{OPE}}(p^{\prime
2},p^{2}).
\end{equation}%
The Borel transformed amplitude $\Pi ^{\mathrm{OPE}}\left( M_{1}^{2},\
M_{2}^{2}\right) $ can be computed using recipes of Ref.\ \cite%
{Agaev:2016srl}, and written down in form of a double dispersion integral.
To simplify following operations, it is convenient to relate parameters $%
M_{1}^{2}$ and $M_{2}^{2}$ to each other by employing the relation $%
M_{1}^{2}/M_{2}^{2}=m_{Y}^{2}/m_{f}^{2}$ and introducing  the common parameter $%
M^{2}$ through the relation
\begin{equation}
\frac{1}{M^{2}}=\frac{1}{M_{1}^{2}}+\frac{1}{M_{2}^{2}}.
\end{equation}%
This implies replacements
\begin{equation}
M_{1}^{2}=\frac{m_{f}^{2}+m_{Y}^{2}}{m_{f}^{2}}M^{2},\ \ M_{2}^{2}=\frac{%
m_{f}^{2}+m_{Y}^{2}}{m_{Y}^{2}}M^{2},
\end{equation}%
which allows us to carry out an integration over one of variables in the
double dispersion integral. The expression obtained in this phase of
computations depends also on the parameter $u_{0}$
\begin{equation}
u_{0}=\frac{M_{1}^{2}}{M_{1}^{2}+M_{2}^{2}}=\frac{m_{Y}^{2}}{%
m_{f}^{2}+m_{Y}^{2}}.
\end{equation}%
As a result, we get a single integral representation for $\Pi ^{\mathrm{OPE}%
}\left( M^{2}\right) $ which simplifies the continuum subtraction procedure:
Formulas required to fulfil the subtraction are collected in Appendix B of
Ref.\ \cite{Agaev:2016srl}.

Distribution amplitudes of the $\phi $ meson contain a lot of parameters.
Thus, the leading twist DAs of the longitudinally and transversely polarized
$\phi $ meson have the forms
\begin{equation}
\phi _{2}^{\parallel (\perp )}(u)=6u\overline{u}\left[ 1+\sum_{n=1,2\ldots
}^{\infty }a_{2n}^{\parallel (\perp )}C_{2n}^{3/2}(2u-1)\right] ,
\label{eq:LTDAPhi}
\end{equation}%
which are general expressions for $\phi _{2}^{\parallel }(u)$ and $\phi
_{2}^{\perp }(u)$. In computations we use DAs with only one nonzero
coefficients $a_{2}^{\parallel (\perp )}\neq 0$. \ Analytic forms of higher
twist DAs of the $\phi $ meson are borrowed from Refs. \cite%
{Ball:2007rt,Ball:2007zt}, where one can find also parameters of these
functions (see Tables 1 and 2 in Ref.\ \cite{Ball:2007zt}).

The sum rule for $G_{Y\phi f}$ depends on various condensates, and on mass
of $s$ quark presented already in Eq.\ (\ref{eq:VCond}). The masses and
decay constants (couplings) of the particles $\widetilde{Y}$, $\phi $, and $%
f_{0}(980)$ are input information of computations as well. The spectroscopic
parameters of $\widetilde{Y}$ \ have been evaluated in the previous
subsection. For mass and decay constant of the $\phi $ meson, we employ
experimental data $m_{\phi }=(1019.461\pm 0.019)~\mathrm{MeV}$ and $f_{\phi
}=(215\pm 5)~\mathrm{MeV}$. The mass of the meson $f_{0}(980)$ is known from
measurements, whereas the coupling $F_{f}$ of the meson $f_{0}(980)$ is
taken from Ref.\ \cite{Agaev:2017cfz}
\begin{equation}
F_{f}\equiv F_{\mathbf{H}}=\left( 1.35\pm 0.34\right) \times 10^{-3}~\mathrm{%
GeV}^{4}.
\end{equation}%
Let us remind that in Ref.\ \cite{Agaev:2017cfz}  the meson $f_{0}(980)$ was
considered as a scalar diquark-antidiquark state. The sum rule depends also
on the Borel and continuum threshold parameters $M^{2}$ and $s_{0}$. We fix
working windows for the $M^{2}$ and $s_{0}$
\begin{equation}
M^{2}\in \lbrack 2.4,\ 3.4]~\mathrm{GeV}^{2},\ s_{0}\in \lbrack 6,\ 6.5]~%
\mathrm{GeV}^{2},  \label{eq:Wind2Y}
\end{equation}%
which satisfy constraints of sum rule computations.

For the strong coupling $G_{Y\phi f}$ our computations yield
\begin{equation}
G_{Y\phi f}=(1.62\pm 0.41)\ \mathrm{GeV}^{-1}.
\end{equation}%
The width of the decay $\widetilde{Y}\rightarrow \phi f_{0}(980)$ is found
by employing Eq.\ (\ref{eq:DW}), in which one has to make the substitutions $%
g_{Z\psi \pi }$, $m_{\psi }$, $\lambda \left( m_{Z},\ m_{\psi },m_{\pi
}\right) \rightarrow G_{Y\phi f}$, $m_{\phi }$, $\lambda (m_{Y},m_{\phi
},m_{f})$.

For the partial width of the decay $\widetilde{Y}\rightarrow \phi f_{0}(980)$%
, we get
\begin{equation}
\Gamma (\widetilde{Y}\rightarrow \phi f)=(49.2\pm 17.6)~\mathrm{MeV}.
\label{eq:DW0}
\end{equation}%
The result for $\Gamma (\widetilde{Y}\rightarrow \phi f)$ is the principal
output of this subsection, and we are going to use it to evaluate the full
width of the tetraquark $\widetilde{Y}$.


\subsection{The decays $\widetilde{Y}\rightarrow $ $\protect\phi \protect%
\eta $ and $\widetilde{Y}\rightarrow $ $\protect\phi \protect\eta ^{\prime }$%
}

\label{subsec:Decay2}

The decay modes $\widetilde{Y}\rightarrow $ $\phi \eta $ and $\widetilde{Y}%
\rightarrow $ $\phi \eta ^{\prime }$ are next two processes which will be
analyzed in this section. We are going to concentrate here on the channel $%
\widetilde{Y}\rightarrow $ $\phi \eta $, and provide final predictions for
the process $\widetilde{Y}\rightarrow $ $\phi \eta ^{\prime }$.

In the context of the LCSR method the vertex $\widetilde{Y}\phi \eta $ can
be examined using the correlation function
\begin{equation}
\Pi _{\mu \nu }(p,q)=i\int d^{4}xe^{ipx}\langle \eta (q)|\mathcal{T}\{J_{\mu
}^{\phi }(x)J_{\nu }^{Y\dagger }(0)\}|0\rangle ,  \label{eq:CF5Y}
\end{equation}%
where $J_{\mu }^{\phi }(x)$ is the interpolating current for the vector $%
\phi $ meson (\ref{eq:Curr3Phi}).

The phenomenological side of the required sum rule for the strong coupling $%
g_{Y\phi \eta }$ is equal to
\begin{equation}
\Pi _{\mu \nu }^{\mathrm{Phys}}(p,q)=g_{Y\phi \eta }\frac{f_{\phi }m_{\phi
}f_{Y}m_{Y}}{\left( p^{2}-m_{\phi }^{2}\right) \left( p^{\prime
2}-m_{Y}^{2}\right) }\varepsilon _{\mu \nu \alpha \beta }p^{\alpha }q^{\beta
}+\cdots .  \label{eq:Phys4Y}
\end{equation}%
In deriving Eq.\ (\ref{eq:Phys4Y}), we have introduced the vertex
\begin{equation}
\langle \phi \left( p\right) \eta (q)|Y(p^{\prime })\rangle =g_{Y\phi \eta
}\varepsilon _{\mu \nu \alpha \beta }p^{\mu }q^{\nu }\epsilon ^{\ast \alpha
}\epsilon ^{\prime \beta },  \label{eq:ME3}
\end{equation}%
with $\epsilon ^{\prime \beta }$ being the polarization vector of the
tetraquark $\widetilde{Y}$.

It is evident, that the correlation function $\Pi _{\mu \nu }^{\mathrm{Phys}%
}(p,q)$ has a simple Lorentz structure. The invariant amplitude $\Pi ^{%
\mathrm{Phys}}(p^{\prime 2},p^{2})$, which is necessary to obtain the sum
rule for $g_{Y\phi \eta }$, can be extracted from Eq.\ (\ref{eq:Phys4Y}) by
factoring out the structure $\varepsilon _{\mu \nu \alpha \beta }p^{\alpha
}q^{\beta }$.

We calculate the invariant amplitude $\Pi ^{\mathrm{OPE}}(p^{\prime
2},p^{2}) $ from the correlation function $\Pi _{\mu \nu }^{\mathrm{OPE}%
}(p,q)$. In our case $\Pi _{\mu \nu }^{\mathrm{OPE}}(p,q)$ is determined by
the formula
\begin{eqnarray}
&&\Pi _{\mu \nu }^{\mathrm{OPE}}(p,q)=-i\int d^{4}xe^{ipx}\left[ \gamma _{5}%
\widetilde{S}_{s}^{ib}(x){}\gamma _{\mu }\widetilde{S}_{s}^{bi}(-x)\gamma
_{5}\gamma _{\nu }+\gamma _{\nu }\gamma _{5}\widetilde{S}_{s}^{ib}(x){}%
\gamma _{\mu }\widetilde{S}_{s}^{bi}(-x)\gamma _{5}\right] _{\alpha \beta }
\notag \\
&&\times \langle \eta (q)|\overline{u}_{\alpha }^{a}(0)u_{\beta
}^{a}(0)|0\rangle.  \label{eq:CF6Y}
\end{eqnarray}

The correlation function $\Pi _{\mu \nu }^{\mathrm{OPE}}(p,q)$ is expressed
using $s$ quark propagators and local matrix elements of the $\eta $ meson.
The local matrix elements\ $\langle \eta (q)|\overline{u}_{\alpha
}^{a}u_{\beta }^{a}|0\rangle $ should be transformed in accordance with Eq.\
(\ref{eq:MatEx}). Our analysis proves that $\Pi _{\mu \nu }^{\mathrm{OPE}%
}(p,q)$ receives a contribution from the matrix element $\langle \eta (q)|%
\overline{u}\gamma _{\mu }\gamma _{5}u|0\rangle $ of the $\eta $ meson
\begin{equation}
\langle \eta (q)|\overline{u}\gamma _{\mu }\gamma _{5}u|0\rangle =-i\frac{%
f_{\eta }^{q}}{\sqrt{2}}q_{\mu },  \label{eq:ME}
\end{equation}%
where $f_{\eta }^{q}$ is the decay constant of the $\eta $ meson's $q$
component. Here, some comments are in order concerning the matrix element (%
\ref{eq:ME}). It differs from matrix elements of other pseudoscalar mesons,
and this is related to the  mixing in the $\eta -\eta ^{\prime }$ system.
Relevant problems have been discussed in Sec.\ \ref{sec:4s}, where one can
find further details.

Using Eqs.\ (\ref{eq:CF6Y}) and (\ref{eq:ME}), we find the invariant
amplitude $\Pi ^{\mathrm{OPE}}(p^{\prime 2},p^{2})$. This amplitude has to
be equated to $\Pi ^{\mathrm{Phys}}(p^{\prime 2},p^{2})$ which allows us to
extract the sum rule for the strong coupling $g_{Y\phi \eta }$.

Because the correlation function $\Pi _{\mu \nu }^{\mathrm{OPE}}(p,q)$
depends on local matrix elements of the $\eta $ meson, we apply technical
tools of the soft-meson approximation. In the soft limit $p^{\prime }=p$,
and for the strong coupling $g_{Y\phi \eta }$ we get the sum rule
\begin{equation}
g_{Y\phi \eta }=\frac{1}{f_{\phi }m_{\phi }f_{Y}m_{Y}}\mathcal{P}%
(M^{2},m^{2})\Pi ^{\mathrm{OPE}}\left( M^{2},s_{0}\right),
\label{eq:Coupl1g}
\end{equation}%
where $\Pi ^{\mathrm{OPE}}(M^{2},s_{0})$ is the invariant amplitude $\Pi ^{%
\mathrm{OPE}}(p^{2})$ after the Borel transformation and subtraction
procedures. The amplitude $\Pi ^{\mathrm{OPE}}\left( M^{2},s_{0}\right) $
computed by including into analysis nonperturbative terms up to dimension-$5
$ is
\begin{equation}
\Pi ^{\mathrm{OPE}}\left( M^{2},s_{0}\right) =\frac{f_{\eta }^{q}m_{s}}{8%
\sqrt{2}\pi ^{2}}\int_{4m_{s}^{2}}^{s_{0}}dse^{-s/M^{2}}+\frac{f_{\eta
}^{q}m_{s}^{2}}{6\sqrt{2}M^{2}}\langle \overline{s}s\rangle +\frac{f_{\eta
}^{q}}{12\sqrt{2}M^{2}}\langle \overline{s}g_{s}\sigma Gs\rangle.
\label{eq:CF7}
\end{equation}%
The width of the process $\widetilde{Y}\rightarrow $ $\phi \eta $ is
determined by the formula
\begin{equation}
\Gamma (\widetilde{Y}\rightarrow \phi \eta )=\frac{g_{Y\phi \eta
}^{2}\lambda ^{3}(m_{Y},m_{\phi },m_{\eta })}{12\pi }.  \label{eq:DWGen}
\end{equation}

For the strong coupling $g_{Y\phi \eta }$ and width of the decay $\widetilde{%
Y}\rightarrow $ $\phi \eta $ numerical computations yield
\begin{equation}
g_{Y\phi \eta }=(1.85\pm 0.38)~\mathrm{GeV}^{-1},\ \Gamma (\widetilde{Y}%
\rightarrow \phi \eta )=(35.8\pm 10.4)~\mathrm{MeV}.  \label{eq:DW1Y2175}
\end{equation}%
Let us note that in calculations of $g_{Y\phi \eta }$, we have varied $M^{2}$
and $s_{0}$ within the intervals
\begin{equation}
M^{2}\in \lbrack 1.3,\ 1.8]~\mathrm{GeV}^{2},\ s_{0}\in \lbrack 6,\ 6.5]~%
\mathrm{GeV}^{2}.  \label{eq:Wind3Y}
\end{equation}

The partial width of the decay $\widetilde{Y}\rightarrow \phi \eta ^{\prime
} $ can be found by using expressions obtained for the first process. To
this end, we take into account the mass of the $\eta ^{\prime }$ meson, the
new coupling $f_{\eta ^{\prime }}^{q}$, and function $\lambda (m_{Y},m_{\phi
},m_{\eta ^{\prime }})$
\begin{equation}
f_{\eta ^{\prime }}^{q}=f_{q}\sin \varphi ,\ \lambda \rightarrow \lambda
(m_{Y},m_{\phi },m_{\eta ^{\prime }}),
\end{equation}
which can be easily implemented into analysis. For the parameters of the
second decay channel, we get
\begin{equation}
g_{Y\phi \eta ^{\prime }}=(1.59\pm 0.31)~\mathrm{GeV}^{-1},\ \Gamma (%
\widetilde{Y}\rightarrow \phi \eta ^{\prime })=(6.1\pm 1.7)~\mathrm{MeV}.
\label{eq:DW2Y2175}
\end{equation}

Saturating the full width of the $\widetilde{Y}$ resonance by three decay
channels analyzed in the present section, we find%
\begin{equation}
\Gamma _{\mathrm{full}}=(91.1\pm 20.5)~\mathrm{MeV}.  \label{eq:FullWY2175}
\end{equation}

The result for the mass $m_{Y}$ obtained in this section by treating $%
\widetilde{Y}$ as the vector tetraquark $\widetilde{Y}=[su][\overline{s}%
\overline{u}]$ agrees with the BaBar data, but is consistent with BESIII
measurements as well. The full width $\Gamma _{\mathrm{full}}$ has some
overlapping region with $\Gamma =(58\pm 16\pm 20)~\mathrm{MeV}$ extracted in
Ref.\ \cite{Aubert:2006bu}, but agreement with data of BESIII is
considerably better.

Encouraging is also our prediction for the ratio
\begin{equation}
\frac{\Gamma (\widetilde{Y}\rightarrow \phi \eta )}{\Gamma (\widetilde{Y}%
\rightarrow \phi f)}\approx 0.73,
\end{equation}%
which is almost identical to its experimental value $\approx 0.74$. The
latter has been obtained from analysis of experimental information on the
ratios
\begin{equation}
\frac{\Gamma (\widetilde{Y}\rightarrow \phi \eta )\times \Gamma (\widetilde{Y%
}\rightarrow e^{+}e^{-})}{\Gamma _{\mathrm{total}}}=1.7\pm 0.7\pm 1.3,
\end{equation}%
and%
\begin{equation}
\frac{\Gamma (\widetilde{Y}\rightarrow \phi f)\times \Gamma (\widetilde{Y}%
\rightarrow e^{+}e^{-})}{\Gamma _{\mathrm{total}}}=2.3\pm 0.3\pm 0.3,
\end{equation}%
from Ref. \cite{Tanabashi:2018oca} [ $\widetilde{Y}$ is denoted there $\phi
(2170)$].

In calculations of $\Gamma _{\mathrm{full}}$, we have included into analysis
only three strong decays of the resonance $\widetilde{Y}$. \ Decay channels $%
\widetilde{Y}\rightarrow \phi \pi \pi ,\ K^{+}K^{-}\pi ^{+}\pi ^{-},\
K^{\ast }(892)^{0}\overline{K}^{\ast }(892)^{0}$ of $\widetilde{Y}$
(observed in experiments and/or theoretically allowed) and other possible
modes have not been taken into account. Partial width of these decays may
improve our prediction for $\Gamma _{\mathrm{full}}$. Theoretical analyses
of these channels, as well as their detailed experimental investigations can
help to answer open questions about the structure of the resonance $Y(2175)$%
.

\section{Concluding notes}

\label{sec:Conc}

In this article we have reviewed our works devoted to investigations of
resonances observed by different collaborations, and which are considered as
candidates to exotic four-quark mesons. As usual, experimental measurements
provide valuable information on masses, widths and quantum numbers of these
states. Corresponding theoretical studies start from interpretations of
observed resonances as ground-state or excited conventional mesons, from
assumptions on their dynamical or exotic nature. These suggestions should be
supported by successful confrontation of theoretical predictions for their
spectroscopic parameters and decay widths with experimental data. It is
worth noting that existing theoretical computations use all diversity of
available methods and schemes.

In our articles we studied the resonances $Z_{c}(3900)$, $\,Z_{c}(4430)$, $%
Z_{c}^{-}(4100)$, $X(4140)$, $X(4274)$, $a_{1}(1420)$, $Y(4660)$, $X(2100)$,
$X(2239)$, and $Y(2175)$ by assuming that they are exotic four-quark mesons
with diquark-antidiquark structures. We constructed relevant interpolating
currents for these states and calculated their masses and couplings. All
resonances considered here are strong-interaction unstable particles, and
decay to a pair of ordinary mesons. We computed partial widths of their
dominant decay modes. Obtained results allowed us to interpret these
resonances as diquark-antidiquark states with different spin-parities and
quark contents or pose additional questions on their nature.

Some of our predictions deserves to be mentioned here. Thus, we interpreted
the resonance $Z_{c}(4430)$ as the radial excitation of $Z_{c}(3900)$, and
computed the masses and couplings, as well as estimated full widths of these
states. It seems experimental data do not contradict to this assumption, and
resonances $Z_{c}(3900)$ and $Z_{c}(4430)$ are the ground-state and radial
excitation of the same tetraquark. Another interesting result is connected
with an assumption about quark content of $Y(2175)$. In fact, despite
widespread $4s$ picture of $Y(2175)$, we argued that this vector resonance
may have a content $[sq][\overline{s}\overline{q}]$. Predictions for
parameters of this tetraquark agree with available measurements. Interesting
was also our suggestion about different internal color structures of the
axial-vector resonances $X(4140)$ and $X(4274)$. We interpreted them as
tetraquarks with identical quark contents and spin-parities, but built of
color-triplet and -sextet diquarks, respectively. While results for masses
of these states confirm our assignments, prediction for the full width of $%
X(4274)$ overshoots experimental data considerably. Alternative ideas on the
structure $X(4274)$ seem may be helpful to solve this problem.

In our calculations we used the QCD sum rules approach, which is a powerful
tool to explore features not only of conventional, but also exotic hadrons.
The spectral parameters of four-quark states were computed by employing
two-point correlation functions, and sum rules extracted from their
analysis. To explore numerous strong decays of particles under discussion,
we applied either three-point or light-cone sum rule methods. From these sum
rules it is possible to extract numerical values of strong couplings
corresponding to vertices of involved particles. The three-point sum rule
method is effective for computations in the case of heavy final-state
mesons. The light-cone sum rules are applicable to situations when at least
one of final mesons is a particle with well-known distribution amplitudes or
local matrix elements. Let us note that tetraquark-tetraquark-meson vertices
can be explored by means of standard methods of LCSRs, whereas treatment of
tetraquark-meson-meson vertices requires additionally a soft-meson technique.

There are a lot of results left beyond the scope of the present review.
Thus, we did not consider our papers on the structure of the resonance $%
X(5568)$ and its charmed partner, which presumably are particles made of
four quarks of different flavors \cite%
{Agaev:2016mjb,Agaev:2016ijz,Agaev:2016lkl,Agaev:2016urs}. Very interesting
investigations of tetraquarks stable against strong and electromagnetic
decays were not included into this review as well. Because stable
tetraquarks can dissociate to conventional mesons only through weak
transformations, lifetimes of these states are approximately $%
10^{-12}-10^{-13}~\mathrm{s}$ and significantly longer than that of unstable
tetraquarks. The famous member of this class is the axial-vector tetraquark $%
T_{bb;\overline{u}\overline{d}}^{-}$ which is one of candidates to stable
heavy exotic mesons \cite{Navarra:2007yw,Karliner:2017qjm,Eichten:2017ffp}.
We calculated the full width of $T_{bb;\overline{u}\overline{d}}^{-}$
through its semileptonic decays \cite{Agaev:2018khe}. It is remarkable that
family of stable exotic mesons is wider than one might suppose: some of
these states were studied in our articles \cite%
{Sundu:2019feu,Agaev:2019wkk,Agaev:2019lwh,Agaev:2020dba}. Problems of the
resonance $X(5568)$, in general tetraquarks built of four quarks of
different flavors, and stable heavy tetraquarks deserve detailed
investigations and separate reviews.


\section*{ACKNOWLEDGEMENTS}

We are grateful to our colleagues working on same subjects for
correspondence and criticism. This research, at various stages, was
supported by grants of TUBITAK such as Grant 2221-"Fellowship Program For
Visiting Scientists and Scientists on Sabbatical Leave", as well as the
Grant No. 115F183.

\appendix*

\section{ The quark propagators}

\renewcommand{\theequation}{\Alph{section}.\arabic{equation}} \label{sec:App}
The light and heavy quark propagators are necessary to find QCD side of the
different correlation functions. In the present work we use the light quark
propagator $S_{q}^{ab}(x)$ which is given by the following formula
\begin{eqnarray}
&&S_{q}^{ab}(x)=i\delta _{ab}\frac{\slashed x}{2\pi ^{2}x^{4}}-\delta _{ab}%
\frac{m_{q}}{4\pi ^{2}x^{2}}-\delta _{ab}\frac{\langle \overline{q}q\rangle
}{12}+i\delta _{ab}\frac{\slashed xm_{q}\langle \overline{q}q\rangle }{48}%
-\delta _{ab}\frac{x^{2}}{192}\langle \overline{q}g_{s}\sigma Gq\rangle
\notag \\
&&+i\delta _{ab}\frac{x^{2}\slashed xm_{q}}{1152}\langle \overline{q}%
g_{s}\sigma Gq\rangle -i\frac{g_{s}G_{ab}^{\alpha \beta }}{32\pi ^{2}x^{2}}%
\left[ \slashed x{\sigma _{\alpha \beta }+\sigma _{\alpha \beta }}\slashed x%
\right] -i\delta _{ab}\frac{x^{2}\slashed xg_{s}^{2}\langle \overline{q}%
q\rangle ^{2}}{7776}  \notag \\
&&-\delta _{ab}\frac{x^{4}\langle \overline{q}q\rangle \langle
g^2_{s}G^{2}\rangle }{27648}+\cdots .  \label{eq:A1}
\end{eqnarray}%
For the heavy quarks $Q$ we utilize the propagator $S_{Q}^{ab}(x)$
\begin{eqnarray}
&&S_{Q}^{ab}(x)=i\int \frac{d^{4}k}{(2\pi )^{4}}e^{-ikx}\Bigg \{\frac{\delta
_{ab}\left( {\slashed k}+m_{Q}\right) }{k^{2}-m_{Q}^{2}}-\frac{%
g_{s}G_{ab}^{\alpha \beta }}{4}\frac{\sigma _{\alpha \beta }\left( {\slashed %
k}+m_{Q}\right) +\left( {\slashed k}+m_{Q}\right) \sigma _{\alpha \beta }}{%
(k^{2}-m_{Q}^{2})^{2}}  \notag  \label{eq:A2} \\
&&+\frac{g_{s}^{2}G^{2}}{12}\delta _{ab}m_{Q}\frac{k^{2}+m_{Q}{\slashed k}}{%
(k^{2}-m_{Q}^{2})^{4}}+\frac{g_{s}^{3}G^{3}}{48}\delta _{ab}\frac{\left( {%
\slashed k}+m_{Q}\right) }{(k^{2}-m_{Q}^{2})^{6}}\left[ {\slashed k}\left(
k^{2}-3m_{Q}^{2}\right) +2m_{Q}\left( 2k^{2}-m_{Q}^{2}\right) \right] \left(
{\slashed k}+m_{Q}\right) +\cdots \Bigg \}.  \notag \\
&&
\end{eqnarray}

The light-cone propagator of the light quark is given by the expression
\begin{eqnarray}
&&\mathcal{S}_{q}^{ab}(x)=\frac{i\slashed x}{2\pi ^{2}x^{4}}\delta _{ab}-%
\frac{m_{q}}{4\pi ^{2}x^{2}}\delta _{ab}-\frac{\langle \overline{q}q\rangle
}{12}\left( 1-i\frac{m_{q}}{4}\slashed x\right) \delta _{ab}-\frac{x^{2}}{192%
}\langle \overline{q}g_{s}\sigma Gq\rangle \left( 1-i\frac{m_{q}}{6}\slashed %
x\right) \delta _{ab}  \notag \\
&&-ig_{s}\int_{0}^{1}du\left\{ \frac{\slashed x}{16\pi ^{2}x^{2}}G_{ab}^{\mu
\nu }(ux)\sigma _{\mu \nu }-\frac{iux_{\mu }}{4\pi ^{2}x^{2}}G_{ab}^{\mu \nu
}(ux)\gamma _{\nu }-\frac{im_{q}}{32\pi ^{2}}G_{ab}^{\mu \nu }(ux)\sigma
_{\mu \nu }\left[ \ln \left( \frac{-x^{2}\Lambda ^{2}}{4}\right) +2\gamma
_{E}\right] \right\} ,  \label{eq:A3}
\end{eqnarray}%
For the heavy $Q$ quark's light-cone propagator we have
\begin{eqnarray}
&&\mathcal{S}_{Q}^{ab}(x)=\frac{m_{Q}^{2}}{4\pi ^{2}}\frac{K_{1}\left( m_{Q}%
\sqrt{-x^{2}}\right) }{\sqrt{-x^{2}}}\delta _{ab}+i\frac{m_{Q}^{2}}{4\pi ^{2}%
}\frac{{\slashed x}K_{2}\left( m_{Q}\sqrt{-x^{2}}\right) }{\left( \sqrt{%
-x^{2}}\right) ^{2}}\delta _{ab}  \notag \\
&&-\frac{g_{s}m_{Q}}{16\pi ^{2}}\int_{0}^{1}dvG_{ab}^{\mu \nu }(vx)\left[
(\sigma _{\mu \nu }{\slashed x}+{\slashed x}\sigma _{\mu \nu })\frac{%
K_{1}\left( m_{Q}\sqrt{-x^{2}}\right) }{\sqrt{-x^{2}}}+2\sigma ^{\mu \nu
}K_{0}\left( m_{Q}\sqrt{-x^{2}}\right) \right] .  \label{eq:A4}
\end{eqnarray}%
In the expressions above
\begin{equation}
G_{ab}^{\alpha \beta }=G_{A}^{\alpha \beta }t_{ab}^{A},\,\,~~G^{2}=G_{\alpha
\beta }^{A}G_{\alpha \beta }^{A},\ \ G^{3}=\,\,f^{ABC}G_{\mu \nu }^{A}G_{\nu
\delta }^{B}G_{\delta \mu }^{C},
\end{equation}%
where $a,\,b=1,2,3$ are color indices and $A,B,C=1,\,2\,\ldots 8$. Here $%
t^{A}=\lambda ^{A}/2$ , where $\lambda ^{A}$ are the Gell-Mann matrices, and
the gluon field strength tensor is fixed at $x=0$, i.e., $G_{\alpha \beta
}^{A}\equiv G_{\alpha \beta }^{A}(0)$. In Eq.\ (\ref{eq:A4}) $K_{\nu }(z)$
are Bessel functions of the second kind.



\begin{thebibliography}{999}

\bibitem{Jaffe:1976ig} R.~L.~Jaffe,
Phys.\ Rev.\ D \textbf{15}, 267 (1977). 


\bibitem{Kim:2017yvd} H.~Kim, K.~S.~Kim, M.~K.~Cheoun and M.~Oka,
Phys.\ Rev.\ D \textbf{97}, 094005 (2018). 


\bibitem{Agaev:2017cfz} S.~S.~Agaev, K.~Azizi and H.~Sundu,
Phys.\ Lett.\ B \textbf{781}, 279 (2018). 


\bibitem{Agaev:2018sco} S.~S.~Agaev, K.~Azizi and H.~Sundu,
Phys.\ Lett.\ B \textbf{784}, 266 (2018). 


\bibitem{Agaev:2018fvz} S.~S.~Agaev, K.~Azizi and H.~Sundu,
Phys.\ Lett.\ B \textbf{789}, 405 (2019). 


\bibitem{Jaffe:1976yi} R.~L.~Jaffe, 
Phys.\ Rev.\ Lett.\ \textbf{38}, 195 (1977) Erratum: [Phys.\ Rev.\ Lett.\
\textbf{38}, 617 (1977)].


\bibitem{Farrar:2003qy} G.~R.~Farrar and G.~Zaharijas,
Phys.\ Rev.\ D \textbf{70}, 014008 (2004).


\bibitem{Farrar:2017ysn} G.~R.~Farrar, 
PoS ICRC \textbf{2017}, 929 (2018).


\bibitem{Farrar:2018hac} G.~R.~Farrar,
arXiv:1805.03723 [hep-ph]. 


\bibitem{Azizi:2019xla} K.~Azizi, S.~S.~Agaev, and H.~Sundu,
arXiv:1904.09913 [hep-ph].


\bibitem{Ader:1981db} J.~P.~Ader, J.~M.~Richard and P.~Taxil,
Phys.\ Rev.\ D \textbf{25}, 2370 (1982).


\bibitem{Lipkin:1986dw} H.~J.~Lipkin,
Phys.\ Lett.\ B \textbf{172}, 242 (1986).


\bibitem{Zouzou:1986qh} S.~Zouzou, B.~Silvestre-Brac, C.~Gignoux and
J.~M.~Richard, 
Z.\ Phys.\ C \textbf{30}, 457 (1986).


\bibitem{Carlson:1987hh} J.~Carlson, L.~Heller and J.~A.~Tjon,
Phys.\ Rev.\ D \textbf{37}, 744 (1988).


\bibitem{Manohar:1992nd} A.~V.~Manohar and M.~B.~Wise,
Nucl.\ Phys.\ B \textbf{399}, 17 (1993). 


\bibitem{Balitsky:1982ps} I.~I.~Balitsky, D.~Diakonov and A.~V.~Yung,
Phys.\ Lett.\ B \textbf{112}, 71 (1982). 


\bibitem{Govaerts:1984hc} J.~Govaerts, L.~J.~Reinders, H.~R.~Rubinstein and
J.~Weyers, 
Nucl.\ Phys.\ B \textbf{258}, 215 (1985).


\bibitem{Govaerts:1985fx} J.~Govaerts, L.~J.~Reinders and J.~Weyers,
Nucl.\ Phys.\ B \textbf{262}, 575 (1985).


\bibitem{Balitsky:1986hf} I.~I.~Balitsky, D.~Diakonov and A.~V.~Yung,
Z.\ Phys.\ C \textbf{33}, 265 (1986). 


\bibitem{Braun:1985ah} V.~M.~Braun and A.~V.~Kolesnichenko,
Phys.\ Lett.\ B \textbf{175}, 485 (1986). 


\bibitem{Braun:1988kv} V.~M.~Braun and Y.~M.~Shabelski,
Sov.\ J.\ Nucl.\ Phys.\ \textbf{50}, 306 (1989). 


\bibitem{Choi:2003ue} S.~K.~Choi \textit{et al.} [Belle Collaboration],
Phys.\ Rev.\ Lett.\ \textbf{91}, 262001 (2003).


\bibitem{Abazov:2004kp} V.~M.~Abazov \textit{et al.} [D0 Collaboration],
Phys.\ Rev.\ Lett.\ \textbf{93}, 162002 (2004).


\bibitem{Acosta:2003zx} D.~Acosta \textit{et al.} [CDF Collaboration],
Phys.\ Rev.\ Lett.\ \textbf{93}, 072001 (2004).


\bibitem{Aubert:2004ns} B.~Aubert \textit{et al.} [BaBar Collaboration],
Phys.\ Rev.\ D \textbf{71}, 071103 (2005).


\bibitem{Choi:2007wga} S.~K.~Choi \textit{et al.} [Belle Collaboration],
Phys.\ Rev.\ Lett.\ \textbf{100}, 142001 (2008).


\bibitem{Mizuk:2009da} R.~Mizuk \textit{et al.} [Belle Collaboration],
Phys.\ Rev.\ D \textbf{80}, 031104 (2009).


\bibitem{Chilikin:2013tch} K.~Chilikin \textit{et al.} [Belle
Collaboration],
Phys.\ Rev.\ D \textbf{88}, 074026 (2013).


\bibitem{Chilikin:2014bkk} K.~Chilikin \textit{et al.} [Belle
Collaboration],
Phys.\ Rev.\ D \textbf{90}, 112009 (2014).


\bibitem{Aaij:2014jqa} R.~Aaij \textit{et al.} [LHCb Collaboration],
Phys.\ Rev.\ Lett.\ \textbf{112}, 222002 (2014).


\bibitem{Aaij:2015zxa} R.~Aaij \textit{et al.} [LHCb Collaboration],
Phys.\ Rev.\ D \textbf{92}, 112009 (2015).


\bibitem{Ablikim:2013mio} M.~Ablikim \textit{et al.} [BESIII Collaboration],
Phys.\ Rev.\ Lett.\ \textbf{110}, 252001 (2013).


\bibitem{Liu:2013dau} Z.~Q.~Liu \textit{et al.} [Belle Collaboration],
Phys.\ Rev.\ Lett.\ \textbf{110}, 252002 (2013).


\bibitem{Xiao:2013iha} T.~Xiao, S.~Dobbs, A.~Tomaradze and K.~K.~Seth,
Phys.\ Lett.\ B \textbf{727}, 366 (2013). 


\bibitem{Ablikim:2015tbp} M.~Ablikim \textit{et al.} [BESIII Collaboration],
Phys.\ Rev.\ Lett.\ \textbf{115}, 112003 (2015).


\bibitem{D0:2016mwd} V.~M.~Abazov \textit{et al.} [D0 Collaboration],
Phys.\ Rev.\ Lett.\ \textbf{117}, 022003 (2016).


\bibitem{Aaij:2016iev} R.~Aaij \textit{et al.} [LHCb Collaboration],
Phys.\ Rev.\ Lett.\ \textbf{117}, 152003 (2016).


\bibitem{Aaij:2016iza} R.~Aaij \textit{et al.} [LHCb Collaboration],
Phys.\ Rev.\ Lett.\ \textbf{118}, 022003 (2017).


\bibitem{Aaij:2016nsc} R.~Aaij \textit{et al.} [LHCb Collaboration],
Phys.\ Rev.\ D \textbf{95}, 012002 (2017).


\bibitem{Aaltonen:2009tz} T.~Aaltonen \textit{et al.} [CDF Collaboration],
Phys.\ Rev.\ Lett.\ \textbf{102}, 242002 (2009). 


\bibitem{Chatrchyan:2013dma} S.~Chatrchyan \textit{et al.} [CMS
Collaboration],
Phys.\ Lett.\ B \textbf{734}, 261 (2014).


\bibitem{Abazov:2013xda} V.~M.~Abazov \textit{et al.} [D0 Collaboration],
Phys.\ Rev.\ D \textbf{89}, 012004 (2014).


\bibitem{Wang:2007ea} X.~L.~Wang \textit{et al.} [Belle Collaboration],
Phys.\ Rev.\ Lett.\ \textbf{99}, 142002 (2007).


\bibitem{Wang:2014hta} X.~L.~Wang \textit{et al.} [Belle Collaboration],
Phys.\ Rev.\ D \textbf{91}, 112007 (2015).


\bibitem{Pakhlova:2008vn} G.~Pakhlova \textit{et al.} [Belle Collaboration],
Phys.\ Rev.\ Lett.\ \textbf{101}, 172001 (2008).


\bibitem{Lees:2012pv} J.~P.~Lees \textit{et al.} [BaBar Collaboration],
Phys.\ Rev.\ D \textbf{89}, 111103 (2014).


\bibitem{Aaij:2018bla} R.~Aaij \textit{et al.} [LHCb Collaboration],
Eur.\ Phys.\ J.\ C \textbf{78}, 1019 (2018).


\bibitem{Shifman:1978bx} M.~A.~Shifman, A.~I.~Vainshtein and V.~I.~Zakharov,
Nucl.\ Phys.\ B \textbf{147}, 385 (1979).


\bibitem{Shifman:1978by} M.~A.~Shifman, A.~I.~Vainshtein and V.~I.~Zakharov,
Nucl.\ Phys.\ B \textbf{147}, 448 (1979).


\bibitem{Balitsky:1989ry} I.~I.~Balitsky, V.~M.~Braun and
A.~V.~Kolesnichenko,
Nucl.\ Phys.\ B \textbf{312}, 509 (1989). 


\bibitem{Belyaev:1994zk} V.~M.~Belyaev, V.~M.~Braun, A.~Khodjamirian and
R.~Ruckl, 
Phys.\ Rev.\ D \textbf{51}, 6177 (1995).


\bibitem{Ioffe:1983ju} B.~L.~Ioffe and A.~V.~Smilga,
Nucl.\ Phys.\ B \textbf{232}, 109 (1984).


\bibitem{Agaev:2016dev} S.~S.~Agaev, K.~Azizi and H.~Sundu,
Phys.\ Rev.\ D \textbf{93}, 074002 (2016).


\bibitem{Agaev:2016srl} S.~S.~Agaev, K.~Azizi and H.~Sundu,
Phys.\ Rev.\ D \textbf{93}, 114036 (2016).


\bibitem{Jaffe:2004ph} R.~L.~Jaffe, 
Phys.\ Rept.\ \textbf{409}, 1 (2005).


\bibitem{Swanson:2006st} E.~S.~Swanson,
Phys.\ Rept.\ \textbf{429}, 243 (2006). 


\bibitem{Klempt:2007cp} E.~Klempt and A.~Zaitsev,
Phys.\ Rept.\ \textbf{454}, 1 (2007). 


\bibitem{Godfrey:2008nc} S.~Godfrey and S.~L.~Olsen,
Ann.\ Rev.\ Nucl.\ Part.\ Sci.\ \textbf{58}, 51 (2008).


\bibitem{Esposito:2014rxa} A.~Esposito, A.~L.~Guerrieri, F.~Piccinini,
Int.\ J.\ Mod.\ Phys.\ A \textbf{30}, 1530002 (2014).


\bibitem{Chen:2016qju} H.~X.~Chen, W.~Chen, X.~Liu and S.~L.~Zhu,
Phys.\ Rept.\ \textbf{639}, 1 (2016).


\bibitem{Chen:2016spr} H.~X.~Chen, W.~Chen, X.~Liu, Y.~R.~Liu and S.~L.~Zhu,
Rept.\ Prog.\ Phys.\ \textbf{80}, 076201 (2017).


\bibitem{Esposito:2016noz} A.~Esposito, A.~Pilloni and A.~D.~Polosa,
Phys.\ Rept.\ \textbf{668}, 1 (2017).


\bibitem{Ali:2017jda} A.~Ali, J.~S.~Lange and S.~Stone,
Prog.\ Part.\ Nucl.\ Phys.\ \textbf{97}, 123 (2017).


\bibitem{Olsen:2017bmm} S.~L.~Olsen, T.~Skwarnicki and D.~Zieminska,
Rev.\ Mod.\ Phys.\ \textbf{90}, 015003 (2018).


\bibitem{Albuquerque:2018jkn} R.~M.~Albuquerque, J.~M.~Dias,
K.~P.~Khemchandani, A.~Martinez Torres, F.~S.~Navarra, M.~Nielsen and
C.~M.~Zanetti, 
J.\ Phys.\ G \textbf{46}, 093002 (2019).


\bibitem{Brambilla:2019esw} N.~Brambilla, S.~Eidelman, C.~Hanhart,
A.~Nefediev, C.~P.~Shen, C.~E.~Thomas, A.~Vairo and C.~Z.~Yuan,
arXiv:1907.07583 [hep-ex]. 



\bibitem{Liu:2008qx} X.~H.~Liu, Q.~Zhao and F.~E.~Close,
Phys.\ Rev.\ D \textbf{77}, 094005 (2008).


\bibitem{Ebert:2008kb} D.~Ebert, R.~N.~Faustov and V.~O.~Galkin,
Eur.\ Phys.\ J.\ C \textbf{58}, 399 (2008).


\bibitem{Bracco:2008jj} M.~E.~Bracco, S.~H.~Lee, M.~Nielsen and R.~Rodrigues
da Silva, 
Phys.\ Lett.\ B \textbf{671}, 240 (2009).


\bibitem{Maiani:2008zz} L.~Maiani, A.~D.~Polosa and V.~Riquer,
New J.\ Phys.\ \textbf{10}, 073004 (2008).


\bibitem{Wang:2010rt} Z.~G.~Wang,
Eur.\ Phys.\ J.\ C \textbf{70}, 139 (2010).


\bibitem{Maiani:2014} L.~Maiani, F.~Piccinini, A.~D.~Polosa and V.~Riquer,
Phys.\ Rev.\ D \textbf{89}, 114010 (2014).


\bibitem{Wang:2014vha} Z.~G.~Wang,
Commun.\ Theor.\ Phys.\ \textbf{63}, 325 (2015).


\bibitem{Agaev:2017tzv} S.~S.~Agaev, K.~Azizi and H.~Sundu, Phys.\ Rev.\ D
\textbf{96}, 034026 (2017).


\bibitem{Lee:2007gs} S.~H.~Lee, A.~Mihara, F.~S.~Navarra and M.~Nielsen,
Phys.\ Lett.\ B \textbf{661}, 28 (2008).


\bibitem{Liu:2008xz} X.~Liu, Y.~R.~Liu, W.~Z.~Deng and S.~L.~Zhu,
Phys.\ Rev.\ D \textbf{77}, 094015 (2008).


\bibitem{Braaten:2007xw} E.~Braaten and M.~Lu,
Phys.\ Rev.\ D \textbf{79}, 051503 (2009).


\bibitem{Branz:2010sh} T.~Branz, T.~Gutsche and V.~E.~Lyubovitskij,
Phys.\ Rev.\ D \textbf{82}, 054025 (2010).


\bibitem{Goerke:2016hxf} F.~Goerke, T.~Gutsche, M.~A.~Ivanov, J.~G.~Korner,
V.~E.~Lyubovitskij and P.~Santorelli,
Phys.\ Rev.\ D \textbf{94}, 094017 (2016).


\bibitem{Rosner:2007mu} J.~L.~Rosner,
Phys.\ Rev.\ D \textbf{76}, 114002 (2007).


\bibitem{Dubynskiy:2008mq} S.~Dubynskiy and M.~B.~Voloshin,
Phys.\ Lett.\ B \textbf{666}, 344 (2008).


\bibitem{Dias:2013xfa} J.~M.~Dias, F.~S.~Navarra, M.~Nielsen and
C.~M.~Zanetti, 
Phys.\ Rev.\ D \textbf{88}, 016004 (2013). 


\bibitem{Wang:2013vex} Z.~G.~Wang and T.~Huang,
Phys.\ Rev.\ D \textbf{89}, 054019 (2014). 


\bibitem{Deng:2014gqa} C.~Deng, J.~Ping and F.~Wang,
Phys.\ Rev.\ D \textbf{90}, 054009 (2014).


\bibitem{Wang:2013daa} Z.~G.~Wang and T.~Huang,
Phys.\ J.\ C \textbf{74}, 2891 (2014).


\bibitem{Wilbring:2013cha} E.~Wilbring, H.-W.~Hammer and U.-G.~Meisner,
Phys.\ Lett.\ B \textbf{726}, 326 (2013).


\bibitem{Dong:2013iqa} Y.~Dong, A.~Faessler, T.~Gutsche and
V.~E.~Lyubovitskij,
Phys.\ Rev.\ D \textbf{88}, 014030 (2013).


\bibitem{Ke:2013gia} H.~W.~Ke, Z.~T.~Wei and X.~Q.~Li,
Eur.\ Phys.\ J.\ C \textbf{73}, 2561 (2013).


\bibitem{Gutsche:2014zda} T.~Gutsche, M.~Kesenheimer and V.~E.~Lyubovitskij,
Phys.\ Rev.\ D \textbf{90}, 094013 (2014).


\bibitem{Esposito:2014hsa} A.~Esposito, A.~L.~Guerrieri and A.~Pilloni,
Phys.\ Lett.\ B \textbf{746}, 194 (2015).


\bibitem{Chen:2015igx} D.~Y.~Chen and Y.~B.~Dong,
Phys.\ Rev.\ D \textbf{93}, 014003 (2016).


\bibitem{Gong:2016hlt} Q.~R.~Gong, Z.~H.~Guo, C.~Meng, G.~Y.~Tang,
Y.~F.~Wang and H.~Q.~Zheng,
Phys.\ Rev.\ D \textbf{94}, 114019 (2016).


\bibitem{Ke:2016owt} H.~W.~Ke and X.~Q.~Li,
Eur.\ Phys.\ J.\ C \textbf{76}, 334 (2016).


\bibitem{Swanson:2014tra} E.~S.~Swanson,
Phys.\ Rev.\ D \textbf{91}, 034009 (2015).


\bibitem{Ikeda:2016zwx} Y.~Ikeda \textit{et al.} [HAL QCD Collaboration],
Phys.\ Rev.\ Lett.\ \textbf{117}, 242001 (2016).


\bibitem{Tanabashi:2018oca} M.~Tanabashi \textit{et al.} [Particle Data
Group], Phys.\ Rev.\ D \textbf{98}, 030001 (2018).


\bibitem{Negash:2015rua} H.~Negash and S.~Bhatnagar,
Int.\ J.\ Mod.\ Phys.\ E \textbf{25}, 1650059 (2016).


\bibitem{Ball:1998ff} P.~Ball and V.~M.~Braun,
Nucl.\ Phys.\ B \textbf{543}, 201 (1999).


\bibitem{Ball:2007zt} P.~Ball, V.~M.~Braun and A.~Lenz,
JHEP \textbf{0708}, 090 (2007).


\bibitem{Wang:2018ntv} Z.~G.~Wang,
Eur.\ Phys.\ J.\ C \textbf{78}, 933 (2018).


\bibitem{Wu:2018xdi} J.~Wu, X.~Liu, Y.~R.~Liu and S.~L.~Zhu,
Phys.\ Rev.\ D \textbf{99}, 014037 (2019).


\bibitem{Voloshin:2018vym} M.~B.~Voloshin,
Phys.\ Rev.\ D \textbf{98}, 094028 (2018).


\bibitem{Cao:2018vmv} X.~Cao and J.~P.~Dai,
Phys.\ Rev.\ D \textbf{100}, 054004 (2019).


\bibitem{Sundu:2018nxt} H.~Sundu, S.~S.~Agaev and K.~Azizi,
Eur.\ Phys.\ J.\ C \textbf{79}, 215 (2019).


\bibitem{Wang:2017lbl} Z.~G.~Wang,
Eur.\ Phys.\ J.\ A \textbf{53}, 192 (2017).


\bibitem{Chilikin:2017evr} K.~Chilikin \textit{et al.} [Belle
Collaboration],
Phys.\ Rev.\ D \textbf{95}, 112003 (2017).


\bibitem{Meissner:1995ra} T.~Meissner,
Phys.\ Rev.\ C \textbf{52}, 3386 (1995).


\bibitem{Maltman:1997jb} K.~Maltman,
Phys.\ Rev.\ C \textbf{57}, 69 (1998).


\bibitem{Bracco:2006xf} M.~E.~Bracco, A.~Cerqueira, Jr., M.~Chiapparini,
A.~Lozea and M.~Nielsen,
Phys.\ Lett.\ B \textbf{641}, 286 (2006).


\bibitem{Cerqueira:2015vva} A.~Cerqueira, B.~Osorio Rodrigues, M.~E.~Bracco
and M.~Nielsen,
Nucl.\ Phys.\ A \textbf{936}, 45 (2015).

\bibitem{Liu:2008tn} X.~Liu, Z.~G.~Luo, Y.~R.~Liu and S.~L.~Zhu,
Eur.\ Phys.\ J.\ C \textbf{61}, 411 (2009).


\bibitem{Wang:2009ue} Z.~G.~Wang,
Eur.\ Phys.\ J.\ C \textbf{63}, 115 (2009).


\bibitem{Albuquerque:2009ak} R.~M.~Albuquerque, M.~E.~Bracco and M.~Nielsen,
Phys.\ Lett.\ B \textbf{678}, 186 (2009).


\bibitem{Wang:2009ry} Z.~G.~Wang, Z.~C.~Liu and X.~H.~Zhang,
Eur.\ Phys.\ J.\ C \textbf{64}, 373 (2009).


\bibitem{Wang:2011uk} Z.~G.~Wang,
Int.\ J.\ Mod.\ Phys.\ A \textbf{26}, 4929 (2011).


\bibitem{Liu:2010hf} X.~Liu, Z.~G.~Luo and S.~L.~Zhu,
Phys.\ Lett.\ B \textbf{699}, 341 (2011,) Erratum: [Phys.\ Lett.\ B \textbf{%
707}, 577 (2012)].


\bibitem{He:2011ed} J.~He and X.~Liu,
Eur.\ Phys.\ J.\ C \textbf{72}, 1986 (2012).


\bibitem{Finazzo:2011he} S.~I.~Finazzo, M.~Nielsen and X.~Liu,
Phys.\ Lett.\ B \textbf{701}, 101 (2011).


\bibitem{HidalgoDuque:2012pq} C.~Hidalgo-Duque, J.~Nieves and
M.~P.~Valderrama,
Phys.\ Rev.\ D \textbf{87}, 076006 (2013).


\bibitem{Stancu:2009ka} F.~Stancu,
J.\ Phys.\ G \textbf{37}, 075017 (2010).


\bibitem{Patel:2014vua} S.~Patel, M.~Shah and P.~C.~Vinodkumar,
Eur.\ Phys.\ J.\ A \textbf{50}, 131 (2014).


\bibitem{Molina:2009ct} R.~Molina and E.~Oset,
Phys.\ Rev.\ D \textbf{80}, 114013 (2009).


\bibitem{Branz:2010rj} T.~Branz, R.~Molina and E.~Oset,
Phys.\ Rev.\ D \textbf{83}, 114015 (2011).


\bibitem{Danilkin:2009hr} I.~V.~Danilkin and Y.~A.~Simonov,
Phys.\ Rev.\ D \textbf{81}, 074027 (2010).


\bibitem{Bhardwaj:2015rju} V.~Bhardwaj \textit{et al.} [Belle
Collaboration],
Phys.\ Rev.\ D \textbf{93}, 052016 (2016).


\bibitem{Aubert:2008bl} B.~Aubert \textit{et al.} [BaBar Collaboration],
Phys.\ Rev.\ D \textbf{77}, 111102 (2008).


\bibitem{Chen:2016iua} D.~Y.~Chen, 
Eur.\ Phys.\ J.\ C \textbf{76}, 671 (2016).


\bibitem{Liu:2016onn} X.~H.~Liu,
Phys.\ Lett.\ B \textbf{766}, 117 (2017).


\bibitem{Chen:2010ze} W.~Chen and S.~L.~Zhu,
Phys.\ Rev.\ D \textbf{83}, 034010 (2011).


\bibitem{Chen:2016oma} H.~X.~Chen, E.~L.~Cui, W.~Chen, X.~Liu and S.~L.~Zhu,
Eur.\ Phys.\ J.\ C \textbf{77}, 160 (2017).


\bibitem{Wang:2016tzr} Z.~G.~Wang,
Eur.\ Phys.\ J.\ C \textbf{76}, 657 (2016).


\bibitem{Wang:2016dcb} Z.~G.~Wang,
Eur.\ Phys.\ J.\ C \textbf{77}, 174 (2017).


\bibitem{Wang:2016gxp} Z.~G.~Wang,
Eur.\ Phys.\ J.\ C \textbf{77}, 78 (2017).


\bibitem{Agaev:2017foq} S.~S.~Agaev, K.~Azizi and H.~Sundu,
Phys.\ Rev.\ D\textbf{95}, 114003 (2017). 


\bibitem{Abe:2004zs} K.~Abe \textit{et al.} [Belle Collaboration],
Phys.\ Rev.\ Lett.\ \textbf{94}, 182002 (2005).


\bibitem{Uehara:2009tx} S.~Uehara \textit{et al.} [Belle Collaboration],
Phys.\ Rev.\ Lett.\ \textbf{104}, 092001 (2010).


\bibitem{Aubert:2007vj} B.~Aubert \textit{et al.} [BaBar Collaboration],
Phys.\ Rev.\ Lett.\ \textbf{101} 082001 (2008).


\bibitem{Lebed:2016yvr} R.~F.~Lebed and A.~D.~Polosa,
Phys.\ Rev.\ D \textbf{93}, 094024 (2016).


\bibitem{Stancu:2006st} F.~Stancu,
hep-ph/0607077.


\bibitem{Maiani:2016wlq} L.~Maiani, A.~D.~Polosa and V.~Riquer,
Phys.\ Rev.\ D \textbf{94}, 054026, (2016).


\bibitem{Zhu:2016arf} R.~Zhu,
Phys.\ Rev.\ D \textbf{94}, 054009 (2016).


\bibitem{Adolph:2015pws} C.~Adolph \textit{et al.} [COMPASS Collaboration],
Phys.\ Rev.\ Lett.\ \textbf{115}, 082001 (2015).


\bibitem{Weinstein:1990gu} J.~D.~Weinstein and N.~Isgur,
Phys.\ Rev.\ D \textbf{41}, 2236 (1990). 


\bibitem{Alford:2000mm} M.~G.~Alford and R.~L.~Jaffe,
Nucl.\ Phys.\ B \textbf{578}, 367 (2000).


\bibitem{Amsler:2004ps} C.~Amsler and N.~A.~Tornqvist,
Phys.\ Rept.\ \textbf{389}, 61 (2004). 


\bibitem{Bugg:2004xu} D.~V.~Bugg, 
Phys.\ Rept.\ \textbf{397}, 257 (2004). 


\bibitem{Maiani:2004uc} L.~Maiani, F.~Piccinini, A.~D.~Polosa and V.~Riquer,
Phys.\ Rev.\ Lett.\ \textbf{93}, 212002 (2004).


\bibitem{Hooft:2008we} G.~'t Hooft, G.~Isidori, L.~Maiani, A.~D.~Polosa and
V.~Riquer, 
Phys.\ Lett.\ B \textbf{662}, 424 (2008).


\bibitem{Latorre:1985uy} J.~I.~Latorre and P.~Pascual,
J.\ Phys.\ G \textbf{11}, L231 (1985). 


\bibitem{Narison:1986vw} S.~Narison,
Phys.\ Lett.\ B \textbf{175}, 88 (1986). 


\bibitem{Brito:2004tv} T.~V.~Brito, F.~S.~Navarra, M.~Nielsen and
M.~E.~Bracco,
Phys.\ Lett.\ B \textbf{608}, 69 (2005).


\bibitem{Wang:2005cn} Z.~G.~Wang and W.~M.~Yang,
Eur.\ Phys.\ J.\ C \textbf{42}, 89 (2005).


\bibitem{Chen:2007xr} H.~X.~Chen, A.~Hosaka and S.~L.~Zhu,
Phys.\ Rev.\ D \textbf{76}, 094025 (2007). 


\bibitem{Lee:2005hs} H.~J.~Lee,
Eur.\ Phys.\ J.\ A \textbf{30}, 423 (2006). 


\bibitem{Sugiyama:2007sg} J.~Sugiyama, T.~Nakamura, N.~Ishii, T.~Nishikawa
and M.~Oka,
Phys.\ Rev.\ D \textbf{76}, 114010 (2007). 


\bibitem{Kojo:2008hk} T.~Kojo and D.~Jido,
Phys.\ Rev.\ D \textbf{78}, 114005 (2008). 


\bibitem{Wang:2015uha} Z.~G.~Wang,
Eur.\ Phys.\ J.\ C \textbf{76}, 427 (2016).


\bibitem{Achasov:2020fee} N.~N.~Achasov, 
arXiv:2002.01354. 


\bibitem{Wang:2014bua} Z.~G.~Wang,
arXiv:1401.1134 [hep-ph]. 


\bibitem{Chen:2015fwa} H.~X.~Chen, E.~L.~Cui, W.~Chen, T.~G.~Steele, X.~Liu
and S.~L.~Zhu,
Phys.\ Rev.\ D \textbf{91}, 094022 (2015). 


\bibitem{Sundu:2017xct} H.~Sundu, S.~S.~Agaev and K.~Azizi,
Phys.\ Rev.\ D \textbf{97}, 054001 (2018).


\bibitem{Gutsche:2017oro} T.~Gutsche, V.~E.~Lyubovitskij and I.~Schmidt,
Phys.\ Rev.\ D \textbf{96}, 034030 (2017). 


\bibitem{Ketzer:2015tqa} M.~Mikhasenko, B.~Ketzer and A.~Sarantsev,
Phys.\ Rev.\ D \textbf{91}, 094015 (2015). 


\bibitem{Liu:2015taa} X.~H.~Liu, M.~Oka and Q.~Zhao,
Phys.\ Lett.\ B \textbf{753}, 297 (2016).


\bibitem{Aceti:2016yeb} F.~Aceti, L.~R.~Dai and E.~Oset,
Phys.\ Rev.\ D \textbf{94}, 096015 (2016). 


\bibitem{Basdevant:2015wma} J.~L.~Basdevant and E.~L.~Berger,
Phys.\ Rev.\ Lett.\ \textbf{114}, 192001 (2015).


\bibitem{Wang:2015cis} W.~Wang and Z.~X.~Zhao,
Eur.\ Phys.\ J.\ C \textbf{76}, 59 (2016).


\bibitem{Gutsche:2017twh} T.~Gutsche, M.~A.~Ivanov, J.~G.~K\"orner,
V.~E.~Lyubovitskij and K.~Xu,
Phys.\ Rev.\ D \textbf{96}, 114004 (2017).


\bibitem{Braun:1989iv} V.~M.~Braun and I.~E.~Filyanov,
Z.\ Phys.\ C \textbf{48}, 239 (1990).


\bibitem{Ball:1998je} P.~Ball,
JHEP \textbf{9901}, 010 (1999).


\bibitem{Agaev:2010aq} S.~S.~Agaev, V.~M.~Braun, N.~Offen and F.~A.~Porkert,
Phys.\ Rev.\ D \textbf{83}, 054020 (2011).


\bibitem{Agaev:2012tm} S.~S.~Agaev, V.~M.~Braun, N.~Offen and F.~A.~Porkert,
Phys.\ Rev.\ D \textbf{86}, 077504 (2012).


\bibitem{Braun:2015axa} V.~M.~Braun, S.~Collins, M.~G\"ockeler,
P.~Perez-Rubio, A.~Sch\"afer, R.~W.~Schiel and A.~Sternbeck,
Phys.\ Rev.\ D \textbf{92}, 014504 (2015).


\bibitem{Ding:2007rg} G.~J.~Ding, J.~J.~Zhu and M.~L.~Yan,
Phys.\ Rev.\ D \textbf{77}, 014033 (2008).


\bibitem{Li:2009zu} B.~Q.~Li and K.~T.~Chao,
Phys.\ Rev.\ D \textbf{79}, 094004 (2009).


\bibitem{Guo:2008zg} F.~K.~Guo, C.~Hanhart and U.~G.~Meissner,
Phys.\ Lett.\ B \textbf{665}, 26 (2008).


\bibitem{Wang:2009hi} Z.~G.~Wang and X.~H.~Zhang,
Commun.\ Theor.\ Phys.\ \textbf{54}, 323 (2010).


\bibitem{Albuquerque:2011ix} R.~M.~Albuquerque, M.~Nielsen and R.~Rodrigues
da Silva, 
Phys.\ Rev.\ D \textbf{84}, 116004 (2011).


\bibitem{Qiao:2007ce} C.~F.~Qiao,
J.\ Phys.\ G \textbf{35}, 075008 (2008).


\bibitem{Cotugno:2009ys} G.~Cotugno, R.~Faccini, A.~D.~Polosa and
C.~Sabelli, 
Phys.\ Rev.\ Lett.\ \textbf{104}, 132005 (2010).


\bibitem{Zhang:2010mw} J.~R.~Zhang and M.~Q.~Huang,
Phys.\ Rev.\ D \textbf{83}, 036005 (2011).


\bibitem{Albuquerque:2008up} R.~M.~Albuquerque and M.~Nielsen,
Nucl.\ Phys.\ A \textbf{815}, 53 (2009) Erratum: [Nucl.\ Phys.\ A \textbf{857%
}, 48 (2011)].


\bibitem{Wang:2013exa} Z.~G.~Wang,
Eur.\ Phys.\ J.\ C \textbf{74}, 2874 (2014).


\bibitem{Wang:2016mmg} Z.~G.~Wang,
Eur.\ Phys.\ J.\ C \textbf{76}, 387 (2016).


\bibitem{Wang:2018rfw} Z.~G.~Wang,
Eur.\ Phys.\ J.\ C \textbf{78}, 518 (2018).


\bibitem{Sundu:2018toi} H.~Sundu, S.~S.~Agaev and K.~Azizi,
Phys.\ Rev.\ D \textbf{98}, 054021 (2018).


\bibitem{Aubert:2006bu} B.~Aubert \textit{et al.} [BaBar Collaboration],
Phys.\ Rev.\ D \textbf{74}, 091103 (2006).


\bibitem{Ablikim:2007ab} M.~Ablikim \textit{et al.} [BES Collaboration],
Phys.\ Rev.\ Lett.\ \textbf{100}, 102003 (2008).


\bibitem{Shen:2009zze} C.~P.~Shen \textit{et al.} [Belle Collaboration],
Phys.\ Rev.\ D \textbf{80}, 031101 (2009).


\bibitem{Ablikim:2014pfc} M.~Ablikim \textit{et al.} [BESIII Collaboration],
Phys.\ Rev.\ D \textbf{91}, 052017 (2015).


\bibitem{Ablikim:2018iyx} M.~Ablikim \textit{et al.} [BESIII Collaboration],
Phys.\ Rev.\ D \textbf{99}, 032001 (2019).


\bibitem{Ablikim:2018xuz} M.~Ablikim \textit{et al.} [BESIII Collaboration],
Phys.\ Rev.\ D \textbf{99}, 112008 (2019). 


\bibitem{Ding:2007pc} G.~J.~Ding and M.~L.~Yan,
Phys.\ Lett.\ B \textbf{657}, 49 (2007).


\bibitem{Wang:2012wa} X.~Wang, Z.~F.~Sun, D.~Y.~Chen, X.~Liu and T.~Matsuki,
Phys.\ Rev.\ D \textbf{85}, 074024 (2012).


\bibitem{MartinezTorres:2008gy} A.~Martinez Torres, K.~P.~Khemchandani,
L.~S.~Geng, M.~Napsuciale and E.~Oset,
Phys.\ Rev.\ D \textbf{78}, 074031 (2008).


\bibitem{AlvarezRuso:2009xn} L.~Alvarez-Ruso, J.~A.~Oller and J.~M.~Alarcon,
Phys.\ Rev.\ D \textbf{80}, 054011 (2009).


\bibitem{Wang:2006ri} Z.~G.~Wang,
Nucl.\ Phys.\ A \textbf{791}, 106 (2007).


\bibitem{Chen:2008ej} H.~X.~Chen, X.~Liu, A.~Hosaka and S.~L.~Zhu,
Phys.\ Rev.\ D \textbf{78}, 034012 (2008).


\bibitem{Chen:2018kuu} H.~X.~Chen, C.~P.~Shen and S.~L.~Zhu,
Phys.\ Rev.\ D \textbf{98}, 014011 (2018).


\bibitem{Agaev:2019coa} S.~S.~Agaev, K.~Azizi, and H.~Sundu,
Phys.\ Rev.\ D \textbf{101}, 074012 (2020).



\bibitem{Lu:2019ira} Q.~F.~Lu, K.~L.~Wang and Y.~B.~Dong,
arXiv:1903.05007 [hep-ph]. 


\bibitem{Cui:2019roq} E.~L.~Cui, H.~M.~Yang, H.~X.~Chen, W.~Chen and
C.~P.~Shen,
Eur.\ Phys.\ J.\ C \textbf{79}, 232 (2019).


\bibitem{Wang:2019nln} Z.~G.~Wang, 
arXiv:1901.04815 [hep-ph]. 


\bibitem{Wang:2019qyy} L.~M.~Wang, S.~Q.~Luo and X.~Liu,
arXiv:1901.00636 [hep-ph]. 


\bibitem{Azizi:2019ecm} K.~Azizi, S.~S.~Agaev, and H.~Sundu, Nucl.\ Phys.\ B
\textbf{948}, 114789 (2019).


\bibitem{Agaev:2014wna} S.~S.~Agaev, V.~M.~Braun, N.~Offen, F.~A.~Porkert
and A.~Sch\"{a}fer,
Phys.\ Rev.\ D \textbf{90}, 074019 (2014).


\bibitem{Agaev:2015faa} S.~S.~Agaev, K.~Azizi and H.~Sundu,
Phys.\ Rev.\ D \textbf{92}, 116010 (2015).


\bibitem{Agaev:2016dsg} S.~S.~Agaev, K.~Azizi and H.~Sundu,
Phys.\ Rev.\ D \textbf{95}, 034008 (2017).



\bibitem{Ball:1996tb} P.~Ball and V.~M.~Braun,
Phys.\ Rev.\ D \textbf{54}, 2182 (1996).


\bibitem{Ball:1998sk} P.~Ball, V.~M.~Braun, Y.~Koike and K.~Tanaka,
Nucl.\ Phys.\ B \textbf{529}, 323 (1998).


\bibitem{Ball:2007rt} P.~Ball and G.~W.~Jones,
JHEP \textbf{0703}, 069 (2007).




\bibitem{Agaev:2016mjb} S.~S.~Agaev, K.~Azizi and H.~Sundu,
Phys.\ Rev.\ D\ \textbf{93}, 074024 (2016).


\bibitem{Agaev:2016ijz} S.~S.~Agaev, K.~Azizi and H.~Sundu,
Phys.\ Rev.\ D \textbf{93}, 114007 (2016).


\bibitem{Agaev:2016lkl} S.~S.~Agaev, K.~Azizi and H.~Sundu, Phys.\ Rev.\ D\
\textbf{93}, 094006 (2016).


\bibitem{Agaev:2016urs} S.~S.~Agaev, K.~Azizi and H.~Sundu,
Eur.\ Phys.\ J.\ Plus \textbf{131}, 351 (2016).


\bibitem{Navarra:2007yw} F.~S.~Navarra, M.~Nielsen, and S.~H.~Lee,
Phys.\ Lett.\ B \textbf{649}, 166 (2007).


\bibitem{Karliner:2017qjm} M.~Karliner and J.~L.~Rosner,
Phys.\ Rev.\ Lett.\ \textbf{119}, 202001 (2017).


\bibitem{Eichten:2017ffp} E.~J.~Eichten and C.~Quigg,
Phys.\ Rev.\ Lett.\ \textbf{119}, 202002 (2017).


\bibitem{Agaev:2018khe} S.~S.~Agaev, K.~Azizi, B.~Barsbay, and H.~Sundu,
Phys.\ Rev.\ D \textbf{99}, 033002 (2019).


\bibitem{Sundu:2019feu} H.~Sundu, S.~S.~Agaev and K.~Azizi,
Eur.\ Phys.\ J.\ C \textbf{79}, 753 (2019).


\bibitem{Agaev:2019wkk} S.~S.~Agaev, K.~Azizi and H.~Sundu,
Phys.\ Rev.\ D \textbf{100}, 094020 (2019).


\bibitem{Agaev:2019lwh} S.~S.~Agaev, K.~Azizi, B.~Barsbay, and H.~Sundu,
arXiv:1912.07656 [hep-ph].


\bibitem{Agaev:2020dba} S.~S.~Agaev, K.~Azizi, B.~Barsbay, and H.~Sundu,
arXiv:2001.01446 [hep-ph].
\end{thebibliography}
\end{document}